\documentclass[phd,12pt]{psuthesis}
\usepackage[T1]{fontenc}
\usepackage[utf8x]{inputenc}
\usepackage{lmodern}
\usepackage{textcomp}
\usepackage{microtype}

\usepackage{amsmath}
\usepackage{amsfonts}
\usepackage{amssymb}
\usepackage{subfigure}
\usepackage{eqlist} % Makes for a nice list of symbols.
\usepackage[nosepfour,warning,np,debug,autolanguage]{numprint}
\usepackage{acro}
\usepackage{booktabs}
\usepackage[final]{graphicx}
\usepackage[dvipsnames]{color}

\usepackage{amsmath}
\usepackage{amssymb}
\usepackage{graphicx}
\usepackage{tikz-network}
\usepackage{amsmath}
\usepackage{amssymb}
\usepackage{amsthm}
\usepackage{xcolor}
\usepackage{bbm}
\usepackage{algorithm}
\usepackage{algcompatible}
\DeclareGraphicsExtensions{.pdf, .jpg}

\usepackage{url}

\usepackage{cite}

\usepackage{titlesec}

\newcommand{\ppg}[1]{p_{#1}^G}
\newcommand{\ppgt}[1]{p_{#1}^{G^T}}
\newcommand{\ppgs}[1]{p_{#1}^{G|_{\sigma}}}
\newcommand{\ppgst}[1]{p_{#1}^{(G|_{\sigma})^T}}
\newcommand{\pppg}[2]{p_{#1 , #2}^G}
\newcommand{\pppgs}[2]{p_{#1 , #2}^{G|_{\sigma}}}

\newtheorem{ddd}{Definition}
\newtheorem{lem}{Lemma}
\newtheorem{thm}{Theorem}
\newtheorem{corr}{Corollary}

\newtheorem{prp}{Proposition}
\newtheorem{rmk}{Remark}

\title{Modeling Bias in Decision-Making Attractor Networks}

\author{Syed Safaan Sadiq}
\dept{Mathematics}
\degreedate{August 2025}
\copyrightyear{2025}

\documenttype{Dissertation}
\submittedto{The J. Jeffrey and Ann Marie Fox Graduate School}

\collegesubmittedto{ }%Eberly College of Science}

\numberofreaders{5}

\honorsadvisor{Honors P. Advisor}
{Associate Professor of Engineering Science and Mechanics}
\honorsadvisortwo{Honors P. Advisor, Jr.}
{Professor of Engineering Science and Mechanics}

\secondthesissupervisor{Second T. Supervisor}

\escdepthead{Department Q. Head}
\escdeptheadtitle{P. B. Breneman Chair and Professor 
of Engineering Science and Mechanics
}

\readerone[Co-Chair of Committee]
		{Carina Curto}
		{Professor of Applied Mathematics and Brain Science, Brown University \newline Dissertation Co-Advisor}
		
\advisor[Dissertation Co-Advisor][Co-Chair of Committee]
		{Leonid Berlyand}
		{Professor of Mathematics}

\readertwo
			{Vladimir Itskov}
			{Associate Professor of Mathematics}

\readerthree
			{Jessica Conway}
			{Associate Professor of Mathematics and Biology}

\readerfour
			{Reka Albert}
			{Distinguished Professor of Physics and Biology}

\readerfive
			{Pierre-Emmanuel Jabin}
			{Director of Graduate Studies \\ Distinguished Professor of Mathematics}

\definecolor{gray75}{gray}{0.75}
\newcommand{\hsp}{\hspace{15pt}}
\titleformat{\chapter}[display]{\fontsize{30}{30}\selectfont\bfseries\sffamily}{Chapter \thechapter\hsp\textcolor{gray75}{\raisebox{3pt}{|}}}{0pt}{}{}

\titleformat{\section}[block]{\Large\bfseries\sffamily}{\thesection}{12pt}{}{}
\titleformat{\subsection}[block]{\large\bfseries\sffamily}{\thesubsection}{12pt}{}{}

\usepackage{listings}
\begin{document}
\pagestyle{fancy}
\fancyhead[L,C,R]{}
\fancyfoot[L,R]{}
\fancyfoot[C]{\thepage}
\renewcommand{\headrulewidth}{0pt}
\renewcommand{\footrulewidth}{0pt}
%%%%%%%%%%%%%%%%%%%%%%%%
% Preliminary Material %
%%%%%%%%%%%%%%%%%%%%%%%%
% This command is needed to properly set up the frontmatter.
\frontmatter

%%%%%%%%%%%%%%%%%%%%%%%%%%%%%%%%%%%%%%%%%%%%%%%%%%%%%%%%%%%%%%
% IMPORTANT
%
% The following commands allow you to include all the
% frontmatter in your thesis. If you don't need one or more of
% these items, you can comment it out. Most of these items are
% actually required by the Grad School -- see the Thesis Guide
% for details regarding what is and what is not required for
% your particular degree.
%%%%%%%%%%%%%%%%%%%%%%%%%%%%%%%%%%%%%%%%%%%%%%%%%%%%%%%%%%%%%%
% !!! DO NOT CHANGE THE SEQUENCE OF THESE ITEMS !!!
%%%%%%%%%%%%%%%%%%%%%%%%%%%%%%%%%%%%%%%%%%%%%%%%%%%%%%%%%%%%%%

% Generates the title page based on info you have provided
% above.
\psutitlepage

% Generates the committee page -- this is bound with your
% thesis. If this is an baccalaureate honors thesis, then
% comment out this line.
\psucommitteepage

% Generates the abstract. The argument should point to the
% file containing your abstract. 
%
%\pagestyle{plain}
\chapter*{Abstract}
    \begin{singlespace}
        % Place abstract below.

\vspace{-0.3in}

Attractor neural network models of cortical decision-making circuits represent them as dynamical systems in the state space of neural firing rates with the attractors of the network encoding possible decisions.  While the attractors of these models are well studied, far less attention is paid to the basins of attraction even though their sizes can be said to encode the biases towards the corresponding decisions.  The parameters of an attractor network control both the attractors and the basins of attraction.  However, findings in behavioral economics suggest that the framing of a decision-making task can affect preferences even when the same choices are being offered.  This suggests that the circuit encodes both choices and biases separately, that preferences can be changed without disrupting the encoding of the choices themselves.  In the context of attractor networks, this would mean that the parameters can be adjusted to reshape the basins of attraction without changing the attractors themselves.  How can this be realized and how do the parameters shape decision-making biases?

We study this question in the context of threshold linear networks (TLNs), a common model with recurrent dynamics.  In Chapter 2, we rigorously prove in the case of two competing neural populations how the parameters of the network shape the basins of attraction and, consequently, bias.  In Chapter 3, we raise the problem into larger networks in a class of TLNs called CTLNs, where network parameters are derived from a directed graph structure.  We explore and find the challenges of computer-assisted approaches to the problem.  Starting in Chapter 4, we focus on CTLNs derived from directed acyclic graphs.  We prove how the dynamics of the network can be derived from the combinatorics of the directed acyclic graph.  While falling short of determining the full basins of attraction, in Chapter 5, we demonstrate numerically how connectivity shapes the basins’ encoding of bias under assumptions of low dimensional dynamics.  In Chapter 6, we consider the existence of trajectories beginning in a state of excitatory/inhibitory balance and prove how their existence can be inferred from the directed graph structure.  Finally, in Chapter 7, we generalize results from Chapters 4 and 6 to a class of TLNs we call heterogeneous CTLNs (hCTLNs).

    \end{singlespace}
\newpage

% Generates the Table of Contents
\thesistableofcontents

% Generates the List of Figures
\begin{singlespace}
\renewcommand{\listfigurename}{\sffamily\Huge List of Figures}
\setlength{\cftparskip}{\baselineskip}
\addcontentsline{toc}{chapter}{List of Figures}
%\fancypagestyle{plain}{%
%\fancyhf{} % clear all header and footer fields
%\fancyfoot[C]{\thepage}} % except the center
\listoffigures
\end{singlespace}
\clearpage

% Generates the List of Tables
%\begin{singlespace}
%\renewcommand{\listtablename}{\sffamily\Huge List of Tables}
%\setlength{\cftparskip}{\baselineskip}
%\addcontentsline{toc}{chapter}{List of Tables}% !BIB program = 
%\listoftables
%\end{singlespace}
%\clearpage

% Generates the List of Symbols. The argument should point to
% the file containing your List of Symbols. 
%\thesislistofsymbols{SupplementaryMaterial/ListOfSymbols}

% Generates the Acknowledgments. The argument should point to
% the file containing your Acknowledgments. 
%
\chapter{Acknowledgments}
%\addcontentsline{toc}{chapter}{Acknowledgments}
\begin{singlespace}
    % Place acknowledgments below.
The journey to this thesis was a long and winding road and I have many people to thank for reaching here. 

First I thank my parents, Syed and Sahira Sadiq, for being an endless inspiration to me and undoubtedly being responsible for the best parts of who I am.  Also I thank my sister, Saena Sadiq, for her constant encouragement and for diligently executing her responsibility as a younger sibling to keep me humble (perhaps to a fault).  Finally, I would like to acknowledge my extended family as a whole who have always been my supporters and well-wishers.  I am blessed with a family beyond what I deserve and for that I am immensely grateful.

I would like to thank my thesis supervisor Carina Curto without whom this research would not have been possible.  Her mentorship, guidance, and support are responsible for my growth as a researcher and for shaping this project into what it is today.  She has been a model for me in both what it means to be an exceptional researcher and an effective mentor.  

I am also grateful to Jessica Conway, Vladimir Itskov, and Reka Albert for offering their time to serve on my committee and for their feedback and guidance.  Additionally I would like to thank Leonid Berlyand for co-chairing my committee and acting as thesis advisor in my final year while this project reached its conclusion.

I want to thank my peers during my time at Penn State, Stephen White, Neel Raha, and Kieran Cavanagh, for their friendship over the years.  Outside of Penn State, I want to thank Sameer Kailasa for the many helpful conversations and great times we shared as roommates in Providence.  

Finally, I am immensely grateful for the funding provided by the Department of Mathematics, the NIH grant R01 EB022862 and NSF grant DMS-1951165, without which this research would not have been possible.  Additionally, many of the findings in Chapter 3 are based upon work supported by the NSF under grant DMS-1929284 while I was in residence at the Institute for Computational and Experimental
Research in Mathematics (ICERM) in Providence, RI, during the program: "Math + Neuroscience: Strengthening the Interplay Between Theory and Mathematics."  The findings and conclusions in this dissertation do not necessarily reflect the view of the funding agencies.

\end{singlespace}
 %Dont forget this!!!!!!!!

% Generates the Epigraph/Dedication. The first argument should
% point to the file containing your Epigraph/Dedication and
% the second argument should be the title of this page. 
%\thesisdedication{SupplementaryMaterial/Dedication}{Dedication}

%%%%%%%%%%%%%%%%%%%%%%%%%%%%%%%%%%%%%%%%%%%%%%%%%%%%%%
% This command is needed to get the main part of the %
% document going.                                    %
%%%%%%%%%%%%%%%%%%%%%%%%%%%%%%%%%%%%%%%%%%%%%%%%%%%%%%
\thesismainmatter

%%%%%%%%%%%%%%%%%%%%%%%%%%%%%%%%%%%%%%%%%%%%%%%%%%
% This is an AMS-LaTeX command to allow breaking %
% of displayed equations across pages. Note the  %
% closing the "}" just before the bibliography.  %
%%%%%%%%%%%%%%%%%%%%%%%%%%%%%%%%%%%%%%%%%%%%%%%%%%
\allowdisplaybreaks{
%\pagestyle{fancy}
%\fancyhead{}
%
%%%%%%%%%%%%%%%%%%%%%%
% THE ACTUAL CONTENT %
%%%%%%%%%%%%%%%%%%%%%%
% Chapters
\chapter{Introduction}

The course of our lives is charted by how we navigate a gauntlet of decisions.  At each stage, our brain must select from a menu of options our next action.  How does it do this?  How does the brain select an option off the menu?  Naively, we might assign each item on the menu a "value" and then claim that the highest value choice will be the decision.  Such an approach is how decision-making is conceived by standard economics \cite{utils}.  However, studies in behavioral economics have demonstrated convincingly that this cannot be the whole story and that actual decision-making is often \textit{menu-variant}.  

One such example is the Decoy Effect \cite{huber}.  Consider two choices, call them X and Y, which when compared with each other have various strengths and weaknesses that would make it so that either one would be chosen with equal likelihood when offered together on a menu.  If offered to a room of people each choice will be selected at a roughly $50\%-50\%$ split.  Now, let there then be a third, irrelevant, choice Z added to the menu which is clearly inferior to one of the two original choices, say X.  It has been found that the mere presence of this third choice on the menu skews preferences toward its superior option, in this case X.  As there is never a reason to pick Z, this new menu of three choices still has the same two ultimate choices, X and Y, but now more people in the room will be drawn to X at perhaps a $60\%-40\%$ split.  Figure~\ref{fig:fig1} illustrates how the presence of an irrelevant choice can shift the decision-making boundary between choices.  Under the standard economics paradigm, the framing of the decision should not bias our choices, but the Decoy Effect clearly demonstrates that this is not the case.

%Fig 1
\begin{figure}[!ht]
\begin{center}
\vspace{.1in}
\includegraphics[width=5.75in]{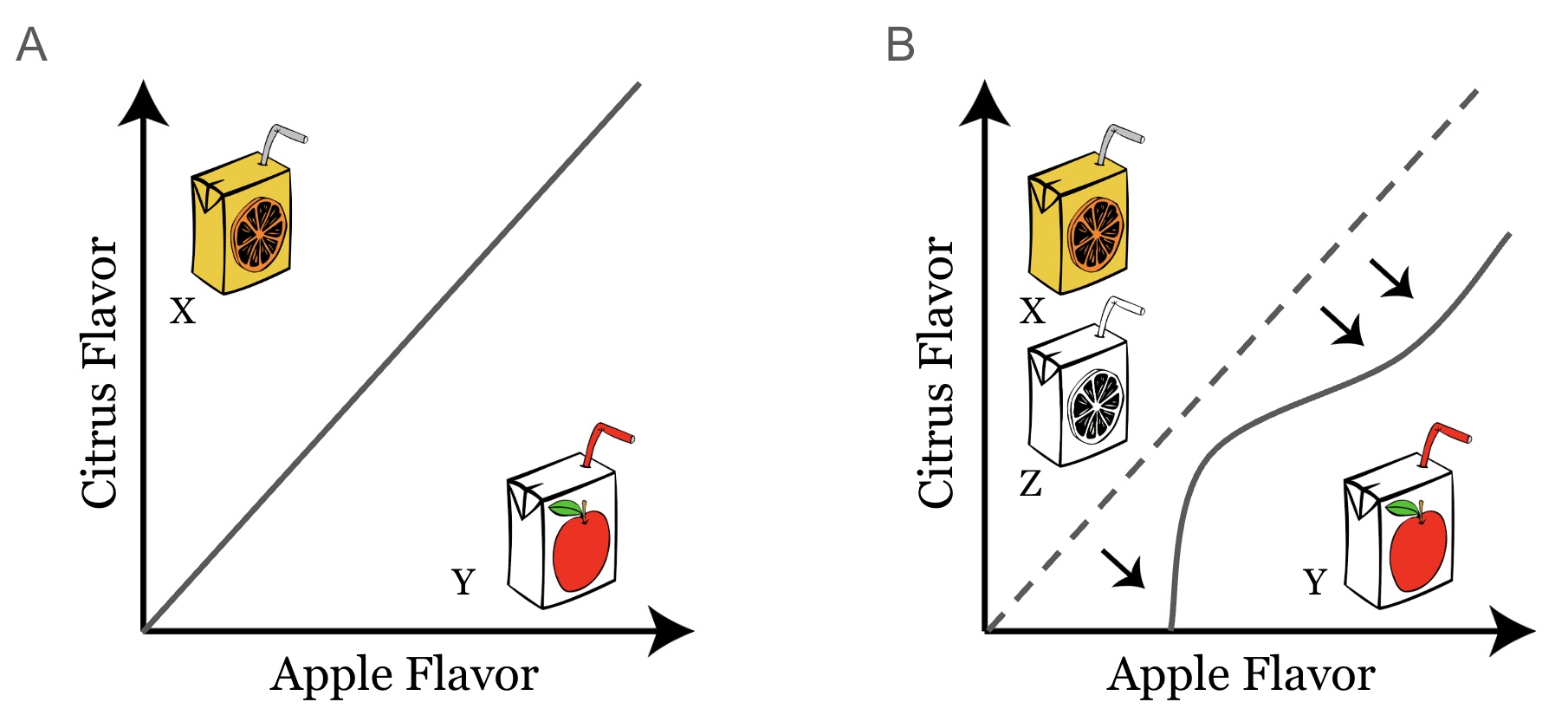}
\vspace{.1in}
\caption[The Decoy Effect]{{\bf The Decoy Effect.}  (A) Orange juice "X" and apple juice "Y" have distinct flavors and depending upon preferences toward one or the other, either may be chosen roughly equivalently.  The diagonal decision boundary reflects the $50\%-50\%$ split.  (B) The presence of a poorer quality orange juice "Z" does not add a true choice, but it increases the number of situations where "X" is the preferred choice, shifting the decision boundary.}
\label{fig:fig1}
\end{center}
\vspace{-.2in}
\end{figure}

These phenomenological observations also find biological support in the activity of neurons in the lateral intraparietal area (LIP), a region of the cerebral cortex in primate brains, which does appear to be menu-variant \cite{DG}.  The LIP is of substantial interest in the study of decision-making and these findings lend credence to the idea that decision-making is a complex process and that the biases that shape it are nontrivial to understand.  Moreover, these findings suggest a natural question which has been the foundation of my doctoral work.  

\textbf{Question}: How is the brain able to encode biases separately from the decisions themselves?

We will study this mathematically through a dynamical systems approach, applying the theory of attractor neural networks to study how a network model of a neural circuit encodes decision-making bias.

\section{Decision-Making as an Attractor Network}

Attractor networks are a popular computational framework for studying neural circuits.  In the vein of the Hopfield model\cite{hops} for associative memory, we conceive of a circuit of neurons (Fig~\ref{fig:fig2}A) as a dynamical system in the state space of neural activity.  For a given input (an initial condition) the activity converges towards an attractor.  The attractors of the network represent the various outcomes of neural computation for the circuit and generally correspond to some phenomena of interest.  In the case of the Hopfield model the attractors are stable fixed points representing the memories which are being retrieved.  Other models make use of continuous attractors.  Examples include ring-shaped attractors which describe the maintenance of heading directional information \cite{ring} and line attractors which keep track of eye position \cite{seung}.  Attractor networks have also been used in the modeling of decision-making circuits with the attractors representing the decisions themselves \cite{wang}.

Much of the literature surrounding attractor networks is focused on how many attractors can be encoded into a network and the types of attractors \cite{attrev}.  Far less attention is paid to the basins of attraction.  As the basins of attraction are the sets of initial conditions converging to respective attractors, they are representative of the bias towards those corresponding decisions as depicted in Fig~\ref{fig:fig2}B.  Shaping the basins then shapes the bias.  So then, we can frame our question in the context of attractor network models.

%Fig 2
\begin{figure}[!ht]
\begin{center}
\vspace{.1in}
\includegraphics[width=6.5in]{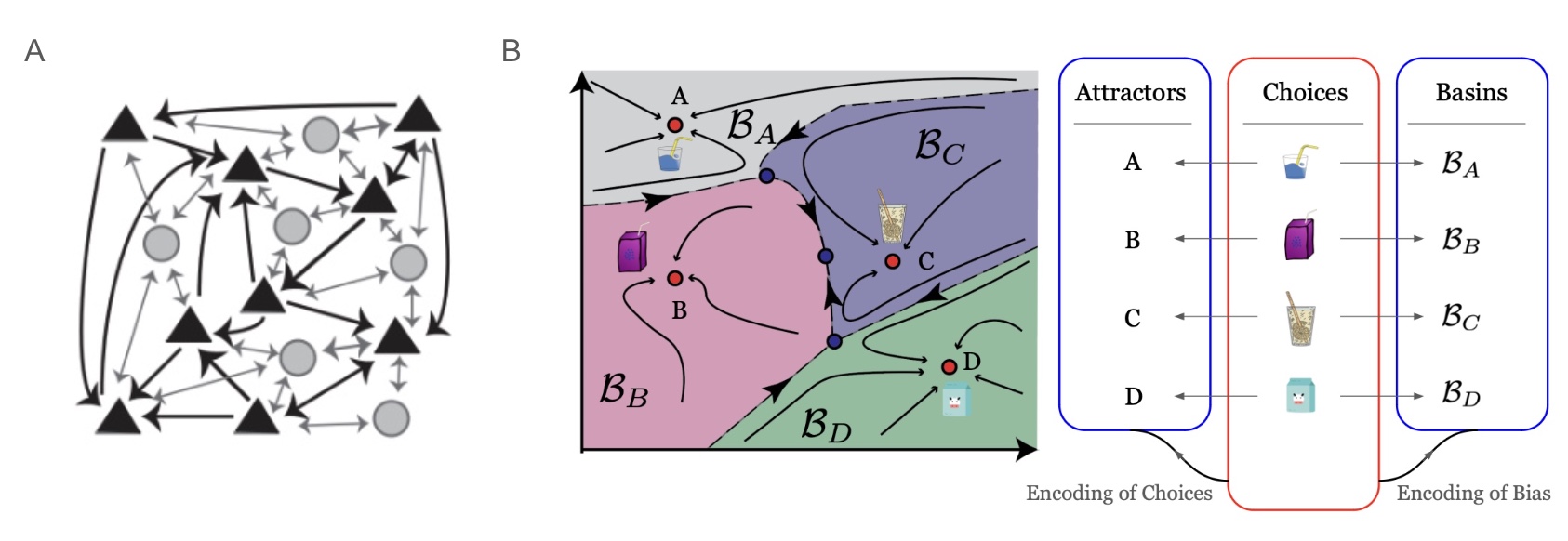}
\vspace{.1in}
\caption[Attractor networks and decision-making]{{\bf Attractor networks and decision-making.}  (A) Diagram of a cortical circuit with triangles representing excitatory neurons and circles representing inhibitory interneurons.  (B) This schematic illustrates the attractor network interpretation of a decision-making circuit.  The point attractors in the state space correspond to the choices of the decision-making task, in this case drink selection, and the sizes of the basins of attraction correspond to the bias towards those choices.  There are additionally unstable fixed points which lie on the decision boundaries.}
\label{fig:fig2}
\end{center}
\vspace{-.2in}
\end{figure}

\textbf{Question:} How can the parameters of a network be used shape the basins of attraction (the bias) without changing the attractors (the decisions) themselves?

Specifically, we will investigate this question in the setting of firing-rate models of network activity.

\section{Firing Rate Models and TLNs}

The rate coding hypothesis suggests that neurons communicate through their firing rates rather than by the particulars of individual action potentials \cite{rate}.  Firing rate models are dynamical systems operating in the state space of neural firing rates where each state variable corresponds to the firing rate of a neuron or neural population.  To construct such a model, the first thing that must be considered is the relationship between the firing rate of the neurons and the synaptic input current it receives.  The following differential equation captures this \cite{AppendixF}:

\begin{center}
    $\tau \dfrac{dx}{dt} = -x + \Phi (I_{total})$ 
\end{center}

In this equation, $x$ is the firing rate of the neuron and $I_{total}$ is the total synaptic current being received by the neuron.  The function $\Phi (z)$ is a firing rate function which describes the relationship between $I_{total}$ and $x$.  Finally, the parameter $\tau$ is a time constant.

The input current received by a neuron $i$ in a recurrent neural network can be decomposed as the sum of some external input current, $\theta_{i} (t)$, plus a weighted sum, $\sum_{j=1}^n W_{ij}x_j $, of the firing rates of the neurons in the network.  The weights $W_{ij}$ form the connectivity matrix $W$ and represent the strength of each synapse.  The firing rate differential equation can then be rewritten as:

\begin{center}
    $\tau \dfrac{dx_i}{dt} =-x_i + \Phi \left( \sum_{j=1}^n W_{ij}x_j +\theta_i(t)\right)$, $i\in [n]$
\end{center}

where $x_i(t)$ is the firing rate of the $i$-th neuron in a network of $n$ neurons.  As a point of notation, we use $[n]$ to denote the set of indices (i.e. $[n]=\{1,2,\dots , n\}$).  The choice of firing rate function $\Phi (z)$ makes a significant difference in the mathematical tractability of a problem.  While the linear firing rate function $\Phi (z)=z$ remains popular in neuroscience modeling, it would be of no help in our study of decision-making because a linear dynamical system cannot demonstrate multistability.  If we want to study bias in decision-making circuits, we need to encode at least two decisions (attractors) in the circuit, so we need a nonlinear firing rate function.

We will use threshold linear networks (TLNs) \cite{tlns}, a recurrent neural network model governed by the following differential equations:

$$\dfrac{dx_i}{dt} = -x_i + \left[\sum_{j=1}^n W_{ij}x_j +\theta_i \right]_+, i \in [n] $$ .

where $[\cdot]_+=\max\{0,\cdot\}$ is the ReLU activation function.  Often, the actual neural activity at fixed points is of less interest than which neurons are firing, which is called the \textit{support} of a fixed point.  The set of fixed point supports for a TLN with connectivity matrix $W$ and external input current vector $\Vec{\theta}$ is written as:

$$\operatorname{FP}(W,\Vec{\theta}):=\{\sigma \subseteq [n] \mid \sigma \text{ supports a fixed point } x^* \text{ of the TLN with parameters $W$ and $\Vec{\theta}$}\}.$$

TLNs are a rich class of firing rate models which display a wealth of dynamical properties periodic attractors and, crucially for our purposes, multistability \cite{gr}.

\section{Neuroscience of decision-making}

Neuroscience experiments in decision-making generally involve a subject being presented with a task where, upon some stimulus, the subject must decide which among various alternatives is in accordance with the stimulus.  One of the canonical decision-making tasks is simian motion discrimination \cite{church}.  Monkeys are fixed in head position and presented with a visual field of moving dots.  They must then make a saccadic eye movement in the net direction of the motion to receive rewards.  A benefit of this kind of highly constrained task, where the decision is as specific as an eye movement, is to allow for concentrating on a smaller region of the brain.  Saccadic eye movement in particular is associated with the LIP.  A more general decision-making task may engage more parts of the brain and require recording of various regions to get a meaningful picture of the dynamics.

One of the common problems within decision-making studies is replicability across labs \cite{ibo}.  A 2021 paper by the International Brain Laboratory, a consortium of labs, attempts to resolve this by presenting an assay of decision-making tasks involving mice which produced consistent results across the member labs.  The core task involves a head-fixed mouse being presented a screen with a grating on either the left or the right.  The mouse must turn a steering wheel to the left or right to bring the grating toward the center of the screen to receive the reward of sweetened water.  The assay consists of variations on this core task which allow for investigating different aspects of the decision-making process, such as modifying the probability with which the grating appears on the right or left side and the contrast of the grating to see how prior experience affects the mouse's choices \cite{ibo}.

An important distinction arises in the two tasks that we have discussed.  The grating task merely requires a decision between right or left, a forced set of two choices.  On the other hand, a motion discrimination task could be decided in any direction, theoretically an infinite set of choices.  The reason this distinction is significant for our purposes is that the second might cause is to think we need a continuous attractor to describe the results.  However, in practice, a continuous choice is often approximated with a discrete set of choices.  For example, instead of treating each possible direction of saccadic eye motion as different, they may grouped into upper right, lower right, upper left, and lower left motions, reducing a continuous set of choices into just four \cite{church}. 

The key point here is that tasks with continuous choices are often approximated to have a forced set of choices.  For this reason, our analysis will primarily focus on TLNs with discrete point attractors.  While TLNs are know to have continuous, and even dynamic, attractors, there fortunately exists a class of TLNs whose attractors are known to be only discrete point attractors.

\section{DAG CTLN Models for Decision-Making Circuits}

The role of connectivity structure in shaping basins of attraction can be emphasized using a special subclass of TLNs called combinatorial TLNs (CTLNs).  In CTLNs, the weight matrix is derived from a directed graph (e.g. Fig~\ref{fig:fig3a}) and the external input currents are made uniform i.e. $\Vec{\theta}=\theta \mathbbm{1}$ where $\mathbbm{1}$ is the all ones vector.  We take a directed graph (with no self-edges) and derive weights according to the following rule \cite{gr}:

 \begin{center}
$W_{ij} = \begin{cases} 0 & \text{if } i=j \\ -1-\delta & \text{if } j \not\rightarrow i \\ -1+\varepsilon & \text{if } j \rightarrow i \end{cases}$

with $\varepsilon, \delta > 0$ and $0<\varepsilon<\frac{\delta}{1+\delta}$.
\end{center}.

%Fig 3
\begin{figure}[!ht]
\begin{center}
\vspace{.1in}
\includegraphics[width=4in]{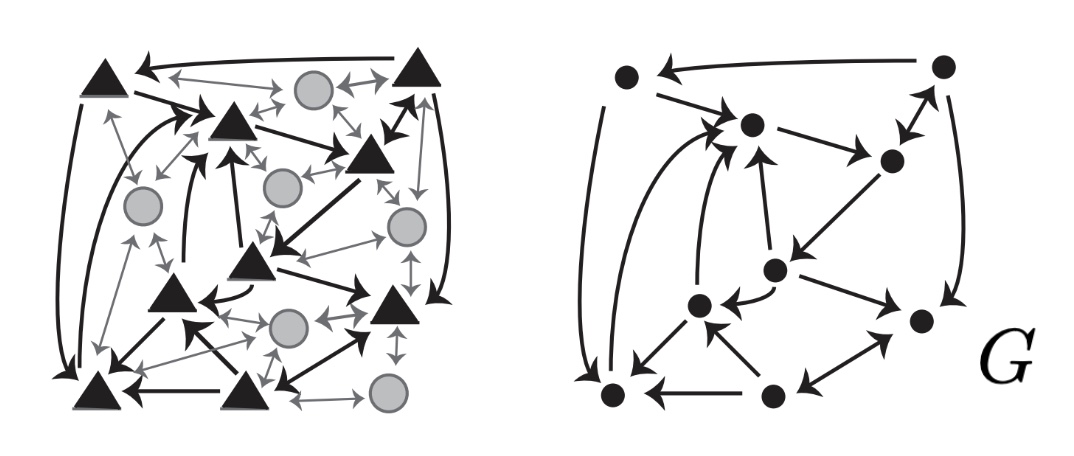}
\vspace{.1in}
\caption[Deriving a CTLN directed graph from a neural circuit]{{\bf Deriving a CTLN directed graph from a neural circuit.} An example of a directed graph representation of a neural circuit.  The inhibitory interneurons are removed, so we are left with a network of just the excitatory neurons.  The model incorporates their role in making the excitatory neurons effectively inhibitory toward one another.  For a neuron $i$, $j \not\rightarrow i$ indicates \textit{strong inhibition} as there are no excitatory connections to reduce the effective inhibition.  Alternatively, $j \rightarrow i$ indicates \textit{weak inhibition}, with the excitatory connection reducing the effective inhibition.}
\label{fig:fig3a}
\end{center}
\vspace{-.2in}
\end{figure}

We effectively create two kinds of synaptic weights.  A neuron $i$ is said to be \textit{strongly inhibited} by another neuron $j$ if $j \not\rightarrow i$ and \textit{weakly inhibited} if $j\rightarrow i$.  An excitatory connection corresponding to an edge is reducing the inhibition \cite{gr}.  This class takes as a modeling assumption that the excitatory neurons are effectively inhibiting one another indirectly through inhibitory interneurons with the excitatory connections merely reducing the inhibition as depicted in Fig~\ref{fig:fig3a}.

A remarkable property of CTLNs is that there exist correspondences between the combinatorial properties of the graph and the fixed points of the dynamical system \cite{gr}.   This enables us to talk about the attractors of a CTLN in terms of the directed graph $G$.  In many cases, the attractor supports are controlled entirely by the directed graph structure and the fixed point support set can be defined as a function of the graph.  We then use the notation:

$$\operatorname{FP}(G):=\{\sigma \subseteq [n] \mid \sigma \text{ is the support for some fixed point } x^* \text{ of a CTLN derived from $G$}\}$$

%Fig 3_2
\begin{figure}[!ht]
\begin{center}
\vspace{.1in}
\includegraphics[width=5.25in]{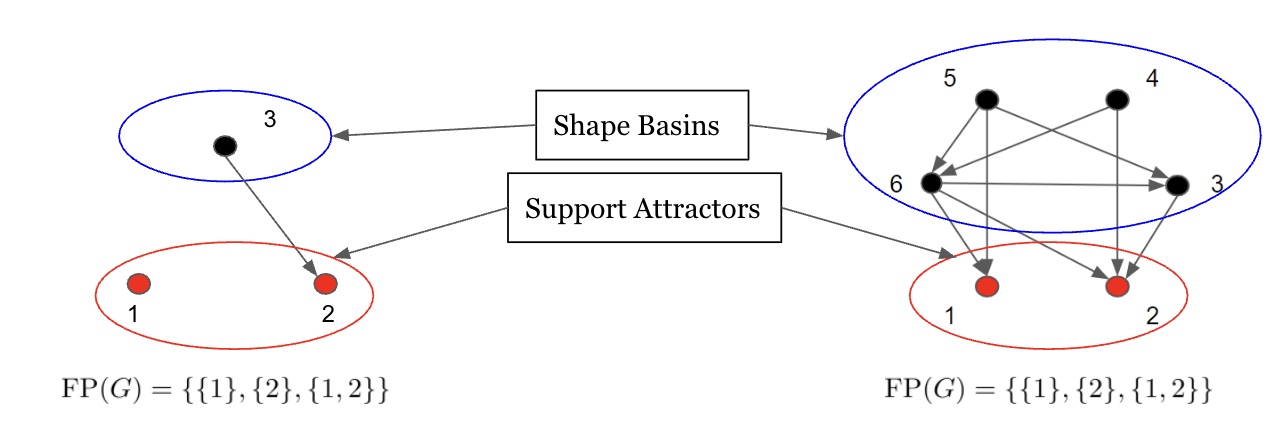}
\vspace{.1in}
\caption[CTLNs of directed acyclic graphs (DAGs)]{{\bf CTLNs of directed acyclic graphs (DAGs).} Directed acyclic graphs have fixed points supported only on sinks and the unions of sinks.  Only the fixed points supported on the sinks themselves yield attractors, one for each sink ($\{1\}$ and $\{2\}$).  The union of sinks ($\{1,2\}$) supports the saddle point which lies on the separatrix}
\label{fig:fig3}
\end{center}
\vspace{-.2in}
\end{figure}

In particular, CTLNs derived from directed acyclic graphs (DAGs) have very predictable stable fixed point attractors, one for each sink, with unions of sinks being the unstable fixed points \cite{gr}.  Moreover, DAG CTLNs have no dynamic attractors \cite{dagdyn}, making it very easy to generate a variety of CTLNs with the same attractors by fixing the sinks and altering the graph structure of the non-sink neurons (Fig~\ref{fig:fig3}).  Doing so will not change the attractors, but it will change the basins of attraction!  

%Fig 4
\begin{figure}[!ht]
\begin{center}
\vspace{.1in}
\includegraphics[width=6.5in]{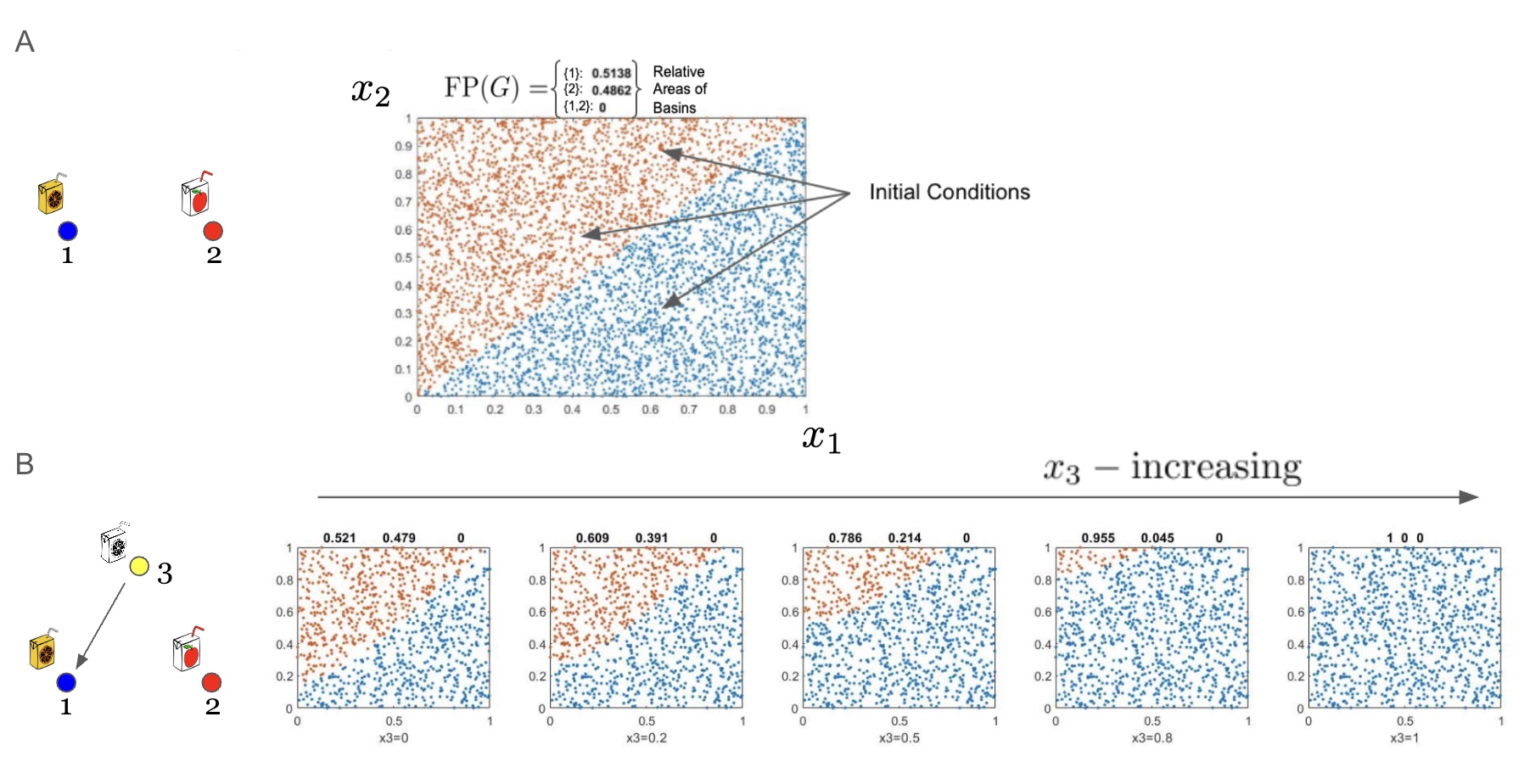}
\vspace{.1in}
\caption[Non-sink neurons shape basins of attraction in DAG CTLNs]{{\bf Non-sink neurons shape basins of attraction in DAG CTLNs.} We show in this figure the results of Monte Carlo simulation to determine the basins of attraction.  We randomly sample initial conditions $(x_1(0), x_2(0), \dots , x_n(0))$ for a CTLN and we color code that point in state space according to the sink attractor to which it converges.  (A) In this DAG, there exist only the two sinks.  The figure color codes initial conditions in the $x_1, x_2$ plane to numerically capture the basins of attraction.  We see that the basins of attraction are symmetric.  (B) This DAG corresponds to the Decoy Effect scenario where the third neuron represent the irrelevant choice.  It is never part of the support of an attractor, but nonetheless it shapes the dynamics.  Here we have taken cross sections of the state space for different values of $x_3(0)$ in the initial condition.  We see that the activity of the third neuron skews the basins of attraction making the one corresponding to $x_1$ larger.  While the attractors themselves remain unchanged from A, the basins have been altered.  Notably, the shifting of the decision boundary is even present in the $x_3(0)=0$ cross section.}
\label{fig:fig3b}
\end{center}
\vspace{-.2in}
\end{figure}

We can construct a DAG which mirrors the aforementioned Decoy Effect.  We begin with two sinks corresponding to the main choices.  By randomly sampling initial conditions and tracking the attractor to which they converge, we can numerically reconstruct the basins of attraction for the CTLN (Fig~\ref{fig:fig3b}A) and see that they are equivalent.  We can then compare this to a case where we add the asymmetrically dominated, irrelevant, choice as a non-sink neuron, neuron 3 (Fig~\ref{fig:fig3b}B).  Both of these circuits have the same attractors, but different basins of attraction.  Notice that in the case of the Decoy Effect, the basin of attraction of the weakly inhibited neuron is larger, indicating that there are more initial conditions converging towards its attractor relative to the symmetric case.  This captures how the bias has been skewed towards that choice.  Notice that in both cases there is a fixed point supported on both sinks (marked $\{1,2\}$) which has no trajectories converging to it.  This is because this fixed point is a saddle point.

From this case of DAGs with two sinks, we can build an intuitive picture of binary choice decision-making dynamics.  We have two basins of attraction separated by a decision boundary of codimension 1 and this decision boundary is the stable manifold associated with a saddle point supported on the union of the sinks.  When we have a decision boundary \cite{carnevale}, to see how the basins of attraction are shaped, we would need to understand the dynamics of the network relative to it.  Consider Fig~\ref{fig:fig4}, where we have a stable manifold marked in green for a saddle point which separates two basins of attraction.  That stable manifold represents the decision boundary.

\section{Basins of Attraction and Decision-Making Bias: Choosing Initial Conditions}

An assumption up until now has been that all of the initial conditions are equally relevant (Fig~\ref{fig:fig4}A).  This certainly is implausible biologically, but then this begs the question of what initial conditions should we be focused on?  

% Fig 4_2
\begin{figure}[!ht]
\begin{center}
\vspace{.1in}
\includegraphics[width=6.25in]{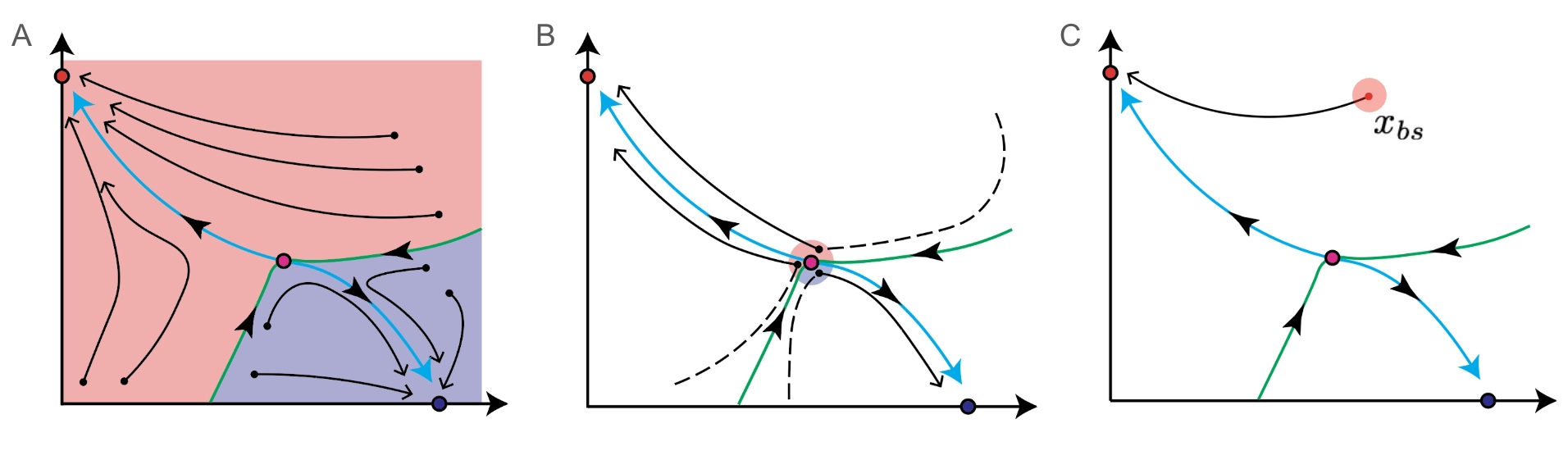}
\vspace{.1in}
\caption[Bias, basins of attraction, and three hypotheses of neural dynamics]{{\bf Bias, basins of attraction, and three hypotheses of neural dynamics}  A two dimensional schematic depicting three ways of relating basins of attraction to a decision-making circuit.  (A) Determining bias as the relative sizes of the full basins of attraction, where all initial conditions matter equally, aligns with the high-dimensional reservoir dynamics hypothesis of neural dynamics.  (B) Focusing on initial conditions near the saddle points emphasizes trajectories which approximate the branching stable (green) and unstable (cyan) manifolds of the saddle point, hewing along the decision boundary.  This aligns with the hypothesis of low-dimensional subspace dynamics and associates bias with the relative sizes of the basins of attraction within the region around the saddle point.  (C) Prioritizing the trajectory using the balanced state as the initial condition is in accordance with path-following dynamics and understands bias of the network as being towards the attractor to which the balanced state trajectory converges.}
\label{fig:fig4}
\end{center}
\vspace{-.2in}
\end{figure}

We could instead restrict ourselves to trajectories close to the decision boundary.  Even if we don't quite know where the decision boundary is located, as DAG CTLNs are continuous dynamical systems, trajectories beginning sufficiently close to the stable manifold of the saddle point will eventually be near the fixed point before following the unstable manifold toward one of the attractors.  We could then understand these trajectories by concentrating on initial conditions in the neighborhood of the saddle point, as in Fig~\ref{fig:fig4}B, and determining which basin is larger in this neighborhood.

Another initial condition which is of interest is informed by the biology of neural dynamics.  There exists a broad literature arguing that neural circuits exist in a state of balance between inhibition and excitation \cite{bs1,bs2,bs3}.  This means that the input received by every neuron, both from within the network and from outside of it, should by zero for every neuron.  The excitatory/inhibitory balance would then arise at the state $\Vec{x}_{bs}$ such that $W \Vec{x}_{bs} + \Vec{\theta} = 0$.  This \textit{balanced state} typically lies in one of the basins of attraction and we could then see to which attractor it converges (Fig~\ref{fig:fig4}C).  

A notable concern with balanced states is that their trajectories often do not make sense biologically, presenting a challenge to this approach.  We call a CTLN where these issues do not arise \textit{balanced}.

To summarize, we have suggested three paradigms.  The first treats all initial conditions equally, the second focuses on initial conditions in the neighborhood of the saddle point as a means of studying trajectories near the decision boundary, and the third studies the singular initial condition of the balanced state.  

Notably these three paradigms map on to three hypotheses of neural dynamics \cite{hyp}.

\textbf{H1: } High-dimensional reservoir dynamics

\textbf{H2: } Low-dimensional subspace structured dynamics

\textbf{H3: } Path-following dynamics

In the case of \textbf{H1}, where neural dynamics happens in the full state space of the network, we would need to consider initial conditions generally and this corresponds to the approach of looking at the size of the full basins of attraction relative to one another.  If we were to take \textbf{H2}, where neural dynamics are believed to primarily operate on low dimensional manifolds, we would focus on initial conditions near the lower dimensional decision boundaries which relates to the paradigm of considering initial conditions near the saddle point.  Finally, if we were to accept \textbf{H3}, where we are privileging a particular trajectory this would align with the idea of seeing to which attractor a trajectory beginning at the balanced state converges. 

\section{Summary of Results}

The organization of this dissertation is as follows:

In Chapter 2 we discuss a two neuron TLN model for decision-making.  We conduct a bifurcation analysis to prove the conditions required for them to be useful in the study of decision-making i.e. for there to be two stable attractors.  We will then rigorously prove how the basins of attraction evolve through parameter space by determining bifurcations on the qualitative, coarse-grained dynamics of the network.  This will show fully how the parameters can be manipulated to adjust the sizes of the basins of attraction while preserving the attractors.  We conclude by explicitly calculating the basins and detailing the relationship between their sizes and the model parameters.  As determining the basins of attraction \textit{a fortiori} tells us about the basins of attraction near the saddle point and the attractor to which the balanced state converges, this constitutes a full solution to the problem in the two dimensional case.

As solving a general TLN in higher dimensions is impractical, in Chapter 3 we explore computer assisted approaches drawing on Conley Index Theory.  We offer an algorithm to rigorously compute coarse-grained dynamics for TLNs while also showing the challenges it faces in determining the basins of attraction in higher dimensions.  

In Chapter 4 we restrict our attention to CTLNs. Specifically, we focus on DAG CTLNs and give theoretical results demonstrating a deep association between the properties of these networks and generating function constructions on the DAG that we refer to as \textit{localized path polynomials}.  Using them, we will not only offer a novel proof for the relationship between the sinks of the DAG and the attractors of its CTLNs, but we will also use them to rigorously prove analytic solutions for the dynamics of a subclass of DAG CTLNs.  After offering an analytic approach to resolving the associated initial value problem however, we realize that the expressions end up being too cumbersome to allow us to resolve the full basins of attraction as we did in Chapter 2.  Still, we press forward with studying the basins of attraction in the vicinity of the saddle point and of the balanced state.

In Chapter 5 we show how the theoretical properties of DAG CTLNs suggest a relationship between sink in-degree and the sizes of basins of attraction near the saddle point.  We conjecture a relationship between sink in-degree and the local basin size, and we find numerical evidence to support this relationship.  

Chapter 6 is focused on the path-following, balanced state paradigm and we tackle head on the problem of determining when a CTLN is balanced.  We rigorously prove results for general CTLNs and then demonstrate how incorporating the properties of DAG CTLNs and their localized path polynomials enables us to prove stronger results in that case.  We will then discussed the notion of \textit{balanced graphs}, graphs such that any CTLN derived from them is balanced, and how localized path polynomials can be used to build such graphs.  We conclude by presenting an algorithm which has had partial success at predicting the attractor to which the balanced state converges and we provide numerical evidence demonstrating its effectiveness.

The last chapter will weaken the CTLN conditions by allowing the external input currents received by the neurons to vary, what will be referred to as an hCTLN (heterogeneous CTLN).  We look at hCTLNs derived from DAGs and will generalize some of the theoretical results from DAG CTLNs to this setting.

\subsection*{Major Original Contributions}

\begin{itemize}

\item The main theoretical contributions in Chapter 2 are Theorem~\ref{thm:bcmcb} and Theorem~\ref{thm:bcmbasin}.  Together, these provide a comprehensive description of how the combinatorial dynamics and basins of attraction of competitive TLNs evolve with respect to each other in a four dimensional parameter space.  Also, while Theorem~\ref{thm:paradox} is primarily an illustrative exercise, it is nonetheless an original result.

\item The novel contributions in Chapter 3 are Proposition~\ref{prp:sep}, Proposition~\ref{prp:simpsep}, and Algorithm 1.  They allow the efficient generation of a state transition graph on any TLN subject to a convex polytope partition of the state space generated by a hyperplane arrangement and respecting the piecewise linearity of the TLN.

\item Chapter 4 defines localized path polynomials and contains a number of original theoretical results related to DAG CTLNs, notably Corollary~\ref{corr:charpolyctln}, Corollary~\ref{corr:evs}, and Proposition~\ref{prp:dagfp}.  Applied together in the context of a special class of DAG CTLNs, they produce Theorem~\ref{thm:final} determining general solutions for solving the underlying linear systems composing DAGs.  We also give a way of resolving the initial value problem for these systems.

\item Chapter 5 is primarily computational, numerically detecting correlation between the sizes of the basins of attraction for DAG CTLNs in the neighborhood of saddle points and the indegrees of the sinks relative to one another.  The analysis is motivated by Proposition~\ref{prp:stab} and Proposition~\ref{prp:stab2}.

\item Chapter 6 again contains a number of theoretical results.  With respect to determining when a CTLN is balanced, the notable results are Theorem~\ref{thm:bs} which gives a sufficient condition on parameters for any CTLN to be balanced and Theorem~\ref{thm:dagbs} which is an improved result specifically for DAGs.  Additionally, Theorem~\ref{thm:ufdbs} and Corollary~\ref{corr:outtree} describes classes of balanced graphs.  Lastly we contribute Algorithm 2 which has had partial success at predicting the attractor to which the balanced state trajectory converges.

\item The key contributions of Chapter 7 are Lemma~\ref{lem:DAGG}, Proposition~\ref{prp:fpg}, Proposition~\ref{prp:minone}, and Proposition~\ref{prp:dagbsg}.  Proposition~\ref{prp:minone} is used to obtain a stronger form of Theorem~\ref{thm:final} whereas the remainder generalize results from Chapter 4 and Chapter 6 into the setting of hCTLNs.

\end{itemize}

\chapter{Combinatorial Dynamics of Two-Dimensional TLNs: The Binary Competition Model}

%\section{TLNs: A Patchwork of Linear Systems}
One of the main benefits of working with TLN models is in the piecewise linearity of their differential equations.  In this chapter we will show how this can be exploited in a simple two dimensional model of  competing neural populations to demonstrate how the parameters can shape the basins of attraction while preserving the attractors.  

Before we delve into the details of the model, we introduce some of the fundamental tools used to study TLN dynamics.

Recall that the TLN differential equations are of the form:

$$\dfrac{dx_i}{dt} = -x_i + \left[\sum_{j=1}^n W_{ij}x_j +\theta_i \right]_+, i=1,...,n.$$ 

Note that the term inside the ReLU function is a linear function of $x$, 

$$y_i(x):=\sum_{j=1}^n W_{ij}x_j +\theta_i.$$

Because of the threshold linearity of the ReLU function, $\dfrac{dx_i}{dt}$ will be linear on either side of the hyperplane:

$$H_i: y_i(x)= 0.$$

The hyperplanes $\{H_i\}_{i=1}^n$ can be used to divide the state space into chambers where the system is linear \cite{gr}.  This shows that TLNs are a very special case of a continuous Filippov system\cite{fil} and are patchworks of linear systems.  Within each chamber of the hyperplane arrangement, the dynamics are governed by a linear dynamical system.  

The chambers created by the partitioning of $H_i$ (Fig~\ref{fig:chambersRL}) and their corresponding linear systems of ODEs will be referenced in the following manner:

\begin{ddd}
    $R_\sigma=\{x\in \mathbb{R}^n \mid y_i(x) > 0, \forall i \in \sigma$ and $y_k(x)\leq 0, \forall k \not \in \sigma\}$
\end{ddd}

\begin{rmk}
We will occasionally use the symbol $R_{\sigma}^+$ to indicate the restriction of $R_{\sigma}$ to the positive orthant i.e. $R_{\sigma}^+ = \{ x\in R_{\sigma} \mid x\geq 0 \}$.
\end{rmk}

\begin{ddd}
     $L_\sigma=\left\{\dfrac{dx_i}{dt} = -x_i +\sum_{j=1}^n W_{ij}x_j+b_i\mid i\in\sigma\right\} \bigcup \left\{\dfrac{dx_k}{dt}=-x_k \mid  k \not\in \sigma\right\}$
\end{ddd}

\begin{figure}[!ht]
\begin{center}
\vspace{.1in}
\includegraphics[width=2.75in]{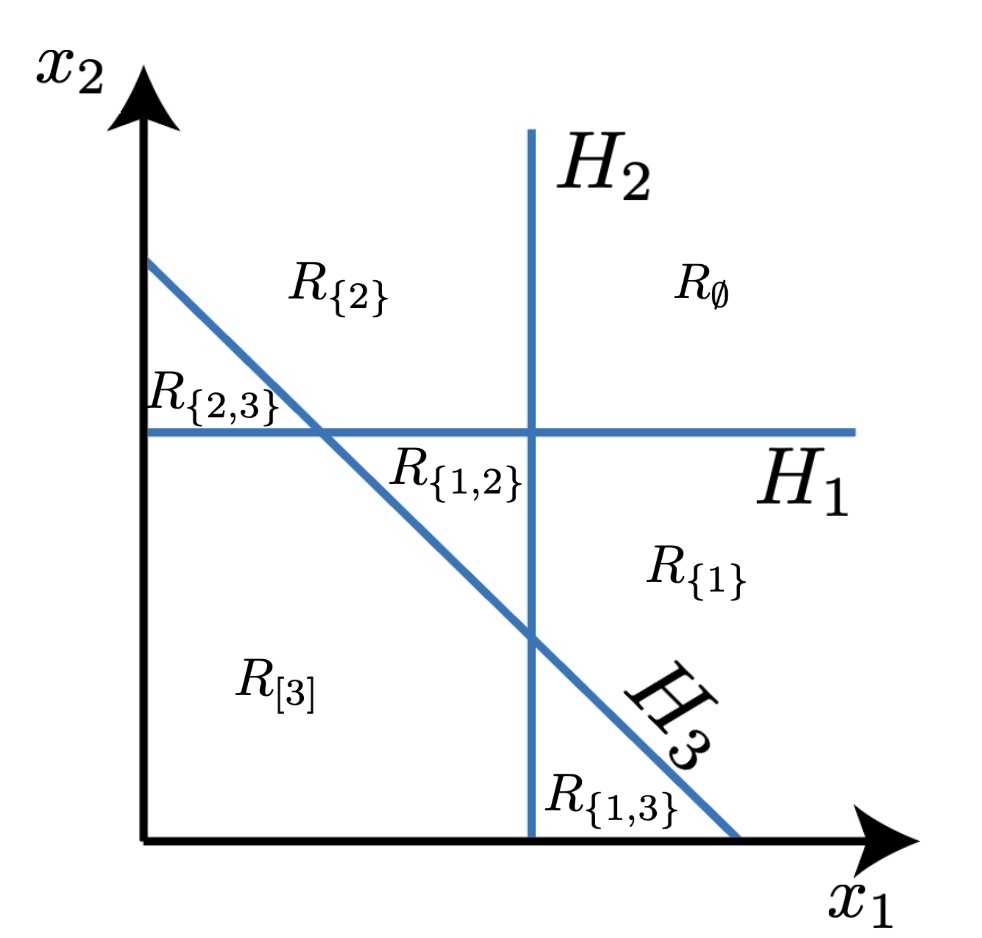}
\vspace{.1in}
\caption[TLN state space partition using $H_i$ hyperplane arrangement]{{\bf TLN state space partition using $H_i$ hyperplane arrangement.} An example cross section of the hyperplane arrangement for a three neuron TLN.  Each neuron $i$ produces a hyperplane $H_i$ and the full hyperplane arrangement creates a partition of the state space into chambers $R_{\sigma}$ governed by linear dynamics $L_{\sigma}$.} 
\label{fig:chambersRL}
\end{center}
\vspace{-.2in}
\end{figure}

The local linearity of TLNs make analyses of the dynamics far more tractable than for most nonlinear dynamical systems.  To illustrate this, we look now at an application of TLNs to the so-called "paradoxical effect" problem.

\section{Example: Paradoxical Effect}

The paradoxical effect is a term used to describe how increasing input current to an inhibitory neuron can, in certain contexts, reduce its steady state activity \cite{dox}.  The mathematical question that this introduces is under what conditions can a mathematical network model recreate this phenomena.

The phenomenon is most easily understood through using a \textit{nullcline} analysis.

\begin{ddd}

Let $\dfrac{dx}{dt}=f(x)$ be an autonomous differential equation.  The \textbf{nullcline} of this differential equation is the curve given by the relation $\mathcal{N}: f(x)=0$.

\end{ddd}

In the context of a system of an $n$-dimensional system of differential equations, each differential equation contributes a nullcline along which the derivative component in the direction of the associated state variable is zero.  Taking the nullcline of each differential equation in the system we obtain the collection, $\{\mathcal{N}_i\}_{i=1}^n$.

\begin{rmk}
The intersection of the nullclines gives the steady states of the dynamical system.  In the event that they intersect at isolated points, those will be the fixed points of the system.
\end{rmk}

One of the great benefits of working with TLNs is that they have piecewise linear nullclines.  For the differential equation:

$$\dfrac{dx_i}{dt} = -x_i + \left[\sum_{j=1}^n W_{ij}x_j +\theta_i \right]_+$$

the associated nullcline will be:

$$\mathcal{N}_i : x_i = \left[\sum_{j=1}^n W_{ij}x_j +\theta_i \right]_+$$

We now demonstrate the paradoxical effect in TLNs.  Consider an E-I network, as depicted in Fig~\ref{fig:ei2}A, consisting of a single excitatory and a single inhibitory neuron with weight matrix:

$$W=\left[
\begin{array}{cc}
    w_e & w_i^e \\
    w_e^i & w_i
\end{array}
\right]
$$

% Fig 4
\begin{figure}[!ht]
\begin{center}
\vspace{.1in}
\includegraphics[width=5.75in]{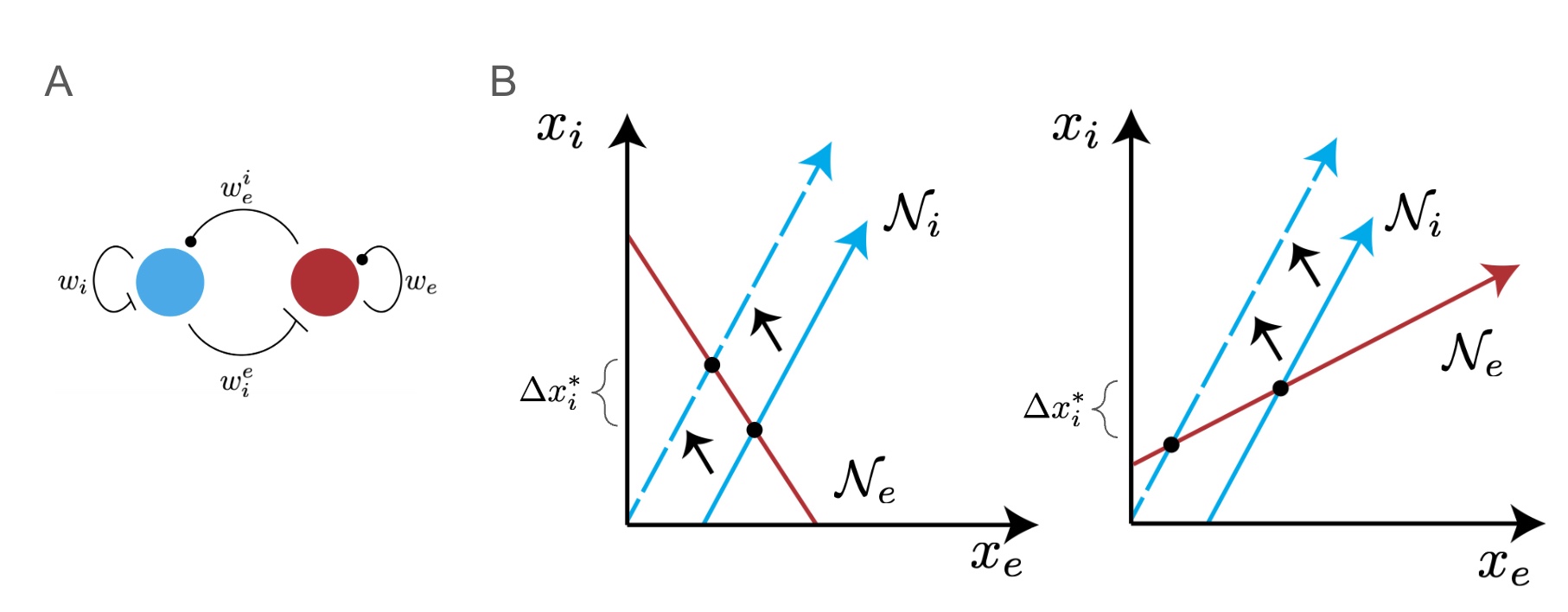}
\vspace{.1in}
\caption[E-I Networks and the paradoxical effect]{{\bf E-I Networks and the paradoxical effect.} (A) An E-I network with two neurons.  The parameters $w_i, w_e$ are the self-inhibition and self-excitation of the neurons whereas $w_i^e$ and $w^e_i$ represent the excitation to the inhibitory neuron and the inhibition to the excitatory neuron respectively.  (B) A nullcline schematic illustrating the cause of the paradoxical effect in two dimensions where $\mathcal{N}_{e,i}$ represent the nullclines for the neurons.  The first image shows the "non-paradoxical" case.  Increasing the external input current to the $x_i$ shifts its nullcline upward and increases its steady state activity.  However, under the proper orientation of the nullclines, this upward shift $\mathcal{N}_i$ decreases the steady state activity of $x_i$.} 
\label{fig:ei2}
\end{center}
\vspace{-.2in}
\end{figure}

where $w_e$ is the self-excitation of the excitatory neuron, $w_i^e$ is the inhibition of the inhibitory neuron to the excitatory neuron, $w_e^i$ is the excitation from the excitatory neuron to the inhibitory neuron, and $w_i$ is the self-inhibition of the inhibitory neuron.  The ODE system is then:

$$\dfrac{dx_e}{dt}=-x_e+[w_e x_e +w_i^e x_i +\theta_e]_+$$
$$\dfrac{dx_i}{dt}=-x_i+[w_e^i x_e +w_i x_i +\theta_i]_+$$

In particular, focus on the chamber $R_{\left[2\right]}$ with the corresponding ODE system $L_{\left[2\right]}$:

$$\dfrac{dx_e}{dt}=-x_e+w_e x_e +w_i^e x_i +\theta_1=(w_e-1)x_e+w_i^e x_i+\theta_e$$
$$\dfrac{dx_i}{dt} =-x_i+w_e^i x_e +w_i x_i +\theta_2=w_e^i x_e +(w_i-1) x_i +\theta_i$$

The fixed point of this system is:

$$
x^*=\left(\dfrac{\theta_i w_i^e +\theta_e(1-w_i)}{(1-w_i)(1-w_e)-w_e^i w_i^e}, \dfrac{\theta_e w_e^i +\theta_i(1-w_e)}{(1-w_i)(1-w_e)-w_e^i w_i^e} \right)
$$

Notice that if:

$$
\dfrac{1-w_e}{(1-w_i)(1-w_e)-w_e^i w_i^e}<0
$$

then $x_i^*$ decreases as $\theta_i$ increases.  That is to say that increasing the external drive on the inhibitory neuron will reduce its steady state value.  

The fixed point of $L_{\left[2\right]}$ is only a fixed point of the the system as a whole if it lies in $R_{\left[2\right]}$.  %It can be shown that $y_e(x^*),y_i(x^*)>0$ precisely when $(1-w_i)(1-w_e)-w_e^i w_i^e>0$.

It is clear then that if $x^*$ is a fixed point of the TLN and $(1-w_i)(1-w_e)-w_e^i w_i^e>0$, then the paradoxical effect arises when $w_e>1$.  This realizes the paradoxical effect. 

To understand what is happening consider how the nullclines change.  For this two-dimensional system $L_{[2]}$ we have the nullclines:

$$\mathcal{N}_e : x_i = \dfrac{1-w_e}{w_i^e} x_e - \dfrac{\theta_e}{w_i^e}$$

$$\mathcal{N}_i : x_i = \dfrac{w_e^i}{1-w_i} x_e + \dfrac{\theta_i}{1-w_i}$$

The condition $w_e>1$ change the sign of the slope for $\mathcal{N}_e$ and produces the paradoxical effect as depicted in Fig~\ref{fig:ei2}B.  Now let us consider how the nature of TLNs lets us study the paradoxical effect in higher dimensions.

\subsection{The Paradoxical Effect in Larger TLNs}

We now consider TLNs with additional inhibitory neurons identical to the first (see Fig~\ref{fig:ein}).  As the inhibitory neurons are identical we can reuse the four synaptic weight parameters $w_e, w_i^e, w_e^i, w_i$ and add the inhibition between inhibitory neurons $w_i^i$.  As before, $x_1$ will be the firing rate of the excitatory neuron while $x_2, \dots, x_n$ are the inhibitory neurons.

\begin{figure}[!ht]
\begin{center}
\vspace{.1in}
\includegraphics[width=3.75in]{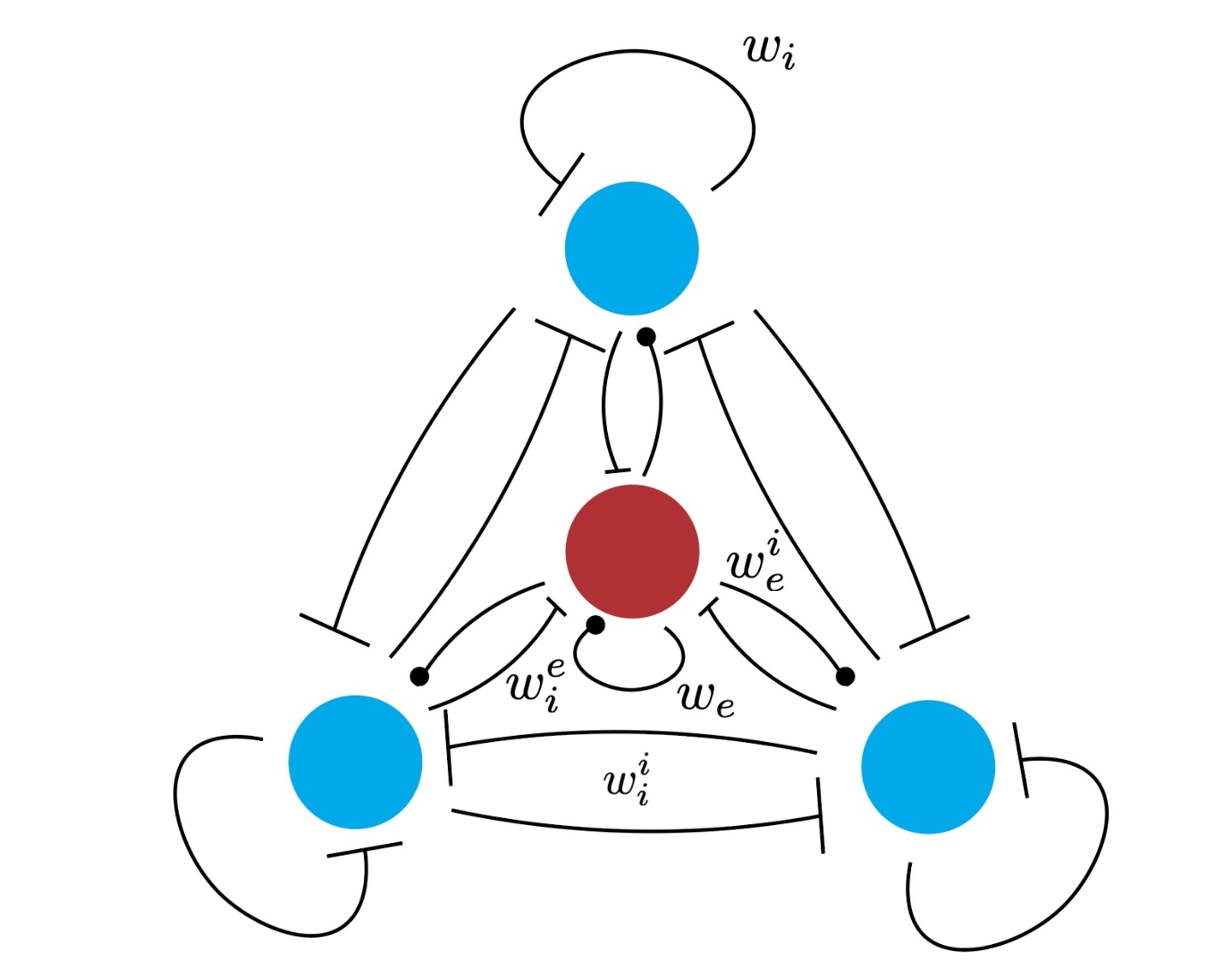}
\vspace{.1in}
\caption[Higher-dimensional E-I networks with identical inhibitory neurons]{{\bf Higher-dimensional E-I networks with identical inhibitory neurons.} An architecture of a four neuron E-I network in higher dimensions with 3 identical inhibitory neurons.  The new parameter $w_i^i$ corresponds to the inhibition between inhibitory neurons and is symmetric among them.}
\label{fig:ein}
\end{center}
\vspace{-.2in}
\end{figure}

We consider the chamber with all the ReLU functions active, $R_{[n]}$, and study the dynamics of the linear system $L_{[n]}$.  The nullclines are then as follows:

$$\mathcal{N}_1: -x_1+w_e x_1+w_i^e \sum_{k=2}^n x_k + \theta_1=0$$
$$\mathcal{N}_j: -x_j +w_e^i x_1 +w_i x_j +w_i^i\sum_{k=2,k\neq j}^n x_k +\theta_j = 0 \text{ if }j>1$$

These can be rewritten as:

$$\mathcal{N}_1: x_1=\dfrac{w_i^e}{1-w_e} \sum_{k=2}^n x_k + \dfrac{\theta_1}{1-w_e}$$
$$\mathcal{N}_{j}: x_1=\dfrac{1-w_i}{w_e^i}x_j -\dfrac{w_i^i}{w_e^i} \sum_{k=2,k\neq j}^n x_k -\dfrac{\theta_j}{w_e^i}\text{ if } j>1$$

The fixed point associated with $L_{\left[n \right]}$ lies at the intersection of the nullclines.  The computations will get a bit involved, so we introduce a useful lemma.

\begin{lem}\label{lem:pe}
    For:

$$\mathcal{N}_{j_1}: x_1=\dfrac{1-w_i}{w_e^i}x_{j_1} -\dfrac{w_i^i}{w_e^i} \sum_{k=2,k\neq j_1}^n x_k -\dfrac{\theta_{j_1}}{w_e^i}$$ 

$$\mathcal{N}_{j_2}: x_1=\dfrac{1-w_i}{w_e^i}x_{j_2} -\dfrac{w_i^i}{w_e^i} \sum_{k=2,k\neq j_2}^n x_k -\dfrac{\theta_{j_2}}{w_e^i}$$

Then $x\in \mathcal{N}_{j_1}\cap \mathcal{N}_{j_2} \implies x_{j_1}=x_{j_2}+\dfrac{\theta_{j_1}-\theta_{j_2}}{1-w_i+w_i^i}$

\end{lem}

\begin{proof}

If $x$ lies on both nullclines, then the two nullcline equations can be set equal to one another.

$$
\dfrac{1-w_i}{w_e^i}x_{j_1} -\dfrac{w_i^i}{w_e^i} \sum_{k=2,k\neq j_1}^n x_k -\dfrac{\theta_{j_1}}{w_e^i}=\dfrac{1-w_i}{w_e^i}x_{j_2} -\dfrac{w_i^i}{w_e^i} \sum_{k=2,k\neq j_2}^n x_k -\dfrac{\theta_{j_2}}{w_e^i}
$$

After multiplying through by $w_e^i$ and eliminating identical terms from both sides, this simplifies to:

$$
(1-w_i)x_{j_1}-w_i^i x_{j_2} -\theta_{j_1}=(1-w_i)x_{j_2}-w_i^i x_{j_1} -\theta_{j_2}
$$

The rearranging of which yields the desired:

$$
E_{j_1 j_2}:x_{j_1}=x_{j_2}+\dfrac{\theta_{j_1}-\theta_{j_2}}{1-w_i+w_i^i}
$$

\end{proof}

Lemma~\ref{lem:pe} is a useful technical tool which we can use to rewrite the entries of the fixed point in terms of one another.  From here the fixed point of $L_{[n]}$ can be calculated.

\begin{prp}
    The fixed point $x^*$ associated with the system $L_{\left[n \right]}$ of a threshold linear network consisting of one excitatory neuron, $x_1$, and multiple identical inhibitory neurons is:

    $$x_1^*=\dfrac{w_i^e\left(\gamma(n-1)+\beta \right)}{1-w_e}\sum_{k=2}^n \theta_k + \dfrac{1-w_i^e\alpha (n-1)}{1-w_e}\theta_1$$
    
   $$ x_{j}^*=\gamma\sum_{k=2}^n \theta_k + \beta \theta_j -\alpha\theta_1\text{ if } (j>1)$$

where:

$\alpha=\dfrac{w_e^i}{w_e^i w_i^e (n-1)+(1-w_e)(-1+w_i +w_i^i (n-2))}$

$\gamma=\dfrac{-w_e^i w_i^e - w_i^i (1-w_e)}{(1-w_i+w_i^i)(w_e^i w_i^e (n-1)+(1-w_e)(-1+w_i+w_i^i(n-2)))}$

$\beta=\dfrac{1}{1-w_i+w_i^i}$

and the synaptic weights as defined in Fig~\ref{fig:ein}.

\end{prp}
\begin{proof}

For the fixed point $x^*$, we can apply Lemma~\ref{lem:pe}, to rewrite each $x^*_j$ such that $j>1$ in the form:

\begin{eqnarray}\label{eq:eq1}
    x^*_j=x^*_n+\dfrac{\theta_j-\theta_n}{1-w_i+w_i^i}
\end{eqnarray}

Then, since $x^*$ lies on the nullcline $\mathcal{N}_1$:

\begin{eqnarray}\label{eq:eq2}
    x^*_1=\dfrac{w^e_i}{1-w_e}\sum_{k=2}^n \left(x^*_n+\dfrac{\theta_k-\theta_n}{1-w_i+w_i^i}\right) +\dfrac{\theta_1}{1-w_e}
\end{eqnarray}

Thus, all $x^*_j$ such that $j\neq n$ can be rewritten in terms of $x^*_n$.

As $x^*$ lies on $\mathcal{N}_n$:

\begin{eqnarray}\label{eq:eq3}
    x^*_1=\dfrac{1-w_i+w_i^i}{w_e^i}x^*_n-\dfrac{w_i^i}{w_e^i}\sum_{k=2}^n x^*_k - \dfrac{\theta_n}{w_e^i}
\end{eqnarray}

Using (2.1) and (2.2), equation (2.3) can be rewritten purely in terms of $x_n^*$:

   $$\dfrac{w^e_i}{1-w_e}\sum_{k=2}^n \left(x^*_n+\dfrac{\theta_k-\theta_n}{1-w_i+w_i^i}\right) +\dfrac{\theta_1}{1-w_e}$$ 
   
   $$=  \dfrac{1-w_i+w_i^i}{w_e^i}x^*_n-\dfrac{w_i^i}{w_e^i}\sum_{k=2}^n \left(x^*_n+\dfrac{\theta_k-\theta_n}{1-w_i+w_i^i}\right) - \dfrac{\theta_n}{w_e^i}$$

Solving for $x_n^*$ yields:

    $$x^*_n=\gamma \sum_{k=2}^n \theta_k + \beta \theta_n - \alpha\theta_1$$

Then, for $j>1$:

    $$x_j^* = x_n^* + \dfrac{\theta_j-\theta_n}{1-w_i+w_i^i}=\gamma\sum_{k=2}^n \theta_k + \beta\theta_j -\alpha\theta_1$$

Similarly, using (2), $x^*_1$ is:

    $$x_1^*=\dfrac{w_i^e\left(\gamma(n-1)+\beta \right)}{1-w_e}\sum_{k=2}^n \theta_k + \dfrac{1-w_i^e\alpha (n-1)}{1-w_e}\theta_1$$

\end{proof}

\begin{thm}\label{thm:paradox}
In a TLN of one excitatory neuron, $x_1$, and $n-1$ identical inhibitory neurons, $x_{j>1}$, let $x^*$ be the fixed point of $L_{[n]}$.  Then, for $j\in [2, n]$, increasing $\theta_j$ will decrease $x^*_j$ when $\dfrac{\tilde{\gamma}}{\tilde{\beta}} <0$ where:

$\tilde{\gamma}=w_e^i w_i^e (n-2)+(1-w_e)(-1+w_i+w_i^i(n-3))$

$\tilde{\beta}=(1-w_i+w_i^i)(w_e^i w_i^e (n-1)+(1-w_e)(-1+w_i+w_i^i(n-2)))$

and the synaptic weights as defined in Fig~\ref{fig:ein}.
\end{thm}

\begin{proof}

Recall that we seek the condition such that $x_j^*$ decreases as $\theta_j$ increases for $j>1$.  Focusing on the role of $\theta_j$ in the expression of $x_j^*$, we have:

$$x_j^*=\gamma\sum_{k=2}^n \theta_k + \beta \theta_j -\alpha\theta_1=(\gamma +\beta)\theta_j+\gamma\sum_{j\neq k=2}^n \theta_k -\alpha\theta_1$$

We would then expect that the paradoxical effect would arise when $\gamma + \beta < 0$.

Now notice that:

$\gamma+\beta = \dfrac{-w_e^i w_i^e - w_i^i (1-w_e)}{(1-w_i+w_i^i)(w_e^i w_i^e (n-1)+(1-w_e)(-1+w_i+w_i^i(n-2)))} + \dfrac{1}{1-w_i+w_i^i} $

$= \dfrac{w_e^i w_i^e (n-2)+(1-w_e)(-1+w_i+w_i^i(n-3))}{(1-w_i+w_i^i)(w_e^i w_i^e (n-1)+(1-w_e)(-1+w_i+w_i^i(n-2)))}= \dfrac{\tilde{\gamma}}{\tilde{\beta}} <0$. 

\end{proof}

We can find a simpler corollary if we assume that inhibitory neurons inhibit themselves to the same extent as they inhibit other neurons, i.e. $w_i = w_i^i$.  Then we have:

\begin{corr}\label{thm:corrdox}
In a TLN of one excitatory neuron, $x_1$, and $n-1$ identical inhibitory neurons, $x_{j>1}$, let $x^*$ be the fixed point of $L_{[n]}$.  Then, for $j\in [2, n]$, increasing $\theta_j$ will decrease $x^*_j$ when $\dfrac{\tilde{\gamma}}{\tilde{\beta}} <0$ where:

$\tilde{\gamma}=w_e^i w_i^e (n-2)+(1-w_e)(-1+w_i(n-2))$

$\tilde{\beta}=w_e^i w_i^e (n-1)+(1-w_e)(-1+w_i(n-1))$

and the synaptic weights as defined in Fig~\ref{fig:ein}.
\end{corr}

Notice that this analysis involved treating the nonlinear TLN as a linear system, which we could do precisely because of the piecewise linearity of the ReLU function.  In a network which used a more complex firing rate function this analysis could have been far more challenging.  The idea of using a chamber by chamber linear dynamical analysis is an approach we will come back to repeatedly and is at the heart of the rich body of theoretical results in TLN literature.

\section{Binary Competition Model}

A strategy that has been used in attractor network models of decision-making is to reduce a larger network of neurons down to a smaller network of neural populations, often with two competing neural populations \cite{wang1,wang2,wang3,bcmarchitect}.  We will treat the populations of excitatory neurons as if they are effectively inhibiting one another, something we already see briefly with the discussion of CTLNs in the introduction.  

Regions of the cerebral cortex associated with decision-making, such as the prefrontal cortex and the LIP, are thought to consist of modular networks of excitatory neurons immersed within a sea of inhibitory interneurons \cite{cpg}.  This means that we can think of a decision-making task as being handled by a modular decision-making circuit of excitatory neurons.  We take as a modeling assumption that the excitatory neurons are effectively inhibiting each other through the interneurons, with excitatory connections merely reducing the inhibition.  What this would mean mathematically is that all the synaptic weights are negative.  This yields the notion of \textit{competitive TLNs}.

\begin{ddd}[Definition 5.1 in \cite{diverse}]
We say that a TLN with weight matrix $W$ and input current vector $\Vec{\theta}$ is \textbf{competitive} if $W_{ij} \leq 0$ and $W_{ii}=0$ $\forall i,j \in [n]$ with $\Vec{\theta} \geq 0$.  The TLN is further said to be \textbf{non-degenerate} if \cite{diverse}: 

$\bullet$ $\theta_i > 0$ for at least one $i \in [n]$

$\bullet$ $\det (I-W|_{\sigma}) \neq 0$ for every $\sigma \subseteq [n].$

$\bullet$ For each $\sigma \subseteq [n]$ such that $\forall i \in \sigma,  \theta_i > 0$, the corresponding Cramer's determinant is nonzero: $\det ((I-W|_{\sigma})_i ; \Vec{\theta}|_{\sigma})$

\end{ddd}  

where the notation $(A_i ; \Vec{b})$ represents the matrix with the $i$-th column removed and replaced by $\Vec{b}$.  The first non-degeneracy condition ensures that the origin is not a fixed point.  The second non-degeneracy condition makes sure that no chamber has a non-degenerate linear system.  The third non-degeneracy condition ensures that the fixed points of linear systems in adjacent chambers do not both lie on the shared wall of the chambers \cite{diverse}.

We will be considering competitive, non-degenerate TLNs of the form:

\begin{figure}[!ht]
\begin{center}
\vspace{.1in}
\includegraphics[width=2.75in]{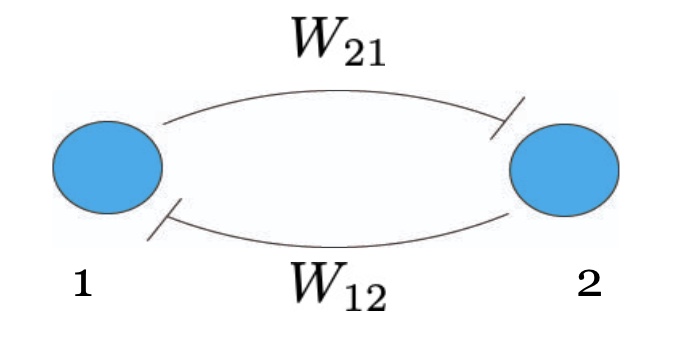}
\vspace{.1in}
\caption[Binary Competition Model]{{\bf Binary Competition Model.} Competitive model of two neural populations.}
\label{fig:bcmarc}
\end{center}
\vspace{-.2in}
\end{figure}

$$\dfrac{dx_1}{dt}=-x_1 +[W_{12}x_2+\theta_1]_+$$
$$\dfrac{dx_2}{dt}=-x_2 +[W_{21}x_1+\theta_2]_+$$

where $W_{12},W_{21}<0$ and $\theta_1,\theta_2>0$.  The network architecture is depicted in Fig~\ref{fig:bcmarc}.  The idea here is that each of the two neural populations corresponds to one choice in a binary choice decision-making task.  We refer to this as the Binary Competition Model.

The qualitative nature of a bistable model is hardly original and there exist multistable models in mathematical biology with similar dynamics (e.g. the competitive exclusion case of the Lotka-Volterra equations \cite{biomod,lv} and cases of genetic toggle switches \cite{toggle}), such models tend to have nonlinearities in their differential equations which make them more challenging to find exact solutions for and for studying the basins of attraction \cite{lvsols}.  Using the piecewise linearity of TLN equations, we will not only be able to calculate the seperatrix, but also integrate beneath it to determine the sizes of the basins of attraction relative to one another.

A challenge with basins of attraction is that they require an understanding of the global dynamics of a dynamical system and to understand how they evolve we need to be able to track those global dynamics through parameter space.  Our approach will be to appeal to \textit{combinatorial dynamics} where the strategy is to impose a combinatorial structure of the state space.  We will partition it into regions with parameter dependent boundaries and then see how trajectories beginning in a region flow into others.  We represent the regions as the vertices on a graph and draw directed edges between them if there exists a trajectory starting in one region that exits into another, creating a state transition graph.  If a state transition graph is such that each vertex has only one outgoing edge, i.e. trajectories beginning in one region flow into only one one other, then we call such a state transition graph a \textbf{trajectory graph}.

The first partitions we make will naturally be using the ReLU hyperplane arrangement $\{H_i\}_{i=1}^2$ (Fig~\ref{fig:empty}).  We have:

$$H_1: x_2=-\dfrac{\theta_1}{W_{12}}$$

$$H_2: x_1=-\dfrac{\theta_2}{W_{21}}$$

\begin{figure}[!ht]
\begin{center}
\vspace{.1in}
\includegraphics[width=4.75in]{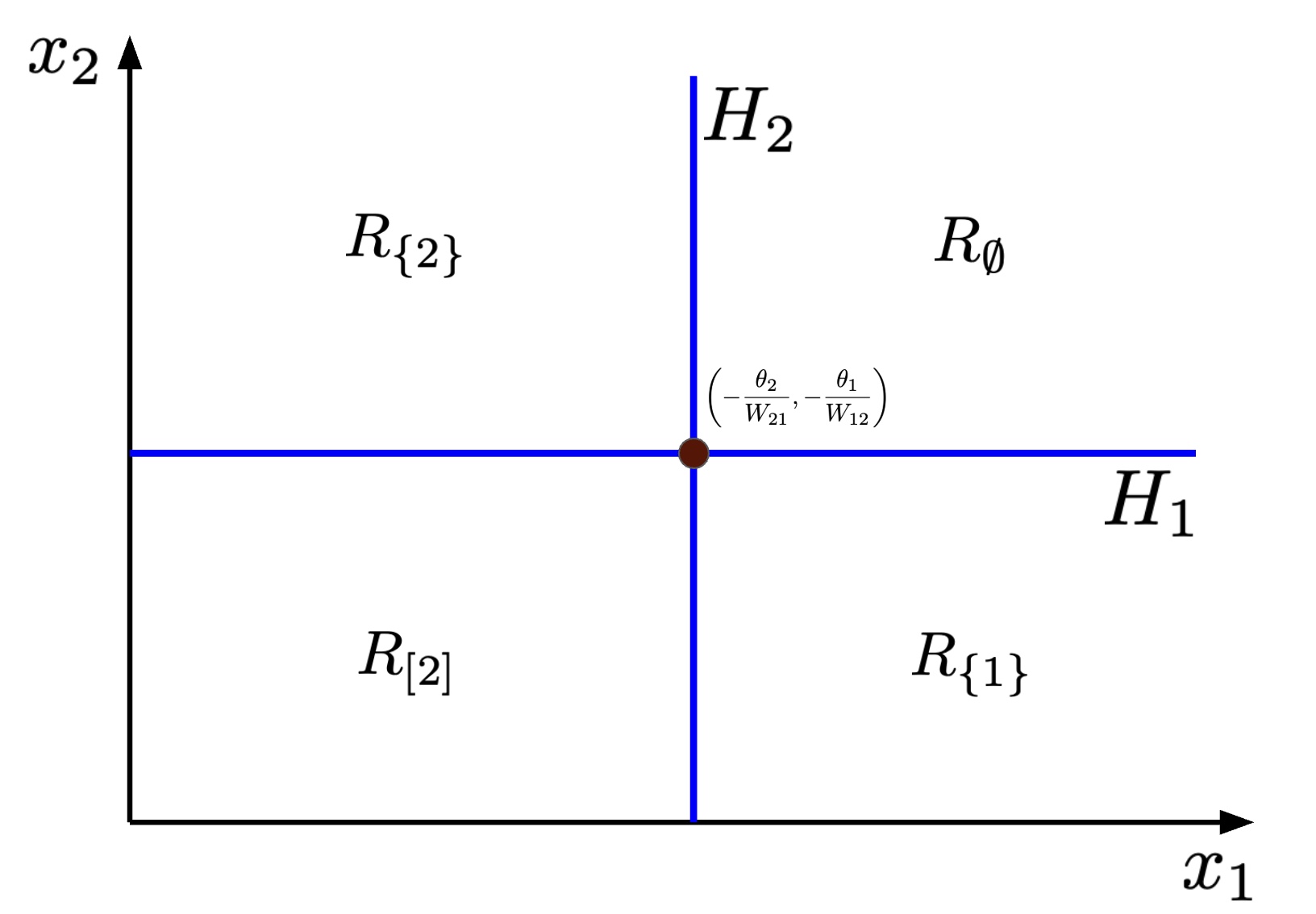}
\vspace{.1in}
\caption[ReLU Partition]{{\bf ReLU Partition.} The hyperplane arrangement $\{H_i\}_{i=1}^2$ divides the state space into regions with linear dynamics.}
\label{fig:empty}
\end{center}
\vspace{-.2in}
\end{figure}

\begin{rmk}
The intersection of the two lines $H_1$ and $H_2$ corresponds to the balanced state, $x_{bs}$, for the Binary Competition Model. 
\end{rmk}

Through standard ODE analysis we solve for the solutions of the linear systems in each of these chambers:

\begin{lem}

For the Binary Competition Model class of TLNs, the chamber by chamber linear ODE systems induced by the ReLU hyperplane partition and given in Figure~\ref{fig:iso_sol},

\begin{figure}[H]
    \begin{center}
    \includegraphics[width=4.75in]{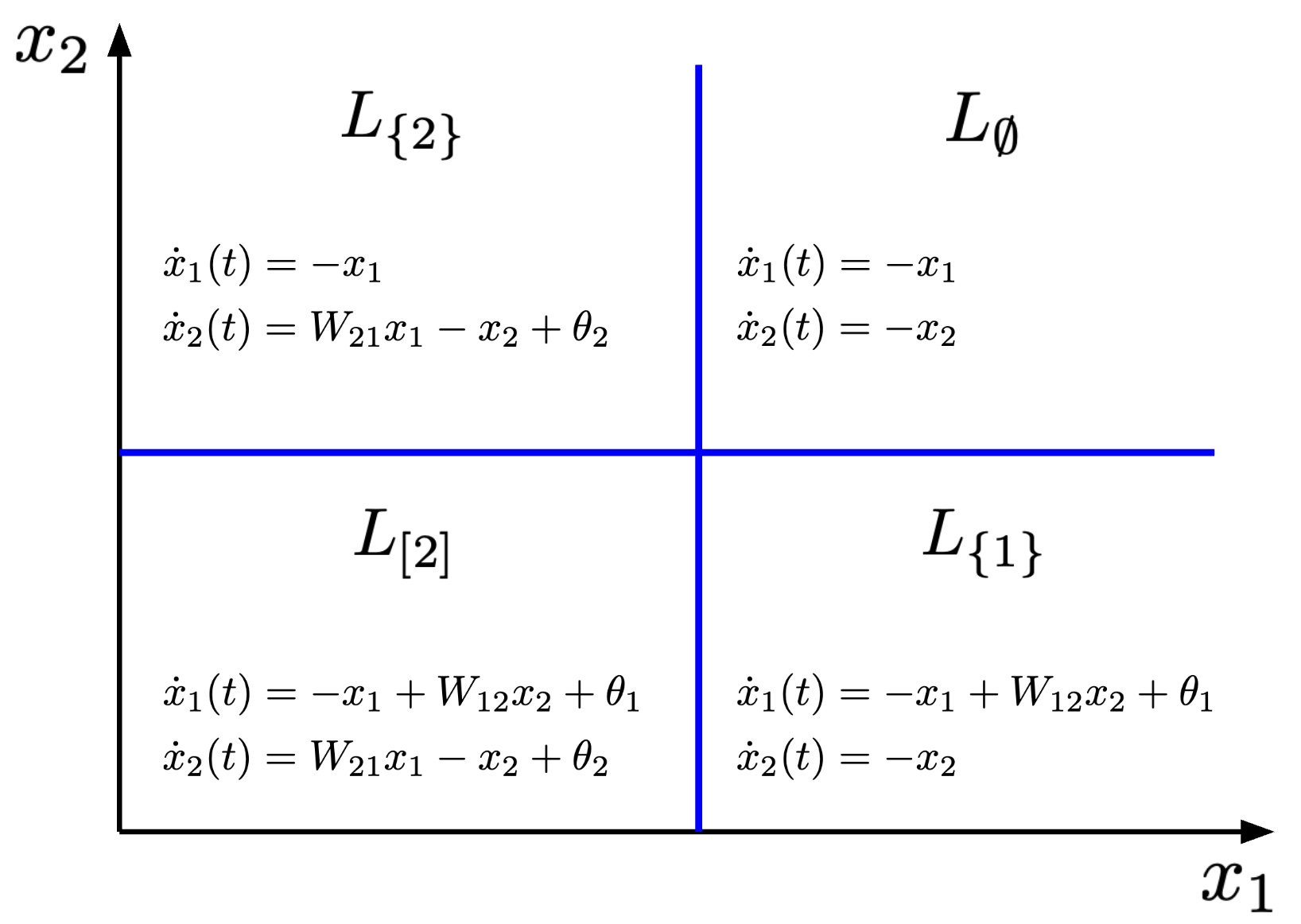}
    \end{center}
    \vspace{-0.1in}
    \caption[Linear ODE systems $L_{\sigma}$ for Binary Competition Model]{{\bf Linear ODE systems $L_{\sigma}$ for Binary Competition Model.}}
    \label{fig:iso_sol}
\end{figure}

have solutions:

\textbf{$L_{\emptyset}$:}

$\text{       }x_1(t)=x_1^0 e^{-t}$

$\text{       }x_2(t)=x_2^0 e^{-t}$

\textbf{$L_{\{1\}}$:}

$\text{       }x_1(t)=x_2^0 W_{12} t e^{-t}+(x_1^0-\theta_1)e^{-t} +\theta_1$

$\text{       }x_2(t)=x_2^0 e^{-t}$

\textbf{$L_{\{2\}}$:}

$\text{       }x_1(t)=x_1^0 e^{-t}$

$\text{       }x_2(t)=x_1^0 W_{21} t e^{-t}+(x_2^0-\theta_2)e^{-t} +\theta_2$

\textbf{$L_{\{1,2\}}$:}

$\text{       }x_1(t)=c_1\sqrt{-W_{12}}e^{(-1-\sqrt{W_{12}W_{21}})t}-c_2\sqrt{-W_{12}}e^{(-1+\sqrt{W_{12}W_{21}})t}+\dfrac{\theta_1+W_{12}\theta_2}{1+|W|}$

$\text{       }x_2(t)=c_1\sqrt{-W_{21}}e^{(-1-\sqrt{W_{12}W_{21}})t}+c_2\sqrt{-W_{21}}e^{(-1+\sqrt{W_{12}W_{21}})t}+\dfrac{W_{21}\theta_1+\theta_2}{1+|W|}$

$\text{       }c_1=\dfrac{(1+|W|)(x_1^0+x_2^0)-(\theta_1+\theta_2+W_{21}\theta_1+W_{12}\theta_2)}{(1+|W|)(\sqrt{-W_{12}}+\sqrt{-W_{21}})}$

$\text{       }c_2=\dfrac{(1+|W|)(x_2^0-x_1^0)-(\theta_2-\theta_1+W_{21}\theta_1-W_{12}\theta_2)}{(1+|W|)(\sqrt{-W_{12}}+\sqrt{-W_{21}})}$

where $|W|=-W_{12}W_{21}$ is the determinant of W.

\end{lem}

We will also introduce a technical lemma which we will use across our calculations.

\begin{lem}\label{lem:continue}
    Let $(a,b)\in \mathbbm{R}_+^2$ and define a vector field according to the linear system $L_{\{1\}}$ of the Binary Competition Model.  Then, the implicitization of the trajectory passing through $(a,b)$ is:

    $$x_1=\left(\dfrac{a-\theta_1}{b}\right)x_2 - W_{12}x_2\ln{\left(\dfrac{x_2}{b}\right)}+\theta_1.$$

    Alternatively, if the vector field were defined according to $L_{\{2\}}$, the implicitization of the alternative trajectory passing through $(a,b)$ would be:

    $$x_2=\left(\dfrac{b-\theta_2}{a}\right)x_1 - W_{21}x_1\ln{\left(\dfrac{x_1}{a}\right)}+\theta_2.$$

\end{lem}

\begin{proof}

The solutions for $L_{\{1\}}$ are of the form:

$x_1(t)=W_{12}x_2^0 t e^{-t} + (x_1^0 -\theta_1)e^{-t}+\theta_1$

$x_2(t)=x_2^0 e^{-t}$

Let $(a,b)\in \mathbbm{R}^2_+$.  For the trajectory of $L_{\{1\}}$ passing through $(a,b)$, $\exists t^*$ such that:

$a=W_{12}x_2^0t^*e^{-t^*}+(x_1^0-\theta_1)e^{-t^*}+\theta_1$

$b=x_2^0 e^{-t^*}$.

The first expression can be manipulated into the following form after multiplying on both sides by $e^{t^*}$:

$$x_1^0=(a-\theta_1)e^{t^*}-W_{12}x_2^0t^*+\theta_1.$$

Manipulating the second expression, we have $e^{t^*}=\dfrac{x_2^0}{b}$ and $t^*=\ln{\left(\dfrac{x_2^0}{b}\right)}$.  Substituting, we have:

$$x_1^0=\left(\dfrac{a-\theta_1}{b}\right)x_2^0 - W_{12}x_2^0\ln{\left(\dfrac{x_2^0}{b}\right)}+\theta_1.$$

This relation defines the set of initial conditions passing through $(a,b)$ in positive or negative time.  But this is exactly the set of points of the trajectory.  Thus we have the implicitization of the trajectory as:

$$x_1=\left(\dfrac{a-\theta_1}{b}\right)x_2 - W_{12}x_2\ln{\left(\dfrac{x_2}{b}\right)}+\theta_1.$$

Performing this same analysis for $L_{\{2\}}$, we obtain the implicitization:

$$x_2=\left(\dfrac{b-\theta_2}{a}\right)x_1 - W_{21}x_1\ln{\left(\dfrac{x_1}{a}\right)}+\theta_2.$$
\end{proof}

For convenience, we define functions for these expressions:

\begin{ddd}
    $$f_1(a,b,x):=\left(\dfrac{a-\theta_1}{b}\right)x - W_{12}x\ln{\left(\dfrac{x}{b}\right)}+\theta_1.$$

    $$f_2(a,b,x):=\left(\dfrac{b-\theta_2}{a}\right)x - W_{21}x\ln{\left(\dfrac{x}{a}\right)}+\theta_2.$$

\end{ddd}

Lemma~\ref{lem:continue} makes it easy to piece trajectories across chambers and its utility will quickly become evident.

\subsection{The Bistable Symmetric Case}

A critical aspect of our strategy for determining trajectory graphs will be nullcline analysis.  We will get a feel for the nullclines in this system by looking at a perfectly symmetric TLN with $\theta_1=\theta_2=\theta$ and $W_{12}=W_{21}=-1-\delta$ taking $\delta>0$.  Note that this corresponds to a CTLN derived from an independent set of two neurons.  This is known to be a bistable system with one attractor supported on each neuron and a saddle point supported on both of them.  The connectivity matrix has the form:

$$W=
\left[
\begin{array}{cc}
0 & -1-\delta \\
-1-\delta & 0 \\
\end{array}
\right].
$$

The next step will be to introduce the nullclines into our partition where the nullclines are given by the equations:

$$\mathcal{N}_1: x_1=[W_{12} x_2 +\theta_1]_+$$
$$\mathcal{N}_2: x_2=[W_{21} x_1 +\theta_1]_+$$

Looking at our partition, we can start labeling our chambers using a canonical labelling scheme.  For a partial partition using just chamber boundaries and nullclines associated with a network of $n$ neurons, the region R will be labelled $S_1(R) S_2(R) \hdots S_{2n}(R)$ derived from the string $H_1 \mathcal{N}_1 H_2 \mathcal{N}_2 \hdots H_n \mathcal{N}_n$ where:

$$
S_i(R)=
\begin{cases}
    0 & \text{ if } R \text{ lies above the corresponding nullcline or chamber boundary}  \\
    1 & \text{ if } R \text{ lies below the corresponding nullcline or chamber boundary}  \\
\end{cases}
$$

So, consider the case for a network of two neurons and in particular the region such that:

$$x_2>\dfrac{-\theta_1}{W_{12}}, x_1\leq\dfrac{-\theta_2}{W_{21}}$$

and

$$x_1>[W_{12}x_2+\theta_1]_+, x_2\leq[W_{21}x_1+\theta_2]_+$$

That is to say, the region lying above $H_1$ and $\mathcal{N}_1$ while lying below $H_2$ and $\mathcal{N}_2$.

The label for this region will then be:

$$\begin{array}{cccc}
     H_1 & \mathcal{N}_1 & H_2 & \mathcal{N}_2  \\
     \vdots & \vdots & \vdots & \vdots \\
     0 & 0 & 1 & 1
\end{array}$$

$$S_1 S_2 S_3 S_4 = 0011$$

Our convention will also be to treat the label as a binary representation read left to right rather than the usual right to left.

So, for the region 0011, the corresponding vertex in the graph structure will be numbered:

$$0 \cdot 2^0 + 0 \cdot 2^1 + 1 \cdot 2^2 + 1 \cdot 2^3 = 0+0+4+8=12$$

So, in the case of the bistable, symmetric TLN, we have the labelling in Figure~\ref{fig:iso_code}

\begin{figure}[H]
    \begin{center}
    \includegraphics[width=.61\textwidth, height=.44\textwidth]{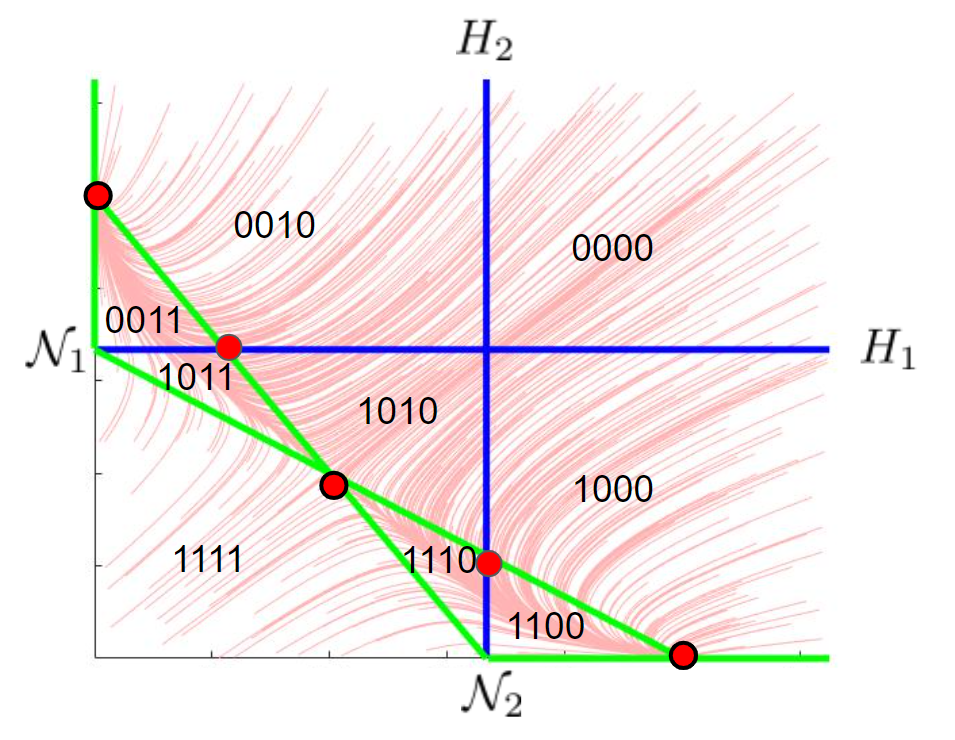}
    \end{center}
    \vspace{-0.1in}
    \caption[ReLU hyperplane and nullcline partition]{{\bf ReLU hyperplane and nullcline partition.} Labelling of $H_i/\mathcal{N}_i$ partition for two neuron independent set CTLN.}
    \label{fig:iso_code}
\end{figure}

From here we can build a state transition graph (Figure~\ref{fig:iso_code_graph}) using a nullcline analysis which tells us when $x_1$ and $x_2$ are increasing and decreasing respectively.  

\begin{figure}[H]
    \begin{center}
    \includegraphics[width=.61\textwidth, height=.44\textwidth]{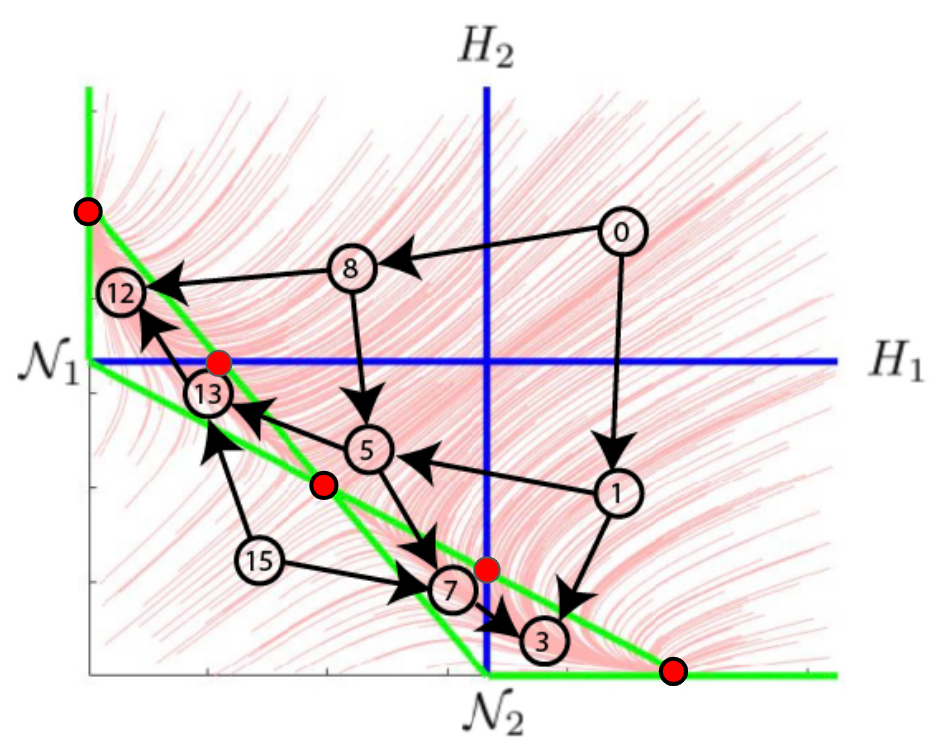}
    \end{center}
    \vspace{-0.1in}
    \caption[Canonical numbering of $H_i/\mathcal{N}_i$ partition]{{\bf Canonical numbering of $H_i/\mathcal{N}_i$ partition.} Canonical numbering scheme for partial trajectory graph of two neuron independent set CTLN}
    \label{fig:iso_code_graph}
\end{figure}

But notice that this state transition graph is not refined enough to give the basins of attraction for each attractor.  We will have to break it down further.  To obtain a fully refined trajectory graph, $R_{\{2\}}$ needs to be subdivided into the trajectories dropping into $R_{[2]}$ and those staying in the chamber with the situation similar for $R_{\{1\}}$.  That division, which corresponds to a trajectory hitting the nullcline/chamber boundary intersection must then be traced through the other chambers as needed.  Additionally, within $R_{[2]}$ and $R_{\emptyset}$ a diagonal separation is required.

\begin{prp}\label{prp:indset} 

The trajectory graph of the CTLN associated with a graph consisting of an independent set with two nodes takes the form of Figure~\ref{fig:iso_tg} where:

The Chamber Boundary equations are:

$$H_1: x_2=\dfrac{\theta}{1+\delta}$$
$$H_2: x_1=\dfrac{\theta}{1+\delta}$$

The nullclines are:

$$\mathcal{N}_1: x_1=[(-1-\delta)x_2+\theta]_+$$
$$\mathcal{N}_2: x_2=[(-1-\delta)x_1+\theta]_+$$

And the refining curves which cross through the nullcline/chamber boundary intersection are given by:

$$\mathcal{M}_1: x_1=
\begin{cases}
f_1 \left( \dfrac{\theta}{1+\delta}, \dfrac{\theta \delta}{(1+\delta)^2}  ,x_2 \right) & \text{if } \dfrac{\theta \delta}{(1+\delta)^2} \leq x_2 \leq \dfrac{\theta}{1+\delta} \\
\dfrac{(1+\delta)}{\theta} f_1\left(  \dfrac{\theta}{1+\delta}, \dfrac{\theta \delta}{(1+\delta)^2}  , \dfrac{\theta}{1+\delta} \right)x_2 & \text{ if } \dfrac{\theta}{1+\delta}< x_2
\end{cases}
$$

$$\mathcal{M}_2: x_2=
\begin{cases}
f_2 \left( \dfrac{\theta \delta}{(1+\delta)^2}, \dfrac{\theta}{1+\delta}  ,x_1 \right) & \text{if } \dfrac{\theta \delta}{(1+\delta)^2} \leq x_1 \leq \dfrac{\theta}{1+\delta} \\
\dfrac{(1+\delta)}{\theta} f_2\left(   \dfrac{\theta \delta}{(1+\delta)^2}, \dfrac{\theta}{1+\delta}  , \dfrac{\theta}{1+\delta} \right)x_1 & \text{ if } \dfrac{\theta}{1+\delta}< x_1
\end{cases}
$$

\begin{figure}[H]
    \begin{center}
    \includegraphics[width=.7\textwidth]{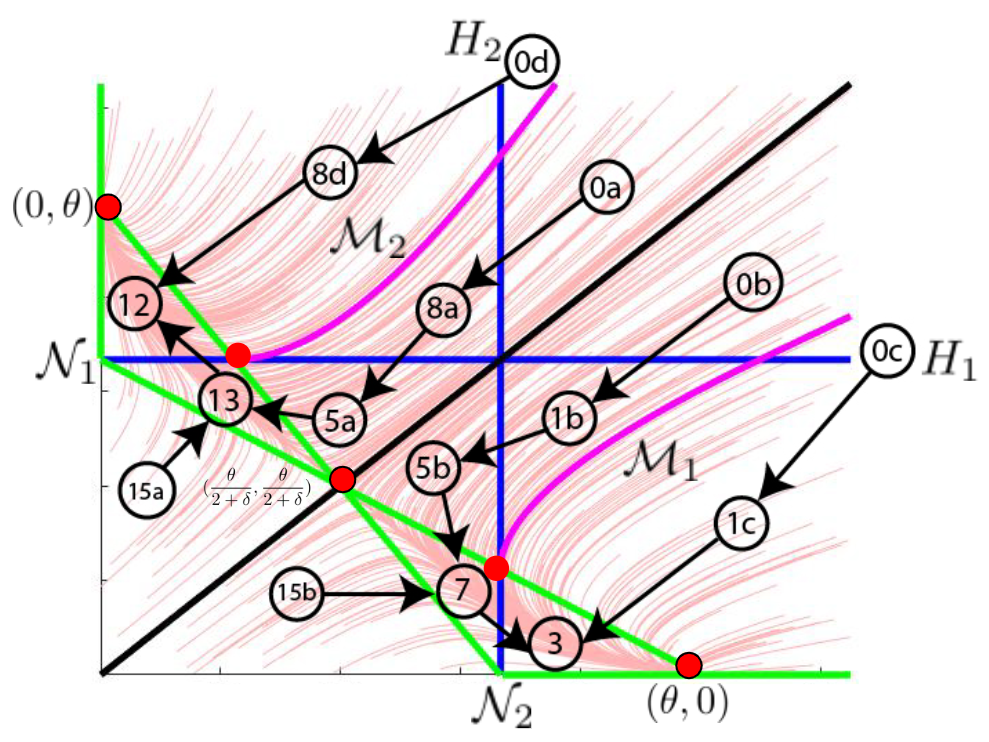}
    \end{center}
    \vspace{-.25in}
    \caption[Trajectory graph for symmetric bistable TLN]{\textbf{Trajectory graph for symmetric bistable TLN.} The shown curves refine the state space into chambers which yield a trajectory graph.  The light red curves in the background are simulated trajectories, confirming the accuracy of our trajectory graph.}
    \label{fig:iso_tg}
\end{figure}

\end{prp}

\begin{proof}

In the chamber $R_{\{2\}}$ the solution to the ODE system is:

$$x_1(t) = x_1^0 e^{-t}$$
$$x_2(t) = (-1-\delta)x_1^0 t e^{-t} + \theta + (x_2^0 - \theta) e^{-t}$$

Since $x_1$ is strictly decreasing, trajectories in this chamber cannot enter $R_{\emptyset}$.

To see which trajectories would enter chamber $R_{[2]}$, we take the point where the nullcline $\mathcal{N}_2$ intersects $H_1$ as this separates the $H_1$ wall of $R_{\{2\}}$ into where trajectories flow beneath it and above it.  This point is $\left( \dfrac{\theta \delta}{(1+\delta)^2}, \dfrac{\theta}{1+\delta} \right)$ and by applying Lemma~\ref{lem:continue} we obtain $\mathcal{M}_2$.

This boundary, restricted to values of $x_1^0$ where $t^*$ (as defined in the proof of Lemma~\ref{lem:continue}) is positive, is the magenta boundary which separates \textbf{8a}.  Thus, $\textbf{8a} \longrightarrow \textbf{5a}$.  Additionally, since \textbf{8d} is above $\mathcal{M}_2$, its trajectories remain in $R_{\{2\}}$,  So trajectories beginning there will go to the fixed point $(0,\theta)$.  To see that they go to the fixed point via \textbf{12} simply resolve $x_2(t^*)<\theta$ and notice that the resulting inequality is trivial.  So, $\textbf{8d} \longrightarrow \textbf{12}$.

Without loss of generality, we can extend these results to $R_{\{1\}}$ as well to conclude $\textbf{1b} \longrightarrow \textbf{5b}$ and $\textbf{1c} \longrightarrow \textbf{3}$.

Now consider chamber $R_{[2]}$.  The solution to the ODE system in this chamber is:

$$
x_1(t)=-\dfrac{1}{2}(x_2^0 - x_1^0)e^{\delta t}+\dfrac{1}{2}\left(x_2^0 + x_1^0 -\dfrac{2\theta}{2+\delta}\right)e^{-(2+\delta)t}+\dfrac{\theta}{2+\delta}
$$
$$
x_2(t)=\dfrac{1}{2}(x_2^0 - x_1^0)e^{\delta t}+\dfrac{1}{2}\left(x_2^0 + x_1^0 -\dfrac{2\theta}{2+\delta}\right)e^{-(2+\delta)t}+\dfrac{\theta}{2+\delta}
$$

Since $e^{\delta t}$ is strictly increasing to $\infty$, this indicates that the trajectories cannot remain in this chamber.  Additionally $x_2(t) > x_1(t)$ precisely when $\dfrac{1}{2}(x_2^0 - x_1^0)e^{\delta t}>-\dfrac{1}{2}(x_2^0 - x_1^0)e^{\delta t}$.  This holds true when $x_2^0 > x_1^0$.  So if a trajectory begins above the diagonal it must remain above the diagonal, and if a trajectory begins below the diagonal it must remain below the diagonal.  Thus, trajectories cannot cross the diagonal as well.

For \textbf{5a}, this means that trajectories cannot move into \textbf{5b} and since the region lies above both nullclines the trajectories must decrease in $x_1$ and $x_2$ into \textbf{13}.  Thus $\textbf{5a} \longrightarrow \textbf{13}$.  By similar reasoning, $\textbf{5b} \longrightarrow \textbf{7}$.

Since \textbf{15a} lies below both nullclines, it must increase into \textbf{13}.  So, $\textbf{15a} \longrightarrow \textbf{13}$ and without loss of generality $\textbf{15b} \longrightarrow \textbf{7}$.

Now, for \textbf{13}, to leave from above through the $x_2$-nullcline it would have to be through a point where $\dfrac{dx_1}{dt}$ is positive, which is impossible since \textbf{13} lies abov\textbf{12}e the $x_1$-nullcline.  Similarly, to leave from below through the $x_1$-nullcline it must be through a point where $\dfrac{dx_2}{dt}$ is negative, which is again impossible as the \textbf{13} lies below the $x_2$-nullcline.  Since trajectories cannot remain within the chamber, $\textbf{13} \longrightarrow \textbf{12}$ and again without loss of generality $\textbf{7} \longrightarrow \textbf{3}$.

We finally turn our attention to $R_{\emptyset}$.  The solution in this chamber is 

$$x_1(t) = x_1^0 e^{-t}$$
$$x_2(t) = x_2^0 e^{-t}$$

Then, $x_2 = \dfrac{x_2^0}{x_1^0} x_1$ which is linear and in particular is above the diagonal if $x_2^0>x_1^0$ and below if $x_2^0<x_1^0$.

So, if $x_1^0 > x_2^0$, then the trajectory hits $H_1$ first and enters $R_{\{1\}}$.  If the other way around, it enters $R_{\{2\}}$.

The only remaining question is which trajectories beginning in the upper diagonal enter \textbf{8a} and which enter \textbf{8d}.  

Since $\mathcal{M}_2$ intersects $H_2$ at $\left(\dfrac{\theta}{1+\delta}, f_2\left( \dfrac{\theta \delta}{(1+\delta)^2},\dfrac{\theta}{1+\delta},\dfrac{\theta}{1+\delta} \right) \right)$ and in $R_{\emptyset}$ $x_2 = \dfrac{x_2^0}{x_1^0} x_1$, we can extend $\mathcal{M}_2$ into $R_{\emptyset}$ accordingly.

So, $\textbf{0d} \longrightarrow \textbf{8d}$ and $\textbf{0a} \longrightarrow \textbf{8a}$.  By similar arguments, we also conclude that $\textbf{0c} \longrightarrow \textbf{1c}$ and $\textbf{0b} \longrightarrow \textbf{1b}$.

\end{proof}

These trajectory graphs are a combinatorial representation of the dynamics and in particular the basins of attraction.  Given an initial condition $(x_1^0,x_2^0)$, by finding which region it falls within and tracing the path through the trajectory graph to the end region, we can determine to which point attractor it converges.  

For the independent set, the basin for the upper-left fixed point is the region above the diagonal while the basin for the bottom-right fixed point is the region below the diagonal.  For $x_1^0=x_2^0$, you have convergence to the unstable fixed point on the diagonal.  The diagonal is the stable manifold separating the basins of attraction for the attractors.

\subsection{Trajectory Graphs of the Binary Competition Model}

The Binary Competition Model is a patchwork of four linear systems: $L_{\emptyset}, L_{1}, L_{2}$, and $L_{12}$.  The $R_{\emptyset}$ chamber will never have a fixed point as the fixed point of $L_{\emptyset}$ is $(0,0)$. This means the TLNs in this class can have up to three fixed points.  The $H_i/\mathcal{N}_i$ arrangements are closely related to the fixed points of the TLN.

%Analysis Starts Here for BCM

\begin{lem}\label{lem:bcmatt}
For the Binary Competition Model, assume without loss of generality that $\theta_2 \geq \theta_1.$  Then we have the bifurcation diagram on fixed point supports depicted in Figure~\ref{fig:att_bif}.

\begin{figure}[H]
    \begin{center}
    \includegraphics[width=.61\textwidth, height=.44\textwidth]{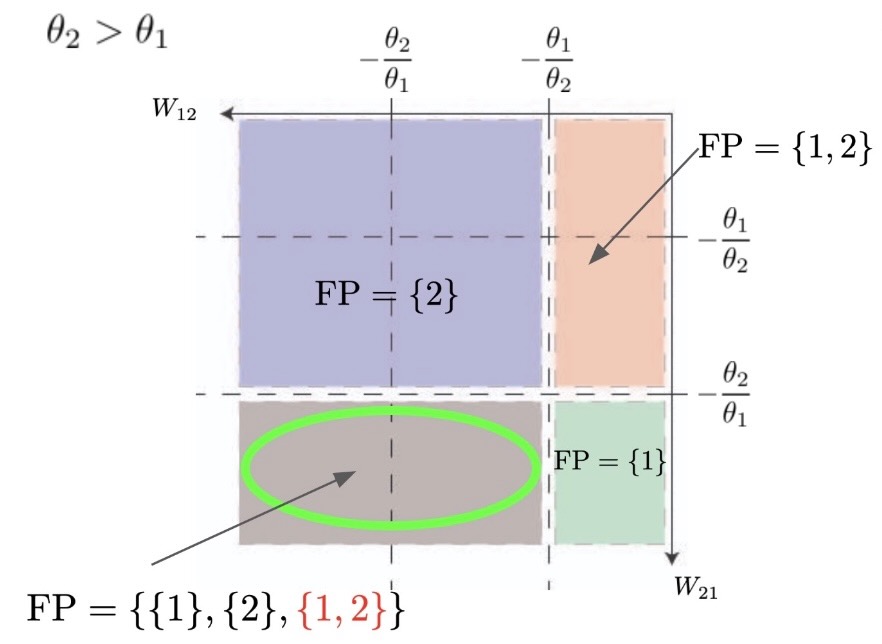}
    \end{center}
    \vspace{-0.1in}
    \caption[Bifurcations on fixed point supports of the Binary Competition Model]{{\bf Bifurcations on fixed point supports of the Binary Competition Model.}. This bifurcation diagram depicts the set of fixed point supports in the various regions of the parameter space.}
    \label{fig:att_bif}
\end{figure}
\end{lem}

\begin{proof}
From standard fixed point analysis we know that the fixed points of linear systems $L_{1}$, $L_{2}$, and $L_{1,2}$ are as follows:

$L_{1}: (\theta_1,0)$

$L_{2}: (0,\theta_2)$

$L_{1,2}: \left( \dfrac{\theta_1 + W_{12}\theta_2}{1-W_{12}W_{21}} , \dfrac{W_{21}\theta_1 + \theta_2}{1-W_{12}W_{21}} \right)$

Each of these will be fixed points of the overall TLN if they lie in the chambers $R_{\{1\}}$, $R_{\{2\}}$, and $R_{[2]}$ respectively.  These chambers are generated by the hyperplane arrangement:

$H_1: W_{12}x_2 + \theta_1 = 0 \implies x_2=-\dfrac{\theta_1}{W_{12}}$

$H_2: W_{21}x_1 + \theta_2 = 0 \implies x_1=-\dfrac{\theta_2}{W_{21}}.$

Now we consider each case separately:

\textbf{Case 1:} $R_{\{1\}}$

In this case, we would need the $L_1$ fixed point to lie inside $H_1$ and outside $H_2$.  Being inside $H_1$ is trivially met as $0<-\dfrac{\theta_1}{W_{12}}$.  Alternatively, to lie outside $H_2$ requires $\theta_1 > -\dfrac{\theta_2}{W_{21}}$.  Bearing in mind that $W_{21}$ is negative, this rearranges to $W_{21}<-\dfrac{\theta_2}{\theta_1}$.  So, we conclude that the $R_{\{1\}}$ fixed point exists when $W_{21}<-\dfrac{\theta_2}{\theta_1}$.

\textbf{Case 2:} $R_{\{2\}}$

Applying the arguments from Case 1 without loss of generality, we conclude that the $R_{\{2\}}$ fixed point exists when $W_{12}<-\dfrac{\theta_1}{\theta_2}$.

\textbf{Case 3:} $R_{[2]}$

For $R_{[2]}$ to have a fixed point, the $L_{1,2}$ fixed point must lie within both  $H_1$ and $H_2$.  So, we must have:

$$\dfrac{\theta_1 + W_{12}\theta_2}{1-W_{12}W_{21}}<-\dfrac{\theta_2}{W_{21}} \text{ and } \dfrac{W_{21}\theta_1 + \theta_2}{1-W_{12}W_{21}}<-\dfrac{\theta_1}{W_{12}}.$$

For the first condition, we can rearrange as follows:

$$ \dfrac{\theta_1 + W_{12}\theta_2}{1-W_{12}W_{21}}<-\dfrac{\theta_2}{W_{21}} \implies \dfrac{\theta_1 + W_{12}\theta_2}{1-W_{12}W_{21}} + \dfrac{\theta_2}{W_{21}}<0 \implies$$

$$\dfrac{W_{21}\theta_1+W_{12}W_{21}\theta_2+\theta_2-W_{12}W_{21}\theta_2}{W_{21}(1-W_{12}W_{21})}<0 \implies \dfrac{W_{21}\theta_1+\theta_2}{1-W_{12}W_{21}}>0.$$

Through similar manipulation the second condition becomes:

$$\dfrac{W_{21}\theta_1 + \theta_2}{1-W_{12}W_{21}}<-\dfrac{\theta_1}{W_{12}} \implies  \dfrac{\theta_1+W_{12}\theta_2}{1-W_{12}W_{21}}>0.$$

Then, it is clear that we have two regions of the parameter space where this dual support fixed point occurs.  If $W_{21} < \dfrac{1}{W_{12}}$ (i.e. $1-W_{12}W_{21} > 0$), then the fixed point exists when $W_{21}>-\dfrac{\theta_2}{\theta_1}$ and $W_{12}>-\dfrac{\theta_1}{\theta_2}$.  Alternatively, if $W_{21} > \dfrac{1}{W_{12}}$, the fixed point exists when $W_{21}<-\dfrac{\theta_2}{\theta_1}$ and $W_{12}<-\dfrac{\theta_1}{\theta_2}$.

Our last consideration is the stability of the $R_{[2]}$ fixed point.  We find the eigenvalues of the $-I+W$ matrix.

$$\operatorname{det}\left(\left[
\begin{array}{cc}
-1-\lambda & W_{12} \\
W_{21} & -1-\lambda
\end{array}
\right]\right)=\lambda^2 +2\lambda + 1 - W_{12}W_{21}=0.$$

The characteristic polynomial has roots $\lambda_1=-1-\sqrt{W_{12}W_{21}}$ and $\lambda_2=-1+\sqrt{W_{12}W_{21}}$.  The eigenvalue $\lambda_1$ is clearly negative, but the sign of $\lambda_2$ is parameter dependent:

$$\lambda_2 > 0 \iff \sqrt{W_{12}W_{21}} > 1.$$ 

Squaring both sides and rearranging, we have:

$$\sqrt{W_{12}W_{21}} > 1 \iff W_{21}>\dfrac{1}{W_{12}}.$$

Thus, we conclude that in the $W_{21}>-\dfrac{\theta_2}{\theta_1}$ and $W_{12}>-\dfrac{\theta_1}{\theta_2}$ regime, the $R_{[2]}$ fixed point is stable whereas in the $W_{21}<-\dfrac{\theta_2}{\theta_1}$ and $W_{12}<-\dfrac{\theta_1}{\theta_2}$ regime it is a saddle point.

\end{proof}

\begin{corr}\label{corr:nullcd}
For the Binary Competition Model, assume without loss of generality that $\theta_2 \geq \theta_1.$  Then we have the bifurcation diagram on $H_i/\mathcal{N}_i$ arrangements depicted in Figure~\ref{fig:NCBCM}.

\begin{figure}[H]
    \begin{center}
    \includegraphics[width=\textwidth, height=.55\textwidth]{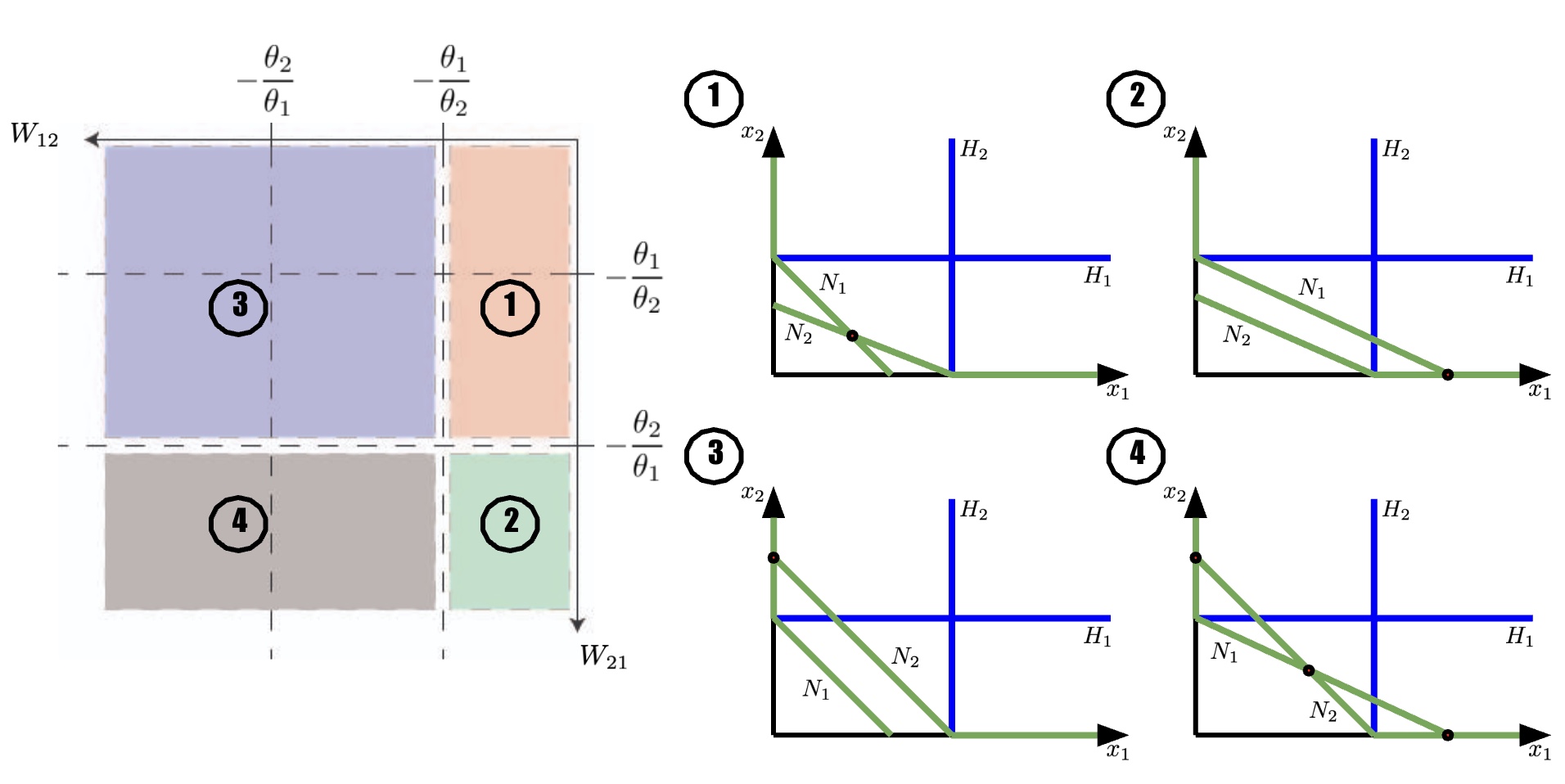}
    \end{center}
    \vspace{-0.1in}
    \caption[Bifurcation of $H_i/\mathcal{N}_i$ arrangement for Binary Competition Model]{{\bf Bifurcation of $H_i/\mathcal{N}_i$ arrangement for Binary Competition Model} Within marked regions of the parameter space, the $H_i/\mathcal{N}_i$ arrangement is as indicated.  By convention, $\theta_2 \geq \theta_1$.}
    \label{fig:NCBCM}
\end{figure}

\end{corr}

%Stab Manifold Lemma

From this we can see that the key difference in trajectory graphs between the CTLN cases and the more general Binary Competition Model is the placement of the positively oriented stable eigenspace of the $L_{1,2}$ fixed point, where it exists.  In the symmetric case it was the diagonal, but now it can be more freely moved.  Our key consideration is whether it continues from $R_{[2]}$ into $R_{\{1\}}$ or $R_{\{2\}}$.

%%%%%%%%%%%%%%%%%%%%%%%%%%%%%%%%%%%%%%%%%%%%%%

\begin{lem}\label{lem:bcmbslem}
For the Binary Competition Model, with parameters in Zone 4 of Fig~\ref{fig:NCBCM}, if $W_{21}<\left( \dfrac{\theta_2}{\theta_1} \right)^2 W_{12}$, then the stable manifold of the saddle point in $R_{[2]}$ intersects the $H_1$ wall of the chamber.  If $W_{21} > \left( \dfrac{\theta_2}{\theta_1} \right)^2 W_{12}$, then it intersects the $H_2$ wall of the chamber.

For parameters in Zone 1 of Fig~\ref{fig:NCBCM}, if $W_{21}>\left( \dfrac{\theta_2}{\theta_1} \right)^2 W_{12}$, then the stable manifold of the saddle point in $R_{[2]}$ intersects the $H_1$ wall of the chamber.  If $W_{21} < \left( \dfrac{\theta_2}{\theta_1} \right)^2 W_{12}$, then it intersects the $H_2$ wall of the chamber.
\end{lem}

\begin{proof}

For the linear system $L_{[2]}$, the stable manifold corresponds to the eigenspace created by the eigenvector $\left[ \begin{array}{c} \sqrt{-W_{12}} \\ \sqrt{-W_{21}} \end{array} \right]$.  Then, the stable manifold in $R_{[2]}$ is a line with slope $\sqrt{\dfrac{W_{21}}{W_{12}}}$ passing through the fixed point $x^*=\left( \dfrac{\theta_1 + W_{12}\theta_2}{1+|W|} , \dfrac{W_{21}\theta_1 + \theta_2}{1+|W|} \right)$.

Now consider the slope of the line passing through both $x^*$ and the intersection of the ReLU lines (the corner of the $R_{[2]}$ chamber), the point $x_{bs}=\left( -\dfrac{\theta_2}{W_{21}} , -\dfrac{\theta_1}{W_{12}} \right)$.  After some calculation we find that:

$$\dfrac{(x_{bs})_2-x_2^*}{(x_{bs})_1 - x_1^*}=\dfrac{W_{21}(\theta_1+W_{12}\theta_2 )}{W_{12}(W_{21}\theta_1+\theta_2)}.$$

Then, if the slope of the stable manifold is less than this slope, the stable manifold intersects the $H_2$ wall of $R_{[2]}$ and continues into $R_{\{1\}}$.  That is to say, it does so if and only if:

$$\sqrt{\dfrac{W_{21}}{W_{12}}} < \dfrac{W_{21}(\theta_1+W_{12}\theta_2 )}{W_{12}(W_{21}\theta_1+\theta_2)} \implies \sqrt{\dfrac{W_{12}}{W_{21}}} < \dfrac{\theta_1+W_{12}\theta_2 }{W_{21}\theta_1+\theta_2} .$$

Alternatively, the stable manifold continues into $R_{\{2\}}$ if and only if:

$$\sqrt{\dfrac{W_{12}}{W_{21}}} > \dfrac{\theta_1+W_{12}\theta_2 }{W_{21}\theta_1+\theta_2} .$$

Consider the first inequality.  Since we are in Zone 4 of Fig~\ref{fig:NCBCM}, we have $W_{21}\theta_1 + \theta_2 <0$.  Thus,

$$\sqrt{\dfrac{W_{12}}{W_{21}}} < \dfrac{\theta_1+W_{12}\theta_2 }{W_{21}\theta_1+\theta_2} \implies \sqrt{-W_{12}}(W_{21}\theta_1 +\theta_2) > \sqrt{-W_{21}}(\theta_1 + W_{12}\theta_2).$$

For the time being, we use the more compact notation $\alpha=\sqrt{-W_{12}}$ and $\beta=\sqrt{-W_{21}}$ and we note that $\alpha, \beta > 0$.  We can then rewrite the inequality as:

$$\alpha (-\beta^2 \theta_1 +\theta_2) > \beta (\theta_1 - \alpha^2 \theta_2) \implies (\beta \theta_2)\alpha^2 + (\theta_2 - \beta^2\theta_1)\alpha - \beta\theta_1 > 0.$$

This is a quadratic inequality in $\alpha$ with a positive leading coefficient.  So, the inequality is satisfied for $\alpha < \alpha_1$ and $\alpha> \alpha_2$ where $\alpha_1,\alpha_2$ are the solutions of $(\beta \theta_2)\alpha^2 + (\theta_2 - \beta^2\theta_1)\alpha - \beta\theta_1 = 0$ such that $\alpha_1 \leq \alpha_2.$

Solving the quadratic, we obtain $\alpha_1=-\beta^{-1}$ and $\alpha_2=\dfrac{\theta_1}{\theta_2}\beta$.  As $\alpha>0$, the inequality $\alpha < -\beta^{-1}$ yields no solutions.  That leaves $\alpha > \dfrac{\theta_1}{\theta_2}\beta$.  This is equivalent to:

$$\sqrt{-W_{12}}>\dfrac{\theta_1}{\theta_2}\sqrt{-W_{21}}.$$

Squaring both sides and rearranging, we obtain:

$$W_{21}>\left( \dfrac{\theta_2}{\theta_1}\right)^2 W_{12}.$$

This is the condition for a TLN with parameters in Zone 4 of Fig~\ref{fig:NCBCM} to have the stable manifold of the $R_{[2]}$ fixed point intersect the $H_2$ wall of the chamber and continue into $R_{\{1\}}$.

Similarly, taking the inequality in the other direction, we get the condition to intersect the $H_1$ wall.

Using the same process, we get the conditions for Zone 1 of Fig~\ref{fig:NCBCM} as well.
\end{proof}

%%%%%%%%%%%%%%%%%%%%%%%%%%%%%%%%%%%%%%%%%%%%%%%%%

Based on this we can see that the bistable regime can yield two possible trajectory graphs.

\begin{prp}\label{prp:bistabtraj}
For the Binary Competition Model, assume without loss of generality that $\theta_2 \geq \theta_1$ and that the conditions $W_{12}<-\dfrac{\theta_1}{\theta_2}$ and $W_{21}<-\dfrac{\theta_2}{\theta_1}$. 

If $W_{21}>\left(\dfrac{\theta_2}{\theta_1}\right)^2 W_{12}$, then the TLN has the trajectory graph in Figure~\ref{fig:bistabt}A.

If $W_{21}<\left(\dfrac{\theta_2}{\theta_1}\right)^2 W_{12}$, then it has the trajectory graph in Figure~\ref{fig:bistabt}B.

\begin{figure}[!ht]
\begin{center}
\vspace{.1in}
\includegraphics[width=6.25in]{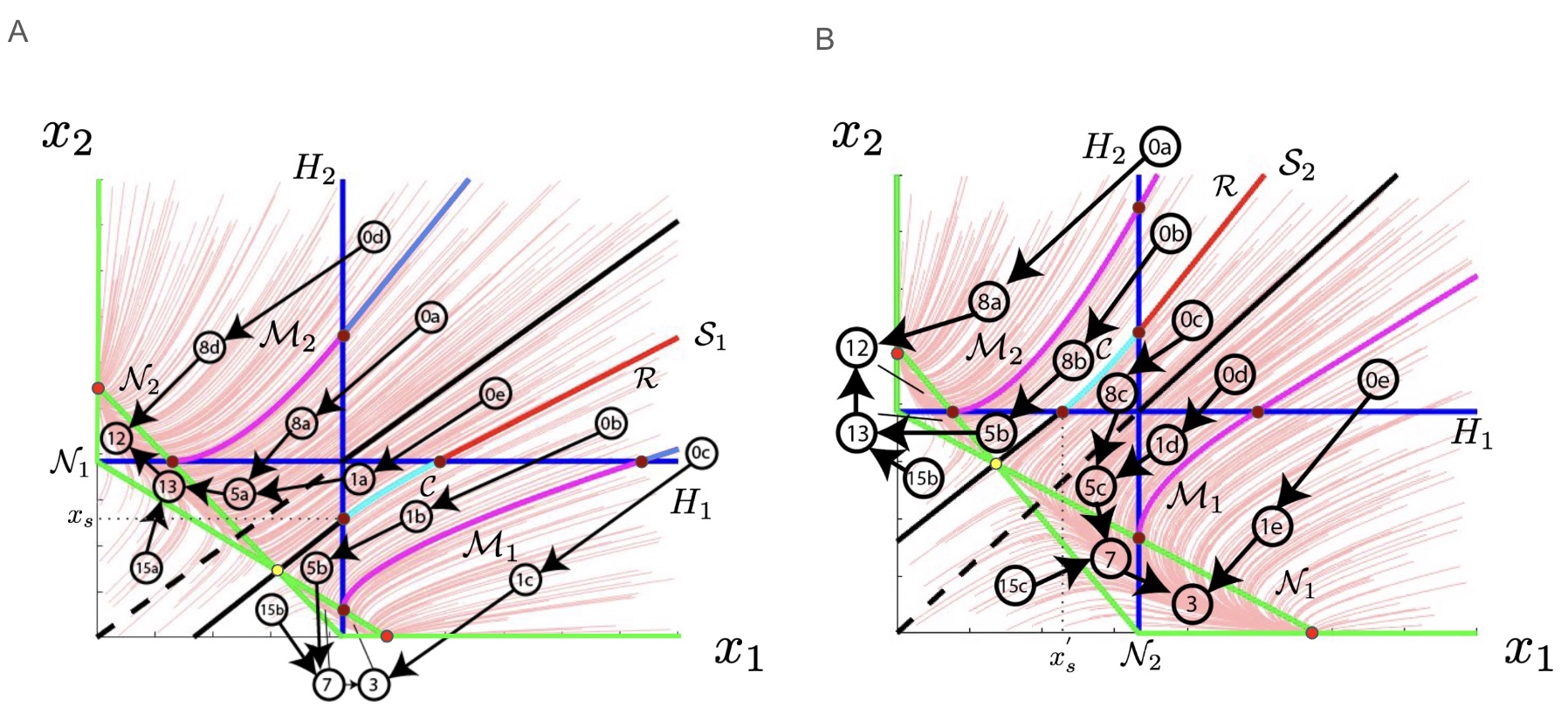}
\vspace{.1in}
\caption[Trajectory graphs under bistable parameter regime]{{\bf Trajectory graphs under bistable parameter regime.} (A) Trajectory graph if stable manifold crosses $R_{\{1\}}$.  (B) Trajectory graph if stable manifold crosses $R_{\{2\}}$.  For clarity, stable fixed points are marked in red, the saddle point in yellow, other points of interest in brown.}
\label{fig:bistabt}
\end{center}
\vspace{-.2in}
\end{figure}

where, for $m=\sqrt{\dfrac{W_{12}}{W_{21}}}$, the refining curves are:

$\mathcal{M}_{(i)}$: Initial conditions for trajectories passing through $H_i/\mathcal{N}_i$ intersections in positive time (\textcolor{magenta}{Magenta})

$$\mathcal{M}_1: x_1=
\begin{cases}
f_1 \left( -\dfrac{\theta_2}{W_{21}}, \dfrac{W_{21}\theta_1+\theta_2}{|W|}  ,x_2 \right) & \text{if } \dfrac{W_{21}\theta_1+\theta_2}{|W|} \leq x_2 \leq -\dfrac{\theta_1}{W_{12}} \\

-\dfrac{W_{12}}{\theta_1} f_1\left(  -\dfrac{\theta_2}{W_{21}}, \dfrac{W_{21}\theta_1+\theta_2}{|W|}  , -\dfrac{\theta_1}{W_{12}} \right)x_2 & \text{ if } -\dfrac{\theta_1}{W_{12}}< x_2
\end{cases}
$$

$$\mathcal{M}_2: x_2=
\begin{cases}
f_2 \left( \dfrac{\theta_1+W_{12}\theta_2}{|W|}, -\dfrac{\theta_1}{W_{12}}  ,x_1 \right) & \text{if } \dfrac{\theta_1+W_{12}\theta_2}{|W|} \leq x_1 \leq -\dfrac{\theta_2}{W_{21}} \\

-\dfrac{W_{21}}{\theta_2} f_2\left(   \dfrac{\theta_1+W_{12}\theta_2}{|W|}, -\dfrac{\theta_1}{W_{12}}  , -\dfrac{\theta_2}{W_{21}} \right)x_1 & \text{ if } -\dfrac{\theta_2}{W_{21}}< x_1
\end{cases}
$$

$\mathcal{S}_{(i)}$: Chamber by chamber breakdown of the positively oriented stable manifold $\mathcal{S}_1/\mathcal{S}_2$ for the $R_{[2]}$ fixed point (Black, \textcolor{cyan}{Cyan}, and \textcolor{red}{Red})

$$\mathcal{S}_1: x_1=\begin{cases}
mx_2+\frac{(1-mW_{21})\theta_1 + (W_{12} - m)\theta_2}{1+|W|} & \text{if } 0\leq x_2 \leq x_s \\
f_1 \left(-\frac{\theta_2}{W_{21}},x_s,x_2 \right) & \text{if }  x_s< x_2 \leq -\frac{\theta_1}{W_{12}} \\
-\frac{W_{12}}{\theta_1}f_1 \left(-\frac{\theta_2}{W_{21}},x_s, -\frac{\theta_1}{W_{12}}  \right)x_2 &  \text{if }  -\frac{\theta_1}{W_{12}} < x_2
\end{cases}
$$

$$\mathcal{S}_2: x_2=\begin{cases}
m^{-1}x_1+\frac{(1-m^{-1}W_{12})\theta_2 + (W_{21} - m^{-1})\theta_1}{1+|W|} & \text{if } 0\leq x_1 \leq x'_s \\
f_2 \left(x'_s,-\frac{\theta_1}{W_{12}},x_1 \right) & \text{if }  x'_s< x_1 \leq -\frac{\theta_2}{W_{21}} \\
-\frac{W_{21}}{\theta_2}f_2 \left(x'_s,-\frac{\theta_1}{W_{12}}, -\frac{\theta_2}{W_{21}}  \right)x_1 &  \text{if }  -\frac{\theta_2}{W_{21}} < x_1
\end{cases}
$$

where $x_s=\dfrac{(W_{21} -m^{-1})(W_{21}\theta_1 +\theta_2)}{W_{21}(1+|W|)}$ and $x'_s=\dfrac{(W_{12} -m)(W_{12}\theta_2 +\theta_1)}{W_{12}(1+|W|)}$.
\end{prp}

\begin{proof}
These cases are very similar to that of Proposition~\ref{prp:indset} and most of the arguments regarding nullclines carry over without loss of generality.  We now find the $H_1/\mathcal{N}_2$ and $H_2/\mathcal{N}_1$ intersections to be $\left( \dfrac{\theta_1+W_{12}\theta_2}{|W|}, -\dfrac{\theta_1}{W_{12}} \right)$ and $\left( -\dfrac{\theta_2}{W_{21}}, \dfrac{W_{21}\theta_1+\theta_2}{|W|} \right)$ respectively.  We then analogously find the curves $\mathcal{M}_{1,2}$ using Lemma~\ref{lem:continue}.

The main difference from the previous case is the stable manifold.  In the earlier case, the diagonal was the stable manifold, but by adjusting the parameters it is now free to move around.  We begin with the $R_{[2]}$ chamber where we have the saddle point $x^*=\left( \dfrac{\theta_1 + W_{12}\theta_2}{1+|W|} , \dfrac{W_{21}\theta_1 + \theta_2}{1+|W|} \right)$.  The eigenvector associated with the negative eigenvalue of $W$ is $\left[ \begin{array}{c} \sqrt{-W_{12}} \\ \sqrt{-W_{21}} \end{array} \right]$.  The stable manifold according to the $L_{[2]}$ system will then be of the form:

$$x_1=mx_2+b$$
 
where $m=\sqrt{\dfrac{W_{12}}{W_{21}}}$ and this line passes through $x^*$.  So, $b=x_1^*-mx_2^*$.  Then, we have:

$$x_1=mx_2+\dfrac{(1-mW_{21})\theta_1 + (W_{12} - m)\theta_2}{1+|W|}.$$

We can also rewrite this as:

$$x_2=m^{-1}x_1+ \dfrac{(1-m^{-1}W_{12})\theta_2 + (W_{21} - m^{-1})\theta_1}{1+|W|}.$$

Applying Lemma~\ref{lem:bcmbslem}, we determine whether to continue it into $R_{\{1\}}$ or $R_{\{2\}}$.  Accordingly we apply Lemma~\ref{lem:continue} to further trace the stable manifold into the appropriate chamber.  

\textbf{Case 1:} The stable manifold proceeds through $R_{\{1\}}$.

We write the stable manifold segment of $R_{[2]}$ as 

$$x_1=mx_2+\dfrac{(1-mW_{21})\theta_1 + (W_{12} - m)\theta_2}{1+|W|}.$$

Then, its intersection with the $H_2$ $\left(\text{recall that this is } x_1=-\dfrac{\theta_2}{W_{21}}\right)$ wall will be when:

$$-\dfrac{\theta_2}{W_{21}}=mx_2+\dfrac{(1-mW_{21})\theta_1 + (W_{12} - m)\theta_2}{1+|W|}.$$

Resolving this, we have:

$$-\dfrac{\theta_2}{W_{21}}=mx_2+\dfrac{(1-mW_{21})\theta_1 + (W_{12} - m)\theta_2}{1+|W|}.$$

Then,

$$mx_2=\dfrac{(W_{21}m-1)(W_{21}\theta_1 +\theta_2)}{W_{21}(1+|W|)}\implies x_2=\dfrac{(W_{21} -m^{-1})(W_{21}\theta_1 +\theta_2)}{W_{21}(1+|W|)}.$$

Applying Lemma~\ref{lem:continue}, we obtain the segment in $R_{\{1\}}$:

$$x_1=f_1 \left(-\dfrac{\theta_2}{W_{21}},\dfrac{(W_{21} -m^{-1})(W_{21}\theta_1 +\theta_2)}{W_{21}(1+|W|)},x_2 \right)$$

It intersects the $H_1$ wall at the point: 

$$\left( f_1 \left(-\dfrac{\theta_2}{W_{21}},\dfrac{(W_{21} -m^{-1})(W_{21}\theta_1 +\theta_2)}{W_{21}(1+|W|)}, -\dfrac{\theta_1}{W_{12}}  \right), -\dfrac{\theta_1}{W_{12}} \right).$$

The linear continuation into $R_{\emptyset}$ is:

$$x_1=-\dfrac{W_{12}}{\theta_1}f_1 \left(-\dfrac{\theta_2}{W_{21}},\dfrac{(W_{21} -m^{-1})(W_{21}\theta_1 +\theta_2)}{W_{21}(1+|W|)}, -\dfrac{\theta_1}{W_{12}}  \right)x_2.$$ 

Thus, we have stable manifold $x_1=\mathcal{S}_1(x_2)$ where:

$$\mathcal{S}_1(x_2)=\begin{cases}
mx_2+\frac{(1-mW_{21})\theta_1 + (W_{12} - m)\theta_2}{1+|W|} & \text{if } 0\leq x_2 \leq \frac{(W_{21} -m^{-1})(W_{21}\theta_1 +\theta_2)}{W_{21}(1+|W|)} \\
f_1 \left(-\frac{\theta_2}{W_{21}},\frac{(W_{21} -m^{-1})(W_{21}\theta_1 +\theta_2)}{W_{21}(1+|W|)},x_2 \right) & \text{if }  \frac{(W_{21} -m^{-1})(W_{21}\theta_1 +\theta_2)}{W_{21}(1+|W|)}< x_2 \leq -\frac{\theta_1}{W_{12}} \\
-\frac{W_{12}}{\theta_1}f_1 \left(-\frac{\theta_2}{W_{21}},\frac{(W_{21} -m^{-1})(W_{21}\theta_1 +\theta_2)}{W_{21}(1+|W|)}, -\frac{\theta_1}{W_{12}}  \right)x_2 &  \text{if }  -\frac{\theta_1}{W_{12}} < x_2
\end{cases}.
$$

There exists a final technical point in confirming that the stable manifold intersects the $x_1$-axis wall of $R_{[2]}$ rather than the $x_2$-axis wall.  This will hold true if, for the segment in $R_{[2]}$:

$$x_1=mx_2+\dfrac{(1-mW_{21})\theta_1 + (W_{12} - m)\theta_2}{1+|W|}.$$

the constant term is positive.  That is:

$$\dfrac{(1-mW_{21})\theta_1 + (W_{12} - m)\theta_2}{1+|W|} > 0.$$

Now, from the proof of Lemma~\ref{lem:bcmatt}, we know that $-1+\sqrt{W_{12}W_{21}}>0$ in this parameter regime and so $1+|W|=1-W_{12}W_{21}<0$.  So, we need to confirm that $(1-mW_{21})\theta_1 + (W_{12} - m)\theta_2<0$ as well.  Again using the compact notation $\alpha=\sqrt{-W_{12}}, \beta=\sqrt{-W_{21}}$, we can rewrite this as:

$$\theta_1 + \dfrac{\alpha \beta^2}{\beta}\theta_1 - \alpha^2 \theta_2 -\dfrac{\alpha}{\beta}\theta_2 <0 \implies (\beta \theta_2)\alpha^2 + (\theta_2 - \beta \theta_1)\alpha - \beta \theta_1 > 0.$$

But as we saw in the proof for Lemma~\ref{lem:bcmbslem}, this condition translates to $W_{21}>\left( \dfrac{\theta_2}{\theta_2} \right)^2 W_{12}$ which is true by assumption.

\textbf{Case 2:} The stable manifold proceeds through $R_{\{2\}}$.

We write the stable manifold segment of $R_{[2]}$ as 

$$x_2=m^{-1}x_1+ \dfrac{(1-m^{-1}W_{12})\theta_2 + (W_{21} - m^{-1})\theta_1}{1+|W|}.$$

By symmetry, we obtain the stable manifold $x_2=\mathcal{S}_2(x_1)$:

$$\mathcal{S}_2(x_1)=\begin{cases}
m^{-1}x_1+\frac{(1-m^{-1}W_{12})\theta_2 + (W_{21} - m^{-1})\theta_1}{1+|W|} & \text{if } 0\leq x_1 \leq \frac{(W_{12} -m)(W_{12}\theta_2 +\theta_1)}{W_{12}(1+|W|)} \\
f_2 \left(\frac{(W_{12} -m)(W_{12}\theta_2 +\theta_1)}{W_{12}(1+|W|)},-\frac{\theta_1}{W_{12}},x_1 \right) & \text{if }  \frac{(W_{12} -m)(W_{12}\theta_2 +\theta_1)}{W_{12}(1+|W|)}< x_1 \leq -\frac{\theta_2}{W_{21}} \\
-\frac{W_{21}}{\theta_2}f_2 \left(\frac{(W_{12} -m)(W_{12}\theta_2 +\theta_1)}{W_{12}(1+|W|)},-\frac{\theta_1}{W_{12}}, -\frac{\theta_2}{W_{21}}  \right)x_1 &  \text{if }  -\frac{\theta_2}{W_{21}} < x_1
\end{cases}.
$$

\end{proof}

From here, it becomes easy to see that we have a bifurcation on the combinatorial dynamics of the Binary Competition Model.

\begin{thm}\label{thm:bcmcb}
For $\theta_2 \geq \theta_1$, the combinatorial structure of the trajectory graphs for a two neuron symmetric TLN with varying inputs bifurcates as depicted in Figure~\ref{fig:combifur} with: 

$\mathcal{H}_i$: Rectifier Chamber Boundary for $x_i$ (\textcolor{blue}{Blue})

$$\mathcal{H}_1: x_2=\dfrac{-\theta_1}{W_{12}}$$
$$\mathcal{H}_2: x_1=\dfrac{-\theta_2}{W_{21}}$$

$\mathcal{N}_i$: Nullclines for $x_i$ (\textcolor{green}{Green})

$$\mathcal{N}_1: x_1=[W_{12}x_2+\theta_1]_+$$
$$\mathcal{N}_2: x_2=[W_{21}x_1+\theta_2]_+$$

$\mathcal{M}_{(j)}$: Initial conditions for trajectories passing through $H_i/\mathcal{N}_j$ intersections in positive time (\textcolor{magenta}{Magenta})

$\mathcal{S}_{(i)}$: Chamber by chamber breakdown of the positively oriented stable manifold $\mathcal{S}_1/\mathcal{S}_2$ for the $R_{[2]}$ fixed point (Black, \textcolor{cyan}{Cyan}, and \textcolor{red}{Red})

\begin{figure}[!ht]
\begin{center}
\vspace{.1in}
\includegraphics[width=6.25in]{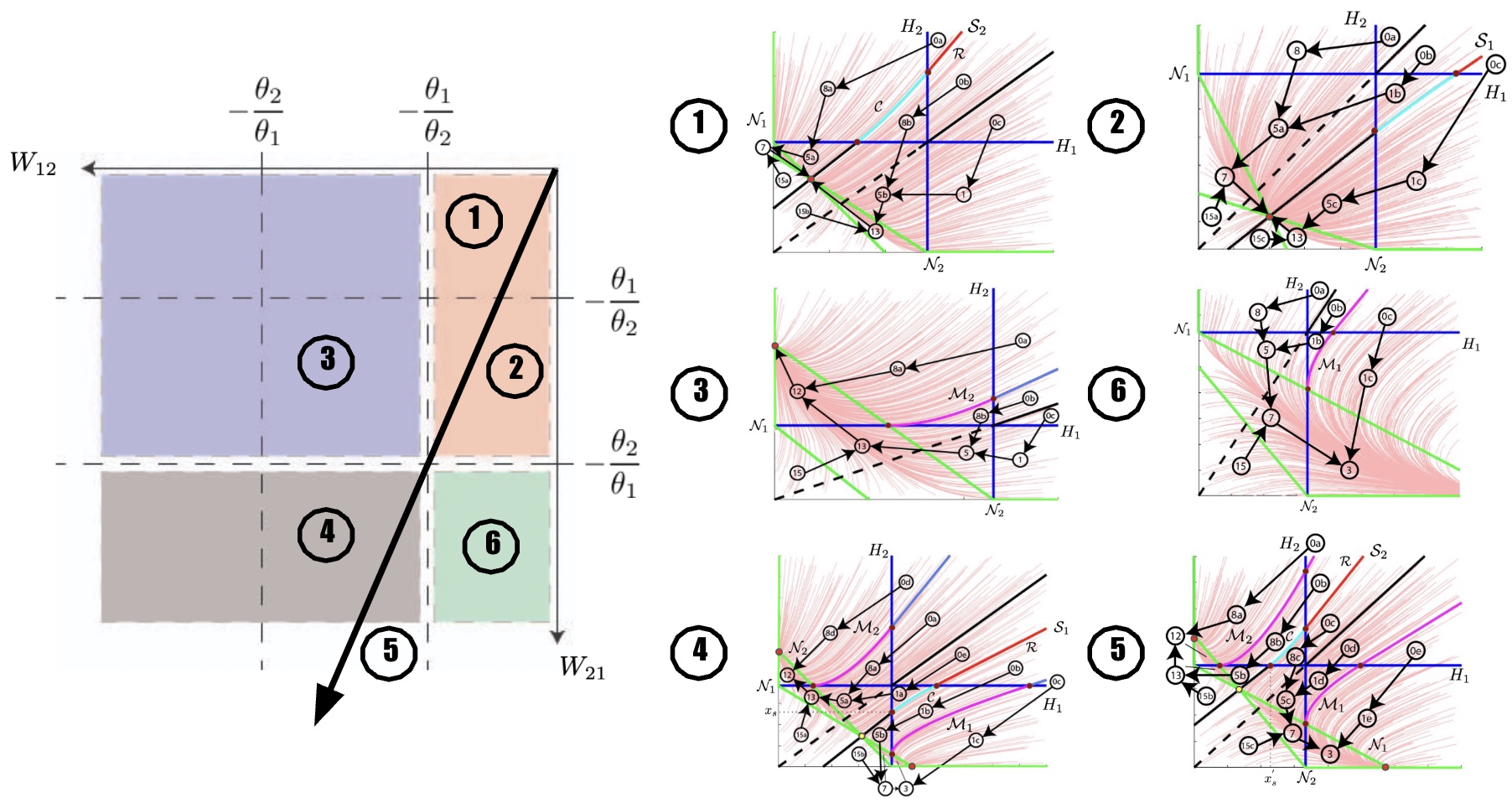}
\vspace{.1in}
\caption[Trajectory graph bifurcation]{{\bf Trajectory graph bifurcation.} Bifurcation diagram on trajectory graph structure of state space.}
\label{fig:combifur}
\end{center}
\vspace{-.2in}
\end{figure}

\end{thm}

\begin{proof}

Drawing on Corollary~\ref{corr:nullcd}, standard nullcline analysis yields most of the trajectory graph upon inspection.  Where there is an $H_i /\mathcal{N}_j$ intersection we draw in the curve $\mathcal{M}_j$.  Additionally, we fill in the Perron-Frobenius stable manifold for Zones 1 and 4 of Figure~\ref{fig:NCBCM} as prescribed by Lemma~\ref{lem:bcmbslem} and Proposition~\ref{prp:bistabtraj}.  The final piece is the diagonal through the intersection of $H_1$ and $H_2$, $x_2=\frac{\theta_1 W_{21}}{\theta_2 W_{12}}x_1$, restricted to $R_{\emptyset}$ and marked in black, which separates trajectories that flow from $R_{\emptyset}$ into $R_{\{1\}}$ and $R_{\{2\}}$ respectively.

\end{proof}

\subsection{Basins of Attraction}

We concern ourselves primarily with the bistable case as it is the only one with distinct basins of attraction.  

In the case where $W_{21} > \left( \dfrac{\theta_2}{\theta_1} \right)^2 W_{12}$, we have the separatrix:

$$x_1=\mathcal{S}_1(x_2).$$

In the case of $W_{21} < \left( \dfrac{\theta_2}{\theta_1} \right)^2 W_{12}$, we have the separatrix:

$$x_2=\mathcal{S}_2(x_1).$$

Both are depicted in Figure~\ref{fig:septx}.

\begin{figure}[!ht]
\begin{center}
\vspace{.1in}
\includegraphics[width=6.25in]{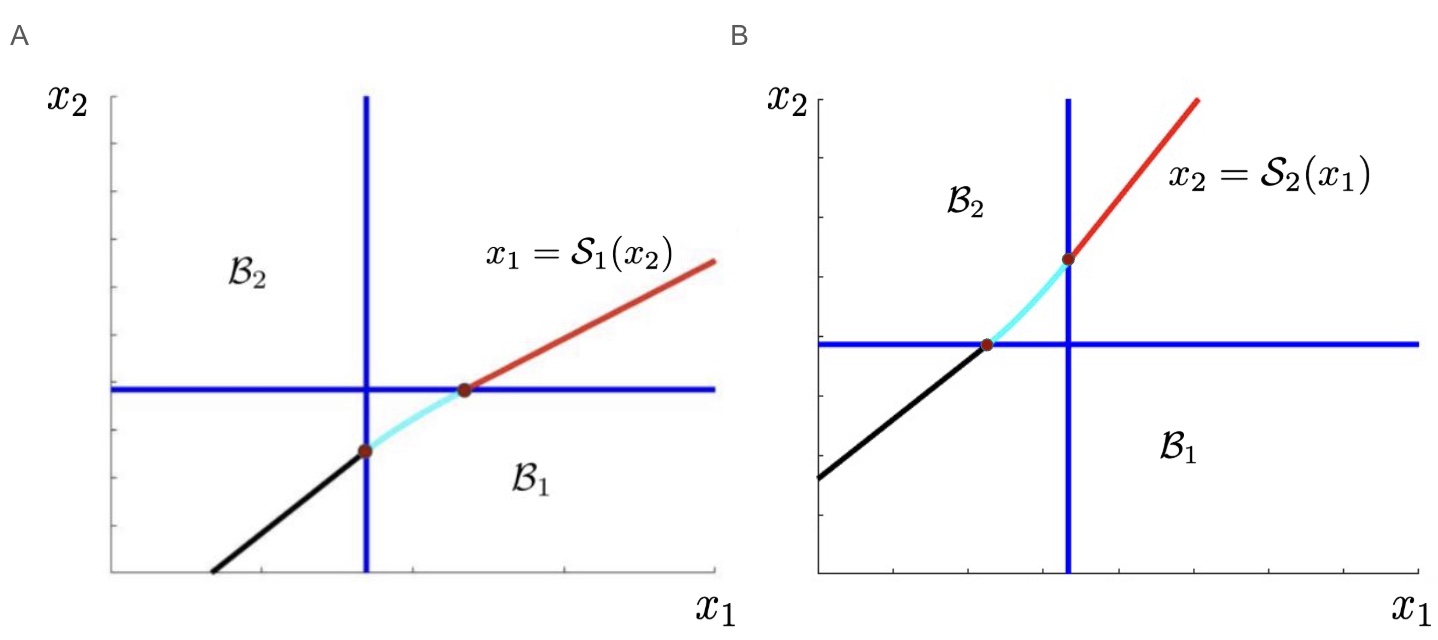}
\vspace{.1in}
\caption[Basins of attraction under bistable parameter regime]{{\bf Basins of attraction under bistable parameter regime.} (A) Separatrix and basins of attraction under Zone 4 parameter regime as indicated in Fig~\ref{fig:combifur}.  (B) Separatrix and basins of attraction under Zone 5 parameter regime.}
\label{fig:septx}
\end{center}
\vspace{-.2in}
\end{figure}

\begin{corr}
For the Binary Competition Model where without loss of generality $\theta_2 \geq \theta_1>0$, the basins of attraction are as follows:

Zones 1 and 2, $W_{21}>-\dfrac{\theta_2}{\theta_1}$ and $W_{12}>-\dfrac{\theta_1}{\theta_2}$: $\operatorname{FP}=\{\{1,2\}\}$, The basin of attraction for the single fixed point supported on $\{1,2\}$ is the entire phase space.

Zone 3, $W_{21}>-\dfrac{\theta_2}{\theta_1}$ and $W_{12}<-\dfrac{\theta_1}{\theta_2}$: $\operatorname{FP}=\{1\}$, The basin of attraction for the single fixed point supported on $\{2\}$ is the entire phase space.

Zone 4, $W_{21}>\left( \dfrac{\theta_2}{\theta_1} \right)^2 W_{12}$ and $W_{21}<-\dfrac{\theta_2}{\theta_1}$: $\operatorname{FP}=\{1,2,\{1,2\}\}$, For the fixed point supported on $\{1\}$, the basin of attraction is:

$$\mathcal{B}_1=\{(x_1^0,x_2^0) \mid x_1^0>\mathcal{S}_1(x_2^0)\}$$

and similarly the basin of attraction for the fixed point supported on $\{2\}$ is given by:

$$\mathcal{B}_2=\{(x_1^0,x_2^0) \mid x_1^0<\mathcal{S}_1(x_2^0)\}$$

Zone 5, $W_{21}< \left( \dfrac{\theta_2}{\theta_1} \right)^2 W_{12}$ and $W_{12}<-\dfrac{\theta_1}{\theta_2}$: $\operatorname{FP}=\{1,2,\{1,2\}\}$, For the fixed point supported on $\{1\}$, the basin of attraction is:

$$\mathcal{B}_1=\{(x_1^0,x_2^0) \mid x_2^0<\mathcal{S}_2(x_1^0)\}$$

and similarly the basin of attraction for the fixed point supported on $\{2\}$ is given by:

$$\mathcal{B}_2=\{(x_1^0,x_2^0) \mid x_2^0>\mathcal{S}_2(x_1^0)\}$$

Zone 6, $W_{21}<-\dfrac{\theta_2}{\theta_1}$ and $W_{12}>-\dfrac{\theta_1}{\theta_2}$: $\operatorname{FP}=\{2\}$, The basin of attraction for the single fixed point supported on $\{2\}$ is the entire phase space.

\end{corr}

While these expressions are unwieldy, notice that it is not especially challenging to integrate beneath these separatrices.  The segments in $R_{[2]}$ and $R_{\emptyset}$ are linear and the segments in $R_{\{1\}}$ $R_{\{2\}}$, of the form $f_1(a,b,x)$ and $f_2(a,b,x)$ respectively, are both easily resolved using integration by parts on the non-linear term.  To find the relative sizes of the basins, we restrict the window to $[0,B]$ x $[0,B]$, find the fractional area of the basin as a function of B, and then take the limit $B \rightarrow \infty$.

\begin{ddd} 

Let $A$ be a measurable set in $\mathbb{R}^N_+$.  Call $A$ an $\mathcal{F}$-\textbf{set} if:

    $$\lim_{B \rightarrow \infty}\dfrac{\lambda(A\cap[0,B]^n)}{B^n}$$

exists (where $\lambda$ is the standard Lebesgue measure).

If $A$ is an $\mathcal{F}$-\textbf{set}, define as its \textbf{fractional area(volume)} $\mathcal{F}(A)$:

    $$\mathcal{F}(A)=\lim_{B \rightarrow \infty}\dfrac{\lambda(A\cap[0,B]^n)}{B^n}.$$

\end{ddd}

\begin{thm} \label{thm:bcmbasin}

For the Binary Competition Model with parameters in Zone 4 of Fig~\ref{fig:combifur}:

\begin{flushleft}
1. If $f_1 \left(-\dfrac{\theta_2}{W_{21}},x_s, -\dfrac{\theta_1}{W_{12}}  \right) > -\dfrac{\theta_1}{W_{12}}$: 
\end{flushleft}

(a) The fractional area of the basin of attraction $\mathcal{B}_1$ for the fixed point supported on $\{1\}$ is:

    $$\mathcal{F}(\mathcal{B}_1)=-\dfrac{\theta_1}{2W_{12} f_1 \left(-\frac{\theta_2}{W_{21}},x_s, -\frac{\theta_1}{W_{12}}  \right)}.$$

(b) The fractional area of the basin of attraction $\mathcal{B}_2$ for the fixed point supported on $\{2\}$ is:

    $$\mathcal{F}(\mathcal{B}_2)= 1+ \dfrac{\theta_1}{2W_{12} f_1 \left(-\frac{\theta_2}{W_{21}},x_s, -\frac{\theta_1}{W_{12}}  \right)}.$$

\begin{flushleft}    
2. However, if $f_1 \left(-\dfrac{\theta_2}{W_{21}},x_s, -\dfrac{\theta_1}{W_{12}}  \right) \leq -\dfrac{\theta_1}{W_{12}}$:
\end{flushleft}

(a) The fractional area of the basin of attraction $\mathcal{B}_1$ for the fixed point supported on $\{1\}$ is:

    $$\mathcal{F}(\mathcal{B}_1)=1+\dfrac{W_{12}}{2\theta_1}f_1 \left(-\frac{\theta_2}{W_{21}},x_s, -\frac{\theta_1}{W_{12}}  \right).$$

(b) The fractional area of the basin of attraction $\mathcal{B}_2$ for the fixed point supported on $\{2\}$ is:

    $$\mathcal{F}(\mathcal{B}_2)=-\dfrac{W_{12}}{2\theta_1}f_1 \left(-\frac{\theta_2}{W_{21}},x_s, -\frac{\theta_1}{W_{12}}  \right).$$

\begin{flushleft}    
For parameters in Zone 5 of Fig~\ref{fig:combifur}:
\end{flushleft}

\begin{flushleft}
1. If $f_2 \left(x'_s,-\dfrac{\theta_1}{W_{12}}, -\dfrac{\theta_2}{W_{21}}  \right) > -\dfrac{\theta_2}{W_{21}}$:
\end{flushleft}

(a) The fractional area of the basin of attraction $\mathcal{B}_1$ for the fixed point supported on $\{1\}$ is:

    $$\mathcal{F}(\mathcal{B}_1)=1+\dfrac{\theta_2}{2W_{21}f_2 \left(x'_s,-\frac{\theta_1}{W_{12}}, -\frac{\theta_2}{W_{21}}  \right)}.$$

(b) The fractional area of the basin of attraction $\mathcal{B}_2$ for the fixed point supported on $\{2\}$ is:

    $$\mathcal{F}(\mathcal{B}_2)=-\dfrac{\theta_2}{2W_{21}f_2 \left(x'_s,-\frac{\theta_1}{W_{12}}, -\frac{\theta_2}{W_{21}}  \right)}.$$

\begin{flushleft}    
2. However, if $f_2 \left(x'_s,-\dfrac{\theta_1}{W_{12}}, -\dfrac{\theta_2}{W_{21}}  \right) \leq -\dfrac{\theta_2}{W_{21}}$:
\end{flushleft}

(a) The fractional area of the basin of attraction $\mathcal{B}_1$ for the fixed point supported on $\{1\}$ is:

    $$\mathcal{F}(\mathcal{B}_1)=-\dfrac{W_{21}}{2\theta_2}f_2 \left(x'_s,-\frac{\theta_1}{W_{12}}, -\frac{\theta_2}{W_{21}}  \right).$$

(b) The fractional area of the basin of attraction $\mathcal{B}_2$ for the fixed point supported on $\{2\}$ is:

    $$\mathcal{F}(\mathcal{B}_2)=1+\dfrac{W_{21}}{2\theta_2}f_2 \left(x'_s,-\frac{\theta_1}{W_{12}}, -\frac{\theta_2}{W_{21}}  \right).$$

where $m=\sqrt{\dfrac{W_{12}}{W_{21}}}$, $x_s=\dfrac{(W_{21} -m^{-1})(W_{21}\theta_1 +\theta_2)}{W_{21}(1+|W|)}$, and $x'_s=\dfrac{(W_{12} -m)(W_{12}\theta_2 +\theta_1)}{W_{12}(1+|W|)}$.
\end{thm}

\begin{proof}

Generally consider a function of the form:

$$\mathcal{S}(x)=\begin{cases}
k_1 x & \text{if } 0 \leq x \leq p \\
f_1(a,b,x) & \text{if } p < x \leq q \\
k_2 x & \text{if } q < x
\end{cases}$$  

\textbf{Case 1: } $k_2>1$

If $k_2>1$, then the segment $k_2 x$ hits the top of the box $[0, B]^2$ for sufficiently large $B$ as depicted in Fig~\ref{fig:intdiag}.  To find the fractional area of the region below $\mathcal{S}(x)$, consider the following limit:

$$\lim_{B\rightarrow \infty} \dfrac{1}{B^2}\left( \int_{0}^{B/k_2} \mathcal{S}(x) dx + B\left(B-\dfrac{B}{k_2} \right) \right)$$

\begin{figure}[!ht]
\begin{center}
\vspace{.1in}
\includegraphics[width=6.25in]{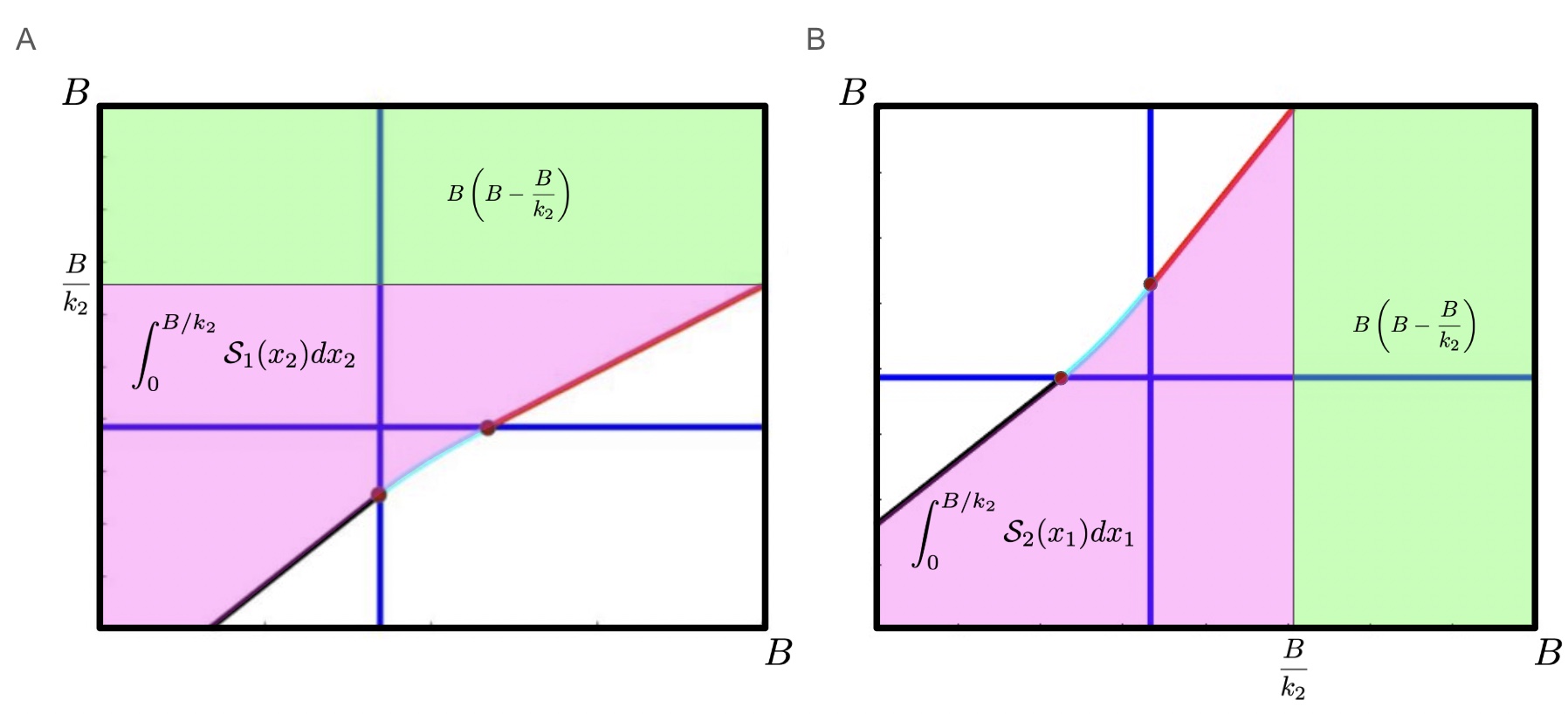}
\vspace{.1in}
\caption[Fractional area computation of basins when $k_2 > 1$]{{\bf Fractional area computation of basins when $k_2 > 1$.} (A) Diagram of two pieces of $\mathcal{B}_2 \cap [0,B]^2$ in Zone 4 of Fig~\ref{fig:combifur}.  When $k_2>1$, the seperatrix hits the right side of the box instead of continuing to the other side, creating the rectangular piece.  (B) Diagram of two pieces of $\mathcal{B}_1 \cap [0,B]^2$ in Zone 5 of Fig~\ref{fig:combifur}.  When $k_2>1$, the seperatrix hits the top of the box instead of continuing to the other side, creating the rectangular piece.}
\label{fig:intdiag}
\end{center}
\vspace{-.2in}
\end{figure}

Notice that:

$$\lim_{B\rightarrow \infty} \dfrac{\int_{0}^{B/k_2} \mathcal{S}(x) dx}{B^2}=\lim_{B\rightarrow \infty}\left( \dfrac{\int_{0}^{p} k_1 x dx}{B^2}+\dfrac{\int_{p}^{q} f_1(a,b,x) dx}{B^2}+\dfrac{\int_{q}^{B/k_2} k_2 x dx}{B^2} \right)$$

$$=0+0+\lim_{B\rightarrow \infty} \dfrac{B^2}{2k_2 B^2}-0=\dfrac{1}{2k_2}.$$

Then,

$$\lim_{B\rightarrow \infty} \dfrac{1}{B^2}\left( \int_{0}^{B/k_2} \mathcal{S}(x) dx + B\left(B-\dfrac{B}{k_2} \right) \right) = \dfrac{1}{2k_2} + 1 -\dfrac{1}{k_2}=1-\dfrac{1}{2k_2}.$$

Now for the key idea.  Notice that, for Zone 4, we have:

$$\mathcal{F}(\mathcal{B}_2)=\lim_{B\rightarrow \infty} \dfrac{1}{B^2}\left( \int_{0}^{B/k_2} \mathcal{S}_1(x_2) dx_2 + B\left(B-\dfrac{B}{k_2} \right) \right)$$

as depicted in Fig~\ref{fig:intdiag}A for $k_2 = -\dfrac{W_{12}}{\theta_1} f_1 \left(-\frac{\theta_2}{W_{21}},\frac{(W_{21} -m^{-1})(W_{21}\theta_1 +\theta_2)}{W_{21}(1+|W|)}, -\frac{\theta_1}{W_{12}}  \right)$.  So, for Zone 4:

$$\mathcal{F}(\mathcal{B}_2)= 1+ \dfrac{\theta_1}{2W_{12} f_1 \left(-\frac{\theta_2}{W_{21}},\frac{(W_{21} -m^{-1})(W_{21}\theta_1 +\theta_2)}{W_{21}(1+|W|)}, -\frac{\theta_1}{W_{12}}  \right)}$$

To obtain $\mathcal{F}(\mathcal{B}_1)$, notice that:

$$\mathcal{F}(\mathcal{B}_1)=\lim_{B\rightarrow \infty} \dfrac{B^2-\int_{0}^B \mathcal{S}_1 (x) dx}{B^2}=1-\mathcal{F}(\mathcal{B}_2).$$

This would also be true if we used $f_2(a,b,x)$ instead of $f_1(a,b,x)$, so we apply this same approach to $\mathcal{S}_2(x)$ to obtain the result for Zone 5 as depicted in Fig~\ref{fig:intdiag}B.

\textbf{Case 2: } $k_2 \leq 1$

Alternatively, if $k_2 \leq 1$, the segment $k_2 x$ would not hit the top of the box, and we may simply integrate beneath it.  We would then have the the fractional area expression:

$$\lim_{B\rightarrow \infty} \dfrac{\int_{0}^{B} \mathcal{S}(x) dx}{B^2}=\lim_{B\rightarrow \infty}\left( \dfrac{\int_{0}^{p} k_1 x dx}{B^2}+\dfrac{\int_{p}^{q} f_1(a,b,x) dx}{B^2}+\dfrac{\int_{q}^{B} k_2 x dx}{B^2} \right)$$

$$=0+0+\lim_{B\rightarrow \infty} \dfrac{k_2 B^2 -k_2 q^2}{2B^2}=\dfrac{k_2}{2}$$.

Then, for Zone 4, we apply this for $\mathcal{S}_1(x)$ to find $\mathcal{F}(\mathcal{B}_2)$.  To obtain $\mathcal{F}(\mathcal{B}_1)$, notice that:

$$\mathcal{F}(\mathcal{B}_1)=\lim_{B\rightarrow \infty} \dfrac{B^2-\int_{0}^B \mathcal{S}_1 (x) dx}{B^2}=1-\mathcal{F}(\mathcal{B}_2).$$

We apply this same approach to $\mathcal{S}_2(x)$ to obtain the result for Zone 5.

\end{proof}

What we see from this is that we are actually able to describe the separatrix and sizes of the basins of attraction quite cleanly for a given set of parameters.  As a final remark, note that the calculations of the fractional area formulas involve the linear segments of the separatrices in $R_{\emptyset}$ dominating and the role of the earlier segments becoming negligible.  This occurs when the limit of the window size is taken to $\infty$.  However, if we were to restrict to initial conditions in a particular bounded set of the state space, perhaps to incorporate biological restrictions, we would still  be able to work out the sizes of the basins relative to one another.  Since the nonlinearities in $f_1(a,b,x)$ and $f_2(a,b,x)$ can be integrated by parts in $x$, our analysis is not overly tied to the limit case.

\subsection{Decision-Making Bias in the Binary Competition Model}

We return now to our three paradigms of how decision-making bias may be encoded in the basins of attraction.  In Fig~\ref{fig:bcmbias}, we demonstrate each of these for this two-dimensional TLN model, assuming that we are in the bistable regime in parameter space.

\begin{figure}[!ht]
\begin{center}
\vspace{.1in}
\includegraphics[width=6.5in]{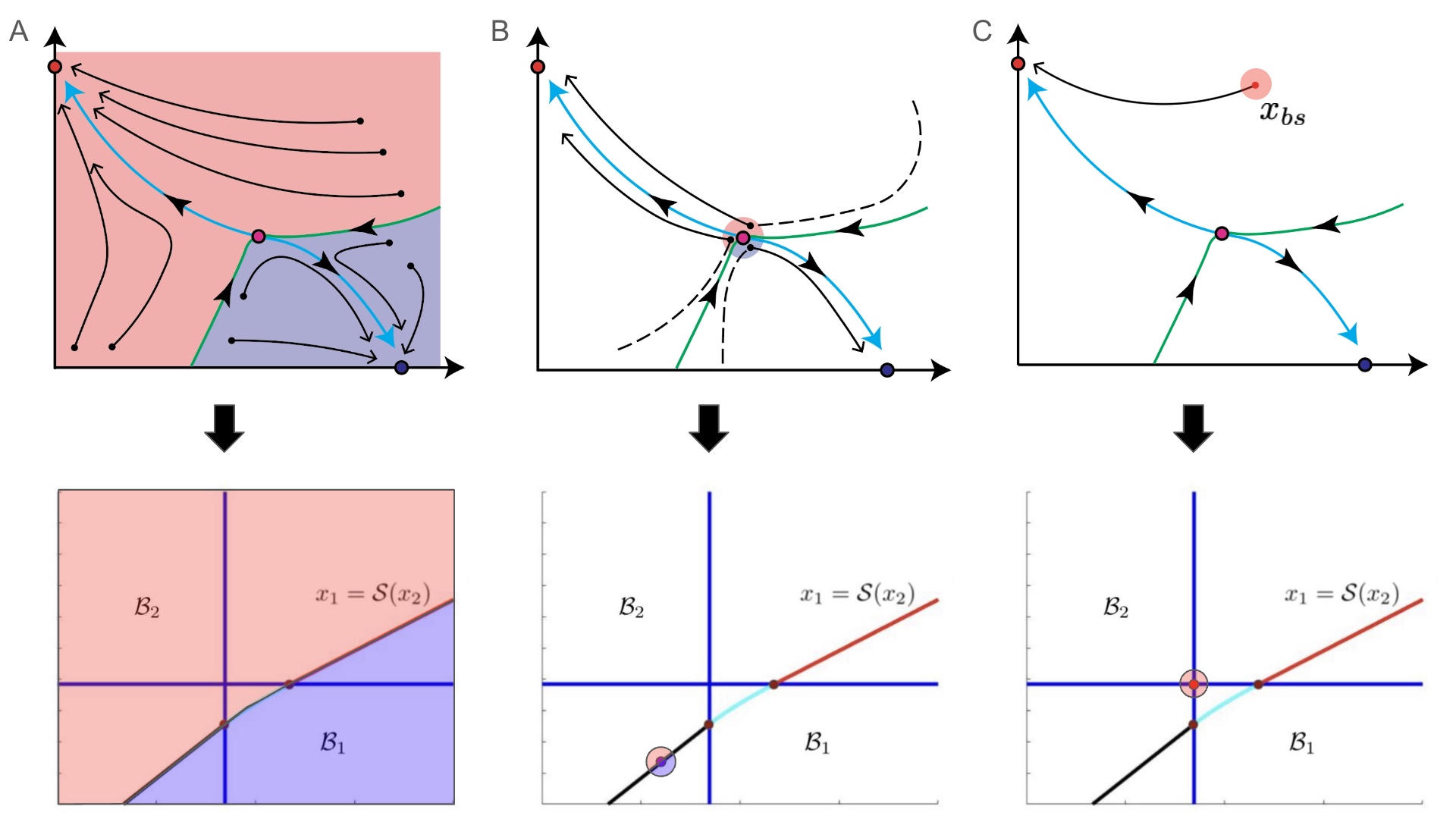}
\vspace{.1in}
\caption[Encoding of decision-making bias in the Binary Competition Model]{{\bf Encoding of decision-making bias in the Binary Competition Model.} (A) Under the assumption that the full basins of attraction are of relevance we take the fractional area of the basins area.  (B) Under the paradigm that the basins in the vicinity of the saddle point matter most, we restrict to the intersection between the basins and a small unit disk around the saddle point, which is problematic as the separatrix is locally linear near the saddle point.  (C) Focusing on the balanced state trajectory, the basin the balanced state lies in is controlled strictly by whether the separatrix exits $R_{[2]}$ through $H_1$ or through $H_2$.}
\label{fig:bcmbias}
\end{center}
\vspace{-.2in}
\end{figure}

As depicted in Fig~\ref{fig:bcmbias}A, using Theorem~\ref{thm:bcmbasin} lets us understand bias under the hypothesis of high dimensional dynamics where we want to compare the relative sizes of the basins of attraction.  While the expressions of the theorem are unwieldy they do precisely quantify this relationship.  However, if we were merely concerned with the balanced state trajectory, we need only use Lemma~\ref{lem:bcmbslem} which indicates on which side of the seperatrix the balanced state lies, as depicted in Fig~\ref{fig:bcmbias}C.

Where problems majorly arise is in the case where we emphasize trajectories along the decision-boundary, sampling initial conditions near the saddle point.  As calculated, the seperatrix is locally linear near the saddle point, so will separate any small disc drawn around it in half as depicted in Fig~\ref{fig:bcmbias}B.  This means that no matter how we modulate the parameters, the relative sizes of the localized basins would always be equivalent.

Another problem with this two-dimensional model is that it is highly simplified and obscures the larger connectivity structure of the network.  While we could simulate a phenomenon like the Decoy Effect by setting $W_{12}=W_{21}$ and making say $\theta_2>\theta_1$, the nuances of a more complex network structure may be destroyed by such a reduction.  This model cannot be the end of the story and analyzing higher dimensional systems will often be necessary to properly understand bias in decision-making circuits.

\chapter{Challenges of Combinatorial Dynamics in Higher Dimensions}

It goes without saying that an analytical approach to working out combinatorial dynamics across higher dimensional TLNs is not tractable.  However, might there be a computer assisted approach that is viable?  Even if we are not able to get a precise trajectory graph, we might still get a state transition graph with gives us rough lower bounds on the sizes of basins.  One challenge of a computer assisted approach is that of attractors.  While in the case of the Binary Competition Model all the attractors were fixed points, this is not true of TLNs in general and it is possible to have dynamic attractors such as limit cycles in higher dimensions \cite{lcexist}.  This means a chamber by chamber linear system analysis is simply not enough, even with computer assistance, as dynamic attractors will span various chambers.  How can we talk about basins of attraction if we do not understand what the attractors are?

One approach to this problem is in the application of Conley Index Theory.  We briefly review the key ideas of applied Conley Index Theory with the exposition being drawn primarily from a review by Konstantin Mischaikow \cite{conley}. 

We will describe how it has been applied to state transition graphs to detect and classify attractors, but also show the challenges that this approach has when dealing with TLNs.  The primary novelty of this chapter is the development of an algorithm which exploits the piecewise linear dynamics of TLNs to computationally determine a state transition graph without requiring brute force methods of simulation to determine edge directions.

\section{Introduction to Conley Index Theory}
Conley Index Theory can be thought of as a coarse graining of traditional dynamical systems theory.  Instead of looking at invariant sets, such as attractors, directly, the objects of focus are instead the \textit{isolating neighborhoods} of invariant sets.

\begin{ddd}
Let $\phi(t,x)$ be a flow in $\mathbbm{R}^n$.  A compact set $N\subset \mathbbm{R}^n$ is an \textbf{isolating neighborhood} if:

$$\operatorname{Inv}(N,\phi):=\{x\in N \mid \phi(\mathbbm{R},x) \subset N\}\subset \operatorname{int} N.$$
\end{ddd} 

Studying attractors directly is mathematically precise whereas studying isolating neighborhoods has more freedom and can be easier.  To be particular, isolating neighborhoods allow us to study \textit{isolated invariant sets} i.e. invariant sets $S$ such that $S=\operatorname{Inv} N$ for some isolating neighborhood $N.$

The next construction to introduce is that of the \textit{index pair} and the \textit{exit set} $L$ of an isolating neighborhood.

\begin{ddd}
Let S be an isolated invariant set.  A pair of compact sets (N,L) with $L \subset N$ is an \textbf{index pair} for S if:

1. $S=\operatorname{Inv}(\operatorname{cl}(N\backslash L))$ and $N\backslash L$ is a neighborhood of S.

2. Given $x\in L$ and $\phi([0,t],x)\subset N$, then $\phi([0,t],x)\subset L$.

3. L is an exit set for N i.e. given $x\in N$ and $t_0>0$ such that $\phi(t_0,x)\not \in N$, then there exists $0\leq t_1 < t_0$ such that $\phi(t_1,x)\in L.$

\end{ddd}

Essentially, these conditions mandate that $L$ be an "outer" component of the isolating neighborhood such that it does not include any of the underlying invariant set, trajectories beginning in $L$ do not enter $N\backslash L$, and that any trajectory leaving $N$ must go through $L$ (Fig~\ref{fig:ci}A).  The Conley index is then a topological invariant assigned to the pointed topological space obtained by collapsing $L$ to a point.  For example the \textit{homotopy Conley index} of an invariant set $S$ with index pair $(N,L)$ is the homotopy type: 

$$h(S)\sim (N/L,[L]).$$

We can use the Betti numbers, $\beta_i$, of the homology groups as a topological signature of the Conley index.

$$\beta_{\bullet}=\operatorname{rank}(H_{\bullet}(N/L,[L])).$$

The Conley Index has three properties that make it a useful tool for studying dynamical systems:

\begin{itemize}

\item If $N$ and $N'$ are isolating neighborhoods such that $\operatorname{Inv}(N)=\operatorname{Inv}(N')$, then Conley Index($N$) = Conley Index($N'$).

\item If Conley Index($N$) is not trivial, then $\operatorname{Inv}(N)\neq \emptyset$.

\item If there are a smooth family of flows $\phi^\lambda (t,x)$ with $\lambda \in [0,1]$. such that $N$ is an isolating neighborhood $\forall \phi^\lambda$.  Then, the Conley Index for $S_\lambda=\operatorname{Inv}(N,\phi^\lambda)$ is independent of $\lambda$.

\end{itemize}

\begin{figure}[!ht]
\begin{center}
\vspace{.1in}
\includegraphics[width=5.75in]{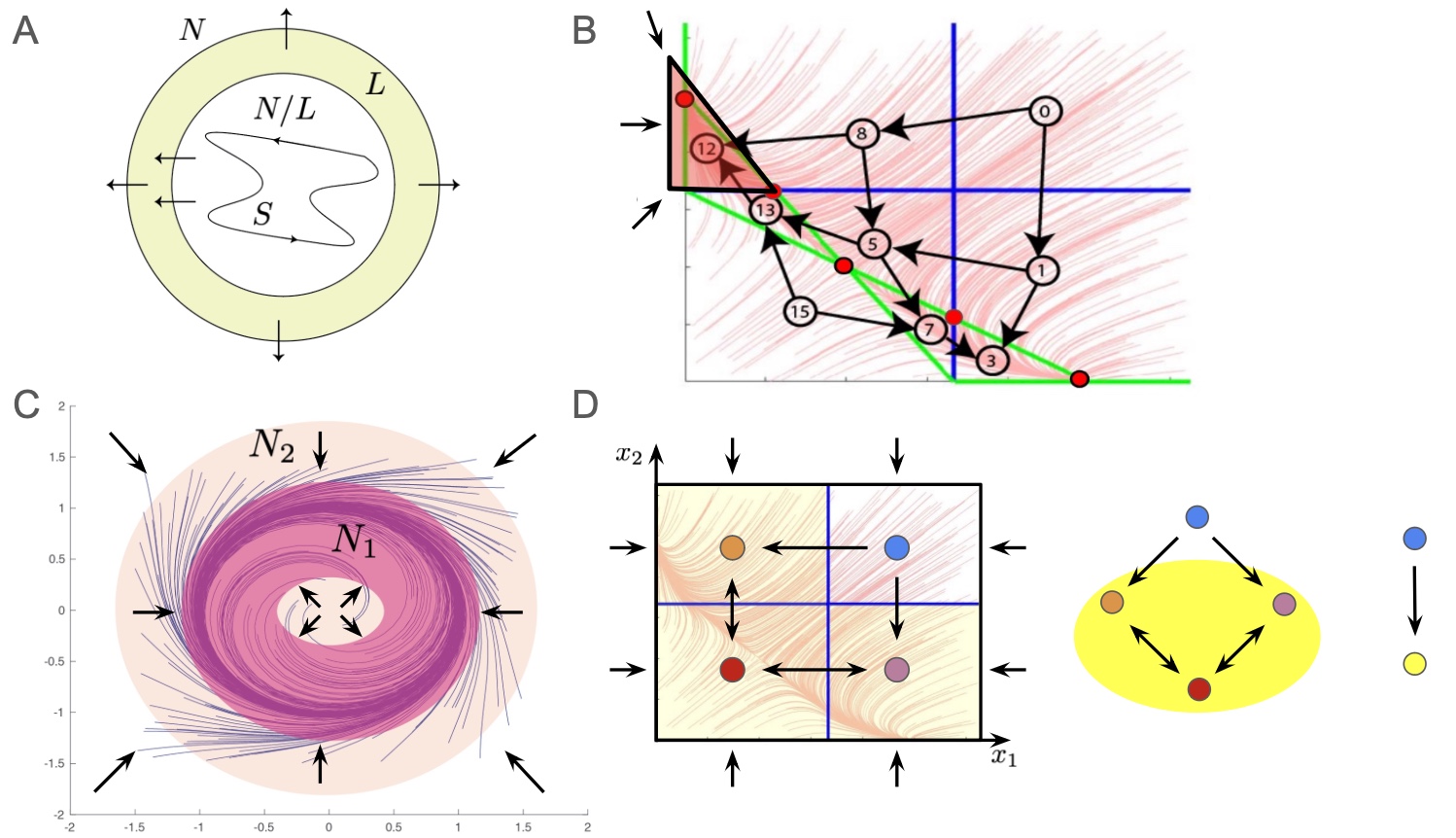}
\vspace{.1in}
\caption[Conley Index Theory and TLNs]{{\bf Conley Index Theory and TLNs.} (A) Schematic showing an invariant set $S$, an isolating neighborhood $N$ with a corresponding exit set $L$. (B) There exist arbitrarily tight compact neighborhoods of chamber 12 which are isolating neighborhoods for the point attractor inside. (C) A limit cycle with two choices of isolating neighborhood $N_1$ and $N_2$.  Even though they are both isolating neighborhoods enclosing the limit cycle, they yield different Conley indices. (D) An example of the state transition graph analysis algorithm being applied to the ReLU hyperplane partition.  Three chambers are collapsed into one strongly connected component which does not separate the two point attractors.}
\label{fig:ci}
\end{center}
\vspace{-.2in}
\end{figure}

This allows us to study the attractors indirectly.  From state transition graphs, we can identify isolating neighborhoods and compute their Conley indices to classify the attractors.  The state transition graph, even if not a full trajectory graph, would then hopefully give us some rough sense of the basins of attraction.

As an exercise for understanding, let us briefly consider the two-dimensional case.  In Fig~\ref{fig:ci}B we have highlighted that arbitrarily tight neighborhoods of chamber 12 are an isolating neighborhood for a point attractor in the independent set CTLN.  Call it $N$.  No trajectories exit this chamber so we have the exit set $L=\emptyset$.  So, the pointed topological space $(N/L,[L])$ is simply $N$ itself.  As $N$ is a contractible set, we have the Betti numbers $\beta_0=1,\beta_1=0$.  Thus, the Conley index would be $(1,0,0)$.  Alternatively, if we had an isolating neighborhood $N_1$ as given in Fig~\ref{fig:ci}C, the Conley Index would be $(1,1,0)$ which characterizes the limit cycle inside.  This is how the Conley index can be used to detect and categorize attractors.  

Once we have a state transition graph, we can collapse strongly connected components to a single vertex, taking the union of the chambers, to allow for attractors that span multiple chambers.  Then we find the sinks of this condensed graph which indicates an attracting set lying within.  Finally, by calculating the Conley Index of that isolating neighborhood, we can try and determine the kind of attractor that lurks inside.  This technique has been used successfully in studying dynamics in systems biology \cite{con2}.

Note that in TLNs fixed points often lie on chamber boundaries in which case these chambers are not isolating neighborhoods in the strictest sense.  However, as discussed, an arbitrarily tight neighborhood of the chamber would be an isolating neighborhood, and so, with this understanding, we informally treat these chambers for now as if they are as well.  A technique of dealing with this issue is that the neighborhood around these fixed points can be "blown-up" into their own chambers \cite{con2}.  Regardless, the first step is the development of a state transition graph.

The two dimensional case also shows us the limitations of Conley Index Theory.  Consider the second isolating neighborhood $N_2$ in Fig~\ref{fig:ci}C.  The set $N_2$ is still an isolating neighborhood where the only attractor inside is the limit cycle, but will have Conley index $(1,0,0)$.  How can this be as the Conley Index is meant to be invariant of the choice of isolating neighborhood?  Notice the subtlety that the Conley index is defined in terms of invariant sets, not attractors.  For $N_2$, the set $\operatorname{Inv}(N_2)$ also includes everything inside of the limit cycle, hence the difference in Conley Index.

What this indicates is that the isolating neighborhood does need to be sufficiently tight to the attractor to properly categorize it.  A unit ball isolating neighborhood could have a point attractor or even multiple chaotic attractors inside and would still yield the same Conley index as long as no trajectories exited the ball.  

The key is then of course is to find a partition scheme that gives us a useable state transition graph.  A problem that can arise is that too coarse a partition will result in too many bidirectional edges in the graph.  Enough bidirectional edges will yield a strongly connected component which will then get collapsed together into a fairly loose isolating neighborhood for the attractors.  Consider the state transition graph which would have arisen in the bistable case of the Binary Choice Model if we had not included the nullclines (Fig~\ref{fig:ci}D) and see how such such a graph structure would tell us very little about the attractors or the basins of attraction.  A key technical point is that our analysis will only be applied to competitive TLNs as the property $W\leq 0$ and $\Vec{\theta}>0$ confines activity in the positive orthant to a bubble within the state space, preventing solutions from blowing up to infinity.  That lets us produce our state transition graph on a finite collection of compact sets (refer to Fig~\ref{fig:ci}D).

\section{Building a State Transition Graph}

Before delving into partition schemes we will first discuss the construction of the state transition graph for a given partition.  

A very naive way to approach this would be to sample initial conditions randomly from partition chambers and track their trajectories into other chambers, drawing edges accordingly.  This is not only inefficient computationally, but it is also not particularly rigorous as we have to hope that our random sampling was large enough to properly approximate the dynamics within the chamber.  Fortunately the linearity of the component dynamical systems of a TLN presents with an alternative approach as long as a partition separates the linear system regions $R_{\sigma}$ and the chambers are convex polytopes generated by a hyperplane arrangement.

As a TLN is a continuous dynamical system, if on a chamber face there is a area where the vector field points inward and an area where it points outward, they are separated by a boundary where the vector field lies within the corresponding hyperplane of the face.

\begin{prp}\label{prp:sep}
    Let $\vec{b}\in\mathbb{R}^n$.  Then the points such that the vector field given by TLN linear ODE system $L_\sigma$ are orthogonal to $\vec{b}$ is $R_\sigma \cap B_\sigma$ where:

    $$B_\sigma:=\sum_{k=1}^n\left(-b_k+\sum_{j\in\sigma}b_j W_{jk}\right)x_k+\sum_{j\in\sigma}b_j \theta_j=0$$
\end{prp}

\begin{proof}

    We seek $\vec{x}\in R_\sigma$ such that $\vec{b}\cdot \frac{d\Vec{x}}{dt}|_{R_\sigma}=0$.

    \begin{eqnarray}
        \vec{b}\cdot \frac{d\Vec{x}}{dt} &=& -\sum_{j\not\in \sigma}b_j x_j+\sum_{j\in\sigma}b_j\dot x_j\\
        &=&-\sum_{j\not\in \sigma}b_jx_j+\sum_{j\in\sigma}b_j(-x_j+\sum_{k=1}^n W_{jk}x_k +\theta_j) \\
        &=&\sum_{k=1}^n\left(-b_k +\sum_{j\in\sigma}b_j W_{jk}\right)x_k +\sum_{j\in\sigma}b_j \theta_j
    \end{eqnarray}
    
\end{proof}

Let a chamber, call it $K^\mu$, be governed by $L_{\sigma}$ and bounded by the hyperplanes $\{K_i\}_{i=1}^m$ where $K_i: k_{i0}+\sum_{j=1}^n k_{ij}x_j=0$.  Then, the chamber is the intersection of half spaces associated with the hyperplanes and its closure can be expressed as the feasible region of the linear program:

$$
\left[
\begin{array}{ccc}
k_{11}&\cdots &k_{1n} \\
\vdots & \ddots &\vdots\\
k_{p1}& \cdots & k_{pn}\\
\hline
-k_{p+1,1}&\cdots & -k_{p+1,n}\\
 \vdots & \ddots & \vdots \\
 -k_{m1}& \cdots & -k_{mn}\\
\end{array}
\right]\Vec{x}\geq \left[ \begin{array}{c} k_{10} \\ \vdots \\ k_{p0} \\ 
\hline  
-k_{p+1,0} \\
\vdots \\
-k_{m0}
\end{array} \right]
$$

where $\Vec{x}\geq 0.$

\begin{figure}[!ht]
\begin{center}
\vspace{.1in}
\includegraphics[width=4.75in]{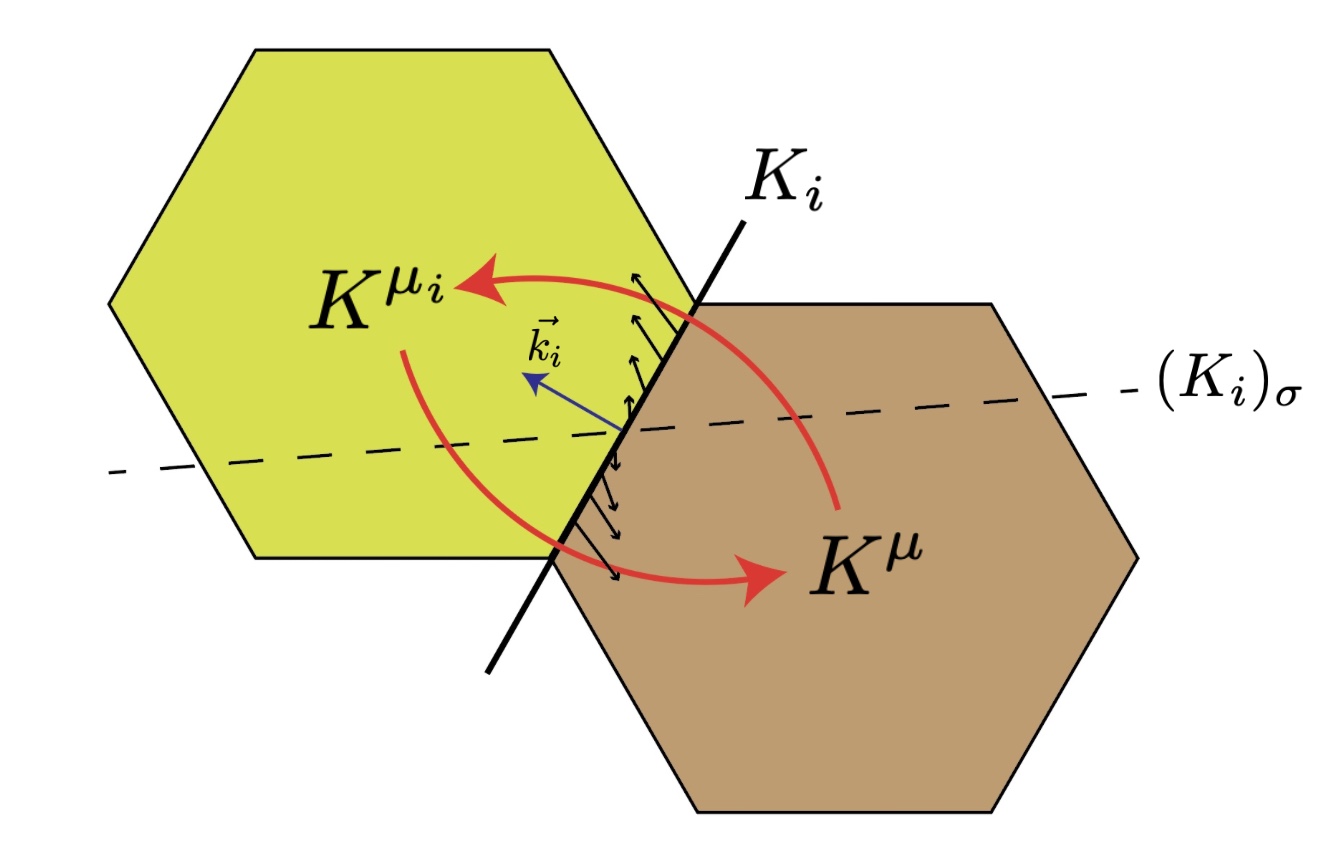}
\vspace{.1in}
\caption[State transition graph construction]{{\bf State transition graph construction.} This diagram depicts the approach towards labeling directed edges of the state transition graph.  The hyperplane $(K_i)_{\sigma}$ divides the hyperplane $K_i$ into regions of inward and outward flow.  In the diagram here outward flows exist on both sides of this chamber wall so edges are drawn in both directions.}
\label{fig:ci2}
\end{center}
\vspace{-.2in}
\end{figure}

To assign an edge through a chamber wall associated with $K_i$, where $K^{\mu_i}$ is the adjacent chamber, we simply turn the associated constraint into an equality to restrict to the face.  We additionally apply Proposition~\ref{prp:sep} to the normal vector of $K_i$, to also add the constraint: 

    $$(K_i)_{\sigma}:=\sum_{\ell=1}^n\left(-k_{i\ell}+\sum_{j\in\sigma}k_{ij} W_{j\ell}\right)x_{\ell}+\sum_{j\in\sigma}k_{ij} \theta_j>0.$$

We see if this linear program has a solution and if so, we have an edge in one direction.  Then we can see if there is a solution when the constraint is changed to:

$$\sum_{\ell=1}^n\left(-k_{i\ell}+\sum_{j\in\sigma}k_{ij} W_{j\ell}\right)x_{\ell}+\sum_{j\in\sigma}k_{ij} \theta_j<0.$$

This would check if there is a directed edge in the other direction.  If both linear programs have a solution, then there is a bidirectional edge over that wall of the chamber.  The bidirectional case is depicted in Fig~\ref{fig:ci2}.  Now, which edge is the one pointing out of the chamber and which is the one pointing in depends on the sign of the normal vector.  Without loss of generality, call the linear program corresponding to the outward pointing edge  $K^{\mu}_{i_+}$.  Using this approach for each wall of each chamber, we can build a state transition graph for the hyperplane arrangement generated by $\{K_i\}_{i=1}^m$.  We start by taking an undirected graph showing adjacency between chambers and then iterate through the adjacent pairs of chambers to determine the directed edges of the state transition graph.

\begin{algorithm}[H]\label{alg:stg}
    \caption{State Transition Graph}
    \begin{algorithmic}[1]
        \STATE{$V(G_1)=\{\mu \mid K^\mu \text{ is a chamber of the partition}\}$}
        \STATE{$V(G_2)=\{\mu \mid K^\mu \text{ is a chamber of the partition}\}$}
        \STATE{$E(G_1)=\{(\mu,\mu') \mid K^\mu , K^{\mu'} \text{ share a face} \}$}
        \FOR{$\mu \in V(G_1)$}
            \FOR{$\mu' \in V(G_1)$}
            	\IF{$(\mu,\mu_{i})\in E(G_1)$}
            		\STATE{$K_i=\mu \cap \mu'$}
			\STATE{$\mu_i=\mu'$}
	        		\IF{$K^\mu_{i_+}$ has a solution}
           			 \STATE{$(\mu,\mu')\in E(G_2)$}
        			\ENDIF
		\ENDIF
            \ENDFOR
        \ENDFOR
            \STATE{\textbf{return} $G_2$}

    \end{algorithmic}
\end{algorithm}

\section{Devising a Partition}

A natural first choice of partition would be our standard $H_i/\mathcal{N}_i$ partition.  An implementation of Algorithm 1for this partition is used for the remainder of this chapter..  Let us begin with our two dimensional case, particularly the independent set CTLN.  We first show an undirected graph of the adjacent chambers.  For ease of comparison we have set aside our "canonical" labeling scheme and use the notation $+/-$ for each hyperplane in the arrangement.  The $+$ symbol is used if the chamber is on the side of the hyperplane containing the origin and the $-$ symbol is used if the chamber is on the other side of the hyperplane.  This label will be written in the form $\mathcal{N}_1  \dots  \mathcal{N}_n , H_1 \dots H_n$.  As an example the chamber lying inside the second and third nullcline and within all three of the ReLU hyperplanes will be labelled $-++,+++$.  Using our algorithm, we compute the state transition graph and then collapse the simply connected components.  

Beginning with the two-dimensional independent set, we recover our analytical results without much issue (Fig~\ref{fig:nullstg}).  However when we try to move to higher dimensional cases, we find things both intriguing and problematic (Fig~\ref{fig:destg} and Fig~\ref{fig:clistg}).  The intriguing aspect is the sheer diversity of state transition graph structures that we find.  

\begin{figure}[!ht]
\begin{center}
\vspace{.1in}
\includegraphics[width=6.25in]{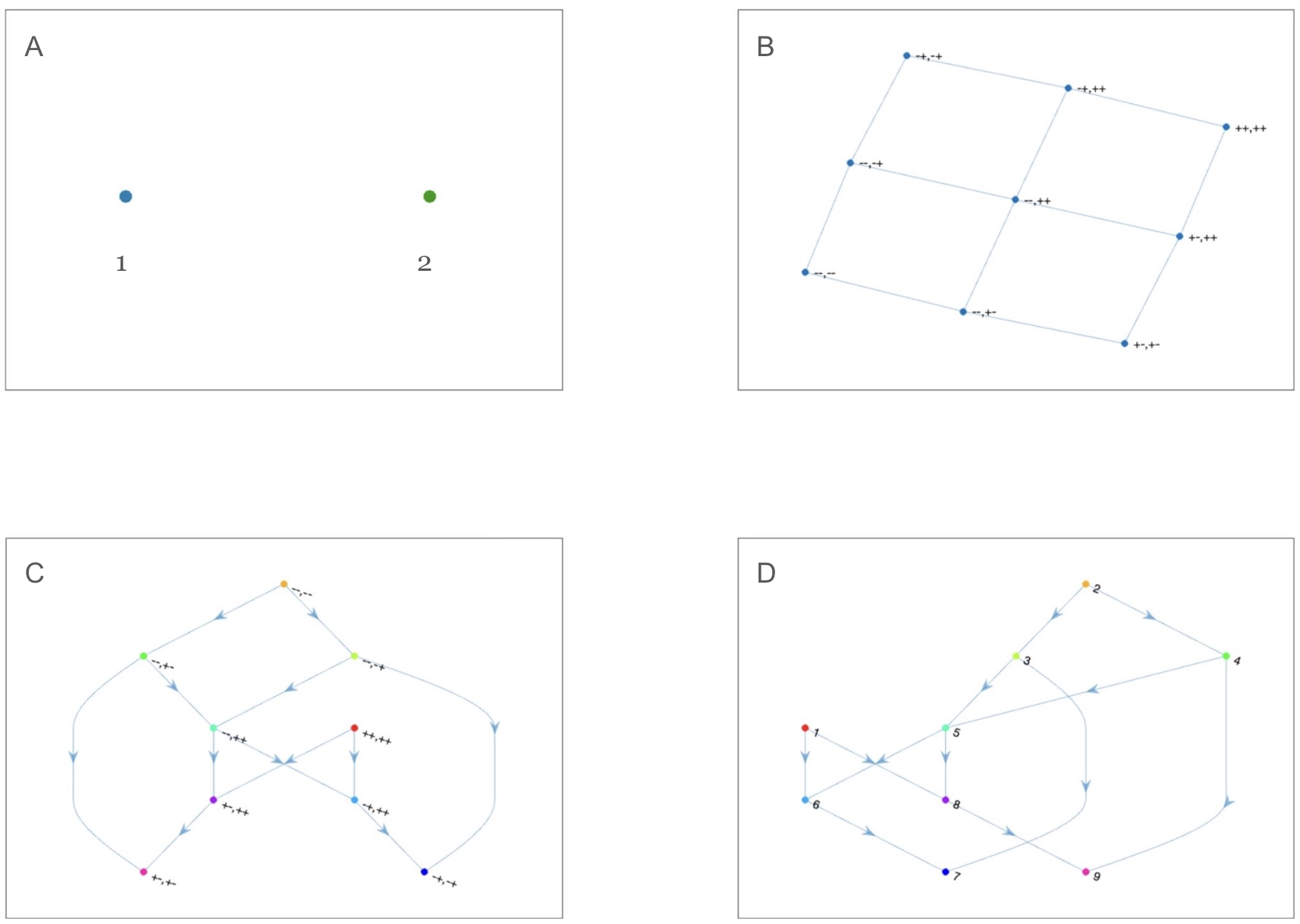}
\vspace{.1in}
\caption[State transition graph of independent set CTLN]{{\bf State transition graph of independent set CTLN.} (A) This CTLN is derived from the independent set of two neurons with parameters $\delta=0.5$, $\varepsilon=0.25$, and $\theta=1$.  (B) This undirected graph shows the the chambers of the $H_i/\mathcal{N}_i$ partition with the undirected edges representing a shared face of the polytope chambers.  (C)  The directed state transition graph color coded to strongly connected components.  (D) The condensed graph of strongly connected components.}
\label{fig:nullstg}
\end{center}
\vspace{-.2in}
\end{figure}

\begin{figure}[!ht]
\begin{center}
\vspace{.1in}
\includegraphics[width=6.25in]{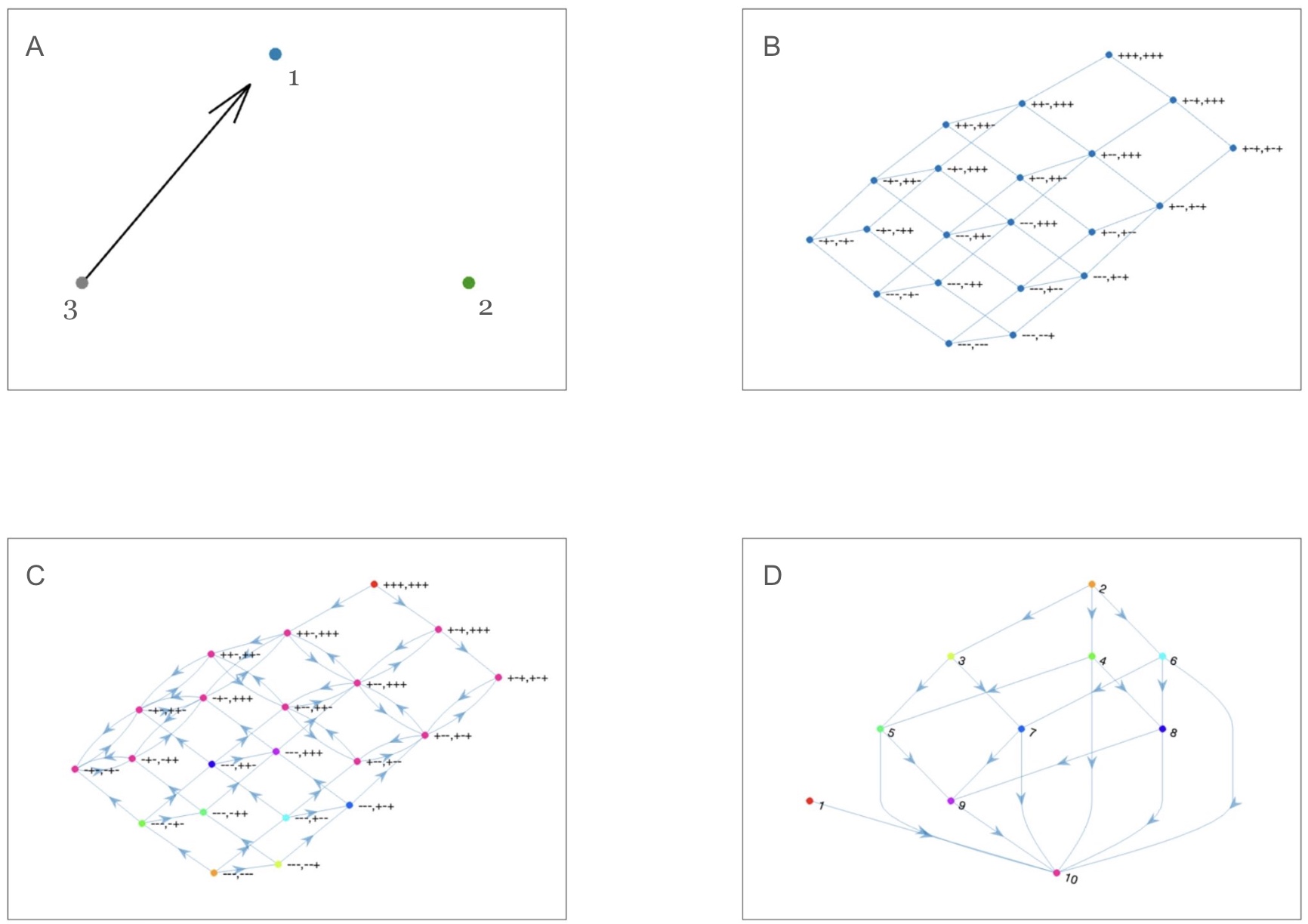}
\vspace{.1in}
\caption[State transition graph of decoy effect CTLN]{{\bf State transition graph of decoy effect CTLN.} As before but with the asymmetric three neuron directed acyclic graph depicted in (A).  Recall that this graph corresponds to the Decoy Effect construction.  Parameters remain $\delta=0.5$, $\varepsilon=0.25$, and $\theta=1$.}
\label{fig:destg}
\end{center}
\vspace{-.2in}
\end{figure}

\begin{figure}[!ht]
\begin{center}
\vspace{.1in}
\includegraphics[width=6.25in]{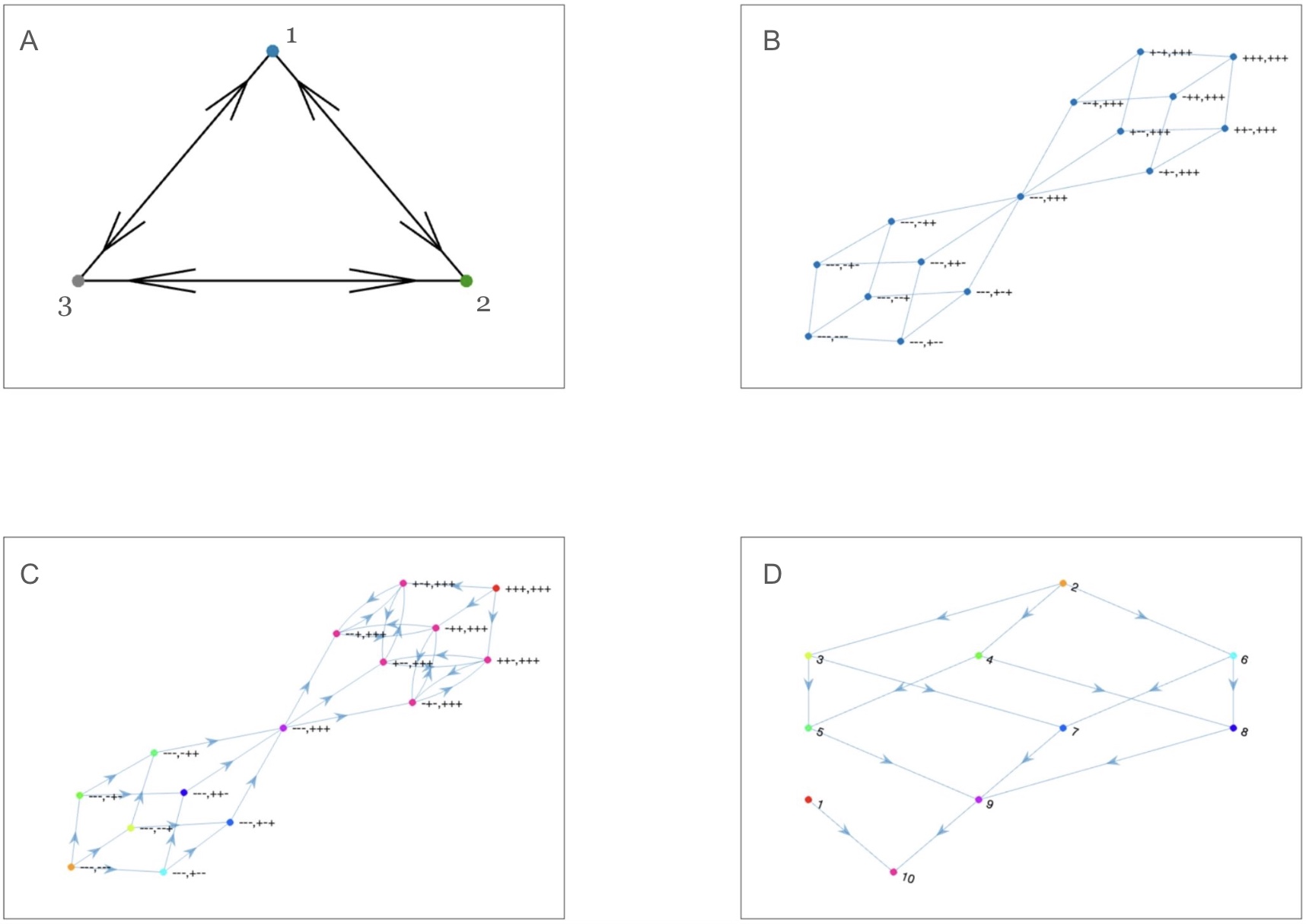}
\vspace{.1in}
\caption[State transition graph of clique CTLN]{{\bf State transition graph of clique CTLN.} As before, but the graph in (A) is now a three neuron clique.  Parameters remain $\delta=0.5$, $\varepsilon=0.25$, and $\theta=1$.}
\label{fig:clistg}
\end{center}
\vspace{-.2in}
\end{figure}

We analytically know that the decoy effect DAG CTLN in Fig~\ref{fig:destg} should have multiple attractors as it is a DAG with multiple sinks, but only a single sink is produced in the condensed state transition graph.  This partition is simply inadequate to distinguish attractors let alone classify them.

Looking at more examples, we find that while the state transition graphs vary considerably and display a rich diversity of structure, they nonetheless only seem to produce one attracting strongly connected component.  Looking at the chambers which compose that sink node in the condensed graph, they are those which are in between nullclines, i.e. which are on the $+$ side of some nullclines and the $-$ side of others.  By and large it appears that these chambers have enough bidirectional edges between that they are collapsed into one strongly connected component which is the single attracting isolating neighborhood in larger CTLNs.  The various attractors of the CTLN seem to lurk within this union and collectively are part of the invariant set for which this union of chambers is an isolating neighborhood.  This is aligned with prior theoretical results on competitive TLNs which indicate that the union of mixed sign nullcline chambers is attracting (Theorem 9.1 in \cite{dagdyn}).  Unfortunately, it seems that even this computational approach often doesn't tell us much more about the attractors of the CTLN than that.  Still, the state transition graphs do capture a kind of combinatorial dynamics and help us to understand how the trajectories beginning outside the attracting set approach it.

While interesting, if our state transition graph is not able to separate out the attractors, it definitely does not give us information about the basins of attraction.  If we want a combinatorial dynamics that guides us to better understanding the basins, we would need something different.

We necessarily need separation into the linear systems.  This means we begin our partition with the hyperplanes $\{H_i\}_{i=1}^n$.  However, as we saw, this is not refined enough even in the two neuron case.  What we need to do is introduce a new set of hyperplanes that that reduce the bidirectional edges over the $R_\sigma$ chamber boundaries.  What we can do is add the hyperplanes which separate the inward and outward flows through the chamber walls $H_i$

\begin{corr}\label{corr:fsep}
   The points in $R_\sigma$ such that the vector field given by the linear ODE system $L_\sigma$ is orthogonal to the normal vector of $H_i$ is $R_\sigma \cap B_i^\sigma$ where:

   $$B_i^\sigma:=\sum_{k=1}^n\left(-W_{ik}+\sum_{j\in\sigma}W_{ij}W_{jk}\right)x_k +\sum_{j\in\sigma}W_{ij}\theta_j=0$$
\end{corr}

So now, our hyperplane arrangement consists of $\{H_i\}_{i=1}^n$ and also for each $H_i\cap R_{\sigma}$ chamber wall with a bidirectional edge, we also include $B_i^{\sigma}$ to separate the wall into a half where the the vector field is "inward" facing and a half where the vector field is "outward" facing.  While all the bidirectional edges with respect to the $H_i$ hyperplanes have been resolved, each $B_i^{\sigma}$ is a new chamber wall which can have a bidirectional edge it its own right.  Of course we can inductively apply the same process to the new hyperplane, terminating when we no longer have bidirectional edges.  But a problem arises.  At a fixed point $x_{\sigma}$, the vector field is zero, and so is always orthogonal to any normal vector.  Therefore, chambers with fixed points will continue to be endlessly partitioned and the process will never terminate.

There remains one hope.  The hyperplane $B_i^{\sigma}$ is unimportant in and of itself, but rather it is the intersection $B_i^{\sigma}\cap H_i$ and $\operatorname{codim}(B_i^{\sigma}\cap H_i)$ is generally 2.  This leaves one dimension of freedom which allows us to draw different hyperplanes which have the same intersection with $H_i$. 

\begin{prp}\label{prp:simpsep}
    Let $B$ be the hyperplane given by $b_0+\sum_{k=1}^n b_k x_k = 0$.  Then $B \cap B_\sigma = B \cap B^*_\sigma$ where:

    $$B^*_\sigma=\sum_{k=1}^n\left(\sum_{j\in\sigma}b_j W_{jk}\right)x_k+b_0+\sum_{j\in\sigma}b_j \theta_j=0$$
    
    and
    
    $$B_\sigma:=\sum_{k=1}^n\left(-b_k+\sum_{j\in\sigma}b_j W_{jk}\right)x_k+\sum_{j\in\sigma}b_j \theta_j=0$$
\end{prp}

\begin{proof}
On $B$, we have the identity $b_0+\sum_{k=1}^n b_k x_k = 0$.  We use this to replace the expression $-\sum_{k=1}^n b_k x_k$ in $B_{\sigma}$ with $b_0$.  This yields the new $B_{\sigma}^*.$
\end{proof}

\begin{corr}\label{corr:ssep}
   For a TLN chamber $R_{\sigma}$, let $H_i\cap R_{\sigma}$ be one of the chamber walls.  Then, $(B_i^{\sigma})^*$ is a hyperplane such that $H_i\cap (B_i^{\sigma})^*$ separates the inward and outward flows of $H_i\cap R_{\sigma}$.

   $$(B_i^\sigma)^*:=\sum_{k=1}^n\left(\sum_{j\in\sigma}W_{ij}W_{jk}\right)x_k + \theta_i +\sum_{j\in\sigma}W_{ij}\theta_j=0$$
\end{corr}

Proposition~\ref{prp:simpsep} gives a new set of hyperplanes which can be used instead of $B_i^{\sigma}$.  This yields a different partition and potentially a different state transition graph.  However, we find that chambers with saddle points still seem to require continual partitioning.  Again, this is not terribly surprising upon some thought because as long as both the stable and unstable manifold of the saddle point are intersecting a hyperplane, then there would be a bidirectional edge through it.  

While an unsatisfying conclusion, this is ultimately where this analysis stands right now.  Hopefully there exists some way to exploit this additional dimension of freedom to obtain a hyperplane arrangement which aligns with either the stable or unstable manifold within a chamber and produces a viable state transition graph, but it remains out of reach at this time.  

To summarize this chapter, we have explored computer assisted ways of extending our combinatorial dynamics into higher dimensions, but what we have primarily found are their limitations.  Our takeaway is that approximating full basins of attraction in higher dimensional TLNs through combinatorial dynamics is a challenging task and now we consider other approaches.

\chapter{Localized Path Polynomials and the Properties of DAG CTLNs}

While the attempts to extend a combinatorial dynamics of TLNs into higher dimensional proved unsuccessful, we did see in the introduction that the attractors of DAG CTLNs can be controlled by fixing the number of sinks.  The following two theorems establish this concretely:

\begin{thm}[Theorem 10.2 in \cite{dagdyn}]
    A CTLN derived from a directed acyclic graph $G$ will have no dynamic attractors.
\end{thm}

\begin{thm}[Rule 7 in \cite{gr}]\label{thm:dagrule}
    The set of fixed points of a CTLN derived from a directed acyclic graph $G$ will be supported on sinks and the unions of sinks.  

    $$\operatorname{FP}(G)=\{ \bigcup s_i \mid  s_i \text{ is a sink in $G$} \}$$
    
    Moreover, each stable fixed point will be supported on exactly one of the sinks.
\end{thm}

\begin{figure}[!ht]
\begin{center}
\vspace{.1in}
\includegraphics[width=6.5in]{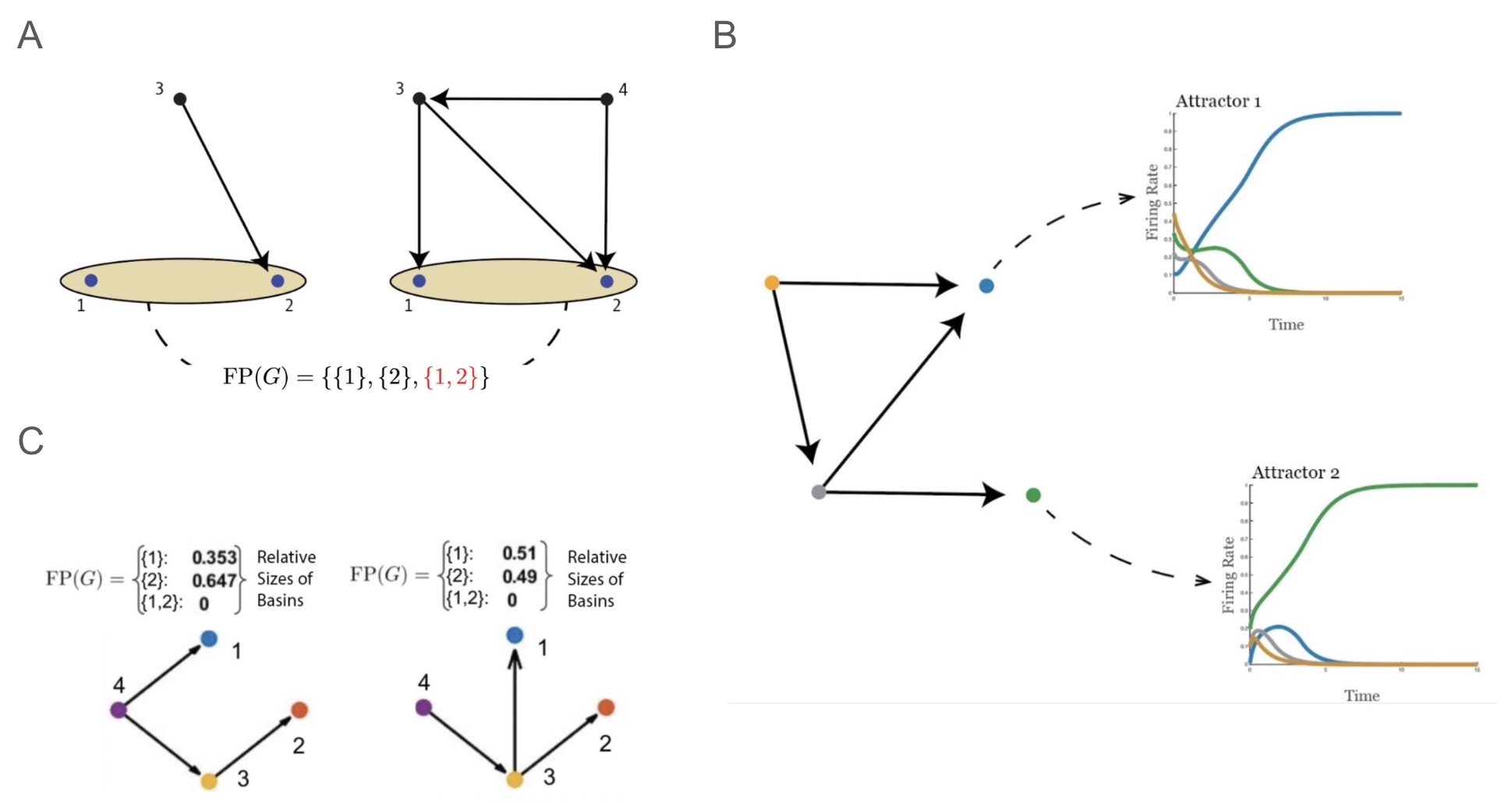}
\vspace{.1in}
\caption[Attractors and basins of attraction in DAG CTLNs]{{\bf Attractors and basins of attraction in DAG CTLNs.}  (A) Directed acyclic graphs have fixed points supported only on sinks and the unions of sinks.  Those which are the unions of sinks will be unstable  (B) Only the fixed points supported on the sinks themselves yield attractors, one for each sink.  (C) As the sinks control the attractors of DAG CTLNs, the only role of the rest of the graph is in shaping the basins of attraction.  The relative sizes of the basins of attraction were found through a Monte Carlo approach by numerically simulating a random set of initial conditions and tracking their trajectories, seeing what fraction converge to each attractor.}
\label{fig:dagrule}
\end{center}
\vspace{-.2in}
\end{figure}

Once the set of sinks is fixed, so are the attractors, with one for each sink.  The unstable saddle points supported on the unions of sinks are associated with the seperatrices which serve as the boundaries between the basins of attraction (Fig~\ref{fig:dagrule}A-B).  Once the sinks are fixed, altering the rest of the network allows us to shape the basins of attraction (Fig~\ref{fig:dagrule}).  

These basins can be highly complex and non-trivial as shown in the numerical simulation in Fig~\ref{fig:MCDAG}.  In Fig~\ref{fig:MCDAG}B, it is clear that the choice of $x_3^0$ and $x_4^0$ are deeply connected to the likelihood of being in either basin.  In addition, even though there is only a single path from the source to each of the sinks, the longer path seems to have a greater biasing effect.  So it seems that the length of the paths from source to sink may have some role in the skewing of basins.  The directed graph in Fig~\ref{fig:MCDAG}A modifies the previous DAG architecture by having node 4 connect directly to the sink 2 rather than indirectly via 3.  A quick look at the relative areas shows that this seems to result in even greater shifting of the basins.  It seems likely from this that the basins of attraction for DAG CTLNs factor in the full extent of the network and understanding them will entail unraveling its global dynamics.

\begin{figure}[!ht]
\begin{center}
\vspace{.1in}
\includegraphics[width=6.5in]{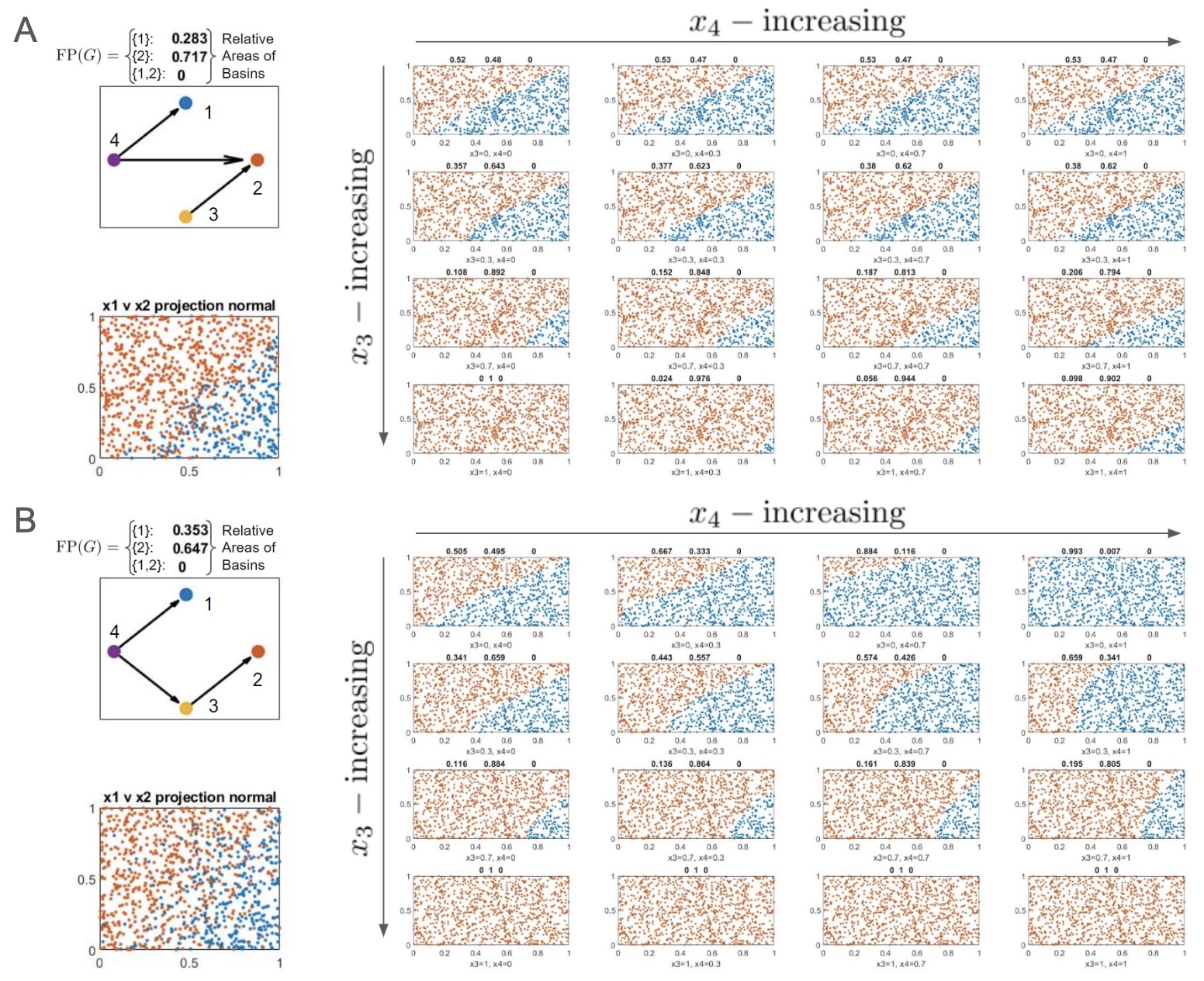}
\vspace{.1in}
\caption[Simulation of basins of attraction in DAG CTLNs]{{\bf Simulation of basins of attraction in DAG CTLNs.} A Monte Carlo simulation of the basins of attraction for two DAG CTLNs.  As before, we randomly sample initial conditions for a CTLN and we color code that point in state space according to the sink attractor to which it converges.  Each slice is a cross section of the state space, and so a cross section of the basins, which fix $x_3^0$ and $x_4^0$.}
\label{fig:MCDAG}
\end{center}
\vspace{-.2in}
\end{figure}

Is there any way that we can take advantage of the structure of DAGs to reveal more about the dynamics of these systems?  What we will show in this chapter is that we can use the combinatorial structure of the DAG to analytically find the solutions of the linear systems, $L_{\sigma}$, composing the CTLN.  The key constructions linking them are \textit{localized path polynomials}.

\begin{ddd}

    For a DAG G of size n, let the vertices be numbered from 1 to n i.e. $V(G)=[n]$.  Then, the \textbf{i-th localized path polynomial}, $p^G_i(z)$, is defined to be: 
    $$p^G_i(z)=1+\sum_{k=1}^n n^i_k z^k$$
    where $n_k^i$ is the number of paths to $i$ of length $k$ (finite because G is acyclic).  Since $G$ is acyclic, $deg(p_i^G(z))$ is finite.
    
\end{ddd}

\begin{figure}[!ht]
\begin{center}
\vspace{.1in}
\includegraphics[width=5.75in]{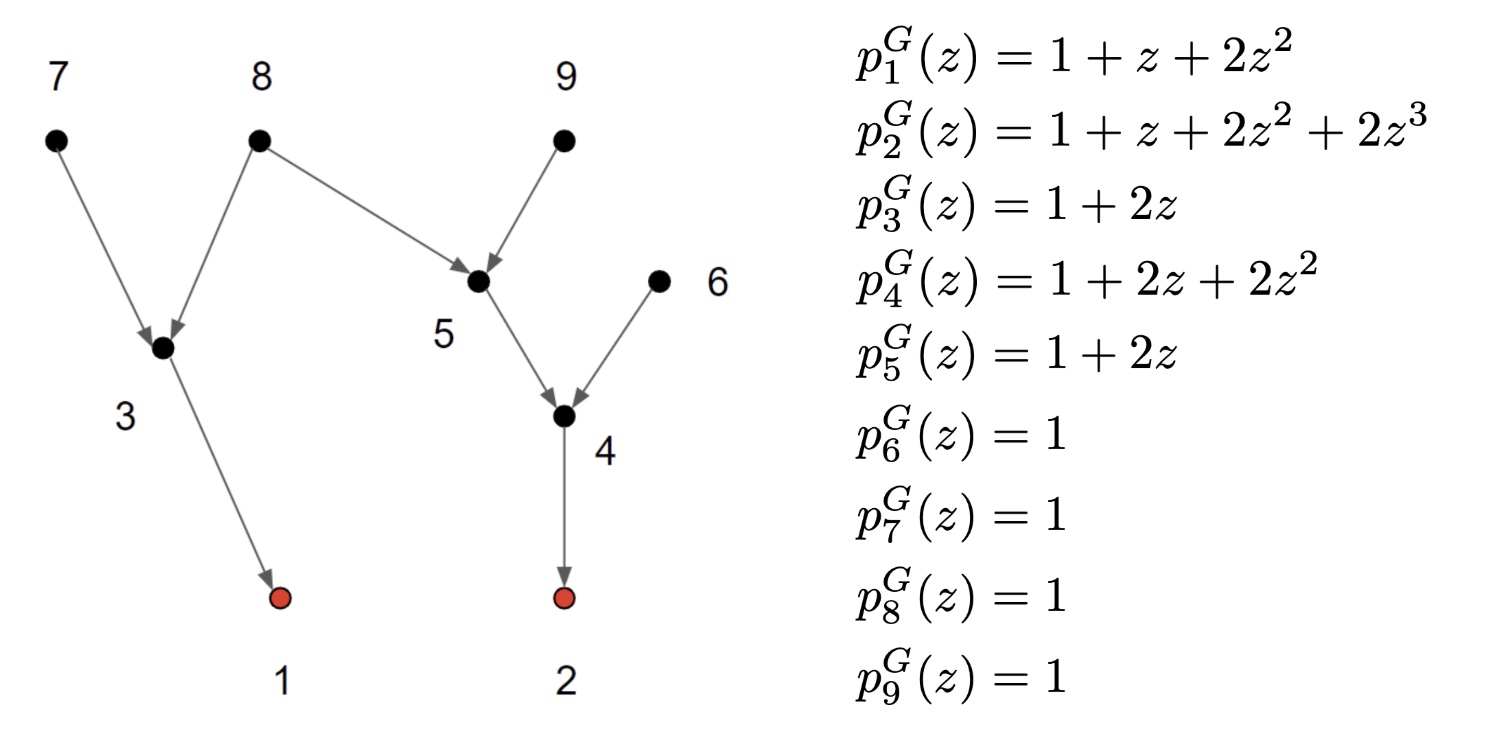}
\vspace{.1in}
\caption[Localized path polynomials]{{\bf Localized path polynomials.}  An example of a DAG CTLN with its associated set of localized path polynomials.}
\label{fig:pathpolys}
\end{center}
\vspace{-.2in}
\end{figure}

\begin{rmk}
The localized path polynomial construction is only of finite degree for DAGs as the presence of a cycle would give the vertices paths of arbitrarily large length.
\end{rmk}

Consider Fig~\ref{fig:pathpolys} where we have depicted a DAG and its associated set of localized path polynomials, one for each vertex. 

\begin{ddd}
Let G be a DAG such that $|G|=n$ and let $\sigma \subseteq [n]$.  Define the \textbf{$\sigma$-path function} $\Vec{p_{\sigma}}: \mathbbm{R}\rightarrow \mathbbm{R}^n$ to be:

$$(\Vec{p_{\sigma}}(z))_i = p_i^{G|_{\sigma}}(z) \text{ } \forall i \in \sigma \text{ and } (\Vec{p_{\sigma}}(z))_i = 0 \text{ } \forall i \not\in \sigma.$$

\end{ddd}

Many of the key properties of the linear systems $L_{\sigma}$ comprising a DAG CTLN can be expressed in terms of localized path polynomials.  To show this, we first transform the matrices $(-I+W)|_{\sigma}$ into a more workable form.  Taking the function $g(x)=-\frac{1}{1+\delta}(x-\delta)$, we have that $B_{\sigma}=g((-I+W)|_{\sigma})=\mathbbm{1}\mathbbm{1}^T-\frac{\varepsilon+\delta}{1+\delta}A|_{\sigma}$ where $A|_{\sigma}$ is the adjacency matrix of the subgraph $G|_{\sigma}$.  By the Spectral Mapping Theorem, the spectrum of $(-I+W)|_{\sigma}$ and $B_{\sigma}$, denoted $\rho ((-I+W)|_{\sigma})$ and $\rho (B_{\sigma})$ respectively, are related in the same way:

\begin{center}
    $\rho (B_{\sigma})= g(\rho ((-I+W)|_\sigma))$.
\end{center}

Additionally, it is not difficult to see that they will share the same eigenvectors.  The insight that undergirds many of the results in this chapter is that through this simple transformation, the CTLN matrices $(-I+W)|_{\sigma}$ can be studied as rank one updates to scaled adjacency matrices.  Recall that the general solution of a non-degenerate, diagonalizable linear system of equations is written in terms of its eigenvectors, eigenvalues, and its fixed point in the following way:

$$\Vec{x}(t)=\sum_{i=1}^n c_i\Vec{v}_i e^{\lambda_i t} +x^*$$

where $\Vec{v_i}$ and $\lambda_i$ are the eigenvectors and eigenvalues of the associated matrix with $x^*$ being the fixed point of the linear system.

The theory developed in this chapter will allow us to find the general solutions of the linear systems $L_{\sigma}$ when the subgraph $G|_{\sigma}$ is an \textit{analytic DAG}.

\begin{ddd}
Let G be a DAG and V(G) its vertex set.  A pair of vertices $(i,j)\in V(G)\times V(G)$ is said to be \textbf{simply embedding} if, $\forall k \in V(G), \text{ } i\rightarrow k \iff j\rightarrow k$.  The \textbf{simply embedding set}, $\operatorname{SE}(G)$, is defined to be: 

$$\operatorname{SE}(G):=\{(i,j)\in V(G) \times V(G) \mid i\neq j \text{ and, } \forall k \in V(G), \text{ } i\rightarrow k \iff j\rightarrow k \}.$$
\end{ddd}

Simply embedding pairs are pairs of vertices which treat the rest of the vertices on the graph identically (Figure~\ref{fig:sepair}).  We associate with each pair $(i,j)$ a vector: $e_i-e_j$.  We are interested in subsets of the simply embedding set where the associated vectors are linearly independent.

\begin{figure}[!ht]
\begin{center}
\vspace{.1in}
\includegraphics[width=3in]{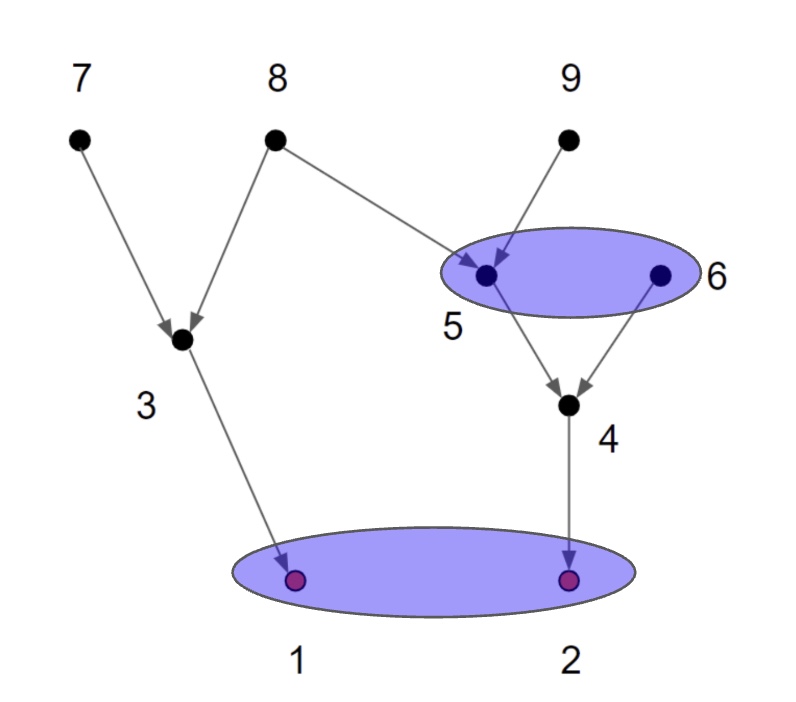}
\vspace{.1in}
\caption[Simply embedding pairs]{{\bf Simply embedding pairs.} Simply embedding pairs are pairs of vertices which treat the rest of the DAG in the same way.  The simply embedding pairs of the earlier DAG are marked and they have edges to precisely the same vertices.}
\label{fig:sepair}
\end{center}
\vspace{-.2in}
\end{figure}

\begin{ddd}

A DAG G is said to be \textbf{analytic} if there exists $\widetilde{\operatorname{SE}}(G)\subseteq \operatorname{SE}(G)$, a subset such that $|\widetilde{SE}(G)|=n-(m+1)$ and $\{e_i-e_j \mid (i,j)\in \operatorname{\widetilde{SE}}(G)\}$ is a linearly independent set where $m$ is the maximum path length in G.
\end{ddd}

The results of this chapter are summarized within the following theorem:

\begin{thm}\label{thm:final}
Let G be a DAG and let W be the weight matrix for an associated CTLN with parameters $\varepsilon, \delta, \theta$.  Let $\sigma \subseteq [n]$ be such that $G|_{\sigma}$ is analytic, $(-I+W)|_{\sigma}$ is diagonalizable, and the polynomial:

$$f(\lambda)=(-\lambda+\delta)^{m+1}-(1+\delta)(|\sigma|(-\lambda+\delta)^{m}+n^{\sigma}_1c(-\lambda+\delta)^{m-1}+...+n^{\sigma}_m c^m)$$

has distinct roots $\{\lambda_k\}_{k=1}^{m+1}$ where $c=-\varepsilon - \delta$, $n^{\sigma}_{j>0}$ is the number of paths of length $j$ in $G|_{\sigma}$ and $m$ is the maximum path length in $G|_{\sigma}$.  

Then, the general solution of $L_{\sigma}$ is of the following form:

$$\Vec{x}(t)=\sum_{k=1}^{m+1} c_k \Vec{p_{\sigma}}(\alpha_k) e^{\lambda_{k} t} +\sum_{(i,j)\in\operatorname{\widetilde{SE}}(G|_{\sigma})} c_{(i,j)}(e_i-e_j)e^{\delta t} +\sum_{k=1}^{n-|\sigma|} c_k \Vec{v}_k e^{-t}+ \Vec{p_{\sigma}}(\beta)\Gamma (\sigma)$$

where $n=|G|$, $\alpha_k=\dfrac{\varepsilon+\delta}{\lambda_k -\delta}$, $\beta=\dfrac{-\varepsilon-\delta}{\delta}$, and $\Gamma (\sigma) = \dfrac{\theta}{-\delta +(1+\delta) \sum_{j\in \sigma} p_j^{G|_{\sigma}} (\beta)}$.
\end{thm}

\section{Eigenvalues and Eigenvectors of $L_{\sigma}$}

We begin by studying the relationship between localized path polynomials of the graph $G|_{\sigma}$ and the eigenvalues/eigenvectors of $(-I+W)|_{\sigma}$.  First, we determine the characteristic polynomial for the matrix.  To do this we will make use of the Matrix Determinant Lemma, which we restate here.

\begin{lem}[Matrix Determinant Lemma: Lemma 1.1 in \cite{MDL}]
Suppose that A is an invertible square matrix and $u$ and $v$ are column vectors.  Then, the determinant of $uv^T + M$ is given by:

$$\det(uv^T + M)=(1+v^T M^{-1} u)\det(M).$$
\end{lem}

We use this lemma to find the characteristic polynomial of matrices of the form $\mathbbm{1}\mathbbm{1}^T +cA$ where $A$ is the adjacency matrix of a DAG.  An important property that we will need to make use of is the nilpotency of DAG adjacency matrices.

\begin{prp}\label{prp:charpoly}
Let B be a matrix derived from directed acyclic graph G with maximum path length $m$ and adjacency matrix A such that:

\begin{center}
    $B=\mathbbm{1}\mathbbm{1}^T+cA$.
\end{center}

Then, the characteristic polynomial of B is:

$$f(\lambda)=\lambda^{n-(m+1)}(\lambda^{m+1}-n_0\lambda^{m}-n_1c\lambda^{m-1}-...-n_{m-1}c^{m-1}\lambda-n_m c^m)$$

where $n_0=|G|$ and $n_{j>0}$ is the number of paths of length $j$ in $G$.

Also, if $c<0$, then the real roots of $\lambda^{m+1}-n_0\lambda^{m}-n_1c\lambda^{m-1}-...-n_{m-1}c^{m-1}\lambda-n_m c^m$ are positive.
\end{prp}

\begin{proof}
The characteristic polynomial is $f(\lambda)=\det(\mathbbm{1}\mathbbm{1}^T +cA - \lambda I)$.  We apply the matrix determinant lemma taking $u,v=\mathbbm{1}$ and $M=cA-\lambda I$.  

So, we have $f(\lambda)=(1+\mathbbm{1}^T (cA-\lambda I)^{-1} \mathbbm{1})\det{(cA-\lambda I)}$.

Note that a DAG can be indexed from sink to source according to its topological ordering i.e. such that if $i\geq j$ then $j \not\rightarrow i$.  In this indexing, the DAG's adjacency matrix is strictly upper triangular.  Then, this means that there exists a matrix $P$ such that $P A P^{-1}$ is strictly upper triangular.  Since $P (\lambda I) P^{-1}=\lambda I$ for any invertible $P$: 

$$\det{(cA-\lambda I)}=\det{(P)}\det{(cA-\lambda I)}\det{(P^{-1})}$$
$$=\det{(cPAP^{-1} -P (\lambda I) P^{-1})}=\det{(cPAP^{-1} -\lambda I)}.$$

Then, $cPAP^{-1} -\lambda I$ is upper triangular with $-\lambda$ on the diagonals, so $\det{(cA - \lambda I)} =-\det{(\lambda I - cA)} =-\lambda^n$.  We conclude from this that:

$$f(\lambda)= -\lambda^n (1+\mathbbm{1}^T (cA-\lambda I)^{-1} \mathbbm{1}).$$

We will now further analyze $((cA-\lambda I)^{-1})= -\frac{1}{\lambda}(I-\frac{c}{\lambda}A)^{-1}$.  As the adjacency matrix of a DAG is nilpotent with index $m+1$, $(\frac{c}{\lambda}A)^{m+1}=0$ so $I-(\frac{c}{\lambda}A)^{m+1}=I$.

Then, since $1-x^{m+1}=(1-x)\sum_{i=0}^m x^i$, we have:

$$I=I-\left(\frac{c}{\lambda}A \right)^{m+1} = \left( I- \left( \frac{c}{\lambda}A \right) \right) \sum_{i=0}^m \left( \frac{c}{\lambda}A \right)^i.$$

Multiplying by $(I-\frac{c}{\lambda}A)^{-1}$ on both sides, we obtain:

$$\left( I-\frac{c}{\lambda}A \right)^{-1} =  \sum_{i=0}^m \left( \frac{c}{\lambda}A \right)^i.$$

So, we can rewrite the expression $1+\mathbbm{1}^T (cA-\lambda I)^{-1} \mathbbm{1}$ as $1-\sum_{i=0}^m \frac{c^i}{\lambda^{i+1}} \mathbbm{1}(A^i)\mathbbm{1}^T.$

Finally, recognizing that $\mathbbm{1}(A^i)\mathbbm{1}^T = n_i$, we conclude:

$$f(\lambda)=-\lambda^n \left(1-\sum_{i=0}^m \frac{c^i}{\lambda^{i+1}}n_i \right)=-\lambda^{n-(m+1)}(\lambda^{m+1}-n_0\lambda^{m}-n_1c\lambda^{m-1}-...-n_{m-1}c^{m-1}\lambda-n_m c^m).$$

Without loss of generality, we change the sign and have the proposed characteristic polynomial.

To find that the real roots of $q(\lambda)=\lambda^{m+1}-n_0\lambda^{m}-n_1c\lambda^{m-1}-...-n_{m-1}c^{m-1}\lambda-n_m c^m$ are positive for $c<0$, we apply Descartes' Rule of Signs.  Notice that for $q(-x)$ there are no variations in this sign (there are two cases here, $m$ even or $m$ odd, but in both cases we end up with no variations in sign).  As there are no negative real roots of $q(\lambda)$, if $\lambda$ is real, $\lambda>0$.

\end{proof}

\begin{corr}\label{corr:charpolyctln}

Let G be a Directed Acyclic Graph and let W be the derived CTLN weight matrix.  Then, let $\sigma \subseteq [n]$ and m the maximum path length of $G|_{\sigma}$.  Then, for $c=-\varepsilon-\delta$, the characteristic polynomial of $(-I+W)|_{\sigma}$ is:

$$f(\lambda)=(-\lambda+\delta)^{|\sigma|-(m+1)}((-\lambda+\delta)^{m+1}-(1+\delta)(|\sigma|(-\lambda+\delta)^{m}+n^{\sigma}_1c(-\lambda+\delta)^{m-1}+...+n^{\sigma}_m c^m)$$

where $n^{\sigma}_{j>0}$ is the number of paths of length $j$ in $G|_{\sigma}$.

Also, the real roots of $(-\lambda+\delta)^{m+1}-|\sigma|(1+\delta)(-\lambda+\delta)^{m}-n^{\sigma}_1c(1+\delta)^2(-\lambda+\delta)^{m-1}-...-n^{\sigma}_m c^m(1+\delta)^{m+1}$ satisfy $\lambda<\delta$.

\end{corr}

Consider what we have demonstrated.  We have made it so that we can read off a characteristic polynomial for the DAG CTLN matrices $(-I+W)|_{\sigma}$ from the combinatorial structure of the subgraph $G|_{\sigma}$.  Moreover, we have also demonstrated that the real eigenvalues are less than or equal to $\delta$.  We will now take this result, and use it to express eigenvectors in terms of the localized path polynomials of $G|_{\sigma}$.

\begin{prp}\label{prp:dagevs}
Let B be a matrix derived from directed acyclic graph G with adjacency matrix A such that:

$$B=\mathbbm{1}\mathbbm{1}^T+cA.$$

Let $\lambda$ be an eigenvalue of B such that $\lambda \neq 0$.  Then, the following is an associated eigenvector $\Vec{v}$ of $\lambda$

$$\vec{v}_j=p_j^G \left(\frac{c}{\lambda}\right).$$ 

\end{prp}

\begin{proof}

If $\lambda \neq 0$, it satisfies: 

$$\lambda^{m+1}-n_0\lambda^{m}-n_1c\lambda^{m-1}-...-n_{m-1}c^{m-1}\lambda-n_m c^m=0$$.  

Now, rearranging this, we have:

$$\lambda^{m+1}=n_0\lambda^{m}+n_1c\lambda^{m-1}+...+n_{m-1}c^{m-1}\lambda+n_m c^m$$.

Then, we divide on both sides by $\lambda^m$ and obtain:

$$\lambda=n_0+n_1\left(\frac{c}{\lambda}\right)+...+n_{m-1}\left(\frac{c}{\lambda}\right)^{m-1}+n_m \left(\frac{c}{\lambda}\right)^m=\sum_{i=1}^n p_i^G \left( \frac{c}{\lambda} \right)$$.

We directly insert and verify that, $\forall i$, $\sum_{j=1}^n B_{ij}\Vec{v}_{j}=\lambda \Vec{v}_i$.

$$\sum_{j=1}^n B_{ij}\Vec{v}_{j}=\sum_{j=1}^n \Vec{v}_j + c\sum_{k\rightarrow i} \Vec{v}_k = \sum_{j=1}^n p_j^G \left( \frac{c}{\lambda} \right) + c\sum_{k\rightarrow i}p_k^G \left( \frac{c}{\lambda} \right)=\lambda + c\sum_{k\rightarrow i}p_k^G \left( \frac{c}{\lambda} \right).$$

If $k\rightarrow i$, a path to $k$ of length $m$ corresponds to a path to $i$ of length $m+1$.  With this we recognize that $c\sum_{k\rightarrow i}p_k^G \left( \frac{c}{\lambda} \right) = \lambda \left(-1+p_i^G \left( \frac{c}{\lambda} \right)\right).$

Then:

$$\sum_{j=1}^n B_{ij}\Vec{v}_{j}=\lambda + \lambda \left( -1+ p_i^G \left( \frac{c}{\lambda} \right) \right)= \lambda p_i^G \left( \frac{c}{\lambda} \right) = \lambda \Vec{v}_i.$$

\end{proof}

\begin{corr}\label{corr:evs}
Let G be a DAG and W be the weight matrix of an associated CTLN.

Let $\lambda$ be an eigenvalue of $(-I+W)|_{\sigma}$ such that $\lambda \neq \delta$.  Then, the following is an associated eigenvector of the matrix for $\lambda$

$$\vec{v}_j=p_j^G \left(\alpha \right).$$ 

where $\alpha=\dfrac{\varepsilon + \delta}{\lambda -\delta}$.
\end{corr}

Now, the localized path polynomials are only able to capture the eigenvectors when $\lambda \neq \delta$.  How much of a problem does this present?  Since we know the characteristic polynomial of the matrices $(-I+W)|_{\sigma}$, we also know that the algebraic multiplicity of $\lambda=\delta$ is $n-(m+1)$ where $m$ is the maximum path length in the DAG subgraph $G|_{\sigma}$ (Fig~\ref{fig:ev_ex}).  While a way to find all of the associated eigenvectors and generalized eigenvectors is lacking,  there is an obvious way to identify some of them. 

\begin{prp}\label{prp:deltavec}
    Let $G|_{\sigma}$ be a subgraph of a DAG with multiple sinks.  Then, $\delta$ is an eigenvalue of $(-I+W)|_{\sigma}$.  Moreover, corresponding eigenvectors of the form $\vec{v}_j=e_i-e_j$ exist if and only if $i,j$ are independent vertices such that if $k\neq i,j$, then $i\rightarrow k \iff j\rightarrow k$. 
\end{prp}

\begin{proof}
    The graph theoretic condition translates to having identical columns $i,j$ in the matrix $-I+W-\delta I$, which means that $e_i-e_j$ is in ker($-I+W-\delta I$).  As the pair of sinks satisfies the graph theoretic condition, then $e_1-e_2$ is an eigenvector corresponding which means that $\delta$ is an eigenvalue.
\end{proof}

\begin{figure}[!ht]
\begin{center}
\vspace{.1in}
\includegraphics[width=5.75in]{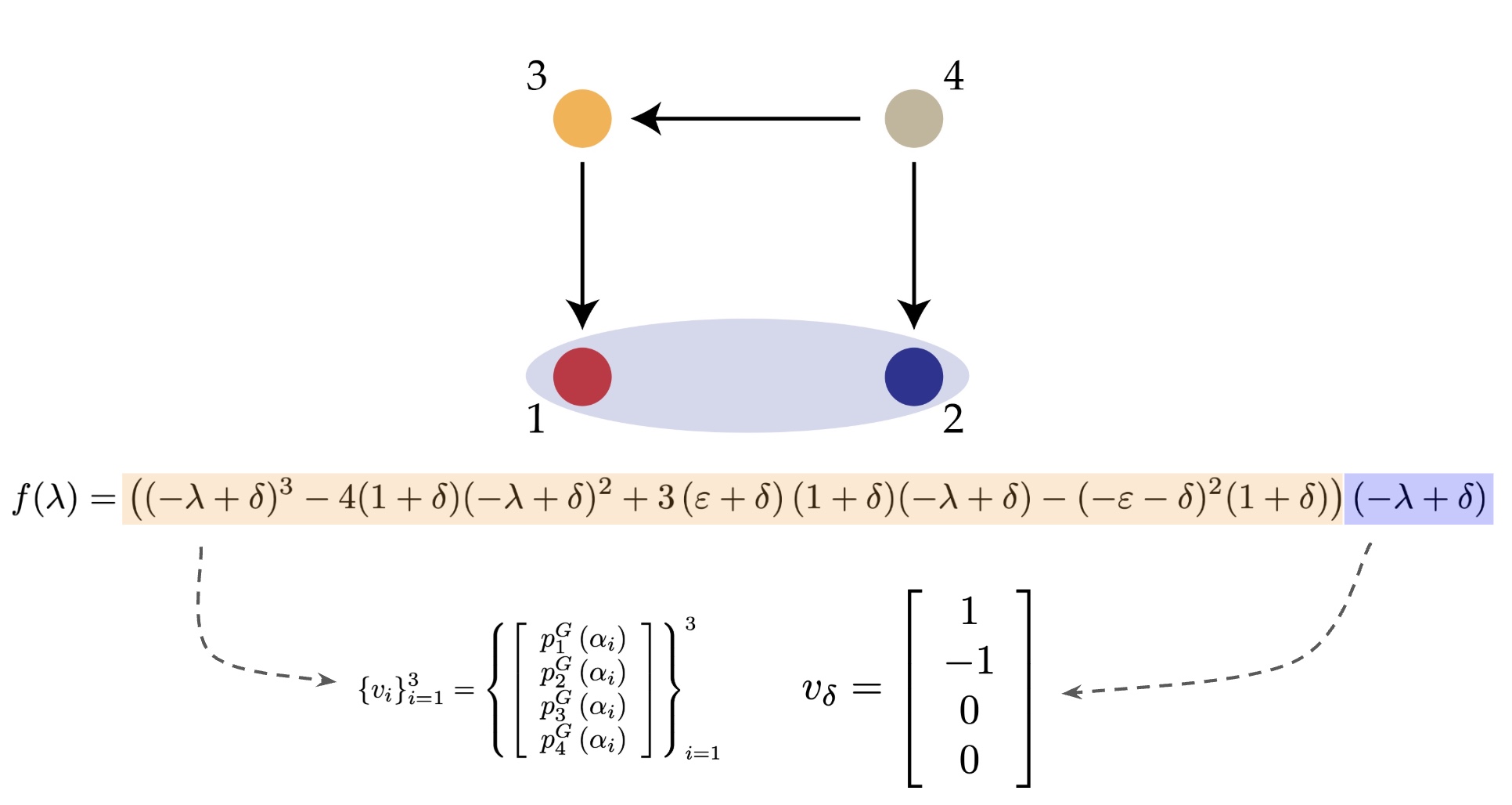}
\vspace{.1in}
\caption[Eigenvectors of DAG CTLNs]{{\bf Eigenvectors of DAG CTLNs.}  An example of a DAG CTLN where four eigenvectors can be found for the $L_{[4]}$ system in the $R_{[4]}$ chamber.  Note that $\alpha_i=\dfrac{\varepsilon +\delta}{\lambda_i - \delta}$.}
\label{fig:ev_ex}
\end{center}
\vspace{-.2in}
\end{figure}

While not a solution in all cases, there are many DAGs where this approach helps us to obtain a full set of linearly independent eigenvectors (Fig~\ref{fig:ev_ex}B).  

\begin{ddd}
We call a CTLN derived from DAG G \textbf{totally analytic} if, $\forall \sigma \subseteq [n]$ such that $|\sigma|>2$, $(-I+W)|_{\sigma}$ is diagonalizable, $G|_{\sigma}$ is analytic, and the polynomial:

$$f(\lambda)=(-\lambda+\delta)^{m+1}-(1+\delta)(|\sigma|(-\lambda+\delta)^{m}+n^{\sigma}_1c(-\lambda+\delta)^{m-1}+...+n^{\sigma}_m c^m)$$

has distinct roots $\{\lambda_k\}_{k=1}^{m+1}$ where $n^{\sigma}_{j>0}$ is the number of paths of length $j$ in $G|_{\sigma}$ and $m$ is the maximum path length in $G|_{\sigma}$.  
\end{ddd}

The observant reader will at this point notice an oversight.  For $\sigma \neq [n]$, the system $L_{\sigma}$ is generally of the form:

$$
\left[
\begin{array}{c}
     \dot x_\sigma  \\
     \dot x_{[n]\backslash\sigma} 
\end{array}
\right]
=
\left[
\begin{array}{c|c}
(-1+W)|_\sigma & C\\
\hline
0 & -I\\
\end{array}
\right]
\left[
\begin{array}{c}
     x_\sigma  \\
     x_{[n]\backslash\sigma} 
\end{array}
\right]
+
\left[
\begin{array}{c}
    \theta \\
     0 
\end{array}
\right]
$$

What we notice here is that there is the additional eigenvalue of $\lambda=-1$ with algebraic multiplicity $n-|\sigma|$.  We have not introduced a way of determining their eigenvectors.  While for $|\sigma|=1$ the solutions are easy to find, this is certainly not true for the other chambers.  There is a way to find these eigenvectors, but to do so will require some more sophisticated machinery.  We will return to this in the final chapter and for now will concern ourselves primarily for the time being with $L_{[n]}$.  In Fig~\ref{fig:comb_resolv}, we present an example of a non-analytic DAG.

\begin{figure}[!ht]
\begin{center}
\vspace{.1in}
\includegraphics[width=4.75in]{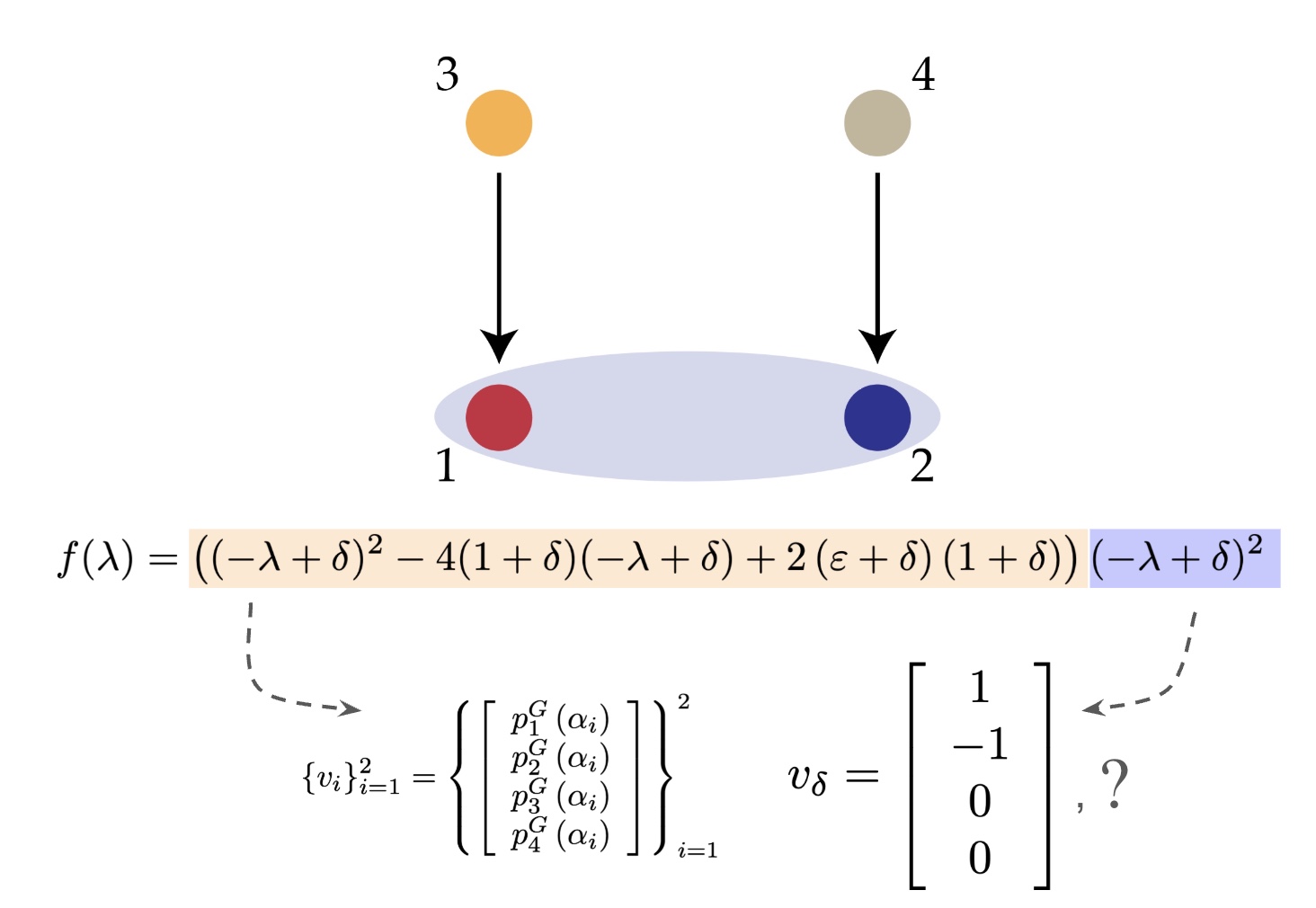}
\vspace{.1in}
\caption[Non-analytic DAGs]{{\bf Non-analytic DAGs.} An example of a DAG CTLN where only three eigenvectors can be found for the $L_{[4]}$ system in the $R_{[4]}$ chamber.}
\label{fig:comb_resolv}
\end{center}
\vspace{-.2in}
\end{figure}

\section{Fixed Points of $L_{\sigma}$: The Chamber Mapping Function}

While eigenvalues and eigenvectors are sufficient to find the homogenous solution for a linear system of ODEs, we will also need a particular solution, i.e. the fixed point.  We will now show the relationship between the localized path polynomial construction and the fixed points of the linear systems $L_{\sigma}$.  The first step in this is introducing a critical lemma 

\begin{lem}[DAG Lemma]\label{lem:DAG}
Let B be a matrix derived from directed acyclic graph G with adjacency matrix A such that:

\begin{center}
    $B=\mathbbm{1}\mathbbm{1}^T+cA$
\end{center}

Then the solution to the following linear system:

\begin{center}
    $B\vec{x}=\gamma\mathbbm{1}+a\vec{x}$
\end{center}

is:

$$x_j=p_j^G\left(\frac{c}{a}\right)\Gamma$$.

where $\Gamma=\dfrac{\gamma}{-a+\sum_{i=1}^n p_i^G(\frac{c}{a})}$
\end{lem}

\begin{proof}
For now, assume $\Gamma$ is defined.  We will show $\forall i\in [n]$, the specified $\Vec{x}$ satisfies the equation $\sum_{j=1}^n B_{ij}x_{j} =\gamma + a x_i$.

First, we expand out the entries of $B$:

$$\sum_{j=1}^n B_{ij}x_{j}=\sum_{j=1}^n x_j + c\sum_{k\rightarrow i} x_k.$$

Next, we will insert our proposed solution $\Vec{x}$:

$$\sum_{j=1}^n x_j + c\sum_{k\rightarrow i} x_k=\Gamma \sum_{j=1}^n p_j^G\left(\frac{c}{a}\right) + c\Gamma \sum_{k\rightarrow i} p_k^G\left(\frac{c}{a}\right).$$

We now process the two sums separately:

$$\Gamma\sum_{j=1}^n p_j^G\left(\frac{c}{a}\right) = \dfrac{\gamma \sum_{j=1}^n p_j^G(\frac{c}{a})}{-a+\sum_{l=1}^n p_l^G(\frac{c}{a})} = \dfrac{\gamma (-a+\sum_{j=1}^n p_j^G(\frac{c}{a}))+a\gamma}{-a+\sum_{l=1}^n p_l^G(\frac{c}{a})}=\gamma + \Gamma a$$ 

$$c\Gamma\sum_{k\rightarrow i} p_k^G\left(\frac{c}{a}\right)= c\Gamma\sum_{k\rightarrow i} \left(1+\sum_{m=1}^n n^k_m \left(\frac{c}{a}\right)^m\right) = \Gamma \sum_{k\rightarrow i} \left(c+\sum_{m=1}^n n^k_m \frac{c^{m+1}}{a^m}\right).$$

Now for the key idea: if $k\rightarrow i$, then a path to $k$ of length $m$ corresponds to a path to $i$ of length $m+1$.  So, then $\sum_{k\rightarrow i} (c+\sum_{m=1}^n n_m^k \frac{c^{m+1}}{a^m}) = \sum_{m=0}^n n_{m+1}^i \frac{c^{m+1}}{a^m}$.  The second sum can then be manipulated further:

$$\Gamma \sum_{k\rightarrow i} \left(c+\sum_{m=1}^n n^k_m \frac{c^{m+1}}{a^m}\right) = \Gamma \sum_{m=0}^n n_{m+1}^i \frac{c^{m+1}}{a^m} = \Gamma a\sum_{m=0}^n n_{m+1}^i \frac{c^{m+1}}{a^{m+1}} = \Gamma a \left(-1+p_i^G\left(\frac{c}{a}\right)\right).$$

At last, we combine the two sums to obtain our desired result.

$$\sum_{j=1}^n B_{ij}x_{j} = \gamma+\Gamma a -\Gamma a  + a p_i^G\left(\frac{c}{a}\right)\Gamma = \gamma + ax_i.$$

We conclude by establishing that $\Gamma$ is defined.  The key issue is whether the denominator is non-zero.  The equation $B\Vec{x}=\gamma \mathbbm{1} + a\Vec{x}$ can be rearranged into $(B-aI)\Vec{x} = \gamma \mathbbm{1}$.  So, we need only confirm that $\operatorname{det}(B-aI) \neq 0 \implies -a+\sum_{i=1}^n p_i^G\left( \frac{c}{a} \right)\neq 0$.  By Lemma~\ref{prp:charpoly}, we see that:

$$\operatorname{det}(B-aI)=a^{n-(m+1)}(a^{m+1}-n_0 a^m - n_1 c a^{m-1}-\hdots-n_{m-1}c^{m-1}a - n_m c^m).$$

Then, this means $a^{m+1}-n_0 a^m - n_1 c a^{m-1}-\hdots-n_{m-1}c^{m-1}a - n_m c^m\neq 0$.  Dividing by $a^m$ on both sides, this yields:

$$a-n_0 - n_1 \left( \frac{c}{a} \right)- \hdots - n_{m-1} \left( \frac{c}{a} \right)^{m-1}-n_m\left( \frac{c}{a} \right)^{m}= a-\sum_{i=1}^n p_i^G \left( \frac{c}{a} \right) \neq 0$$

Multiplying by $-1$ on both sides gives the desired $-a+\sum_{i=1}^n p_i^G\left( \frac{c}{a} \right)\neq 0$.
\end{proof}

We can now use this lemma to solve for the fixed points of the systems $L_{\sigma}$.

\begin{prp}\label{prp:dagfp}
Let G be a DAG of size n, $\sigma\subseteq [n]$, and $\beta=\dfrac{-\varepsilon-\delta}{\delta}$.  Then, for a CTLN associated with G, the fixed point of $L_{\sigma}$ is:

    $$x_{\sigma}^* = \Vec{p_{\sigma}}(\beta) \Gamma (\sigma)$$

where $\Gamma(\sigma)=\dfrac{\theta}{-\delta+(1+\delta)\sum_{i\in \sigma} p_i^{G|_{\sigma}}(\beta)}.$
\end{prp}

\begin{proof}
The matrix corresponding to the system in this chamber is of the form $-I+W$.  Restricting to the set of neurons with active rectifiers, we have $(-I+W)|_\sigma$.  Call this matrix $Z$.  Then, find the fixed point we need to solve $Z\vec{x}=-\theta\mathbbm{1}$.  This is equivalent to $(Z-\delta I)\vec{x}=-\theta\mathbbm{1}-\delta\vec{x}$.  

Notice that $\frac{-1}{1+\delta}(Z-\delta I)=\mathbbm{1}\mathbbm{1}^T+cA|_\sigma=\frac{\theta}{1+\delta}\mathbbm{1}+\frac{\delta}{1+\delta}\vec{x}$ where $c=\frac{-\varepsilon-\delta}{1+\delta}$ and $A$ is the adjacency matrix of $G$.  The result follows by applying Lemma~\ref{lem:DAG}.
\end{proof}

Let us emphasize what was accomplished.  For any chamber $R_{\sigma}$ of the DAG CTLN, we are able to write the fixed point of $L_{\sigma}$ in terms of the combinatorial structure of the DAG subgraph and see precisely how it depends on its paths.  Even though these are often not fixed points of the CTLN as a whole, they nonetheless shape the dynamics of the CTLN within $R_{\sigma}$.  Taking a DAG, we could break it down into its subgraphs $G|_{\sigma}$ and quickly determine the fixed points associated with each $L_{\sigma}$ as shown in Figure~\ref{fig:fp_cham}.

\begin{figure}[!ht]
\begin{center}
\vspace{.1in}
\includegraphics[width=5.75in]{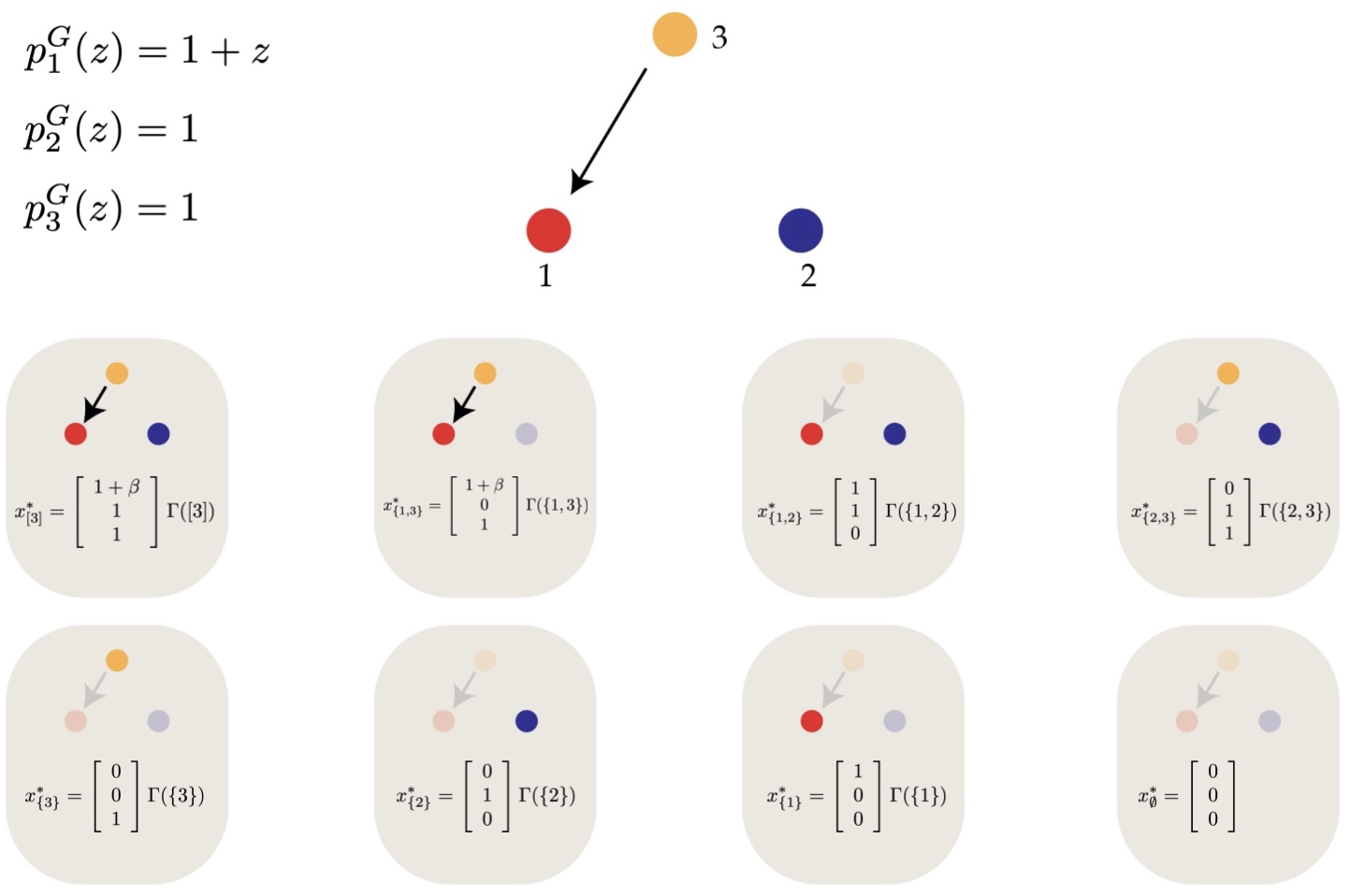}
\vspace{.1in}
\caption[Virtual fixed points of $L_{\sigma}$]{{\bf Virtual fixed points of $L_{\sigma}$.}  Fixed points of the linear systems $L_{\sigma}$.  Most of these are not fixed points of the CTLN as a whole, but nonetheless play a role in shaping the dynamics within the corresponding $R_{\sigma}$.}
\label{fig:fp_cham}
\end{center}
\vspace{-.2in}
\end{figure}

This result begs a natural question.  We know the fixed points of DAG CTLNs are supported on the sinks and the unions of sinks, so most of these fixed points are not located in their own chambers.  Can we find a function that tells us what chamber they do lie in?  A Chamber Mapping Function?

\begin{ddd}
	For a CTLN, define the Chamber Mapping Function to be: 
	
	$$\mathcal{G}:2^{[n]}\rightarrow 2^{[n]}\text{ such that }\mathcal{G}(\sigma)=\rho \iff x^*_{\sigma}\in R_\rho$$
\end{ddd}

\begin{lem}
    For a CTLN, define $x^*_{\sigma}$ to be the fixed point associated with the ODE system $L_\sigma$.  

    Then, $i\in \mathcal{G}(\sigma) \iff y_i(x^*_{\sigma})>0$. 
\end{lem}

\begin{prp}\label{prp:cmf} 
For a CTLN derived from a Directed Acyclic Graph G of size n and $\sigma \subseteq [n]$.  Then, $\mathcal{G}(\sigma)=\rho$ where $\rho$ is a codeword such that $\rho=\{i\in \sigma | g^\sigma_i>0\} \bigcup \{k\not\in \sigma | -g^\sigma_k\geq 0\}$ where:

    $$g^\sigma_i=\dfrac{p_i^{G|_{\sigma \cup \{i\}}}(\beta)}{-\frac{\delta}{1+\delta}+\sum_{k\in \sigma} p^{G|_{\sigma}}_k (\beta)}$$

and $\beta=\frac{-\varepsilon-\delta}{\delta}$.
\end{prp}

\begin{proof}

Let $x^*_{\sigma}$ be the fixed point of $L_\sigma$.

    For $i\in\sigma$, $y_i(x^*_{\sigma})=(x^*_{\sigma})_i$ and the result is trivial.

    For $i\not\in\sigma$:
    
    $y_i(x^*_{\sigma})=(-1-\delta)\sum_{j\in\sigma}x^{\sigma}_j+(\varepsilon+\delta)\sum_{j\in \sigma \rightarrow i}x^{\sigma}_j+\theta$

    Applying Proposition~\ref{prp:dagfp}, this can be rewritten as:

$y_i (x^*_{\sigma})=(-1-\delta)\Gamma \sum_{j\in \sigma} p_j^{G|_{\sigma}}(\beta) + (\varepsilon+\delta)\Gamma \sum_{j\in \sigma \rightarrow i} p_j^{G|_{\sigma}}(\beta) +\theta$

Expanding and combining, we have:

$y_i (x^*_{\sigma})=\left(\dfrac{(-1-\delta)\sum_{j\in \sigma} p_j^{G|_{\sigma}}(\beta) + (\varepsilon + \delta)\sum_{j\in \sigma \rightarrow i} p_j^{G|_{\sigma}}(\beta) -\delta + (1+\delta)\sum_{j\in \sigma} p_j^{G|_{\sigma}}(\beta)}{-\delta+(1+\delta)\sum_{j\in \sigma} p_j^{G|_{\sigma}}(\beta)}\right)\theta$.

This simplifies to:

$y_i (x^*_{\sigma})=\left(\dfrac{(\varepsilon + \delta)\sum_{j\in \sigma \rightarrow i} p_j^{G|_{\sigma}}(\beta) -\delta}{-\delta+(1+\delta)\sum_{j\in \sigma} p_j^{G|_{\sigma}}(\beta)}\right)\theta = \left(\dfrac{-\delta \beta \sum_{j\in \sigma \rightarrow i} p_j^{G|_{\sigma}}(\beta) -\delta}{-\delta+(1+\delta)\sum_{j\in \sigma} p_j^{G|_{\sigma}}(\beta)}\right)\theta$.

We use once again that if $j\rightarrow i$, then a path to $j$ of length $k$ corresponds to a path to $i$ of length $k+1$ to replace $\beta \sum_{j\in \sigma \rightarrow i} p_j^{G|_{\sigma}}(\beta)=p_i^{G|_{\sigma\cup \{i\}}} (\beta) - 1$.

    $y_i(x^*_{\sigma})=\left(\dfrac{-\delta p_i^{G|_{\sigma \cup \{i\}}}(\beta) +\delta -\delta}{-\delta+(1+\delta)\sum_{j\in \sigma} p_j^{G|_{\sigma}}(\beta)} \right)\theta=\dfrac{-\delta \theta p_i^{G|_{\sigma \cup \{i\}}}(\beta)}{-\delta+(1+\delta)\sum_{j\in \sigma} p_j^{G|_{\sigma}}(\beta)}$.
    
    To conclude, we have:
    
    $y_i (x^*_{\sigma})=-g_i^\sigma \left(\dfrac{\delta \theta}{1+\delta}\right)\geq 0 \iff -g_i^\sigma \geq 0$.
    
\end{proof}

\begin{figure}[!ht]
\begin{center}
\vspace{.1in}
\includegraphics[width=5.75in]{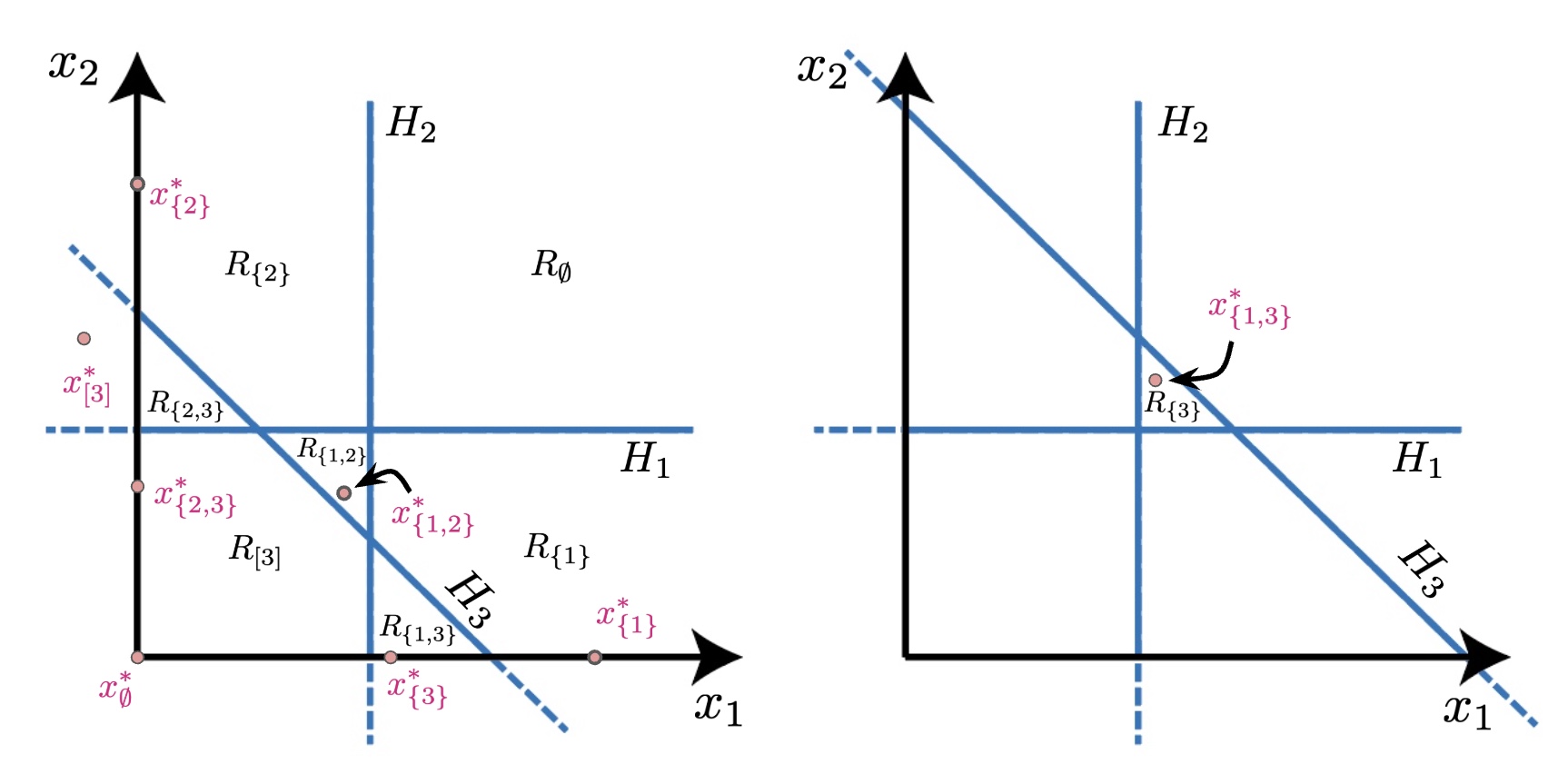}
\vspace{.1in}
\caption[Chamber mapping of virtual fixed points]{{\bf Chamber mapping of virtual fixed points.}  Schematic describing the chambers in which the fixed point solutions of the linear systems $L_{\sigma}$ lie for the Decoy Effect CTLN.  Unless they lie within their own chambers $R_{\sigma}$, these are not fixed points of the CTLN, but their locations still shape dynamics within $R_{\sigma}$.  Each image is a different $x_1,x_2$-cross section of the three dimensional state space and both are needed to account for all chambers of the hyperplane partition (the first lacks $R_{3}$ while the second lacks $R_{1,2}$).  While some effort has been made to give a rough sense of the state space location of the fixed points, this is not entirely accurate.  The accuracy of the schematic is in which chamber the virtual fixed points would lie in.}
\label{fig:cmf_ex}
\end{center}
\vspace{-.2in}
\end{figure}

Figure~\ref{fig:cmf_ex} shows an example of this mapping process.  An additional consequence of this is that we derive an alternative proof for Theorem~\ref{thm:dagrule}, which we restate here:

\begin{flushleft}
\textbf{Theorem~\ref{thm:dagrule}.} (Rule 7 in \cite{gr}) \textit{The set of fixed points of a CTLN derived from a DAG $G$ will be supported on sinks and the unions of sinks.}  
\end{flushleft}

    $$\operatorname{FP}(G)=\{ \bigcup s_i \mid s_i \text{ is a sink in $G$} \}$$
    
    \textit{Moreover, each stable fixed point will be supported on exactly one of the sinks.}

\begin{proof}

First, we show that sinks and unions of sinks are fixed point supports.  Let $\sigma \subseteq [n]$ be a set of sinks.  Then, for $i\in \sigma$, we have:

$$g_i^\sigma=\dfrac{1}{-\frac{\delta}{1+\delta} + |\sigma|}>0$$.

From this we conclude that $\forall i \in \sigma$, we have $i \in \mathcal{G}(\sigma)$.  

Since $\forall i \in \sigma$ are sinks of $G$, for any $k \not\in \sigma$ we have:

$$g_k^\sigma=\dfrac{1}{-\frac{\delta}{1+\delta} + |\sigma|}>0$$.

Then $\forall k \not\in \sigma$, $k \not\in \mathcal{G}(\sigma)$.  So, $\mathcal{G}(\sigma)=\sigma$ and we conclude that all sinks and the unions of sinks are fixed point supports of a DAG CTLN.

Now we show that no other $\sigma \subseteq [n]$ can support a fixed point.  There are two cases.

\textbf{Case 1: } $G|_{\sigma}$ is not an independent set

In the case, there exist a source vertex of $G|_{\sigma}$, $i\in \sigma$, and some $j\in \sigma$ such that there exist paths in $G|_{\sigma}$ to $j$ of length 1 but not length 2.  

$$g_i^\sigma=\dfrac{1}{-\frac{\delta}{1+\delta} + \sum_{l\in \sigma} p_l^{G|_{\sigma}} (\beta)} \text{ , }g_k^\sigma=\dfrac{1+n^k_1 \beta}{-\frac{\delta}{1+\delta} + \sum_{l\in \sigma} p_l^{G|_{\sigma}} (\beta)}$$

Since $\beta<-1$, necessarily $1+n^k_1 \beta < 0$.  Then, if $-\frac{\delta}{1+\delta} + \sum_{l\in \sigma} p_l^{G|_{\sigma}} (\beta) > 0$, we would have $g_k^\sigma < 0$ and so $k\not \in \mathcal{G}(\sigma)$.  Otherwise, we would have $g_i^\sigma < 0$ and $i\not\in \mathcal{G}(\sigma)$.  Thus, $\mathcal{G}(\sigma)\neq \sigma$.

\textbf{Case 2: } $G|_{\sigma}$ is an independent set

Since by assumption $\sigma$ is not a union of sinks, $\exists k \not \in \sigma$ and $i\in \sigma$ such that $i \rightarrow k$.  Then we have:

$$g_k^\sigma = \dfrac{1+n^k_1 \beta}{-\frac{\delta}{1+\delta} + |\sigma|} < 0.$$

Thus, $k\in \mathcal{G}(\sigma)$ so $\mathcal{G}(\sigma) \neq \sigma$.

We conclude from the above that $\mathcal{G}(\sigma)=\sigma \iff \sigma$ is a sink or the union of sinks.

The stability and instability of these fixed points follow from the fact that, using Corollary~\ref{corr:charpolyctln} the characteristic polynomials of the submatrices $\left(-I+W\right)|_{\sigma}$ in these cases are:

$$f(\lambda)=(-\lambda+\delta)^{|\sigma|-1}(-\lambda+\delta-(1+\delta)|\sigma|).$$

These have positive roots if and only if $|\sigma|>1$.
\end{proof}

%%%%%%%%%%%%%%

\section{General Solutions for Linear Systems $L_{\sigma}$}

Bringing our results together, we obtain Theorem~\ref{thm:final} which we restate now.

\begin{flushleft}
\textbf{Theorem~\ref{thm:final}.} \begin{itshape} Let G be a DAG and let W be the weight matrix for an associated CTLN with parameters $\varepsilon, \delta, \theta$.  Let $\sigma \subseteq [n]$ be such that $G|_{\sigma}$ is analytic, $(-I+W)|_{\sigma}$ is diagonalizable, and the polynomial:\end{itshape}
\end{flushleft}

$$f(\lambda)=(-\lambda+\delta)^{m+1}-(1+\delta)(|\sigma|(-\lambda+\delta)^{m}+n^{\sigma}_1c(-\lambda+\delta)^{m-1}+...+n^{\sigma}_m c^m)$$

\begin{itshape}has distinct roots $\{\lambda_k\}_{k=1}^{m+1}$ where $c=-\varepsilon - \delta$, $n^{\sigma}_{j>0}$ is the number of paths of length $j$ in $G|_{\sigma}$ and $m$ is the maximum path length in $G|_{\sigma}$.  

Then, the general solution of $L_{\sigma}$ is of the following form:

$$\Vec{x}(t)=\sum_{k=1}^{m+1} c_k \Vec{p_{\sigma}}(\alpha_k) e^{\lambda_{k} t} +\sum_{(i,j)\in\operatorname{\widetilde{SE}}(G|_{\sigma})} c_{(i,j)}(e_i-e_j)e^{\delta t} +\sum_{k=1}^{n-|\sigma|} c_k \Vec{v}_k e^{-t}+ \Vec{p_{\sigma}}(\beta)\Gamma (\sigma)$$

where $n=|G|$, $\alpha_k=\dfrac{\varepsilon+\delta}{\lambda_k -\delta}$, $\beta=\dfrac{-\varepsilon-\delta}{\delta}$, and $\Gamma (\sigma) = \dfrac{\theta}{-\delta +(1+\delta) \sum_{j\in \sigma} p_j^{G|_{\sigma}} (\beta)}$.
\end{itshape}

\begin{proof}

As $(-I+W)|_{\sigma}$ is diagonalizable, the general solution of the linear system $L_{\sigma}$ is of the form $\Vec{x}(t)=\sum_{i=1}^n c_i\Vec{v}_i e^{\lambda_i t} +x^*$ where $\left\{\lambda_i\right\}_{i=1}^n$ are the eigenvalues of the associated matrix with multiplicity, $\left\{\Vec{v}_i\right\}_{i=1}^n$ are associated eigenvectors, and $x^*$ the particular solution i.e. the fixed point.  By Corollary~\ref{corr:charpolyctln}, we know that the eigenvalue $\lambda=\delta$ has algebraic multiplicity $|\sigma|-(m+1)$, the eigenvalue $\lambda=-1$ has algebraic multiplicity $n-|\sigma|$, and that the remaining eigenvalues satisfy:

$$f(\lambda)=(-\lambda+\delta)^{m+1}-(1+\delta)(|\sigma|(-\lambda+\delta)^{m}+n^{\sigma}_1c(-\lambda+\delta)^{m-1}+...+n^{\sigma}_m c^m).$$

Since, by assumption, $f(\lambda)$ has distinct roots, these eigenvalues will all have algebraic multiplicity $1$ and applying Corollary~\ref{corr:evs} will give us the eigenvector for each of them.  Since $G|_{\sigma}$ is analytic, we can obtain a reduced simply embedding set $\operatorname{\widetilde{SE}(G|_{\sigma})}$ of $|\sigma|-(m+1)$ linearly independent eigenvectors for $\lambda=\delta$, obtained from simply embedding pairs via Proposition~\ref{prp:deltavec}.  All that remains is to determine $x^*$, the fixed point, which is obtained using Proposition~\ref{prp:dagfp}.  Assembling all the pieces gives us our desired general solution.

\end{proof}

%%%%%%%%%%%

\section{Combinatorial Solutions for DAG CTLNs: the Initial Value Problem}

Generally speaking, diagonalizable linear dynamical systems are characterized by their fixed points and the eigenvectors/eigenvalues of their associated matrices.  What we have done in this chapter is show how each of these components can be enumerated from the combinatorial structure of the DAG using the localized path polynomials.  The final piece of the puzzle needed to combinatorially derive the solutions of the systems $L_{\sigma}$ is a way of solving for the coefficients in the initial value problem.  

In fact, the path polynomials do offer a way to do this as well in the case of analytic DAG CTLNs.  However, the results are highly unpleasant most of the time and this show the limitations of this analytical approach.  To illustrate, we will solve the system $L_{[4]}$ in the $R_{[4]}$ chamber for the DAG CTLN depicted in Fig~\ref{fig:ev_ex} with localized path polynomials as depicted in Fig~\ref{fig:iv_ex}.  By convention we number the vertices from sink to source, respecting the topological ordering of the DAG such that the adjacency matrix is strictly upper triangular.  Additionally, in the case we have multiple eigenvectors for $\lambda=\delta$ which share an entry (e.g. $v_1=e_{i_1} - e_{j_1}$ and $v_2=e_{i_2} - e_{j_2}$ such that $i_1=j_2$), we write them so that the $j$'s are distinct.

\begin{figure}[!ht]
\begin{center}
\vspace{.1in}
\includegraphics[width=2.75in]{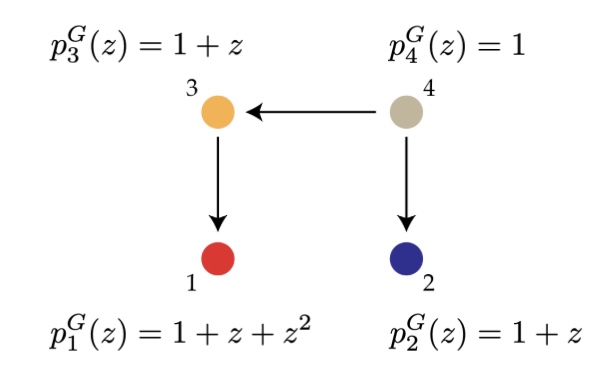}
\vspace{.1in}
\caption[Localized path polynomials for $G$]{{\bf Localized path polynomials for $G$.} Each of the four vertices has a localized path polynomial listed here.}
\label{fig:iv_ex}
\end{center}
\vspace{-.2in}
\end{figure}

\subsection{Fixed Point}

The fixed point is the first component in finding the solution of $L_{\sigma}$, serving as the particular solution.  We simply apply Proposition~\ref{prp:dagfp} and obtain the fixed point:  

$$
\Vec{x}^*(t)=
\left[
\begin{array}{c}
p_1^G(\beta) \\
p_2^G(\beta)\\
p_3^G(\beta) \\
p_4^G(\beta)\\
\end{array}
\right]\Gamma
$$

where, as before, $\beta=\frac{-\varepsilon-\delta}{\delta}$ and $\Gamma=\dfrac{\theta}{-\delta+(1+\delta)\sum_{j=1}^4 p_j (\beta)}$.

\subsection{Eigenvectors}

From Corollary~\ref{corr:charpolyctln}, we know the characteristic polynomial for the matrix $-I+W$ is:

$$f(\lambda) = (-\lambda+\delta)((-\lambda+\delta)^3 - 4(1+\delta)(-\lambda+\delta)^2 -3c(1+\delta)^2(-\lambda+\delta) -c^2(1+\delta)^3).$$

For $\lambda=\delta$, as the algebraic multiplicity is 1 we need only one eigenvector.  Applying Lemma~\ref{prp:deltavec} we have $v_\delta = e_1-e_2$.

The remaining three eigenvalues are the roots of $(-\lambda+\delta)^3 - 4(1+\delta)(-\lambda+\delta)^2 -3c(1+\delta)^2(-\lambda+\delta) -c^2(1+\delta)^3$.  While in this particular case the roots could be obtained through the cubic formula, in general the problem of finding the roots of a sufficiently high degree polynomial is not algebraically tractable.  We will simply refer to the roots as $\lambda_1$, $\lambda_2$, and $\lambda_3$.  We accordingly use the notation $\alpha_i=\frac{\varepsilon+\delta}{\lambda_i-\delta}$.

Then, using Corollary~\ref{corr:evs}, the full set of eigenvectors and eigenvalues are:

$$
\lambda=\delta: \left[\begin{array}{c} 1 \\ -1 \\ 0 \\ 0 \end{array}\right],
\lambda=\{\lambda_i\}_{i=1}^3: \left\{ \left[ \begin{array}{c} p_1^G(\alpha_i) \\ p_2^G(\alpha_i) \\ p_3^G(\alpha_i) \\p_4^G(\alpha_i) \end{array} \right] \right\}_{i=1}^3.
$$

This gives us the general solution:

$$
\Vec{x}(t)=\Vec{x}_H(t)+\Vec{x}_p(t)
$$

$$
=c_1\left[\begin{array}{c} 1 \\ -1 \\ 0 \\ 0 \end{array}\right]e^{\delta t} + c_2 \left[ \begin{array}{c} p_1^G(\alpha_1) \\ p_2^G(\alpha_1) \\ p_3^G(\alpha_1) \\p_4^G(\alpha_1) \end{array}\right] e^{\lambda_1 t} + c_3 \left[ \begin{array}{c} p_1^G(\alpha_2) \\ p_2^G(\alpha_2) \\ p_3^G(\alpha_2) \\p_4^G(\alpha_2) \end{array}\right] e^{\lambda_2 t} + c_4 \left[ \begin{array}{c} p_1^G(\alpha_3) \\ p_2^G(\alpha_3) \\ p_3^G(\alpha_3) \\p_4^G(\alpha_3) \end{array}\right] e^{\lambda_3 t}
+ 
\left[
\begin{array}{c}
p_1^G(\beta) \\
p_2^G(\beta) \\
p_3^G(\beta) \\
p_4^G(\beta) \\
\end{array}
\right]\Gamma.
$$

\subsection{Solving the Initial Value Problem}

We have up until this point the general solution:

$$x_1 (t) = c_1 e^{\delta t} +c_2 p_1^G (\alpha_1) e^{\lambda_1 t} + c_3 p_1^G (\alpha_2) e^{\lambda_2 t}+c_4 p_1^G (\alpha_3) e^{\lambda_3 t} + p_1^G(\beta)\Gamma$$

$$x_2 (t) = -c_1 e^{\delta t} +c_2 p_2^G (\alpha_1) e^{\lambda_1 t} + c_3 p_2^G (\alpha_2) e^{\lambda_2 t}+c_4 p_2^G (\alpha_3) e^{\lambda_3 t} + p_2^G(\beta)\Gamma$$

$$x_3 (t) = c_2 p_3^G (\alpha_1) e^{\lambda_1 t} + c_3 p_3^G (\alpha_2) e^{\lambda_2 t}+c_4 p_3^G (\alpha_3) e^{\lambda_3 t} + p_3^G(\beta)\Gamma$$

$$x_4 (t) = c_2 e^{\lambda_1 t} + c_3 e^{\lambda_2 t}+c_4  e^{\lambda_3 t} + \Gamma.$$

The last step then is to work out the initial value problem for this system.  Taking the initial condition $\Vec{x}_0$, we set up the system:

$$
\left[
\begin{array}{cccc}
	1 & 1+\alpha_1 + \alpha_1^2 & 1+\alpha_2 + \alpha_2^2 & 1+\alpha_3 + \alpha_3^2 \\
	-1 & 1+\alpha_1 & 1+\alpha_2 & 1+\alpha_3\\
	0 & 1+\alpha_1 & 1+\alpha_2 & 1+\alpha_3\\
	0 & 1 & 1 & 1
\end{array}
\right]\Vec{c}
=
\left[
\begin{array}{c}
x_1^0 - p_1^G(\beta)\Gamma \\
x_2^0 - p_2^G(\beta)\Gamma \\
x_3^0 - p_3^G(\beta)\Gamma \\
x_4^0 - \Gamma \\
\end{array}
\right]
$$

We will now employ a three step process to find $\Vec{c}$.  

\textbf{Step 1: } Turn columns from eigenvectors of $\lambda =\delta$ to distinct basis vectors

By construction , all of our eigenvectors for $\lambda=\delta$ are of the form $e_i-e_j$ such that each $j$ is distinct.  For each of these eigenvectors, we add the $j$-th row of the system to the $i$-th row, turning the corresponding column into $-e_j$.  In our current system, we would then have:

$$
\left[
\begin{array}{cccc}
	0 & 2+2\alpha_1 + \alpha_1^2 & 2+2\alpha_2 + \alpha_2^2 & 2+2\alpha_3 + \alpha_3^2 \\
	-1 & 1+\alpha_1 & 1+\alpha_2 & 1+\alpha_3\\
	0 & 1+\alpha_1 & 1+\alpha_2 & 1+\alpha_3\\
	0 & 1 & 1 & 1
\end{array}
\right]\Vec{c}
=
\left[
\begin{array}{c}
x_1^0+x_2^0 - (p_1^G(\beta)+p_2^G(\beta))\Gamma \\
x_2^0 - p_2^G(\beta)\Gamma \\
x_3^0 - p_3^G(\beta)\Gamma \\
x_4^0 - \Gamma\\
\end{array}
\right]
$$

\textbf{Step 2: } Restrict to the last $m+1$ entries of $\Vec{c}$ and solve the Vandermonde Matrix

We will set aside rows that have non-zero entries in the first $n-{m+1}$ columns for the time being.  So, we will have a $(m+1)\times (m+1)$ subsystem.  In our example, we have the following subsystem:

$$
\left[
\begin{array}{ccc}
	2+2\alpha_1 + \alpha_1^2 & 2+2\alpha_2 + \alpha_2^2 & 2+2\alpha_3 + \alpha_3^2 \\
	1+\alpha_1 & 1+\alpha_2 & 1+\alpha_3\\
	1 & 1 & 1
\end{array}
\right]\left[ \begin{array}{c} c_2 \\ c_3 \\ c_4 \end{array} \right]
=
\left[
\begin{array}{c}
x_1^0+x_2^0 - (p_1^G(\beta)+p_2^G(\beta))\Gamma \\
x_3^0 - p_3^G(\beta)\Gamma \\
x_4^0 - \Gamma \\
\end{array}
\right]
$$

Now for the key trick.  We will transform the matrix on the left into the well known Vandermonde Matrix.  By subtracting the bottom row we can eliminate the constant terms.

$$
\left[
\begin{array}{ccc}
	2\alpha_1 + \alpha_1^2 & 2\alpha_2 + \alpha_2^2 & 2\alpha_3 + \alpha_3^2 \\
	\alpha_1 & \alpha_2 & \alpha_3\\
	1 & 1 & 1
\end{array}
\right]\left[ \begin{array}{c} c_2 \\ c_3 \\ c_4 \end{array} \right]
=
\left[
\begin{array}{c}
x_1^0+x_2^0-2x_4^0 - (p_1^G(\beta)+p_2^G(\beta)-2p_4^G(\beta))\Gamma \\
x_3^0-x_4^0 - (p_3^G(\beta)-p_4^G (\beta))\Gamma \\
x_4^0 - \Gamma \\
\end{array}
\right]
$$

We can then repeat this process, with some potential scaling, moving up the rows with $\alpha_i, \alpha_i^2,...,\alpha_i^{m-1}$ respectively until we are left with the Vandermonde Matrix:

$$
\left[
\begin{array}{ccc}
	\alpha_1^2 & \alpha_2^2 & \alpha_3^2 \\
	\alpha_1 & \alpha_2 & \alpha_3\\
	1 & 1 & 1
\end{array}
\right]\left[ \begin{array}{c} c_2 \\ c_3 \\ c_4 \end{array} \right]
=
\left[
\begin{array}{c}
x_1^0+x_2^0-2x_4^0 -2x_3^0 - (p_1^G(\beta)+p_2^G(\beta)-2p_3^G(\beta)-2p_4^G(\beta))\Gamma \\
x_3^0-x_4^0 - (p_3^G(\beta)-p_4^G (\beta))\Gamma \\
x_4^0 - \Gamma \\
\end{array}
\right]
$$

The benefit of this is that the inverse of the Vandermonde matrix of size $(m+1) \times (m+1)$ is known to have the entries:

$$(V^{-1})_{ij}=\dfrac{(-1)^{j-1}E_{j-1}(\{\alpha_1,\dots,\alpha_{m+1}\} \backslash \{\alpha_i\})}{\Pi_{k=1,k\neq i}^{m+1} (\alpha_{i}-\alpha_k)}$$

where $E_m (\{ y_1,...,y_k \})=\sum_{1\leq j_1<\dots<j_m\leq k} y_{j_1}\dots y_{j_m}$ are the elementary symmetric functions.\cite{vand}

So, we can then solve for the coefficients:

$$c_2=\sum_{j=1}^3 (V^{-1})_{1j}\phi_j$$

$$c_3=\sum_{j=1}^3 (V^{-1})_{2j}\phi_j$$

$$c_4=\sum_{j=1}^3 (V^{-1})_{3j}\phi_j$$

where:

$$\Vec{\phi}
=
\left[
\begin{array}{c}
x_1^0+x_2^0-2x_4^0 -2x_3^0 - (p_1^G(\beta)+p_2^G(\beta)-2p_3^G(\beta)-2p_4^G(\beta))\Gamma \\
x_3^0-x_4^0 - (p_3^G(\beta)-p_4^G (\beta))\Gamma \\
x_4^0 - \Gamma \\
\end{array}
\right].$$

\textbf{Step 3: } Solve for the remaining coefficients

We can now also solve for the remaining $n-(m+1)$ coefficients.  As a byproduct of \textbf{Step 1}, each of these coefficients corresponds to exactly one of the rows omitted in \textbf{Step 2}.  In our case, $c_1$ corresponds to row 2.

$$-c_1 + c_2 (1+\alpha_1) + c_3 (1+\alpha_2) + c_4 (1+\alpha_4)=x_2^0 - p_2^G(\beta)\Gamma .$$

Finally we obtain our last coefficient:

$$c_1=c_2 (1+\alpha_1) + c_3 (1+\alpha_2) + c_4 (1+\alpha_4) - x_2^0+p_2^G(\beta)\Gamma.$$

\section{Takeaways}

The results from this chapter demonstrate how finding analytical solutions for DAG CTLNs is surprisingly tractable on a theoretical level.  The results in our final example were parameter independent up to some conditions and didn't require particularly difficult calculation to achieve.  However, the obvious problem is that the expressions are simply too unwieldy.  Trying to piece together analytical trajectories across chambers with these coefficients is not practical, especially since finding the points of intersection with the chamber walls will entail the solution of transcendental equations.  Even if we were to do this, notice that the roots of the polynomial in Corollary~\ref{corr:charpolyctln} are sensitive to even slight changes in the graph structure.  This could prove challenging in comparing the basins of attraction after even a slight alternation of graph structure.  Ultimately, our takeaway from this chapter is that, despite the considerable theory that can be developed about DAG CTLNs derived from the graph combinatorics, the limitations in higher dimensions are tough to overcome if the goal is to calculate the full basins of attraction.

\chapter{Decision-Making Bias Near Decision Boundaries}

Our work so far has simultaneously demonstrated both the tractability of DAG CTLNs along with the challenges of even roughly determining their full basins of attraction.  That said, we also had paradigms of decision-making that did not require analyzing the full basins of attraction.  The focus of this chapter will be the paradigm of studying initial conditions near decision boundaries.  This approach is aligned with an ontology of decision-making circuits that sees them as operating along branching manifolds \cite{branch}.  In the case of a binary choice task, this consists of trajectories beginning along a manifold representing the decision boundary before being carried away along another manifold toward one of the attractors encoding a decision.  

\begin{figure}[!ht]
\begin{center}
\vspace{.1in}
\includegraphics[width=6.25in]{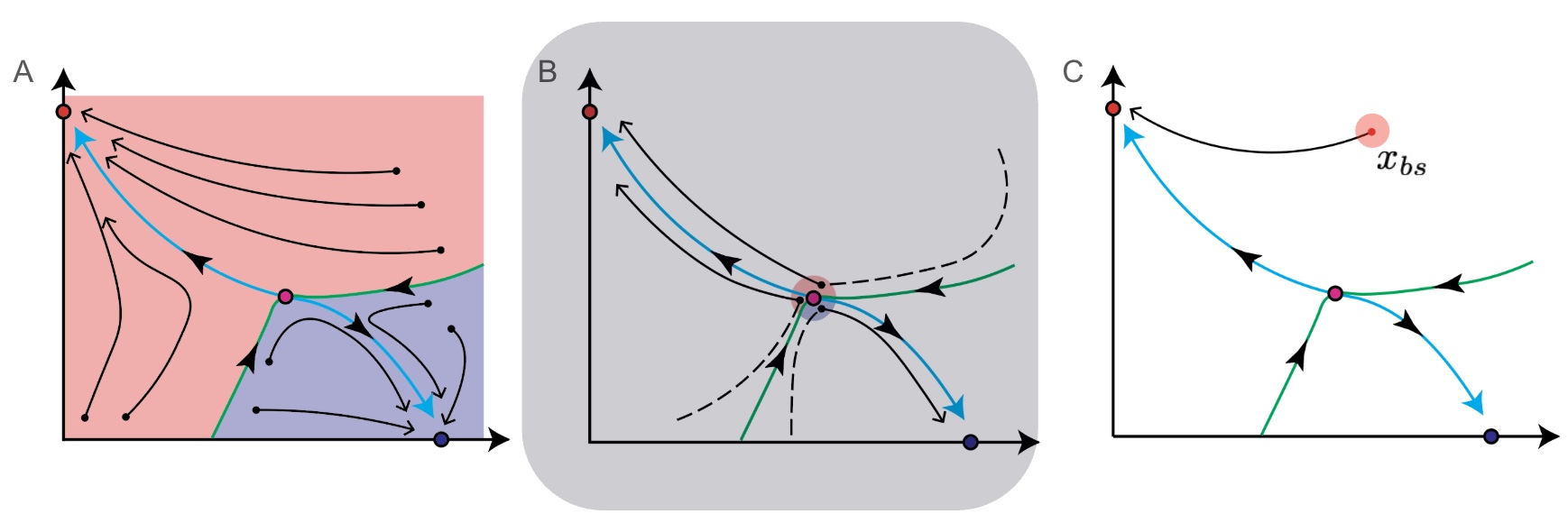}
\vspace{.1in}
\caption[Decision-making dynamics near low dimensional submanifolds]{{\bf Decision-making dynamics near low dimensional submanifolds.}  Having been unsuccessful at analytically determining the basins of attraction for DAG CTLNs, in this chapter we explore the basins of attraction with the neighborhood of an unstable fixed point as a proxy for studying trajectories near the branching stable and unstable manifolds of the network.}
\label{fig:dbfig}
\end{center}
\vspace{-.2in}
\end{figure}

With respect to CTLNs of DAGs with two sinks, our branching manifolds are the codimension 1 stable manifold and the dimension 1 unstable manifold associated with the saddle point supported on the union of sinks.  The question of course is how to study trajectories near decision boundaries without actually knowing the decision boundary itself.  Consider that we are interested then in trajectories which hew close to these manifolds, traveling near the saddle point.  As TLNs are autonomous systems, and we are ultimately only concerned with categorizing trajectories based on which attractors they converge to, we can begin tracking them when they are near the saddle point (Fig~\ref{fig:dbfig}).  Through this thinking, we can move from thinking about initial conditions near the decision boundary to initial conditions near the saddle point.

\section{DAGs with Two Sinks}

\begin{prp}\label{prp:stab}

Let G be a DAG with two sinks.  For a CTLN derived from G such that the sinks correspond to $x_1$ and $x_2$, the stable manifold in the $L_{\{1,2\}}$ system for its saddle point is:

$$-x_1+x_2+a_3 x_3 +...+a_n x_n = 0$$

where:

$$\begin{array}{c}
     a_j=0 \text{ if } j\not\rightarrow 1 \text{ and } j\not\rightarrow 2 \text{ or } j\rightarrow 1 \text{ and } j\rightarrow 2 \\
     a_j=\frac{-\varepsilon-\delta}{1+\delta} \text{ if } j \rightarrow 1 \text{ and } j\not\rightarrow 2 \\
     a_j=\frac{\varepsilon+\delta}{1+\delta} \text{ if } j\not\rightarrow 1 \text{ and } j\rightarrow 2 
\end{array}$$

\end{prp}

\begin{proof}
The matrix corresponding to the linear ODE system $L_{\{1,2\}}$ is:

$
A=
\left[
\begin{array}{cc|ccc}
-1 & -1-\delta & w_{13} & \hdots & w_{1n}\\
-1-\delta & -1 & w_{23} & \hdots & w_{2n}\\
\hline
0 & 0 & -1 & \hdots & 0\\
\vdots & \vdots & \vdots & \ddots & \vdots\\ 
0 & 0 & 0 & \hdots & -1
\end{array}
\right]
$

Which has eigenvalues $\delta$ and $-2-\delta$ from the upper left block and repeated eigenvalue $-1$ from the lower right block.  We claim the eigenvectors are as follows: 

$
\lambda_1=\delta : 
v_1=
\begin{bmatrix}
-1 \\
1 \\
0 \\
\vdots \\
0
\end{bmatrix}
$

$
\lambda_2=-2-\delta : 
v_2=
\begin{bmatrix}
1 \\
1 \\
0 \\
\vdots \\
0
\end{bmatrix}
$

$
\lambda_{3,\hdots,n}=-1: 
v_{3,\hdots,n}=
\begin{bmatrix}
\frac{-w_{23}}{1+\delta} \\
\frac{-w_{13}}{1+\delta}  \\
-1 \\
\vdots \\
0
\end{bmatrix}
\hdots
\begin{bmatrix}
\frac{-w_{2n}}{1+\delta} \\
\frac{-w_{1n}}{1+\delta}  \\
0 \\
\vdots \\
-1
\end{bmatrix}
$

For $\lambda_1$ and $\lambda_2$ this is clear from inspection.  For $\lambda_{k \geq 3}$, see that:

$$Av_k=\left[
\begin{array}{c}
\frac{w_{2k}}{1+\delta}+w_{1k}-w_{1k}\\
w_{2k}+\frac{w_{1k}}{1+\delta}-w_{2k}\\
0 \\
\vdots \\
1 \\
\vdots \\
0
\end{array}
\right]=\left[
\begin{array}{c}
\frac{w_{2k}}{1+\delta}\\
\frac{w_{1k}}{1+\delta}\\
0 \\
\vdots \\
1 \\
\vdots \\
0
\end{array}
\right]=-v_k.
$$

Notice that there are $n-1$ eigenvectors corresponding to negative eigenvalues and so the stable manifold of the system $L_{\{1,2\}}$ will be the hyperplane spanned by those eigenvectors.  We now claim that the normal vector to this stable manifold of $L_{\{1,2\}}$ is:

$$
\vec{n}=
\begin{bmatrix}
-1 \\
1  \\
a_3 \\
\vdots \\
a_n
\end{bmatrix}
\text{ where }
\begin{array}{c}
     a_j=0 \text{ if } j\not\rightarrow 1 \text{ and } j\not\rightarrow 2 \text{ or } j\rightarrow 1 \text{ and } j\rightarrow 2 \\
     a_j=\frac{-\varepsilon-\delta}{1+\delta} \text{ if } j \rightarrow 1 \text{ and } j\not\rightarrow 2 \\
     a_j=\frac{\varepsilon+\delta}{1+\delta} \text{ if } j\not\rightarrow 1 \text{ and } j\rightarrow 2 
\end{array}.
$$

Again, for $\lambda_2$ this is clear from inspection.  We again show via direct computation that this is true for $\lambda_{k\geq 3}$.

$$\Vec{n}\cdot v_k = \frac{w_{2k}}{1+\delta}-\frac{w_{1k}}{1+\delta}-a_k=\frac{w_{2k}-w_{1k}}{1+\delta}-a_k.$$

If $k\not\rightarrow 1 \text{ and } k\not\rightarrow 2 \text{ or } k\rightarrow 1 \text{ and } k\rightarrow 2$:

$$w_{2k}-w_{1k}=0.$$

If $k\rightarrow 1 \text{ and } k\not\rightarrow 2$:

$$w_{2k}-w_{1k}=-\varepsilon-\delta.$$

If $k\not\rightarrow 1 \text{ and } k\rightarrow 2$:

$$w_{2k}-w_{1k}=\varepsilon+\delta.$$

In each of these cases, $\dfrac{w_{2k}-w_{1k}}{1+\delta}=a_k$ as needed to have $\Vec{n}\cdot v_k=0$.  So, we know that the stable manifold of $L_{\{1,2\}}$ is of the form $\Vec{n}\cdot \Vec{x} + c = 0$.

As the saddle point $x^*=\left(\frac{\theta}{2+\delta},\frac{\theta}{2+\delta},0,\hdots,0\right)$ must lie on this manifold, we use it to find $c$.

$$\Vec{n} \cdot x^* + c = 0.$$

Since $\Vec{n} \cdot x^*=-\frac{\theta}{2+\delta}+\frac{\theta}{2+\delta}=0$, we conclude $c=0$ and we are done.

\end{proof}

What Proposition~\ref{prp:stab} indicates is that if we were to ignore the dynamics of other chambers and simply consider the local dynamics in the $L_{\{1,2\}}$ chamber for the saddle point $x^*_{\{1,2\}}$ supported on the union of the two sinks, then the sink with the greater in-degree should have the larger basin of attraction.  Of course the actual stable manifold of the CTLN will be more complex and highly non-linear, even within the $R_{\{1,2\}}$ chamber.  We conjecture that if we restrict our analysis to a small region around the fixed point, $S:=B_{\eta}(x^*_{\{1,2\}}) \cap R^+_{\{1,2\}}$ where $\eta << 1$, then the linear stable manifold of the $L_{\{1,2\}}$ system should approximate this small piece of the actual stable manifold of the CTLN.  For a basin of attraction of a sink, call it $\mathcal{B}_i$, we expect the fractional volume in this region:

$$\mathcal{F}_i:=\dfrac{\lambda(\mathcal{B}_i \cap S)}{\lambda(S)}$$

to be influenced strongly by the sink's indegree.

To be mathematically clear, what we would expect is a strong correlation between $\mathcal{F}_i$ and the fractional indegree of the sink over that of all the sinks:

$$\operatorname{indeg_f} (i):=\dfrac{\operatorname{indeg}(i)}{\sum_{j\in \operatorname{sinks}(G)} \operatorname{indeg}(j)}.$$

Numerically simulating for 1000 random DAGs of 6 neurons with two sinks for $\eta=0.01$, we obtain the results in Fig~\ref{fig:dbbox}A.  Notice the correlation is very strong which provides clear numerical evidence for our conjecture.  Compare this with a weaker relationship if we were to sample initial conditions more broadly from the state space as in Fig~\ref{fig:dbbox}B.  Beyond the correlation coefficients, notice the sheer spread when considering the full basins of attraction

\begin{figure}[!ht]
\begin{center}
\vspace{.1in}
\includegraphics[width=5.75in]{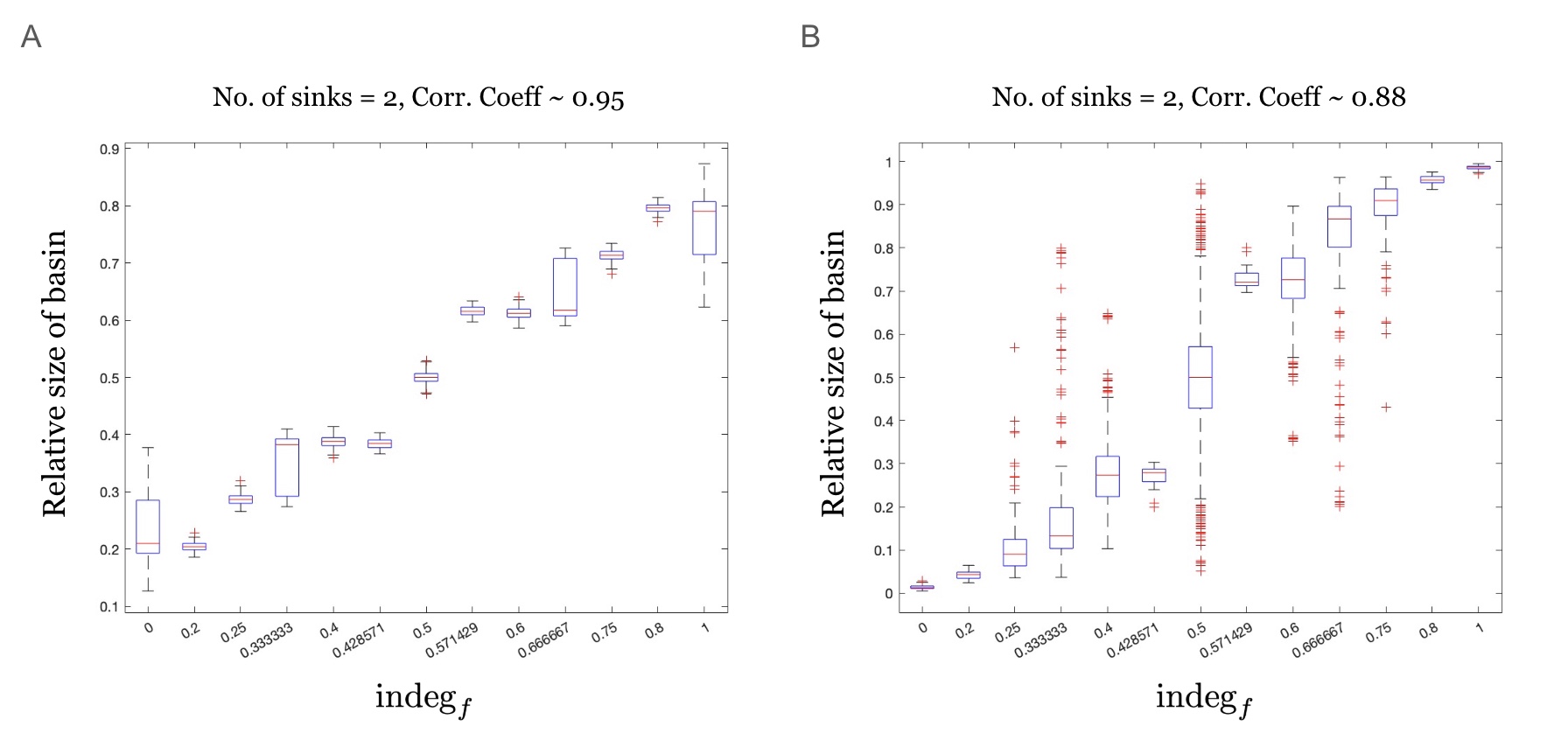}
\vspace{.1in}
\caption[Correlation between basin size and sink fractional indegree]{{\bf Correlation between basin size and sink fractional indegree.} (A)  Boxplots depicting the fraction of trajectories converging to a sink vs. the fractional indegree of the sink for initial conditions sampled near the saddle point.  (B) Boxplots illustrating the same, but with initial conditions sampled from state space more broadly, the box $[0, 1.5]^6$.  Both of these simulations use the same 1000 DAGs of size 6 with parameters $\delta=0.5$, $\varepsilon=0.25$, and $\theta=1$.  The number of initial conditions sampled per DAG was 2500.}
\label{fig:dbbox}
\end{center}
\vspace{-.2in}
\end{figure}

What can we say about DAGs more broadly?

\section{DAGs with Several Sinks}

We can find a result similar to Proposition~\ref{prp:stab} for DAGs with several sinks by making use of the Sherman-Morrison Formula.

\begin{lem}[Sherman-Morrison Formula \cite{smform}]
Suppose $M$ is an $n\times n$ matrix with $u,v$ being $n\times 1$ column vectors.  Then, $uv^T + M$ is invertible if and only if $1+v^T M^{-1} u \neq 0$.  In this case,

$$ (uv^T + M)^{-1} = M^{-1} - \dfrac{M^{-1}uv^T M^{-1}}{1+v^T M^{-1} u}.$$    

\end{lem}

\begin{prp}\label{prp:stab2}
Let G be a DAG of size n and $\sigma \subseteq [n]$ such that $G|_{\sigma}$ is an independent set.  Then, the linear system $L_{\sigma}$ for an associated CTLN has fixed point $x_{\sigma}^*=\dfrac{\theta}{|\sigma|+(|\sigma|-1)\delta}\mathbbm{1}_{\sigma}$ with an unstable manifold of dimension $ |\sigma|-1$ which is :

$$x_{\sigma}^*+\operatorname{span}\left( \{ e_{\sigma_1} - e_{\sigma_j} \mid \sigma_1 , \sigma_j \in \sigma \text{ and } \sigma_j \neq \sigma_1 \} \right)$$

and a stable manifold of codimension $|\sigma|-1$ which is:

$$x_{\sigma}^*+\operatorname{span}\left( \mathbbm{1_{\sigma}} \cup \left\{ \left( \dfrac{1}{|\sigma|-1}-\gamma \dfrac{id_{j}(\sigma)}{|\sigma|-1}\right)\mathbbm{1_{\sigma}} + \gamma \mathbbm{1}_{\sigma \leftarrow j} -e_j \mid j \not\in \sigma \right\} \right)$$

where $\mathbbm{1}_{s}=\sum_{i\in s} e_i$, $\gamma=\dfrac{\varepsilon+\delta}{1+\delta}$, and $id_{j}(\sigma)=|\{k\in\sigma \mid j\rightarrow k \}|$.
\end{prp}

\begin{proof}
Without loss of generality number the vertices in $\sigma$ to be $1,\dots,k$ where $k=|\sigma|$.  Then the matrix for $L_{\sigma}$ is of the form: 

$
A=
\left[
\begin{array}{ccc|ccc}
-1 & \hdots & -1-\delta & w_{1,k+1} & \hdots & w_{1n}\\
\vdots & \ddots & \vdots & \vdots & \hdots & \vdots\\
-1-\delta & \hdots & -1 & w_{k,k+1} & \hdots & w_{kn}\\
\hline
0 & \hdots & 0 & -1 & \hdots & 0\\
\vdots & \ddots & \vdots & \vdots & \ddots & \vdots\\ 
0 & \hdots & 0 & 0 & \hdots & -1
\end{array}
\right]
$

where, because $G|_{\sigma}$ is an independent set, the upper left block is of the form 

$$(-I+W)|_{\sigma} = (-1-\delta)\mathbbm{1}\mathbbm{1}^T + \delta I.$$

Recalling Lemma~\ref{corr:charpolyctln}, we know that the characteristic polynomial of $A$ is:

$$p(\lambda)=(\lambda+1)^{n-k} (-\lambda+\delta)^{k-1} (-\lambda+\delta-(1+\delta)|\sigma|)$$

and so its eigenvalues are $\lambda_1=\delta-(1+\delta)|\sigma|$, $\lambda_2=\delta$, and $\lambda_3=-1$ with algebraic multiplicities 1, $k-1$, and $n-k$ respectively.

The set:

$$\{ e_1 - e_j \mid 1 < j \leq k \}$$

is linearly independent and of size $k-1$ with each element being an eigenvector for $\lambda=\delta$ (Proposition~\ref{prp:deltavec}).  Therefore it is a basis for the unstable manifold.

Now we turn our attention to the stable manifold.  For $\lambda=\delta-(1+\delta)|\sigma|$, applying Corollary~\ref{corr:evs}, we know that $\Vec{p_{\sigma}}(\alpha)$ is the associated eigenvector.  However, since $G|_{\sigma}$ is an independent set, $p_{i}^{G|_{\sigma}} (z) = 1$, $\forall i \in \sigma$.  We will thus refer to this eigenvector as $\mathbbm{1}_{\sigma}$.

Finally, we find eigenvectors for $\lambda=-1$.  We show that there are $n-k$ linearly independent eigenvectors by construction.  We then take as an ansatz vectors of the form:

$$
\Vec{v}=\left[
\begin{array}{c}
c_1 \\
\vdots \\
c_k \\
\hline
0 \\
\vdots \\
-1 \\
\vdots \\
0
\end{array}
\right]
=
\left[
\begin{array}{c}
\Vec{c} \\
\hline
0
\end{array}
\right]-e_j
$$

Then, we have:

$$
A\Vec{v}=\left[
\begin{array}{c}
((-I+W)|_{\sigma})\Vec{c} - \Vec{w}_{*j} \\

\hline

0
\end{array}
\right]+e_j
$$

where

$$
\Vec{w}_{*j}=\left[
\begin{array}{c}
w_{1j} \\
\vdots \\
w_{kj}
\end{array}
\right].
$$

So, if $\Vec{c}$ satisfies $((-I+W)|_{\sigma})\Vec{c} - \Vec{w}_{*j}=-\Vec{c}$ then $\Vec{v}$ is an eigenvector.  Rearranging this system can be rewritten as:

$$((-I+W)|_{\sigma})\Vec{c} + I \Vec{c}= \Vec{w}_{*j} \implies ((-1-\delta)\mathbbm{1}\mathbbm{1}^T + (1+\delta) I)\Vec{c}=\Vec{w}_{*j} \implies (\mathbbm{1}\mathbbm{1}^T - I)\Vec{c}=-\dfrac{\Vec{w}_{*j}}{1+\delta}.$$

Finally, we apply the Sherman-Morrison Formula with $M=-I$ and $u,v=\mathbbm{1}$.  Then $M^{-1}=-I$ and so:

$$(\mathbbm{1}\mathbbm{1}^T - I)^{-1}=-I-\dfrac{1}{1-|\sigma|}\mathbbm{1}\mathbbm{1}^T=\left(\dfrac{1}{|\sigma|-1}\right)\mathbbm{1}\mathbbm{1}^T - I.$$

Finally, we have that:

$$\Vec{c}=-\dfrac{1}{1+\delta}(\mathbbm{1}\mathbbm{1}^T - I)^{-1}\Vec{w}_{*j}=-\dfrac{1}{1+\delta}\left(\dfrac{1}{|\sigma|-1}\sum_{k\in \sigma}w_{kj}\mathbbm{1} -\Vec{w}_{*j} \right) =$$

Substituting the matrix entries, we have:

$$\Vec{c}=\left( \dfrac{|\sigma|}{|\sigma|-1}-\gamma \dfrac{id_{j}(\sigma)}{|\sigma|-1}-1\right)\mathbbm{1} + \gamma \mathbbm{1}_{\sigma \leftarrow j}=\left( \dfrac{1}{|\sigma|-1}-\gamma \dfrac{id_{j}(\sigma)}{|\sigma|-1}\right)\mathbbm{1} + \gamma \mathbbm{1}_{\sigma \leftarrow j}.$$

The value of the fixed point follows from Proposition~\ref{prp:dagfp}.
\end{proof}

What we see from Proposition~\ref{prp:stab2} is that in larger DAGs with multiple saddle points, stable manifolds of the linear systems $L_{\sigma}$ are of codimension $|\sigma|-1$ whereas the unstable manifold is of dimension $|\sigma|-1$.  Moreover, the expression of each depends only on neurons which depends only on the activity of nonsink neurons which have contribute edges toward the sinks.  While of course the actual stable and unstable manifolds will be more complex and connected, we might still expect that sink indegree continues to be a strong factor in shaping dynamics along the decision boundaries.  

It is worth noting that up until this point we have somewhat been referring to the stable manifolds of saddle points and the decision boundaries interchangeably because in the case of two sink DAGs they are equivalent.  It is clear that this relationship does not hold quite so directly when there are $k>2$ sinks, i.e. $k$ basins of attraction, and $2^k-k-1$ saddle points.  Instead we should expect that the trajectories composing the decision boundaries will pass near the unique saddle point supported on the union of the sinks, $x^*_{\operatorname{sinks}(G)}$.  

\begin{figure}[!ht]
\begin{center}
\vspace{.1in}
\includegraphics[width=5.75in]{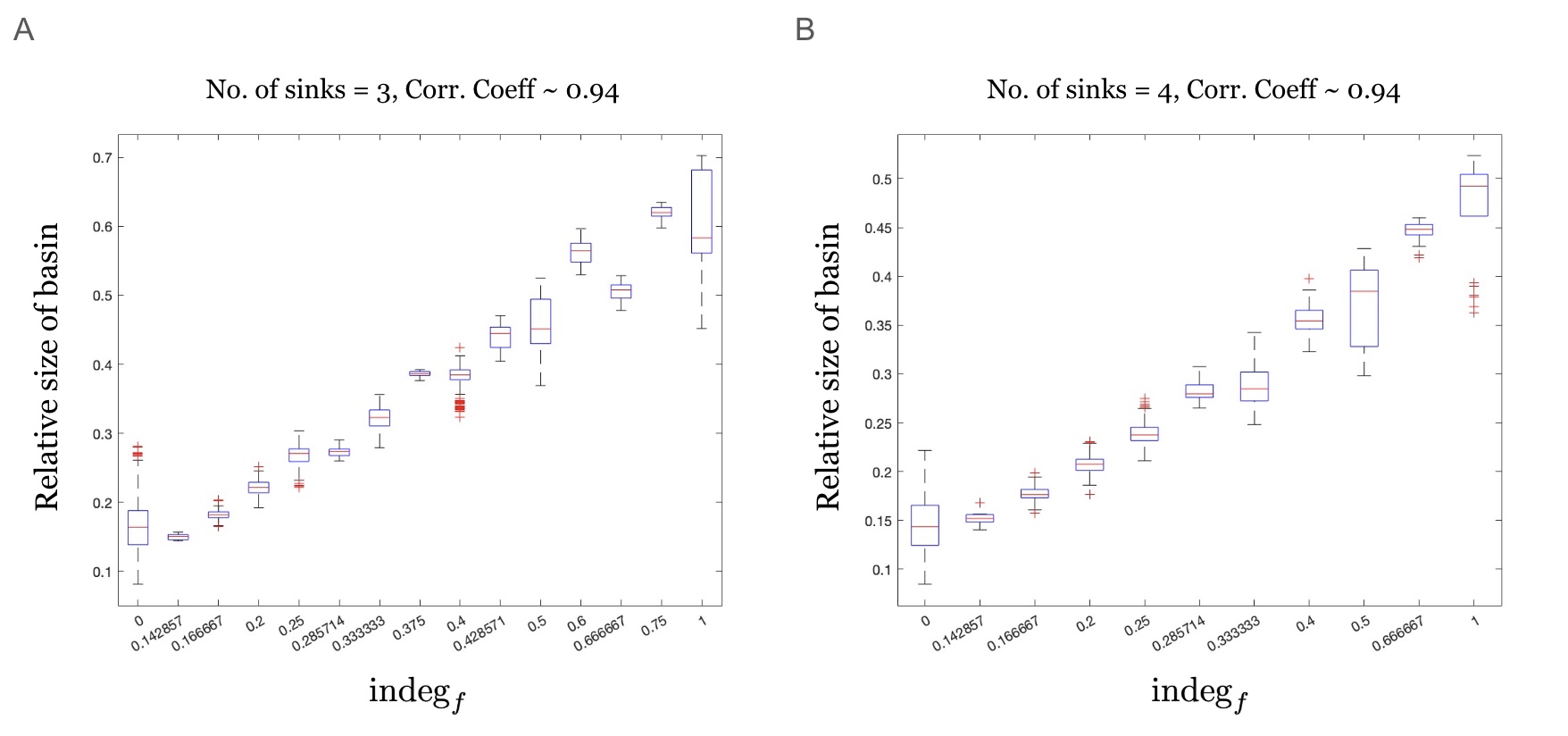}
\vspace{.1in}
\caption[Correlation between basin size and sink fractional indegree]{{\bf Correlation between basin size and sink fractional indegree.} Simulation similar to Fig~\ref{fig:dbfig}A sampling in the neighborhood of the fixed points supported on all sinks.  1000 DAGs were simulated of size $n=6$ where 500 had 3 sinks and 500 had 4 sinks with parameters $\delta=0.5$, $\varepsilon=0.25$, and $\theta=1$.  The number of initial conditions sampled per DAG was 2500. (A) Restricted to the 500 DAGs with 3 sinks.  (B) Restricted to the 500 DAGs with 4 sinks.}
\label{fig:db3}
\end{center}
\vspace{-.2in}
\end{figure}

What we conjecture now is that, the fractional volume of a basin of attraction for a sink $i$:

$$\mathcal{F}_i:=\dfrac{\lambda(\mathcal{B}_i \cap S)}{\lambda(S)}$$ 

where $S:=B_{\eta}\left(x^*_{\operatorname{sinks}(G)}\right) \cap R^+_{\operatorname{sinks}(G)}$, correlates with $\operatorname{indeg_{f}}(i)$.

Numerically simulating this for 1000 random DAGs of six neurons for $\eta=0.01$ split between 3 and 4 sinks, we again find a reasonably strong correlation as depicted in Fig~\ref{fig:db3}A and Fig~\ref{fig:db3}B respectively.  While the spread is noticeably less tame than in the two dimensional case, and the monotonicity does not quite seem to hold, it is still a far superior than the relationship in Fig~\ref{fig:dbbox}B.  A point to note is that, in all the cases we have looked at, the accuracy drops considerably if we consider sinks where $\operatorname{indeg_{f}}=0$.  This is likely because of a limitation in the $\operatorname{indeg_{f}}$ construction where in this case it makes no distinction between a neuron having indegree 0 whether $\sum_{j\in \operatorname{sinks}(G)} \operatorname{indeg}(j)$ is large or small.  This could create considerable variance in this case.

In the above analysis we found strong computational results, justified by some theoretical results, that, in a DAG CTLN, if we restrict our basins of attraction to a neighborhood of the saddle point $x^*_{\operatorname{sinks}(G)}$, the volumes of the basins relative to one another are strongly related to the fractional indegree, $\operatorname{indeg_{f}}$, received by the sinks relative to one another.  The significance of this is that, assuming a circuit architecture comparable to that of a DAG CTLN, under the hypothesis that the dynamics of a decision-making circuit operate along their decision-boundaries before arriving at a decision, we expect bias to be strongly affected by the direct excitation received by the neural population encoding a choice relative to that received by populations encoding other choices.

\chapter{Balanced States and their Decision-Making Dynamics in DAG CTLNs}

We turn our attention now back to balanced states.  As a reminder, the balanced state of a TLN is when the internal inhibition of the circuit is cancelled out by the excitation of external input current i.e. when $\sum_{i=1}^n W_{ij}x_j +\theta_j=0$ for all $i \in [n]$.  Combining each of these conditions, the balanced state is the state $x_{bs}$ such that $W x_{bs} + \Vec{\theta}=0$.

\begin{rmk}
Notice that the balanced state $\Vec{x}_{bs}$ corresponds to the intersection of the hyperplanes $\{H_i\}_{i=1}^n$.  It is the unique point adjacent to all chambers of the TLN.
\end{rmk}

Balanced states are of great interest in the theory of attractor networks, but are often computed without considering further dynamics and are at times thought to be related to the stable states of the network.\cite{bs3}. This is often not true in many attractor network models.  Certainly this is not generally the case in TLNs.  What we will consider here is the possibility that the balanced state represents the appropriate starting point of the network as it begins computation.  Since the balanced state does not typically lie on a separatrix, it will generally converge to one of the attractors.  In this paradigm, we will consider the bias of the network to be toward that attractor.  What we seek to determine is which basin of attraction the balanced state lies in.  To be more precise, as the balanced state is a computation on the external input received by a network, the vector $\Vec{\theta}$, we might consider that the appropriate way of measuring the bias of the decision-making circuit is to evaluate to which attractor the balanced state converges in the event of uniform external input i.e. $\Vec{\theta}=\theta \mathbbm{1}$, which is the case in CTLNs.  As shown in Fig~\ref{fig:bdhyp}, privileging this particular trajectory this aligns with the paradigm of path-following dynamics.  

\begin{figure}[!ht]
\begin{center}
\vspace{.1in}
\includegraphics[width=6.25in]{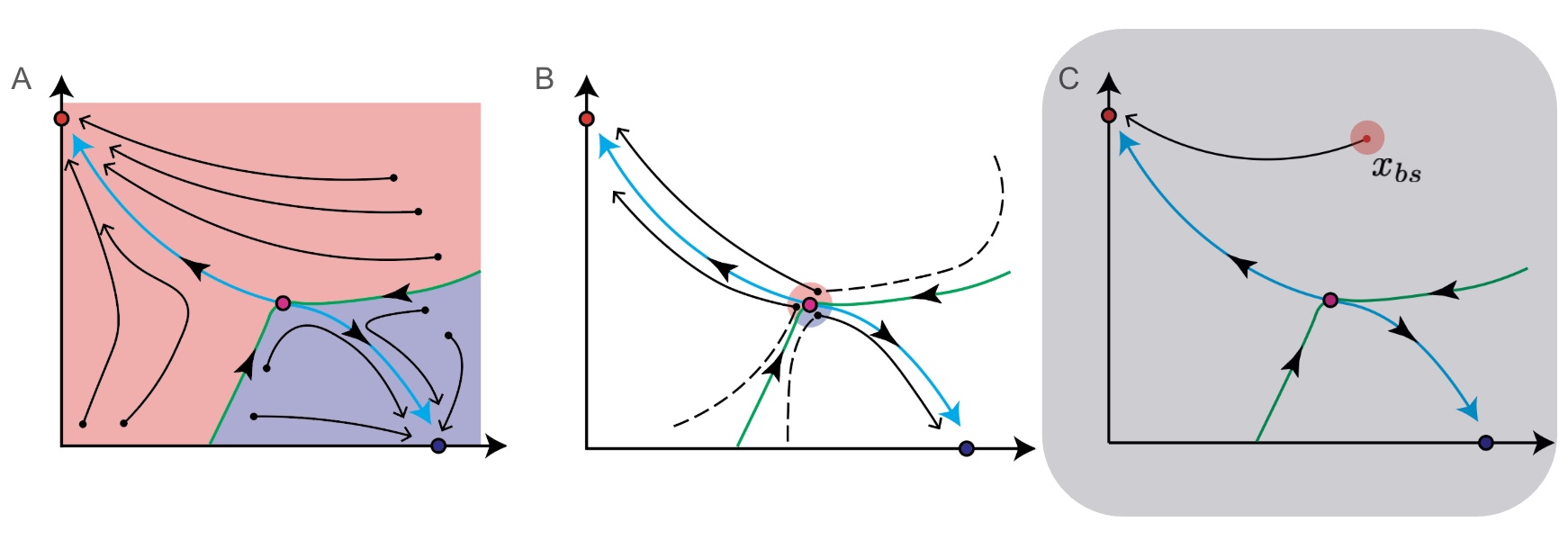}
\vspace{.1in}
\caption[Decision-making dynamics along the balanced state trajectory]{{\bf Decision-making dynamics along the balanced state trajectory.}  Within this paradigm, the bias of the decision-making circuit is aligned with the attractor to which the balanced state trajectory converges.}
\label{fig:bdhyp}
\end{center}
\vspace{-.2in}
\end{figure}

As seen in our discussion of the Binary Competition Model, two-dimensional competitive TLNs always have a balanced state in the positive quadrant, but this is certainly not the case in general.  In higher dimensional TLNs, $\Vec{x}_{bs}$ frequently lies outside of the positive orthant (Fig~\ref{fig:X0}).  This presents a challenge as it is nonsensical to have negative firing rates.

\begin{figure}[!ht]
\begin{center}
\vspace{.1in}
\includegraphics[width=5.75in]{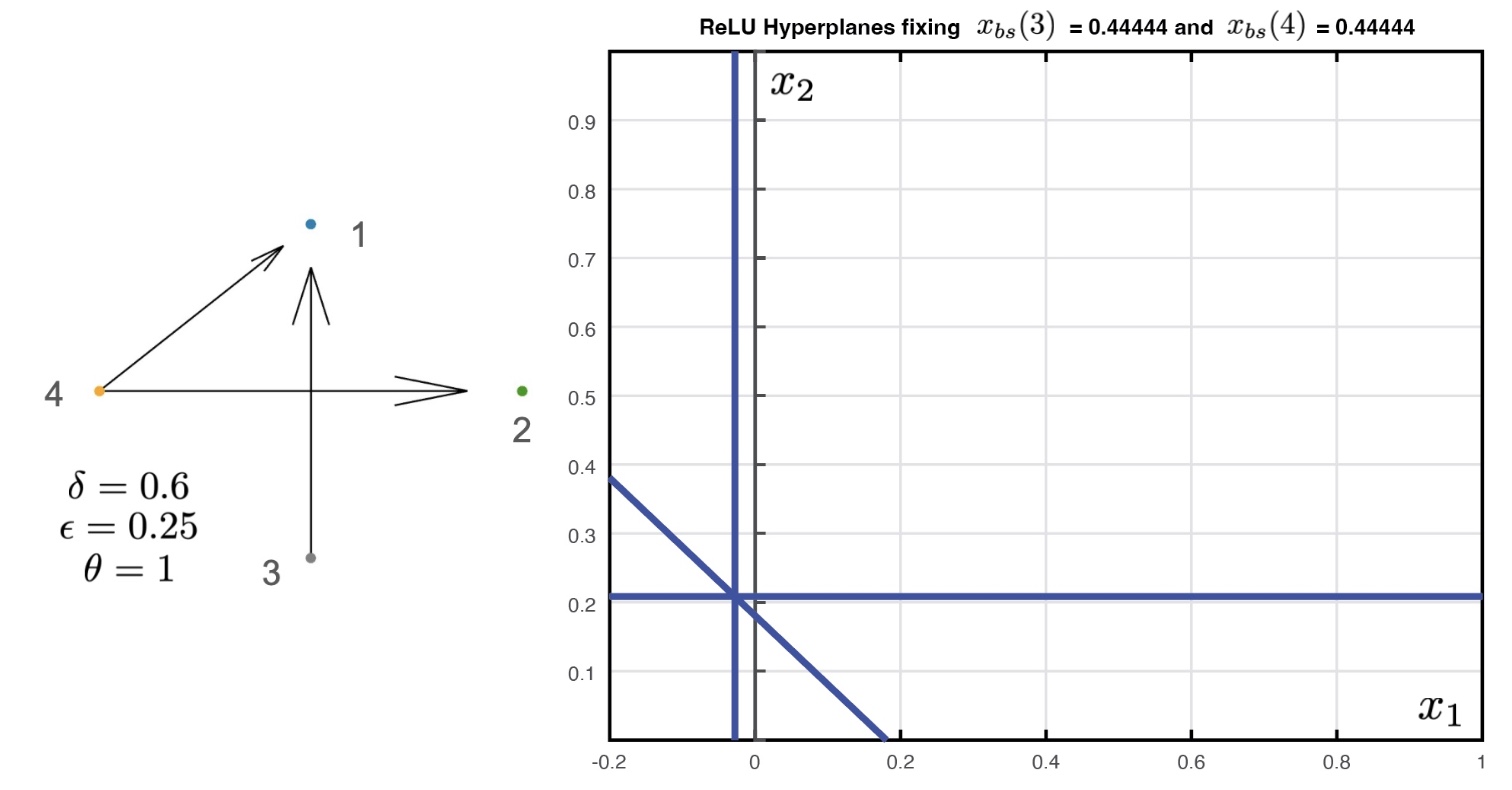}
\vspace{.1in}
\caption[Unbalanced CTLN]{{\bf Unbalanced CTLN.}  For the depicted graph and parameter values, the associated CTLN is unbalanced with $x_{bs}(1)$ having a negative value.  This is biologically nonsensical as the state variables are meant to represent firing rates of neurons.}
\label{fig:X0}
\end{center}
\vspace{-.2in}
\end{figure}

\begin{ddd}
A TLN is said to be \textbf{balanced} if the point $x_{bs}=-W^{-1}\Vec{\theta}$ lies in the positive orthant.
\end{ddd}

Our goal in this chapter is twofold.  First, we want to determine when we can use the balanced state as a reasonable initial condition.  In other words, we want to determine when a CTLN is balanced.  Our second goal will be to better understand to what attractor a trajectory beginning at the balanced state converges.  We prove that there exists a sufficient condition for a CTLN to be balanced derived from the maximum in-degree of a graph.  Employing again localized path polynomials, we find a sharper result for CTLNs derived from DAGs and use it to show that there exist graphs such that their CTLNs are always balanced, regardless of the choice of parameters.  We then return to the question of basins of attraction, presenting an algorithm which aims to predict within which basin the balanced state trajectory lies.

\section{Balanced CTLNs}

We begin with a very general theorem to determine that a CTLN is balanced that works for any CTLN with no assumptions on graph structure.  The proof for this result makes use of Farkas' Lemma which we restate here in a slightly adapted form.

\begin{lem}[Farkas' Lemma: Theorem 4.6 in \cite{fark}]\label{lem:fark}
    Let $W\in \mathbbm{R}^{m\times n}$ and $\Vec{b}\in \mathbbm{R}^m$.  Then exactly one of the following two assertions is true:

    1. There exists an $\Vec{x}\in \mathbbm{R}^n$ such that $W\Vec{x}=\Vec{b}$ and $\Vec{x}\geq 0$.

    2. There exists a $\Vec{y}\in \mathbbm{R}^m$ such that $W^T \Vec{y}\geq 0$ and $\Vec{b}\cdot \Vec{y} < 0$.
\end{lem}

Using Farkas' Lemma with $\Vec{b}=-\theta \mathbbm{1}$, we obtain the following result:

\begin{thm}\label{thm:bs}
(Balanced State Theorem) Let G be a directed graph with maximum in-degree $d_{\max}$.  Any CTLN satisfying: 

$$\dfrac{\varepsilon+\delta}{1+\delta} \leq \dfrac{1}{d_{\max}}$$ 

\noindent associated with G is balanced.
\end{thm}

\begin{proof}
    The CTLN has a balanced state if alternative (1) of Lemma~\ref{lem:fark} holds true for $\Vec{b}=-\theta \mathbbm{1}$.  We will assume alternative (2) of Lemma~\ref{lem:fark} and aim for a contradiction.  We will show that if $\dfrac{\varepsilon+\delta}{1+\delta} \leq \dfrac{1}{d_{\max}}$, there does not exist $\Vec{y} \in \mathbbm{R}^n$ satisfying:

    a. $-\theta\mathbbm{1} \cdot \Vec{y}<0$
    
    b. $(W^T \Vec{y})_i \geq 0$, $\forall i \in [n]$
    
    Dividing through by $-\theta$, (a) can be rewritten as:
    
    a. $\sum_{i=1}^n y_i > 0$

    We will show that if (a) is true, (b) is contradicted.

    Since (b) requires the inequality to hold $\forall i\in [n]$, it would follow that $\forall S \subseteq [n]$:

    $$\sum_{i\in S} (W^T \Vec{y})_i \geq 0$$

    Recall that $W=(-1-\delta) \mathbbm{1}\mathbbm{1}^T+(1+\delta)I+(\varepsilon+\delta)A$ where $A$ is the adjacency matrix of $G$. It follows that $W^T=(-1-\delta)\mathbbm{1}\mathbbm{1}^T+(1+\delta)I+(\varepsilon+\delta)A^T$.  

    This allows for the above inequality to be rewritten as:

    $$\sum_{i\in S} (W^T \Vec{y})_i = (-1-\delta)|S| \sum_{i=1}^n y_i + (1+\delta)\sum_{i\in S} y_i + (\varepsilon + \delta)\sum_{i=1}^n d^S_i y_i \geq 0$$

    where $d_i^S=|\{ j \in S | j \rightarrow i \}|$

    Dividing by $(-1-\delta)$ on both sides yields the equivalent:

    $$|S| \sum_{i=1}^n y_i -\sum_{i\in S} y_i - \alpha \sum_{i=1}^n d^S_i y_i \leq 0$$

    where $\alpha = \dfrac{\varepsilon + \delta}{1+\delta}$.

    Fix $\Vec{y}$ and construct the sets $P=\{i|y_i \geq 0\}$ and $N=\{j|y_j < 0\}$.  Observe that $P \sqcup N = [n]$ and that since $\sum_{i=1}^n y_i > 0$, $P\neq \emptyset$.

    Then we assume by way of contradiction that:

    $$|N| \sum_{i=1}^n y_i -\sum_{i\in N} y_i - \alpha \sum_{i=1}^n d^N_i y_i \leq 0$$

    By decomposing $[n]$ into $P \sqcup N$, this is equivalent to:

    $$|N|\sum_{i\in P}y_i - \alpha \sum_{i \in P} d^N_i y_i + (|N|-1)\sum_{j \in N} y_j - \alpha \sum_{j \in N} d^N_j y_j \leq 0$$

    Since $0 \leq d^N_i \leq \min(d_{\max},|N|) \leq |N|$, it follows that:

    $$(|N|-\alpha\min(d_{\max},|N|))\sum_{i\in P}y_i + (|N|-1)\sum_{j \in N} y_j \leq 0$$

    Now we show the contradiction.  Taking (a) to be true, $\sum_{i=1}^n y_i= \sum_{i\in P} y_i + \sum_{i \in N} y_j > 0$.  Then it follows that for $A > 0$ and $B \geq 0$ s.t. $A\geq B$, it must also be true that $A \sum_{i\in P} y_i + B \sum_{i \in N} y_j > 0$.

    Clearly $|N|-1 \geq 0$ to avoid a trivial contradiction, and, since $\alpha < 1$, it follows that $|N|-\alpha \min(d_{\max},|N|)\geq |N|(1-\alpha)>0$. So, if $|N|-\alpha \min(d_{\max},|N|)\geq |N|-1$, there is a contradiction.  
    
    If $d_{\max}=1$, this is always true as $|N|-\alpha \min(1,|N|)\geq |N|-\alpha>|N|-1$.

    If $d_{\max}>1$, then the contradiction could potentially be avoided if $|N|-\alpha \min(d_{\max},|N|)<|N|-1$ i.e. if $\dfrac{1}{\min(d_{\max},|N|)}<\alpha$.  However, $\dfrac{1}{d_{\max}}\leq\dfrac{1}{\min(d_{\max},|N|)}$ and, by assumption, $\alpha \leq \dfrac{1}{d_{\max}}$.
\end{proof}

\begin{corr}
Let G be a directed graph with maximum in-degree $d_{\max}$.  If $d_{\max} = 1$, then any CTLN associated with $G$ has a balanced state.  If $d_{\max} > 1$, then any CTLN satisfying: 

$$\delta \leq -1+\sqrt{1+\dfrac{1}{d_{\max}-1}}$$ 

associated with G has a balanced state.
\end{corr}

\begin{proof}
Again, If $d_{\max}>1$, then the contradiction is avoided if $|N|-\alpha \min(d_{\max},|N|)<|N|-1$ i.e. if $\dfrac{1}{\min(d_{\max},|N|)}<\alpha$.  But, $\alpha=\dfrac{\varepsilon+\delta}{1+\delta}<1-\dfrac{1}{(1+\delta)^2}$ and $\dfrac{1}{\min(d_{\max},|N|)} < 1-\dfrac{1}{(1+\delta)^2}$ only if:

    $$\delta > -1+\sqrt{1+\dfrac{1}{\min(d_{\max},|N|)-1}}> -1+\sqrt{1+\dfrac{1}{d_{\max}-1}}$$

    But this contradicts the assumption that $\delta \leq -1+\sqrt{1+\dfrac{1}{d_{\max}-1}}$ in the case of $d_{\max}>1$.
\end{proof}

\begin{corr}
    If G is a directed graph with number of sinks $s>0$, then the CTLN associated with G is balanced for $\delta \leq -1+\sqrt{1+\dfrac{1}{n-s-1}}$.
\end{corr}

\begin{proof}
    If a graph $G$ of size $n$ has $s$ sinks, then, since the sinks have out-degree $0$, $d_{\max}\leq n-s$.
\end{proof}

\begin{figure}[!ht]
\begin{center}
\vspace{.1in}
\includegraphics[width=5.75in]{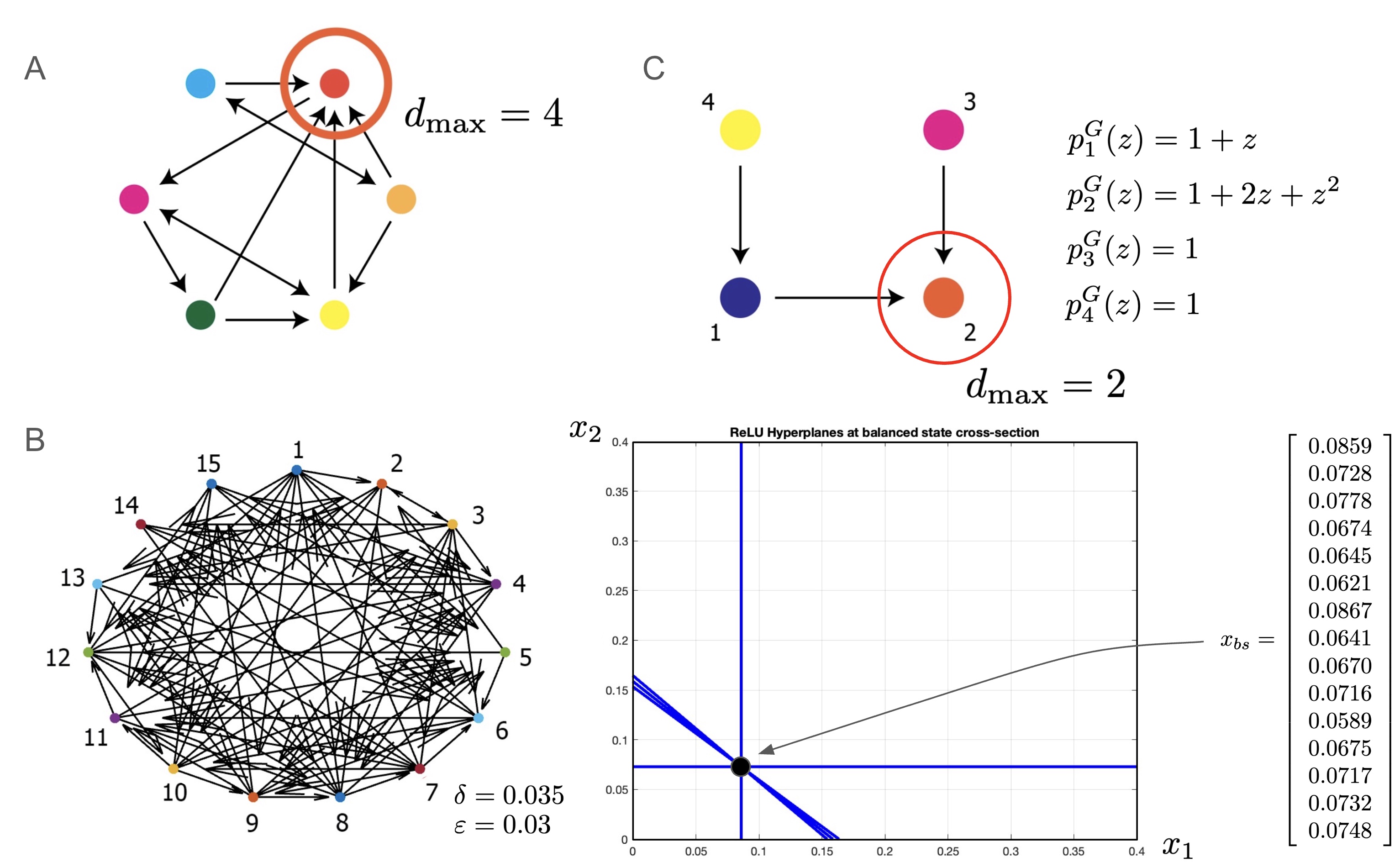}
\vspace{.1in}
\caption[Balanced states of CTLNs]{{\bf Balanced states of CTLNs.}  (A) As $d_{\max}=4$ in this graph, as associated CTLN is balanced for $\dfrac{\varepsilon+\delta}{1+\delta}\leq \dfrac{1}{4}$.  (B) Any directed graph of 15 neurons will have an associated balanced CTLN for $\dfrac{\varepsilon + \delta}{1+\delta} \leq \dfrac{1}{14}$.  (C) Theorem~\ref{thm:bs} is a sufficient but not a necessary condition.  Using Theorem~\ref{thm:dagbs} we know that any CTLN associated with this DAG is balanced.}
\label{fig:X}
\end{center}
\vspace{-.2in}
\end{figure}

There are some interesting takeaways from these results.  The most important of these is that there is no graph such that associated CTLNs are not balanced for sufficiently weak inhibition.  We can easily look at a graph and devise a choice of $\varepsilon$ and $\delta$ such that the CTLN is balanced.  In the case of the example depicted in Fig~\ref{fig:X}A, any choice of parameters such that $\dfrac{\varepsilon+\delta}{1+\delta} \leq 0.25$ will suffice.  Even if we did not know a graph's structure, we could use the fact that $d_{\max} \leq n-1$ and quite generally choose $\varepsilon,\delta$ such that $\dfrac{\varepsilon+\delta}{1+\delta} \leq \dfrac{1}{n-1}$.  In Fig~\ref{fig:X}B we have a random graph of 15 neurons, fairly dense with edges, and have chosen $\delta=0.035$ and $\varepsilon = 0.03$ satisfying $\dfrac{\varepsilon+\delta}{1+\delta} \leq \dfrac{1}{14}$.  The CTLN is balanced as can be seen in the balanced state cross-section of the state space.

While these results are useful and general, a natural concern presents itself.  As the network gets larger and the probability of excitatory connections remains constant, the expected in-degree scales linearly and $d_{\max}$ should rise.  We would need extremely weak inhibition to ensure balance in large networks.

The conditions derived from the above results are sufficient but not always necessary.  Since our interest is mainly in CTLNs derived from DAGs, can we do any better in that case?  In fact we can!

% DAG BS Setup

\begin{lem}\label{lem:dagbsp}
Let G be a DAG.  Letting $\beta=\dfrac{-\varepsilon-\delta}{1+\delta}$, the point $x_{bs}=-W^{-1}(\theta \mathbbm{1})$ of an associated CTLN is:

\begin{center}
    $(x_{bs})_j=\dfrac{p^G_j(\beta)}{-1+\sum_{i=1}^n p^G_i(\beta)}\left(\dfrac{\theta}{1+\delta}\right)$ for $j\in [n]$
\end{center}
\end{lem}

\begin{proof}
We need to solve $W\vec{x}=-\theta\mathbbm{1}$.  This is equivalent to $(W-(1+\delta) I)\vec{x}=-\theta\mathbbm{1}-(1+\delta)\vec{x}$.  

Notice that $\dfrac{-1}{1+\delta}(W-(1+\delta) I)=\mathbbm{1}\mathbbm{1}^T+cA=\dfrac{\theta}{1+\delta}\mathbbm{1}+\vec{x}$ where $c=\dfrac{-\varepsilon-\delta}{1+\delta}$ and $A$ is the adjacency matrix of $G$.  The result follows by applying Lemma~\ref{lem:DAG}.
\end{proof}

The power of using the Lemma~\ref{lem:DAG} here is that the question of whether or not a CTLN is balanced can be transformed from a general problem of matrix inversion and can instead be approached via tracking the sign changes of the path polynomials i.e. a problem of root detection. 

\begin{ddd}
    Define $r_G=\max \{z\in \mathbbm{R} \mid \exists i\in V(G) \text{ s.t. } p^G_i(z)=0\}$ to be the greatest real root for the path polynomials of G.  By convention, if $\{ p_i^G(z) \}_{i\in V(G)}$ has no real roots we say $r_G=-\infty$.
\end{ddd}

\begin{rmk}
    Since $\{ p_i^G(z) \}_{i\in V(G)}$ are polynomials with positive coefficients, $r_G<0$.
\end{rmk}

The relationship between the existence of the balanced state and the roots of the path polynomials is captured in the following result.

% DAG Balanced State Theorem

\begin{thm}\label{thm:dagbs}
    (Balanced State Theorem for DAGs) Let $G$ be a DAG of size $n\geq 2$.  A CTLN associated with G has a balanced state if 
    $$\dfrac{\varepsilon+\delta}{1+\delta}\leq -r_G$$.  
\end{thm}

\begin{proof}
    From Lemma~\ref{lem:dagbsp}, It suffices to show that: 

    $$\dfrac{p^G_j(\beta)}{-1+\sum_{k=1}^n p_k^G(\beta)}>0$$

    for each $j\in [n]$.

    Define $p^G_0(z)=-1+\sum_{k=1}^n p_k^G(z)$.

    For $n\geq 2$, these polynomials have positive coefficients and so their greatest real root, if any real roots exist, are negative.  Call them $r_0, r_1, ..., r_n$ respectively with $r_i=-\infty$ if no real roots exist for $p^G_i(z)$.  Let $K=\sup_{i\in\{0\}\cup [n]} r_i$.  Since $p^G_0(z)=-p^G_s(z)+\sum_{k=1}^n p^G_k(z)=\sum_{i\neq s}^n p_i^G(z)$, where $x_s$ is any one of the source neurons, $r_0 < r_G$ as one of the path polynomials must change sign before $p_0^G(z)$ has a root.  So, $K=\sup_{i\in [n]} r_i=r_G$.  Then, for $\alpha > r_G$, we have $p^G_0(\alpha), p^G_1(\alpha),...,p^G_n(\alpha)>0$.  So, if $\beta=\dfrac{-\varepsilon-\delta}{1+\delta} \geq r_G$, the result holds.
\end{proof}

\begin{corr}
    Let $G$ be a DAG of size $n\geq 2$.  If $r_G\leq -1$, then any CTLN associated with G has a balanced state.  If $r_G>-1$, then a CTLN associated with G has a balanced state if 
    $$\delta<-1+\dfrac{1}{\sqrt{r_G+1}}$$.  
\end{corr}

\begin{proof}
Note that: 
    
    $$\dfrac{-\varepsilon-\delta}{1+\delta}>\dfrac{-\dfrac{\delta}{1+\delta}-\delta}{1+\delta}=\dfrac{-\delta^2-2\delta-1}{\delta^2+2\delta+1}+\dfrac{1}{(1+\delta)^2}=-1+\dfrac{1}{(1+\delta)^2}$$
    
    so it suffices to require that: 
    
    $$-1+\dfrac{1}{(1+\delta)^2}>r_G$$

    Some rearranging yields:

    $$1>(r_G+1)(1+\delta)^2$$

    This yields two cases:

    \textbf{Case 1:} $r_G\leq-1$

    In this case, $r_G+1 \leq 0$ and the inequality is always true.

    \textbf{Case 2:} $r_G>-1$

    In this case, the inequality $1>(r_G+1)(1+\delta)^2$ produces the condition:

    $$\delta<-1+\dfrac{1}{\sqrt{r_G+1}}$$
\end{proof}

% Examples
Consider the example depicted in Fig~\ref{fig:lppbs}.  The use of localized path polynomials allowed us to, without much difficulty, find conditions on CTLN balance, but also note that it was not any better than that obtained by Theorem~\ref{thm:bs}.  

\begin{figure}[!ht]
\begin{center}
\vspace{.1in}
\includegraphics[width=5.25in]{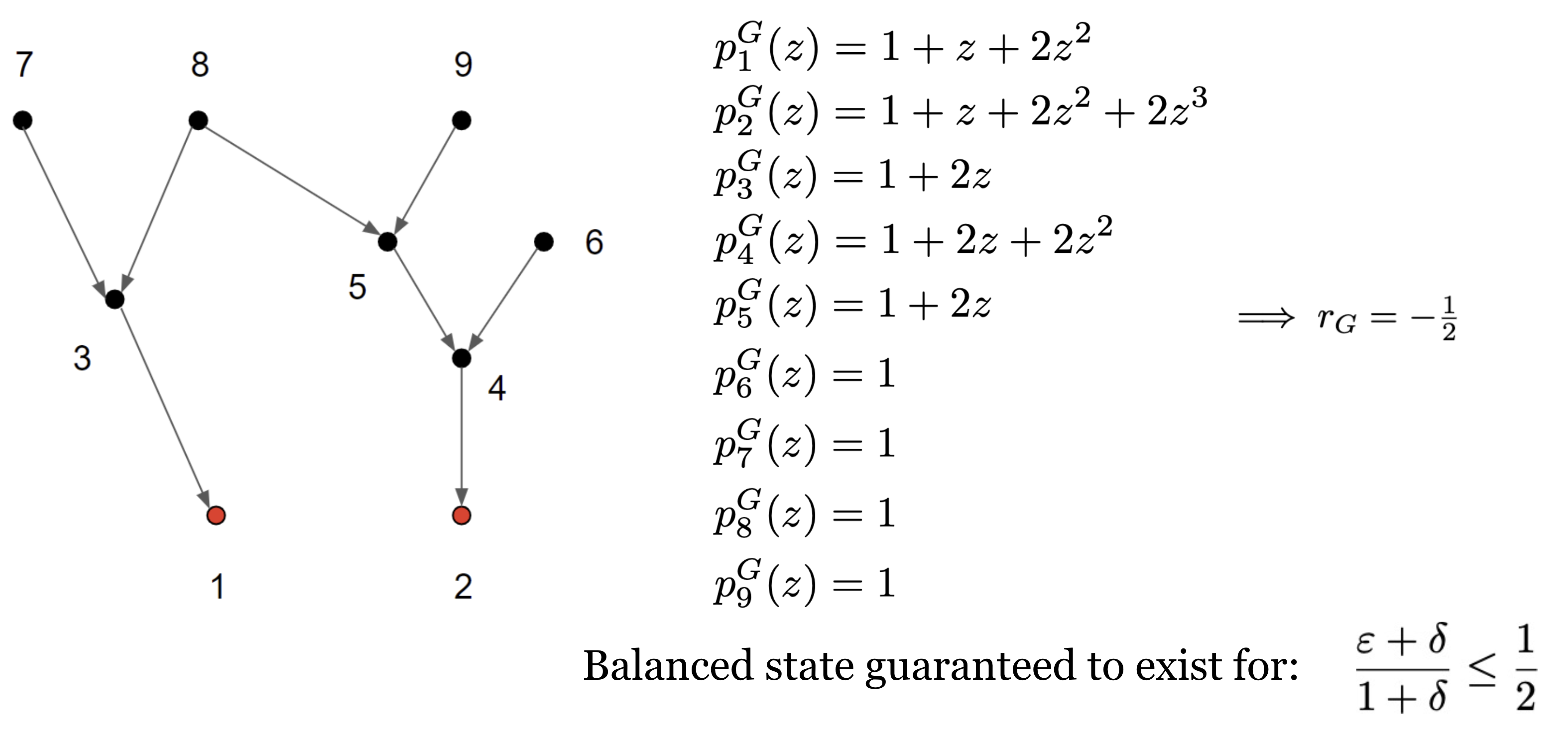}
\vspace{.1in}
\caption[Localized path polynomials and CTLN balance]{{\bf Localized path polynomials and CTLN balance.}  Each vertex of the DAG contributes a localized path polynomial and taking the greatest root $r_G$ among them gives us a condition on CTLN balance.}
\label{fig:lppbs}
\end{center}
\vspace{-.2in}
\end{figure}

That said, this result is indeed stronger and so let us pause briefly to look at an example illustrating the added strength of the above theorem.  We will first evaluate the following example using the original, general Balanced State Theorem and then with the new Theorem~\ref{thm:dagbs}.

In the DAG depicted in Fig~\ref{fig:X}C, $d_{\max}=2$, so Theorem~\ref{thm:bs} guarantees that an associated CTLN will be balanced for $\dfrac{\varepsilon + \delta}{1+\delta} \leq \dfrac{1}{2}$.  

Now, applying the new theorem, we find the following localized path polynomials

$$p_{1}^G (z) = z + 1$$
$$p_{2}^G (z) = z^2 +2z + 1=(z+1)^2$$
$$p_{3}^G (z) = 1$$
$$p_{4}^G (z) = 1$$

It is clear to see for these polynomials that we have $r_G=1$.  Thus, we find that a CTLN derived from this DAG is in fact always balanced!  

This begs another question: for what graphs are all associated CTLNs balanced?

% Balanced Graphs

% Fig XX
\begin{figure}[!ht]
\begin{center}
\vspace{.1in}
\includegraphics[width=5.75in]{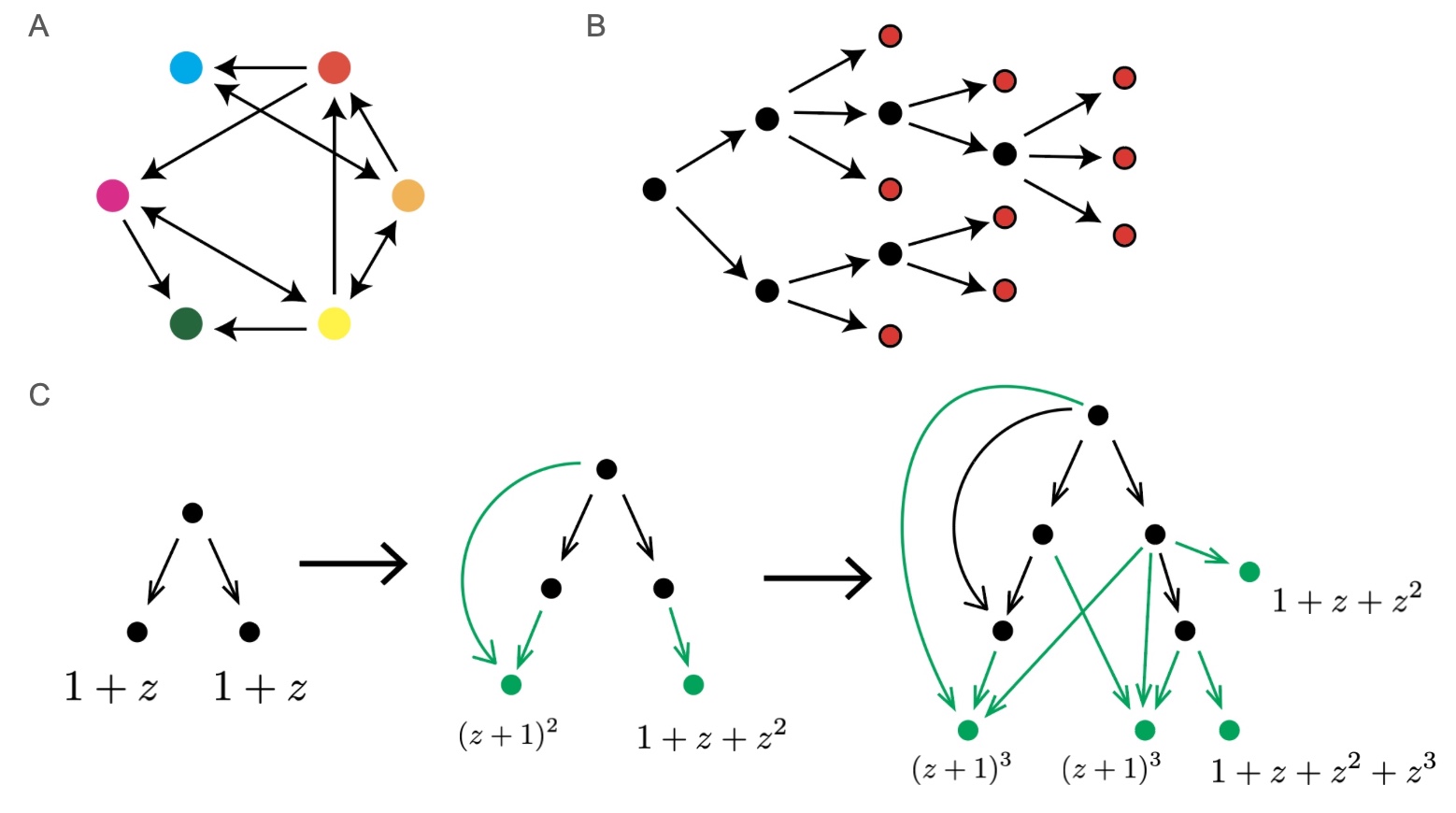}
\vspace{.1in}
\caption[Balanced graphs]{{\bf Balanced graphs.}  (A) Uniform in-degree graphs balanced.  This graph has uniform in-degree 2.  (B) Out-trees are balanced graphs.  (C) Large balanced graphs can be constructed by adding vertices with localized path polynomials known to have no roots in $(-1,0)$.  As no edges are being drawn back to the extant vertices, their localized path polynomials remain unchanged.}
\label{fig:XX}
\end{center}
\vspace{-.2in}
\end{figure}

\begin{ddd}
A directed graph G is said to be \textbf{balanced} if all CTLNs derived from G are balanced.
\end{ddd}

One class of graphs which are balanced is those which are uniform in-degree i.e. all vertices have the same in-degree (Fig~\ref{fig:XX}A).

\begin{thm}\label{thm:ufdbs}
 (UFD Theorem for Balanced States) Let G be a uniform in-degree directed graph.  Then any associated CTLN is balanced. 
\end{thm}

\begin{proof}
    Let $W$ be the weight matrix for a CTLN associated with $G$.
    
    Let $d$ be the uniform in-degree of the vertices.  Observe that the row sum for each row of $W$ will be uniform and equal to $d(-1+\varepsilon)+(n-1-d)(-1-\delta)$.

    Then, construct: 
    
    $$x_{bs}=\dfrac{-\theta \mathbbm{1}}{d(-1+\varepsilon)+(n-1-d)(-1-\delta)}.$$

    Since $d\leq n-1$, both terms of the bottom sum are negative.  As $\theta$ is positive $x_{bs}>0$.  Also observe that since the denominator is the uniform rom sum, $W x_{bs}=-\vec{\theta}$.  Thus, $x_{bs}$ lies in the positive orthant.    
\end{proof}

To be uniform in-degree is a very strong condition on a graph, but fortunately our results suggest other classes of balanced graphs.  Theorem~\ref{thm:bs} suggests that any graph such that $d_{\max} \leq 1$ is balanced.  While this also seems quite narrow, it includes an interesting class of graphs.  Out-trees (Fig~\ref{fig:XX}B), directed trees with in-degree at most 1, satisfy this condition and so are balanced.  

\begin{corr}\label{corr:outtree}

Let G be a directed graph which is an out-tree.  Then, any CTLN associated with G is balanced.

\end{corr}

There have been studies which propose decision-making as potentially a sequence of binary choices\cite{binary} which suggests a DAG network architecture.  All out-trees are DAGs and would fit well with this paradigm.  

Concentrating further on DAGs, an additional consequence of the path polynomial formulation of balanced states is that it gives us the tools for generating balanced DAGs.

\begin{corr}
Let G be a DAG such that the maximum path length in G is 2.  Then, the CTLN associated with G has a parameter independent balanced state if and only if each vertex with maximum path length 1 has in-degree 1 and each vertex $i$ with max path length 2 has incoming paths obeying $n_1^i < 2 \sqrt{n_2^i}$ or, if $n_2^i=1$, $n_1^i \leq 2$.
\end{corr}

\begin{proof}
    The path polynomials of G are of the form $p^G_i(z)=1$,  $p^G_i(z)=n_1^i z + 1$, and $p_i^G(z)=n_2^i z^2 +n_1^i + 1$.  Note that $p^G_i(z)=1$ trivially has no roots in $(-1,0)$ and $p^G_i(z)=n_1^i z + 1$ has no roots in the interval if and only if $n_1^i=1$.  

    Since the constant term of $p_i^G(z)=n_2^i z^2 +n_1^i + 1$ is 1, the roots of the polynomial, $r_1$ and $r_2$, are such that $r_1 r_2=\dfrac{1}{n_2^i} \leq 1$.  This leads to two cases.

    \textbf{Case 1: } $n_2^i=1$
    
    Then, if $n_2^i=1$ the polynomial has a root in $(-1,0)$ if and only if both roots are real and distinct.  Then, there are no roots in the interval if and only if there is a repeated root or if there are imaginary roots.  What is required then is that the discriminant be non-positive i.e. $(n_1^i)^2 - 4 \leq 0$, which is rearranged to $n_1^i \leq 2$.

    \textbf{Case 2: } $n_2^i>1$
    
    If $n_2^i>1$, then the polynomial has a root in $(-1,0)$ if it has a real root.  This is avoided if the roots are imaginary i.e. if $(n_1^i)^2 - 4n_2^i < 0$, which is rearranged to $n_1^i < 2 \sqrt{n_2^i}$.
\end{proof}

If one wished to find similar conditions for parameter independent balance for more complex networks, there exists a vast literature around root detection of polynomials in real intervals, such as Sturm's Theorem (refer to section 8.4 of \cite{sturm} for a discussion of Sturm's Theorem), which could potentially be exploited to these ends.

Additionally, by taking localized path polynomials which are known to not have roots in the interval $(-1,0)$ we can quickly generate balanced DAGs.  Two classes of polynomials which are useful in this enterprise are finite geometric sums of the form $p(z)=\sum_{i=0}^m z^k$ and the binomial expansions $p(z)=(z+1)^m$.  Since localized path polynomials only take into account incoming paths, larger balanced DAGs can be built up from smaller ones.  In Fig~\ref{fig:XX}C, we start by taking a balanced graph of maximum path length 1 and progressively adding neurons in lower and lower topological layers, building larger and larger graphs, all of which are balanced.

% Balanced States as ICs

\section{Balanced States as Initial Conditions}

Now that we have a detailed understanding of balanced states in CTLNs, let us discuss their relevance to the problem at hand.  The particular trajectory is challenging to follow analytically.  It is not hard to see that for any TLN, the first chamber a trajectory beginning at $\Vec{x}_{bs}$ enters is the full support chamber $R_{[n]}$.  This means that even the beginning dynamics are governed by the most complex chamber, making analytical results difficult to obtain, while still being theoretically possible using Theorem~\ref{thm:final} and piecing the trajectory across the chambers $R_{\sigma}$.  What we will now discuss is an imperfect computational algorithm with which we have had considerable success at predicting the attractor to which a trajectory beginning at the balanced state converges.

\begin{ddd}
    $G^{i}=G|_{\sigma}$ where 
    $$\sigma=\{j \in [n] \text{ } | \text{ there does not exist a directed path of length i starting from j} \}$$
\end{ddd}

Notice that $G^i \subseteq G^{i+1}$, $G^1$ is the set of sinks, and $G^{m+1} = G$, where $m$ is the maximum path length in $G$, and so this layering of the DAG is a filtration as depicted in Fig~\ref{fig:filt}. 

\begin{figure}[!ht]
\begin{center}
\vspace{.1in}
\includegraphics[width=3in]{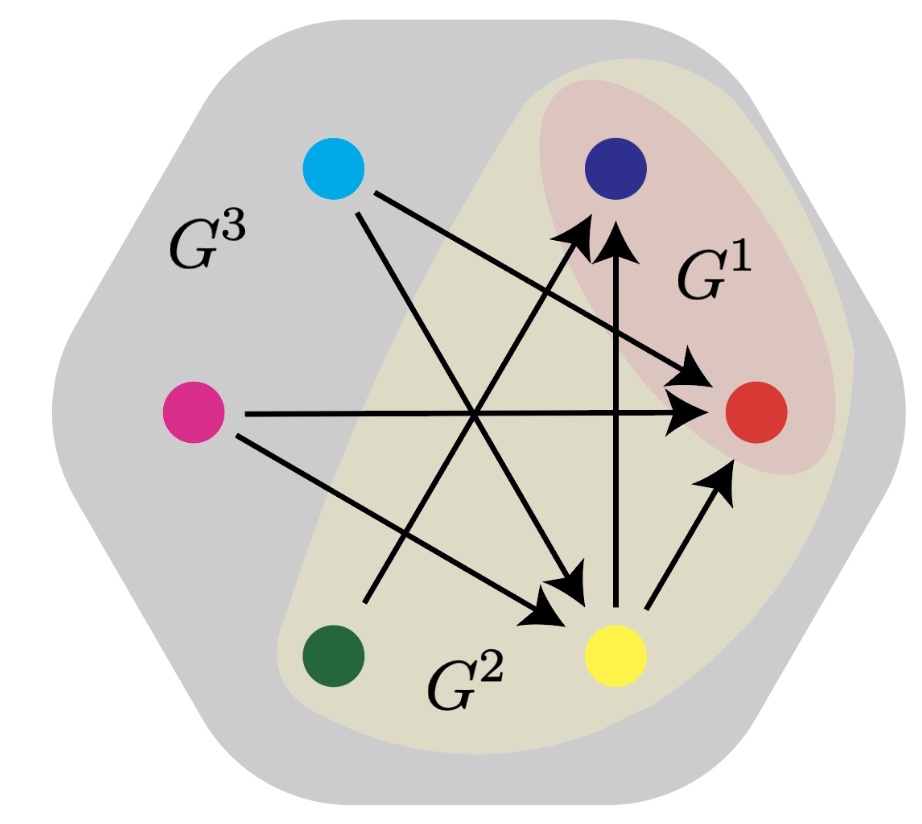}
\vspace{.1in}
\caption[$G^i$ filtration of a DAG]{{\bf $G^i$ filtration of a DAG.} An example of a DAG showing the construction of the subgraph filtration $\{G^i \}_{i=1}^3$.}
\label{fig:filt}
\end{center}
\vspace{-.2in}
\end{figure}

\begin{ddd}
    For a vertex i of a DAG $G$, let $d^i_k$ be the in-degree of vertex i in the subgraph $G^k$.  We call this the \textbf{k-th filtered in-degree of i}.
\end{ddd}

The algorithm begins by assembling a list of the sinks of the DAG as the possible attractor candidates.  We then consider $G^2$ and eliminate any sink from the list if its second filtered in-degree is smaller than that of any of the other sinks.  We then repeat the process, iterating through $G^k$ and comparing values of $d^i_k$ until we are left with one sink, which will be our prediction.  If multiple sinks survive all the way through, the algorithm deems the case inconclusive with the correct fixed point lying among the remaining candidates.  

\begin{algorithm}[H]\label{alg:bs}
    \caption{Balanced State Attractor Prediction}
    \begin{algorithmic}[1]
        \STATE{$C=\{i \in G | \text{ $i$ is a sink of $G$}\}$}
        \FOR{$k \leftarrow$ 2 to $n$}
            \STATE{$D=\{d^i_k | i \in C\}$}
            \STATE{$C \leftarrow \{i \in C | d^i_k =\max{(D)}\}$}
        \ENDFOR
        \IF{length($C$) = 1}
            \STATE{\textbf{return} $C$}
        \ENDIF
    \end{algorithmic}
\end{algorithm}

% Fig XXX
\begin{figure}[!ht]
\begin{center}
\vspace{.1in}
\includegraphics[width=6.25in]{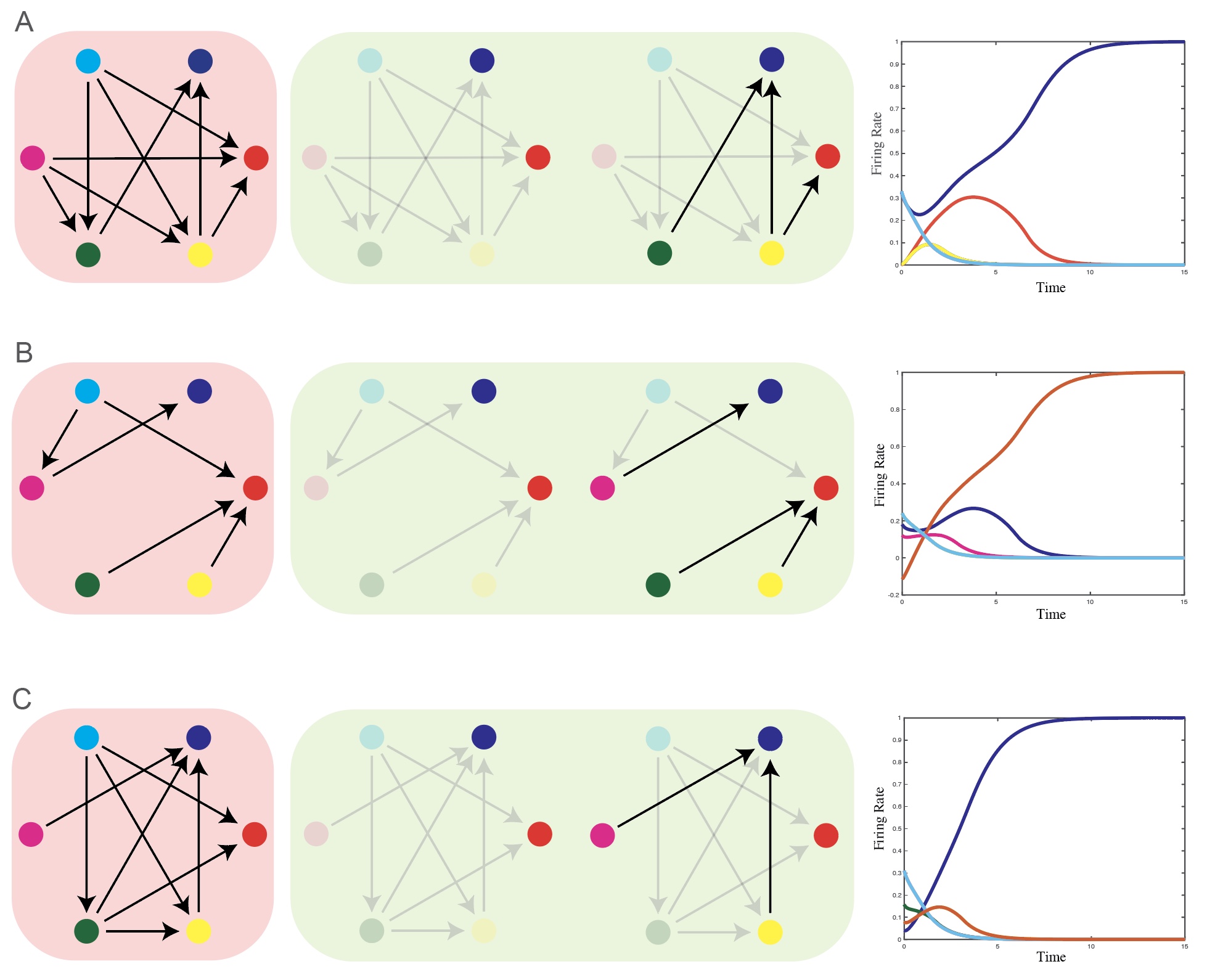}
\vspace{.1in}
\caption[Balanced State Attractor Prediction]{{\bf Balanced state attractor prediction.} An application of the prediction algorithm to three cases with parameters $\delta=0.5$, $\varepsilon=0.24$, and $\theta=1$.  On the left are the graphs, in the center is the filtration of the algorithm, stopping when one of the sink filtered in-degrees is dominant, and on the right is the numerical simulation of the firing rates, colored according to the neuron, confirming the results.  Notably, (B) is even an unbalanced CTLN and yet we have success with the prediction.}
\label{fig:XXX}
\end{center}
\vspace{-.2in}
\end{figure}

Fig~\ref{fig:XXX}A-C illustrate this process for three different DAGs, progressing through the filtration until reaching a prediction.  Comparing each of these with the actual result, we see accurate prediction in each case.  However, there are cases where this algorithm can make an incorrect prediction.  This begs the question: how often does this happen and to what extent does it depend on the parameters $\varepsilon$ and $\delta$?  If we test a set of DAGs at various grid points in parameter space we find the results in Figure~\ref{fig:V}.

\begin{figure}[!ht]
\begin{center}
\vspace{.1in}
\includegraphics[width=5.75in]{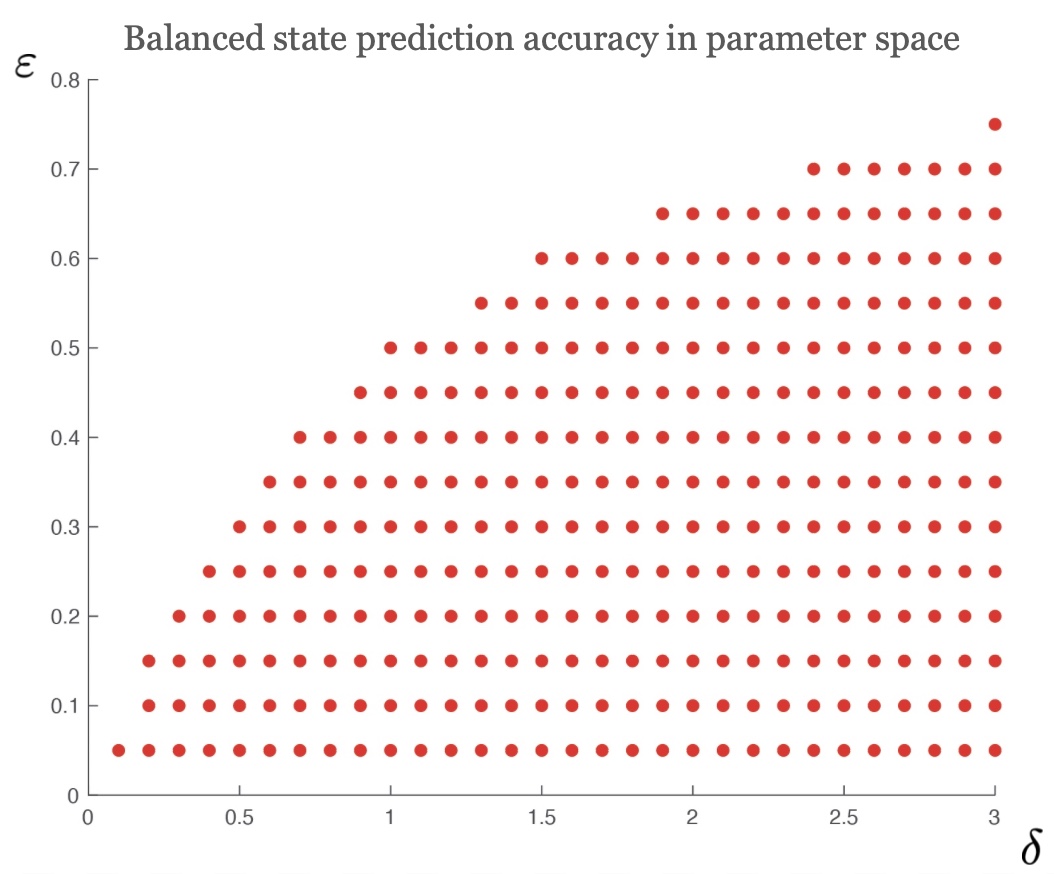}
\vspace{.1in}
\caption[Balanced state attractor prediction accuracy]{{\bf Balanced state attractor prediction accuracy.}  A list of 1000 DAG (n=8) CTLN balanced state trajectories were numerically tested for different parameter values and compared with the prediction from the balanced state attractor prediction algorithm.  The color intensity of the red point in parameter space marks the percentage correct out of the predicted set of the 1000 CTLNs.  We see minimal variation in color because accuracy was similar for various parameter values at approx. $92\%$.}
\label{fig:V}
\end{center}
\vspace{-.2in}
\end{figure}

\chapter{Heterogeneous DAG CTLNs}

We conclude this thesis with a primarily theoretical chapter that will give us tools to generalize some of our earlier results beyond strict CTLNs and also to resolve a shortcoming of our analytical solutions in DAG CTLNs.  We begin by noting that the CTLN conditions are quite strict and it is their strictness that makes the class so tractable.  The easiest condition to weaken is that of external drive symmetry, i.e. $\Vec{\theta}=\theta \mathbbm{1}$.  Instead, we will allow $\Vec{\theta}>0$ to be an arbitrary positive vector.  We will refer to this as a \textit{heterogeneous CTLN} (hCTLN).

The reason this is the easiest condition to weaken is that doing so does not affect the eigenvalues or eigenvectors of the $(-I+W)|_{\sigma}$ matrices.  This means that the general homogeneous solution within the chamber is intact. What is changed however is the hyperplane arrangement and the particular solutions of $L_{\sigma}$ i.e. the fixed points.  A consequence of this is that Theorem 1 no longer applies and we are not even sure what the fixed points of the system are anymore.  What theory can we develop about hCTLNs?

Fortunately for hCTLNs derived from DAGs, many of the results we had for DAG CTLNs in Chapter 7 can be generalized.  To develop them, we need to first introduce the \textit{pinned path polynomials}.

\begin{ddd}
    For a DAG G of size n, let the vertices be numbered from 1 to n i.e. $V(G)=[n]$.  Then, the \textbf{i,j pinned path polynomial}, $p^G_{i,j}(z)$, is defined to be: 
    $$p^G_{i,j}(z)=\delta_{ij}+\sum_{k=1}^n n^{i,j}_k z^k$$
    where $n_k^{i,j}$ is the number of paths from $j$ to $i$ of length $k$ (finite because G is acyclic) and $\delta_{ij}$ is the Kronecker delta.  Since $G$ is acyclic, $deg(p_{i,j}^G(z))$ is finite.
\end{ddd}

\begin{ddd}
    Let G be a labeled directed graph with adjacency matrix A.  Then, $G^T$ is the labeled directed graph induced by the adjacency matrix $A^T$.
\end{ddd}

\begin{rmk}
    Note that $\sum_{j=1}^n p_{i,j}^G(z) = p_i^G(z)$ and $\sum_{i=1}^n p_{i,j}^G(z) = p_j^{G^T}(z)$.  Also, $\sum_{i=1}^n p_i^G (z) = \sum_{j=1}^n p_j^{G^T}(z).$
\end{rmk}

Using this new construction we can both strengthen Theorem~\ref{thm:final} and generalize it to the case of hCTLNs.  We will also develop an appropriate generalization for Lemma~\ref{lem:dagbsp}.

\section{Virtual Fixed Points in hCTLNs}

The component linear systems $L_{\sigma}$ will have the same homogenous solution as $(-I+W)|_{\sigma}$ remains unchanged, but the particular solution is now different as we have the heterogenous external input vector $\Vec{\theta}|_{\sigma}$.

Using pinned path polynomials we have a generalization of Lemma~\ref{lem:DAG}.

\begin{lem}\label{lem:DAGG}
Let B be a matrix derived from a DAG G and adjacency matrix A such that:

\begin{center}
    $B=\mathbbm{1}\mathbbm{1}^T+cA$
\end{center}

Then the solution to the following linear system:

\begin{center}
    $B\vec{x}=\Vec{\phi}+a\vec{x}$
\end{center}

can be written as:

\begin{center}
    $$x_j=-\sum_{i=1}^n \Phi_i p^G_{j,i}\left(\frac{c}{a}\right)+p^G_j\left(\frac{c}{a}\right)\Gamma$$
\end{center}

where $\Gamma=\left(\dfrac{\sum_{i=1}^n \Phi_i p^{G^T}_i\left(\frac{c}{a}\right)}{-a+\sum_{i=1}^n p^G_i\left(\frac{c}{a}\right)}\right)$ and $\Phi_i=\dfrac{\phi_i}{a}$.

\end{lem}

\begin{proof}

The proof will be similar to that of Lemma~\ref{lem:DAG}.  We will again proceed by simply showing that, $\forall i \in [n]$, the specified $\Vec{x}$ satisfies $\sum_{j=1}^n B_{ij}x_j=\phi_i+ax_i$.  For compactness of notation, we will define $\kappa = \dfrac{c}{a}$.

Recall that:

$$\sum_{j=1}^n B_{ij}x_j = \sum_{j=1}^n x_j + c \sum_{j\rightarrow i}x_j.$$

Inserting our proposed solution:

$$\sum_{j=1}^n x_j = \sum_{j=1}^n \left( -\sum_{\ell=1}^n \Phi_\ell p^G_{j,\ell} (\kappa) +p^G_j\left(\kappa\right)\Gamma \right) = - \sum_{\ell=1}^n \Phi_\ell \sum_{j=1}^n p^G_{j,\ell} \left( \kappa \right) + \Gamma \sum_{j=1}^n p_j^G \left( \kappa \right)$$

$$=-\sum_{\ell=1}^n \Phi_\ell p_\ell^{G^T}\left( \kappa \right) + \Gamma \sum_{j=1}^n p_j^G \left( \kappa \right)$$

$$=\dfrac{a\sum_{\ell=1}^n \Phi_\ell p_\ell^{G^T}\left( \kappa \right) - \left(\sum_{j=1}^n p_j^G(\kappa)\right)\left( \sum_{\ell=1}^n \Phi_{\ell} \ppgt{\ell} (\kappa) \right)+\left(\sum_{j=1}^n p_j^G(\kappa)\right)\left( \sum_{\ell=1}^n \Phi_{\ell} \ppgt{i} (\kappa) \right)}{-a+\sum_{\ell=1}^n \ppg{\ell}(\kappa)}.$$

But after simplifying we notice that this is equal to $a\Gamma$.  From this we see that:

$$\sum_{j=1}^n B_{ij}x_j = a\Gamma + c \sum_{j\rightarrow i} \left( -\sum_{\ell=1}^n \Phi_\ell \pppg{j}{\ell} (\kappa) + \ppg{j} (\kappa) \Gamma \right) = a\Gamma -c\sum_{\ell=1}^n \Phi_\ell \sum_{j\rightarrow i} \pppg{j}{\ell} (\kappa) + c\Gamma \sum_{j \rightarrow i} \ppg{j} (\kappa).$$

Recall from previous proofs that $c\sum_{j\rightarrow i} \ppg{j} (\kappa) = a(\ppg{i} (\kappa)-1)$.  So, we can conclude:

$$\sum_{j=1}^n B_{ij}x_j= a\Gamma -c\sum_{\ell=1}^n \Phi_\ell \sum_{j\rightarrow i} \pppg{j}{\ell} (\kappa) -a\Gamma + a \ppg{i} (\kappa) \Gamma = -c\sum_{\ell=1}^n \Phi_\ell \sum_{j\rightarrow i} \pppg{j}{\ell} (\kappa) + a \ppg{i} (\kappa) \Gamma.$$

Now notice similarly that $c\sum_{j \rightarrow i}\pppg{j}{\ell} (\kappa) = a \pppg{i}{\ell} (\kappa)$.  Then:

$$-c\sum_{\ell=1}^n \Phi_{\ell} \sum_{j \rightarrow i} \pppg{j}{\ell} (\kappa) = a\Phi_i \pppg{i}{i} (\kappa) -a\sum_{\ell=1}^n \Phi_{\ell} \pppg{i}{\ell} (\kappa) = \phi_i  -a\sum_{\ell=1}^n \Phi_{\ell} \pppg{i}{\ell} (\kappa).$$

Thus, 

$$\sum_{j=1}^n B_{ij}x_j = \phi_i  -a\sum_{\ell=1}^n \Phi_{\ell} \pppg{i}{\ell} (\kappa) + a \ppg{i} (\kappa) \Gamma = \phi_i + a x_i.$$

\end{proof}

This allows us to generalize the results on the fixed points of $L_{\sigma}$ analogously as we did with Proposition~\ref{prp:dagfp}.

\begin{prp}\label{prp:fpg}
Let G be a DAG of size n, $\sigma\subset [n]$, and $\beta=\frac{-\epsilon-\delta}{\delta}$.  Then, for an hCTLN associated with G, the fixed point of $L_{\sigma}$ is:

\begin{center}
    $$(x^*_{\sigma})_j=-\sum_{i\in \sigma} \Theta_i \pppgs{j}{i} \left( \beta \right)+ \ppgs{j} \left(\beta \right)\Gamma(\sigma) \text{, } \forall j\in \sigma \text{ and } (x^*_{\sigma})_j=0 \text{, } \forall j\not\in \sigma.$$
\end{center}

where $\Gamma(\sigma)=\left(\dfrac{\sum_{i\in \sigma} \Theta_i \ppgst{i} \left( \beta \right)}{-\frac{\delta}{1+\delta}+\sum_{i\in \sigma} \ppgs{i}\left(\beta \right)}\right)$ and $\Theta_i=\dfrac{\theta_i}{\delta}$.

\end{prp}

\begin{proof}
The proof is identical to that of Proposition~\ref{prp:dagfp} with the only change that we use Lemma~\ref{lem:DAGG} rather than Lemma~\ref{lem:DAG}.
\end{proof}

We have nearly reconstructed a version of Theorem~\ref{thm:final} from Chapter 4 in the setting of hCTLNs.  Still, there is one more piece of the puzzle and resolving it will fill in an oversight of our analysis of CTLNs as well.

\section{Revisiting CTLNs}

Recall a gap that was left in the statement of Theorem~\ref{thm:final}.  For the linear systems $L_{\sigma}$, we did not provide the eigenvectors for the eigenvalue $\lambda=-1$ of multiplicity $n-|\sigma|$.  At the time we said that this is something we would revisit.  These eigenvectors can be obtained using Lemma~\ref{lem:DAGG}.

\begin{prp}\label{prp:minone}
Let G be a directed graph such that $|G|=n$ and let $\sigma \subset [n]$.  Then, the matrix for the linear system $L_{\sigma}$ has eigenvalue $\lambda = -1$ with algebraic multiplicity $n-|\sigma|$.  Moreover there are $n-|\sigma|$ distinct eigenvectors of the form:

$$\Vec{v_j}=-e_j+ \Vec{g_j} \text{, } j\not\in \sigma$$

such that, if $k\in \sigma$: 

$$(\Vec{g_j})_k = -\sum_{j \not\rightarrow \ell\in\sigma}\pppgs{k}{\ell} (\alpha) + \dfrac{1-\varepsilon}{1+\delta} \sum_{j \rightarrow \ell\in\sigma} \pppgs{k}{\ell} (\alpha) +\ppgs{k}(\alpha) \Gamma_j (\sigma)$$ 

where $\Gamma_j (\sigma)=\dfrac{\sum_{j\not\rightarrow\ell \in \sigma} (1+\delta) \ppgst{\ell} (\alpha)+\sum_{j\rightarrow\ell \in \sigma} (1-\varepsilon) \ppgst{\ell} (\alpha)}{-(1+\delta)+(1+\delta)\sum_{\ell=1}^n \ppgs{\ell}(\alpha)}$ and $\alpha=\dfrac{-\varepsilon - \delta}{1+\delta}$

and, if $k\not\in \sigma$, $(\Vec{g_j})_k = 0$.

\end{prp}

\begin{proof}

This proof will take a similar approach to that of Proposition~\ref{prp:stab2}, but with the notable modification of using Lemma~\ref{lem:DAGG} rather than the Sherman-Morrison Formula.

Without loss of generality number the vertices in $\sigma$ to be $1,\dots,k$ where $k=|\sigma|$.  Then the matrix for $L_{\sigma}$ is of the form: 

$
B=
\left[
\begin{array}{ccc|ccc}
-1 & \hdots & -1-\delta & w_{1,k+1} & \hdots & w_{1n}\\
\vdots & \ddots & \vdots & \vdots & \hdots & \vdots\\
-1-\delta & \hdots & -1 & w_{k,k+1} & \hdots & w_{kn}\\
\hline
0 & \hdots & 0 & -1 & \hdots & 0\\
\vdots & \ddots & \vdots & \vdots & \ddots & \vdots\\ 
0 & \hdots & 0 & 0 & \hdots & -1
\end{array}
\right]
$

where, taking $A|_{\sigma}$ as the adjacency matrix of $G|_{\sigma}$, the upper left block is of the form 

$$(-I+W)|_{\sigma} = (-1-\delta)\mathbbm{1}\mathbbm{1}^T + \delta I +(\varepsilon +\delta)A|_{\sigma}.$$

Recalling Lemma~\ref{corr:charpolyctln}, we know that $\lambda=-1$ has algebraic multiplicities $n-k$.

We find eigenvectors for $\lambda=-1$.  We show that there are $n-k$ linearly independent eigenvectors by construction.  We then take as an ansatz vectors of the form:

$$
\Vec{v_j}=\left[
\begin{array}{c}
g_1 \\
\vdots \\
g_k \\
\hline
0 \\
\vdots \\
-1 \\
\vdots \\
0
\end{array}
\right]
=
\left[
\begin{array}{c}
\Vec{g} \\
\hline
0
\end{array}
\right]-e_j
$$

Then, we have:

$$
B\Vec{v_j}=\left[
\begin{array}{c}
((-I+W)|_{\sigma})\Vec{g} - \Vec{w}_{*j} \\

\hline

0
\end{array}
\right]+e_j
$$

where

$$
\Vec{w}_{*j}=\left[
\begin{array}{c}
w_{1j} \\
\vdots \\
w_{kj}
\end{array}
\right].
$$

So, if $\Vec{g}$ satisfies $((-I+W)|_{\sigma})\Vec{g} - \Vec{w}_{*j}=-\Vec{g}$ then $\Vec{v_j}$ is an eigenvector.  Rearranging, this system can be rewritten as:

$$((-I+W)|_{\sigma})\Vec{g} + I \Vec{g}= \Vec{w}_{*j} \implies ((-1-\delta)\mathbbm{1}\mathbbm{1}^T +(\varepsilon+\delta) A|_{\sigma})\Vec{g}=\Vec{w}_{*j}-(1+\delta)\Vec{g} $$

$$\implies (\mathbbm{1}\mathbbm{1}^T + \alpha A|_{\sigma})\Vec{g}=-\dfrac{\Vec{w}_{*j}}{1+\delta}+\Vec{g}.$$

where $\alpha=\dfrac{-\varepsilon-\delta}{1+\delta}$.

Applying Lemma~\ref{lem:DAGG} we obtain:

$$g_k = \sum_{\ell\in \sigma} \dfrac{w_{\ell j}}{1+\delta} \pppgs{k}{\ell} (\alpha) + \ppgs{k}(\alpha) \Gamma_j (\sigma).$$

where $\Gamma_j (\sigma)=\dfrac{-\sum_{\ell \in \sigma} w_{\ell j} \ppgst{\ell} (\alpha)}{-(1+\delta)+(1+\delta)\sum_{\ell=1}^n \ppgs{\ell}(\alpha)}$

Then, using that $w_{\ell j}=-1-\delta$ if $j\not \rightarrow \ell$ and $w_{\ell j}=-1+\varepsilon$ if $j\rightarrow \ell$, we finally obtain:

$$g_k=-\sum_{j \not\rightarrow \ell\in\sigma}\pppgs{k}{\ell} (\alpha) + \dfrac{1-\varepsilon}{1+\delta} \sum_{j \rightarrow \ell\in\sigma} \pppgs{k}{\ell} (\alpha) +\ppgs{k}(\alpha) \Gamma_j (\sigma)$$

where $\Gamma_j (\sigma)=\dfrac{\sum_{j\not\rightarrow\ell \in \sigma} (1+\delta) \ppgst{\ell} (\alpha)+\sum_{j\rightarrow\ell \in \sigma} (1-\varepsilon) \ppgst{\ell} (\alpha)}{-(1+\delta)+(1+\delta)\sum_{\ell=1}^n \ppgs{\ell}(\alpha)}.$

\end{proof}

Now we can restate Theorem~\ref{thm:final} in a more complete and comprehensive form.

\begin{flushleft}
\textbf{Theorem~\ref{thm:final}.} \begin{itshape} Let G be a DAG and let W be the weight matrix for an associated CTLN with parameters $\varepsilon, \delta, \theta$.  Let $\sigma \subseteq [n]$ be such that $G|_{\sigma}$ is analytic, $(-I+W)|_{\sigma}$ is diagonalizable, and the polynomial:\end{itshape}
\end{flushleft}

$$f(\lambda)=(-\lambda+\delta)^{m+1}-(1+\delta)(|\sigma|(-\lambda+\delta)^{m}+n^{\sigma}_1c(-\lambda+\delta)^{m-1}+...+n^{\sigma}_m c^m)$$

\begin{itshape} has distinct roots $\{\lambda_k\}_{k=1}^{m+1}$ where $n^{\sigma}_{j>0}$ is the number of paths of length $j$ in $G|_{\sigma}$ and $m$ is the maximum path length in $G|_{\sigma}$. 

Then, the general solution of $L_{\sigma}$ is of the following form:

$$\Vec{x}(t)=\sum_{k=1}^{m+1} c_k \Vec{p_{\sigma}}(\alpha_k) e^{\lambda_{k} t} +\sum_{(i,j)\in\operatorname{\widetilde{SE}}(G|_{\sigma})} c_{(i,j)}(e_i-e_j)e^{\delta t} +\sum_{k\not\in \sigma} c_k (-e_k+\Vec{g_k}) e^{-t}+ \Vec{p_{\sigma}}(\beta)\Gamma (\sigma)$$

where $n=|G|$, $\alpha_k=\dfrac{\varepsilon+\delta}{\lambda_k -\delta}$, $\beta=\dfrac{-\varepsilon-\delta}{\delta}$, and $\Gamma (\sigma) = \dfrac{\theta}{-\delta +(1+\delta) \sum_{j\in \sigma} p_j^{G|_{\sigma}} (\beta)}$.

Additionally, $\Vec{g_j}$ is defined to be: 

$$(\Vec{g_j})_k = -\sum_{j \not\rightarrow \ell\in\sigma}\pppgs{k}{\ell} (\gamma) + \dfrac{1-\varepsilon}{1+\delta} \sum_{j \rightarrow \ell\in\sigma} \pppgs{k}{\ell} (\gamma) +\ppgs{k}(\gamma) \Gamma_j (\sigma) \text{, } \forall k\in\sigma$$ 

where $\Gamma_j (\sigma)=\dfrac{\sum_{j\not\rightarrow\ell \in \sigma} (1+\delta) \ppgst{\ell} (\gamma)+\sum_{j\rightarrow\ell \in \sigma} (1-\varepsilon) \ppgst{\ell} (\gamma)}{-(1+\delta)+(1+\delta)\sum_{\ell=1}^n \ppgs{\ell}(\gamma)}$ and $\gamma=\dfrac{-\varepsilon - \delta}{1+\delta}$

and, if $k\not\in \sigma$, $(\Vec{g_j})_k = 0$.  
\end{itshape}

%%%%%%%%%%%%%%%%%%%%%%%%%%%%%%%%%%%

\subsection{Revised Initial Value Problem}

This new set of eigenvectors does change the initial value problems for the systems $L_{\sigma}$ where $\sigma \subset [n]$, but not considerably.  The same approach can still be employed.  We can use the same CTLN as we did in Chapter 4, but instead of solving the initial value problem for the $L_{[4]}$ system, we will solve it for the $L_{[3]}$ system instead.  In this system we have the localized path polynomials depicted in Figure~\ref{fig:reddag}, but we will also have an eigenvector associated with the eigenvalue $\lambda=-1$.

\begin{figure}[!ht]
\begin{center}
\vspace{.1in}
\includegraphics[width=2.75in]{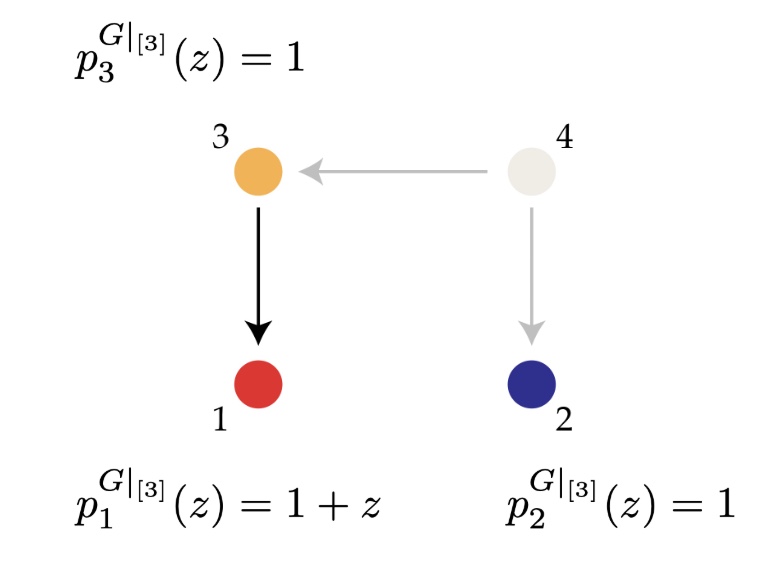}
\vspace{.1in}
\caption[Localized path polynomials for $G|_{[3]}$]{{\bf Localized path polynomials for $G|_{[3]}$.} The localized path polynomials in the subgraph $G|_{[3]}$.}
\label{fig:reddag}
\end{center}
\vspace{-.2in}
\end{figure}

Applying Proposition~\ref{prp:minone}, this eigenvector is:

$$\Vec{v_4}=\left[
\begin{array}{c}
-1 + \gamma \left( \dfrac{1-\varepsilon}{1+\delta} \right) +(1+\gamma) \Gamma_4 ([3]) \\
\left( \dfrac{1-\varepsilon}{1+\delta} \right) + \Gamma_4 ([3]) \\
\left( \dfrac{1-\varepsilon}{1+\delta} \right) + \Gamma_4 ([3]) \\
-1 \\
\end{array}
\right].
$$

Taking $\kappa=\dfrac{1-\varepsilon}{1+\delta}$ and referring to $\Gamma_4([3])=\Gamma_4$, the general solution for this system is of the form:

$$x(t)=
c_1 \left[ \begin{array}{c} 1 \\ -1 \\ 0 \\ 0 \end{array} \right] e^{\delta t} +
c_2 \left[ \begin{array}{c} 1+\alpha_1 \\ 1 \\ 1 \\ 0 \end{array} \right] e^{\lambda_1 t} +
c_3 \left[ \begin{array}{c} 1+\alpha_2 \\ 1 \\ 1 \\ 0 \end{array} \right] e^{\lambda_2 t} +
c_4 \left[
\begin{array}{c}
-1 + \gamma \kappa+(1+\gamma) \Gamma_4 \\
\kappa + \Gamma_4 \\
\kappa + \Gamma_4 \\
-1 \\
\end{array}
\right] e^{-t} +x^*_{[3]}.
$$

Then, the initial value problem can then be set up as:

$$\left[\begin{array}{c} x_1^0 - (x_{[3]}^*)_1 \\ x_2^0 - (x_{[3]}^*)_2 \\ x_3^0 - (x_{[3]}^*)_3 \\ x_4^0  \end{array} \right]=\left[
\begin{array}{cccc}
1 & 1+\alpha_1 & 1+\alpha_2 & -1+\gamma \kappa +(1+\gamma)\Gamma_4 \\
-1 & 1 & 1 & \kappa + \Gamma_4 \\
0 & 1 & 1 & \kappa +\Gamma_4 \\
0 & 0 & 0 & -1 \\
\end{array}
\right]\Vec{c}
$$

Notice that the row corresponding to $x_4$ is empty except for the fourth column.  By construction of the eigenvectors for $\lambda=-1$, this will be true for any of the rows corresponding to $j\not\in \sigma$.  We can then subtract that row to clear out that column.

$$\left[\begin{array}{c} x_1^0 - (x_{[3]}^*)_1+(-1+\gamma \kappa +(1+\gamma)\Gamma_4)x_4^0 \\ x_2^0 - (x_{[3]}^*)_2 + (\kappa + \Gamma_4)x_4^0 \\ x_3^0 - (x_{[3]}^*)_3 +(\kappa + \Gamma_4)x_4^0 \\ x_4^0  \end{array} \right]=\left[
\begin{array}{cccc}
1 & 1+\alpha_1 & 1+\alpha_2 & 0 \\
-1 & 1 & 1 & 0 \\
0 & 1 & 1 & 0 \\
0 & 0 & 0 & -1 \\
\end{array}
\right]\Vec{c}
$$

Then we can conclude that $c_4=-x_4^0$.  Then $c_1,c_2,c_3$ satisfy:

$$\left[\begin{array}{c} x_1^0 - (x_{[3]}^*)_1+(-1+\gamma \kappa +(1+\gamma)\Gamma_4)x_4^0 \\ x_2^0 - (x_{[3]}^*)_2 + (\kappa + \Gamma_4)x_4^0 \\ x_3^0 - (x_{[3]}^*)_3 +(\kappa + \Gamma_4)x_4^0   \end{array} \right]=\left[
\begin{array}{ccc}
1 & 1+\alpha_1 & 1+\alpha_2  \\
-1 & 1 & 1  \\
0 & 1 & 1  \\
\end{array}
\right]\left[ \begin{array}{c} c_1 \\ c_2 \\ c_3 \end{array} \right]
$$

which can be solved using the approach described in Chapter 4.

\section{Balanced hCTLNs}
As Lemma~\ref{lem:DAGG} is an hCTLN analogue for Lemma~\ref{lem:DAG}, we should be able to say something about balanced states in DAG hCTLNs as well.  Using the same process, we obtain a closed form expression of the balanced state.

\begin{prp}\label{prp:dagbsg}
Let G be a DAG.  Letting $\beta=\frac{-\varepsilon-\delta}{1+\delta}$, the point $x_{bs}=-W^{-1}\Vec{\theta}$ of an associated hCTLN can be written as:

\begin{center}
    $$(x_{bs})_j=-\sum_{i=1}^n \left(\dfrac{\theta_i}{1+\delta}\right) p^G_{j,i}(\beta)+p^G_j(\beta)\Gamma$$
    
\end{center}
    
where $\Gamma=\left(\dfrac{\sum_{i=1}^n \theta_i p^{G^T}_i(\beta)}{(1+\delta)(-1+\sum_{i=1}^n p^G_i(\beta))} \right).$

\end{prp}

\begin{proof}
Identical to that of Lemma~\ref{lem:dagbsp} using Lemma~\ref{lem:DAGG} in the place of Lemma~\ref{lem:DAG}.  
\end{proof}

Unfortunately the expressions for the entries of $x_{bs}$ are now far more complicated, but we can still say something about whether an hCTLN is balanced by simply rearranging after requiring $(x_{bs})_j \geq 0$ for all $j \in [n]$.

\begin{prp}\label{prp:bsfinal}
Let G be a DAG.  Let $\beta=\frac{-\varepsilon - \delta}{1+\delta}$ where $\varepsilon$ and $\delta$ are taken from an associated hCTLN with external input vector $\Vec{\theta}$.  Then, that hCTLN is balanced if and only if: 

    $$p^G_j(\beta)\Gamma' \geq \sum_{i=1}^n \theta_i p^G_{j,i}(\beta) \text{, } \forall j\in [n] $$
    
where $\Gamma'=\left(\dfrac{\sum_{i=1}^n \theta_i p^{G^T}_i(\beta)}{-1+\sum_{i=1}^n p^G_i(\beta)} \right).$

\end{prp}
\begin{proof}
For each $j$, multiply $(x_{bs})_j \geq 0$ through by $1+\delta$ and rearrange.

\end{proof}

While this result is a bit more cumbersome than the corresponding result for CTLNs, it nonetheless is a complete answer to the question of when a DAG hCTLN is balanced.  Proposition~\ref{prp:bsfinal} becomes easier to use upon realizing that, given a DAG and a value for $\beta$, each inequality in Proposition~\ref{prp:bsfinal} become linear in $\Vec{\theta}$.  Thus, the problem is reduced to determining the feasible space of a linear program.  We will conclude with an example.

\subsection{Example: Balanced States in DAG hCTLNs}

In the spirit of bringing things full circle, consider again our three neuron DAG illustrating the decoy effect (Figure~\ref{fig:decoydagh}).  Applying Proposition~\ref{prp:bsfinal} yields three inequalities:

1. $p^G_1(\beta)\Gamma' \geq \sum_{i=1}^n \theta_i p^G_{1,i}(\beta)$

2. $p^G_2(\beta)\Gamma' \geq \sum_{i=1}^n \theta_i p^G_{2,i}(\beta)$

3. $p^G_3(\beta)\Gamma' \geq \sum_{i=1}^n \theta_i p^G_{3,i}(\beta)$

\begin{figure}[!ht]
\begin{center}
\vspace{.1in}
\includegraphics[width=5.75in]{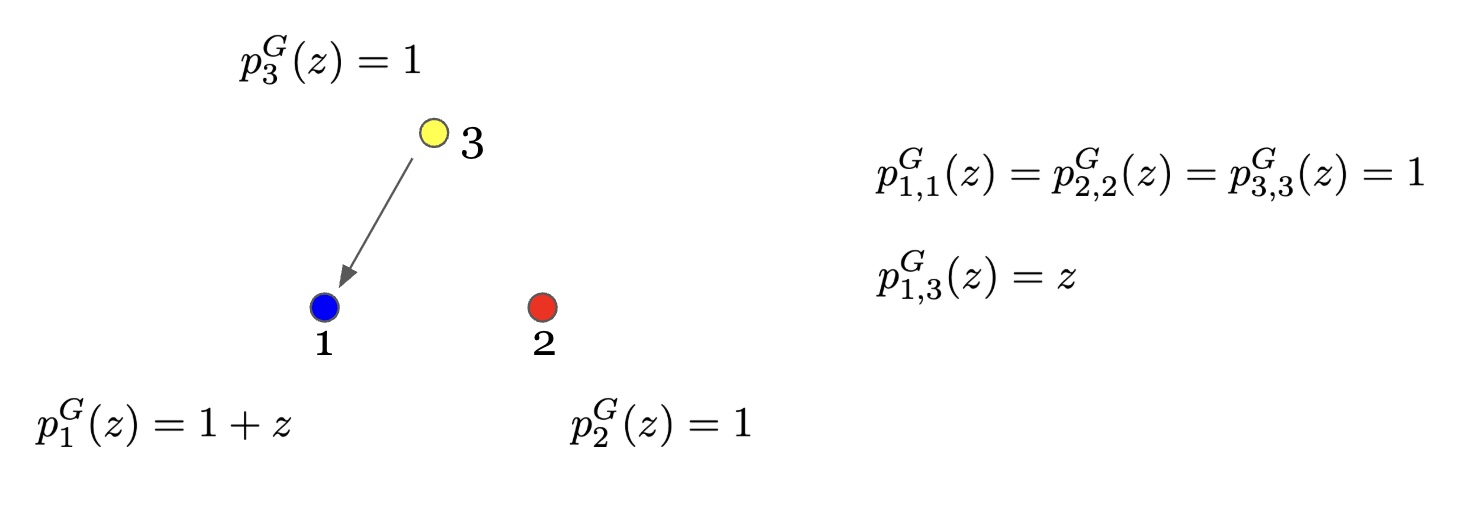}
\vspace{.1in}
\caption[Localized and pinned path polynomials of Decoy Effect DAG]{{\bf Localized and pinned path polynomials of Decoy Effect DAG.} This describes a network corresponding to the Decoy Effect.  Listed are the localized path polynomials and the non-zero pinned path polynomials.}
\label{fig:decoydagh}
\end{center}
\vspace{-.2in}
\end{figure}

Fix $\beta=-0.5$ and then notice the following: $\ppg{1}(\beta)=0.5$, $\ppg{2}(\beta)=1$, and $\ppg{3}(\beta)=1$.  Additionally, $\ppgt{1}(\beta)=1$, $\ppgt{2}(\beta)=1$, and $\ppgt{3}(\beta)=0.5$, so we can conclude: 

$$\Gamma' = \dfrac{\theta_1 + \theta_2 + 0.5\theta_3}{-1+2.5}=\dfrac{2}{3}\theta_1+\dfrac{2}{3}\theta_2+\dfrac{1}{3}\theta_3.$$

Lastly we looked at the pinned path polynomials.  We notice that $\pppg{1}{1}(\beta)=\pppg{2}{2}(\beta)=\pppg{3}{3}(\beta)=1$, $\pppg{1}{3}(\beta)=-0.5$, and the rest vanish.  What remains is a linear system of inequalities describing the potential vectors $\Vec{\theta}$ which produce a balanced hCTLN.  We have the inequalities:

1. $\frac{1}{3}\theta_1+\frac{1}{3}\theta_2+\frac{1}{6}\theta_3 \geq \theta_1 -0.5\theta_3 \rightarrow -2\theta_1+\theta_2+2\theta_3 \geq 0$

2. $\frac{2}{3}\theta_1+\frac{2}{3}\theta_2+\frac{1}{3}\theta_3 \geq \theta_2 \rightarrow 2\theta_1 -\theta_2 + \theta_3 \geq 0$

3. $\frac{2}{3}\theta_1+\frac{2}{3}\theta_2+\frac{1}{3}\theta_3 \geq \theta_3 \rightarrow \theta_1 +\theta_2 -\theta_3 \geq 0$

\begin{figure}[!ht]
\begin{center}
\vspace{.1in}
\includegraphics[width=5.75in]{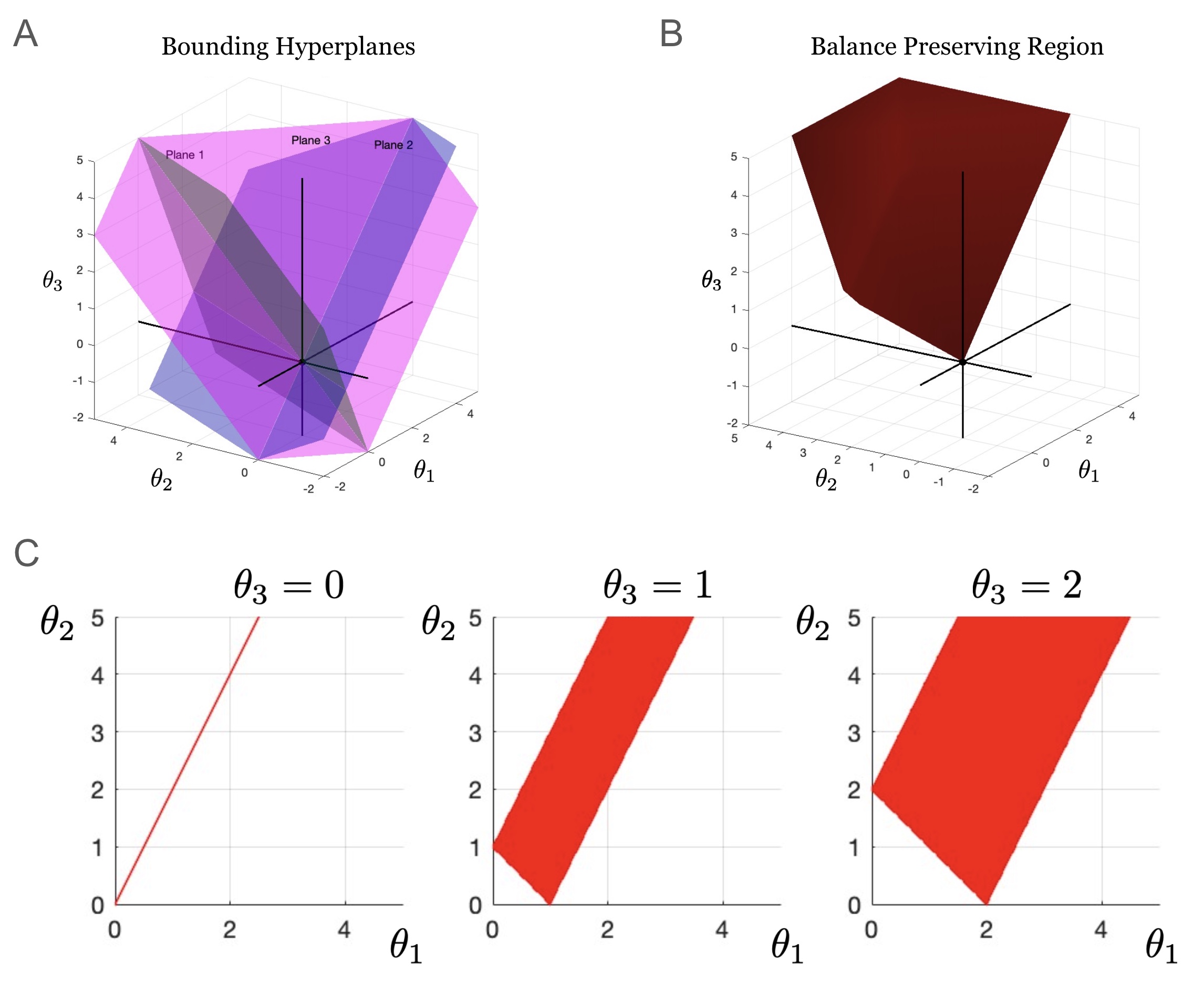}
\vspace{.1in}
\caption[Balance preserving external input vectors]{{\bf Balance preserving input current vectors.} (A) Each inequality produces a plane arrangement.  (B) The inequalities define a region of the space of input current vectors where the an hCTLN derived from the decoy effect DAG, with $\beta=0.5$, is balanced.  (C) Cross-sections of this region for fixed values of $\theta_3$.}
\label{fig:hctlnspace}
\end{center}
\vspace{-.2in}
\end{figure}

Now we have three linear inequalities in $\Vec{\theta}$ that determine whether the hCTLN is balanced.  Together, they define a sort of feasible region within which we have the convex set of $\Vec{\theta}$ producing a balanced hCTLN for the prescribed DAG and connectivity parameters.  This region is illustrated in Figure~\ref{fig:hctlnspace}.  We conclude with a brief remark that the CTLN drive vectors $\theta \mathbbm{1}$ lie within this region.  This is expected as this DAG is balanced because $d_{\max}=1$.

\chapter{Conclusions and Open Questions}

We began this dissertation by asking how parameters of TLN attractor networks shape their basins of attraction while fixing the set of attractors.  Our goal was to understand how decision-making bias was encoded in these models and we considered three ways of relating decision-making bias to basins of attraction.  

1.  The relative sizes of the basins of attraction restricted to the positive orthant.

2.  The relative sizes of the basins within a neighborhood of the saddle point(s).

3.  The basin of attraction within which the balanced state trajectory falls.

Recall that each of these corresponds to a paradigm of how neural dynamics behave.  The first assumes that neural circuits display a high dimensional dynamics with diverse trajectories.  The second assumes that the dynamics are confined to a lower dimensional submanifold on which the saddle point(s) lies.  The third assumes a that the circuit operates along a particular trajectory.

In the context of a two neuron competitive TLN model, we worked the problem out completely, rigorously proving the basins of attraction and analytically calculating their sizes relative to one another.  What we found however is that this model makes too many simplifications to be meaningful in encoding decision-making bias under a low dimensional paradigm.  

We then considered the dynamics of CTLNs, particularly those derived from DAGs.  While we were unable to find ways of determining the relative sizes of the basins of attraction, we did demonstrate how the dynamics could be rigorously worked out in great detail.  Additionally, we were able to numerically demonstrate the relationship between the fractional indegree of sinks and the relative sizes of their basins in the vicinity of the saddle point(s), also providing a partially rigorous justification for the relationship.  We finally explored the existence of balanced state trajectories rigorously in great detail, and offered numerical evidence for how the filtration of a DAG influences the attractor within which the balanced state trajectory lies.  

We will conclude by considering future lines of research which build on the results of this dissertation.

\subsection*{Open Questions}

The following are a series of open questions which invite further inquiry and offer new avenues of study.

\textbf{Question 1: }  What is a viable hyperplane arrangement for building a state transition graph approximating the basins of attraction in a competitive TLN?

This was the problem we left open at the end of Chapter 3.  We showed how there exists a degree of freedom in finding a hyperplane which divides another into regions of inward and outward flow, but we were unable to give a way of exploiting that freedom so that the new hyperplane would not require additional separation.  One approach to this might take inspiration from the two-dimensional trajectory graphs, where the stable manifold was an important separating line.  While the stable manifold is curved, perhaps hyperplanes drawn to be roughly aligned could be used.  In the case of DAG CTLNs, where the eigenvectors of the component systems $L_{\sigma}$ have been worked out in great detail, those corresponding to the stable manifold could be incorporated using the degree of freedom that exists in the partition.

\textbf{Question 2: }  Fixing a DAG and a corresponding CTLN, how does the network structure shape the change in the basins of attraction under a perturbation of parameters?

Among the key challenges with employing Theorem~\ref{thm:final} to determine basins of attraction is tracking the trajectories across chambers and determining what chambers the seperatrix goes through.  However, if a CTLN is fixed with a particular $\varepsilon$ and $\delta$, the chambers can be determined computationally and it may be easier to write out an analytical expression for the separatrix.  Perturbing the parameters slightly will likely not change which chambers the separatrix passes through, so it may be possible to precisely understand how the basins are changing.

\textbf{Question 3: }  How can Balanced State Attractor prediction be improved?

We saw that there were cases where the attractor within which the balanced state trajectory lies was incorrectly predicted by Algorithm 2.  This indicates that the filtration of the DAG is not the full story.  Is there an algorithm with better accuracy?

\textbf{Question 4: }  Can a useful model be obtained by combining a linear recurrent dynamics using the DAG CTLN matrix with a nonlinear readout?

We demonstrated in great detail how the dynamics of the linear systems of a DAG CTLN, $L_{\sigma}$, can be worked out analytically.  Our challenge was in piecing together the trajectories across chambers.  But what if we dispensed with the ReLU nonlinearity and instead had a linear recurrent dynamics?  While we would lose bistability, this could be recovered using a nonlinear readout.  This would be far more tractable, but at the price of losing the elegance that comes with encoding both the neural computation and the attractor end states together in the nonlinear dynamics.  Could this model be meaningfully used to model neural circuits?

\textbf{Question 5: }  How can actual data be incorporated into these frameworks to make testable predictions about decision-making?

The goal of this dissertation was to understand how neural circuits encode decision-making biases.  We found ways in which to theoretically model decision-making biases, but a natural next question is to consider whether the theoretical models can be fit to actual data and tested against experimental outcomes.  The hope of course would be that these models can be used to infer new aspects of decision-making dynamics and inspire experimentation in their own right. 

\section{Numerical Methods Repository}
A package of code for the recreation of numerical experiments and implementation of algorithms is available at: \url{https://github.com/sadiq-safaan/Dissertation-Companion}.

%%%%%%%%%%%%%%%%%%%%%%%%%%%%%%%%%%%%%%%%%%%%%%%%%%%%%%%%%%%%%%%
% Appendices
%
% Because of a quirk in LaTeX (see p. 48 of The LaTeX
% Companion, 2e), you cannot use \include along with
% \addtocontents if you want things to appear the proper
% sequence.

%%%%%%%%%%%%%%%%%%%%%%%%%%%%%%%%%%%%%%%%%%%%%%%%%%%%%%%%%%%%%%%
%\appendix
%\titleformat{\chapter}[display]{\fontsize{30}{30}\selectfont\bfseries\sffamily}{Appendix \thechapter\textcolor{gray75}{\raisebox{3pt}{|}}}{0pt}{}{}
% If you have a single appendix, then to prevent LaTeX from
% calling it ``Appendix A'', you should uncomment the following two
% lines that redefine the \thechapter and \thesection:
%\renewcommand\thechapter{}
%\renewcommand\thesection{\arabic{section}}

%\include{Appendix-A/Appendix-A}
%\include{Appendix-B/Appendix-B}
%\include{Appendix-C/Appendix-C}
%\include{Appendix-D/Appendix-D}
%\include{Appendix-E/Appendix-E}

%%%%%%%%%%%%%%%%%%%%%%%%%%%%%%%%%%%%%%%%%%%%%%%%%%%%%%%%%%%%%%%
% ESM students need to include a Nontechnical Abstract as the %
% last appendix.                                              %
%%%%%%%%%%%%%%%%%%%%%%%%%%%%%%%%%%%%%%%%%%%%%%%%%%%%%%%%%%%%%%%
% This \include command should point to the file containing
% that abstract.
%\include{nontechnical-abstract}
%%%%%%%%%%%%%%%%%%%%%%%%%%%%%%%%%%%%%%%%%%%
} % End of the \allowdisplaybreak command %
%%%%%%%%%%%%%%%%%%%%%%%%%%%%%%%%%%%%%%%%%%%

%%%%%%%%%%%%%%%%
% BIBLIOGRAPHY %
%%%%%%%%%%%%%%%%
% You can use BibTeX or other bibliography facility for your
% bibliography. LaTeX's standard stuff is shown below. If you
% bibtex, then this section should look something like:
	\begin{singlespace}
	\bibliographystyle{GLG-bibstyle}
	\addcontentsline{toc}{chapter}{Bibliography}
	\bibliography{BiblioDatabase}

\begin{thebibliography}{10}
\newcommand{\enquote}[1]{``#1''}
\providecommand{\url}[1]{\texttt{#1}}
\providecommand{\urlprefix}{URL }
\providecommand{\eprint}[2][]{\url{#2}}

\bibitem{utils}
\textsc{Mas-Colell, A.}, \textsc{M.~Whinston}, and \textsc{J.~Green} (1995)
  \emph{Microeconomic Theory}, chap. Preference and Choice, 1 ed., Oxford
  University Press, pp. 6 -- 9.

\bibitem{huber}
\textsc{Huber, J.}, \textsc{J.~Payne}, and \textsc{C.~Puto} (1982)
  \enquote{Adding asymmetrically dominated alternatives: Violations of
  regularity and the similarity hypothesis,} \emph{Journal of Consumer
  Research}, \textbf{9}(1), pp. 90--98.

\bibitem{DG}
\textsc{Dorris, M.} and \textsc{P.~Glimcher} (2004) \enquote{Activity in
  Posterior Parietal Cortex Is Correlated with the Relative Subjective
  Desirability of Action,} \emph{Neuron}, \textbf{44}, pp. 365--378.

\bibitem{hops}
\textsc{Hopfield, J.} (1982) \enquote{Neural networks and physical systems with
  emergent collective computational abilities,} \emph{Proc. Natl. Acad. Sci.
  U.S.A.}, \textbf{79}(8), pp. 2554--2558.

\bibitem{ring}
\textsc{Kim, S.}, \textsc{H.~Rouault}, \textsc{S.~Druckmann}, and
  \textsc{V.~Jayaraman} (2017) \enquote{Ring attractor dynamics in the
  Drosophila central brain,} \emph{Science}, \textbf{356}, pp. 849--853.

\bibitem{seung}
\textsc{Seung, H.} (1996) \enquote{How the brain keeps eyes still,}
  \emph{PNAS}, \textbf{93}, p. 13339–13344.

\bibitem{wang}
\textsc{Wang, X.} (2012) \enquote{Neural dynamics and circuit mechanisms of
  decision-making,} \emph{Current Opinion in Neurobiology}, \textbf{22}(6).

\bibitem{attrev}
\textsc{Khona, M.} and \textsc{I.~R. Fiete} (2022) \enquote{Attractor and
  integrator networks in the brain,} \emph{Nature Reviews Neuroscience},
  \textbf{23}(12), pp. 744--766.

\bibitem{rate}
\textsc{Gerstner, W.}, \textsc{A.~Kreiter}, \textsc{H.~Markram}, and
  \textsc{A.~Herz} (1997) \enquote{Neural codes: firing rates and beyond,}
  \emph{Proc. Natl. Acad. Sci. U.S.A.}, \textbf{94}, p. 12740–12741.

\bibitem{AppendixF}
\textsc{Abbott, L.}, \textsc{S.~Fusi}, and \textsc{K.~Miller} (2012)
  \emph{Principles of Neural Science}, chap. Appendix {F}: Theoretical
  approaches to neuroscience: Examples from single neurons to networks, 5 ed.,
  McGraw-Hill Education/Medical, pp. 1601--1617.

\bibitem{tlns}
\textsc{Hahnloser, R.}, \textsc{S.~Seung}, and \textsc{J.~Slotine} (2003)
  \enquote{Permitted and forbidden sets in symmetric threshold-linear
  networks,} \emph{Neural Comput.}, \textbf{15}(3), pp. 621--638.

\bibitem{gr}
\textsc{Curto, C.} and \textsc{K.~Morrison} (2023) \enquote{Graph Rules for
  Recurrent Neural Network Dynamics,} \emph{Notices of the American
  Mathematical Society}, \textbf{70}(4), pp. 536--551.

\bibitem{church}
\textsc{Churchland, A.}, \textsc{R.~Kiani}, and \textsc{M.~Shadlen} (2008)
  \enquote{Decision-making with multiple alternatives,} \emph{Nature
  Neuroscience}, \textbf{11}(6), pp. 693--702.

\bibitem{ibo}
\textsc{Laboratory, T. I.~B.}, \textsc{V.~Aguillon-Rodriguez},
  \textsc{D.~Angelaki}, \textsc{H.~Bayer}, \textsc{N.~Bonacchi},
  \textsc{M.~Carandini}, \textsc{F.~Cazettes}, \textsc{G.~Chapuis},
  \textsc{A.~K. Churchland}, \textsc{Y.~Dan}, \textsc{E.~Dewitt},
  \textsc{M.~Faulkner}, \textsc{H.~Forrest}, \textsc{L.~Haetzel},
  \textsc{M.~Häusser}, \textsc{S.~B. Hofer}, \textsc{F.~Hu},
  \textsc{A.~Khanal}, \textsc{C.~Krasniak}, \textsc{I.~Laranjeira},
  \textsc{Z.~F. Mainen}, \textsc{G.~Meijer}, \textsc{N.~J. Miska},
  \textsc{T.~D. Mrsic-Flogel}, \textsc{M.~Murakami}, \textsc{J.-P. Noel},
  \textsc{A.~Pan-Vazquez}, \textsc{C.~Rossant}, \textsc{J.~Sanders},
  \textsc{K.~Socha}, \textsc{R.~Terry}, \textsc{A.~E. Urai},
  \textsc{H.~Vergara}, \textsc{M.~Wells}, \textsc{C.~J. Wilson}, \textsc{I.~B.
  Witten}, \textsc{L.~E. Wool}, and \textsc{A.~M. Zador} (2021)
  \enquote{Standardized and reproducible measurement of decision-making in
  mice,} \emph{eLife}, \textbf{10}, p. e63711.

\bibitem{dagdyn}
\textsc{Lienkaemper, C.} (2022), \enquote{Combinatorial geometry of neural
  codes, neural data analysis, and neural networks,} \eprint{2209.07583}.
\newline\urlprefix\url{https://arxiv.org/abs/2209.07583}

\bibitem{carnevale}
\textsc{Carnevale, F.}, \textsc{V.~de Lafuente}, \textsc{R.~Romo},
  \textsc{O.~Barak}, and \textsc{N.~Parga} (2015) \enquote{Dynamic Control of
  Response Criterion in Premotor Cortex during Perceptual Detection under
  Temporal Uncertainty,} \emph{Neuron}, \textbf{86}(4), pp. 1067--1077.

\bibitem{bs1}
\textsc{van Vreeswijk, C.} and \textsc{H.~Sompolinsky} (1996) \enquote{Chaos in
  neuronal networks with balanced excitatory and inhibitory activity,}
  \emph{Science}, \textbf{274}(5293), pp. 1724--1726.

\bibitem{bs2}
\textsc{Deneve, S.} and \textsc{C.~Machens} (2016) \enquote{Efficient codes and
  balanced networks,} \emph{Nature Neuroscience}, \textbf{19}, pp. 375--382.

\bibitem{bs3}
\textsc{Baker, C.}, \textsc{V.~Zhu}, and \textsc{R.~Rosenbaum} (2020)
  \enquote{Nonlinear stimulus representations in neural circuits with
  approximate excitatory-inhibitory balance,} \emph{PLOS Computational
  Biology}, \textbf{16}(9).

\bibitem{hyp}
\textsc{O{\textquoteright}Shea, D.~J.}, \textsc{L.~Duncker}, \textsc{W.~Goo},
  \textsc{X.~Sun}, \textsc{S.~Vyas}, \textsc{E.~M. Trautmann},
  \textsc{I.~Diester}, \textsc{C.~Ramakrishnan}, \textsc{K.~Deisseroth},
  \textsc{M.~Sahani}, and \textsc{K.~V. Shenoy} (2022) \enquote{Direct neural
  perturbations reveal a dynamical mechanism for robust computation,} .

\bibitem{fil}
\textsc{Biák, M.}, \textsc{T.~Hanus}, and \textsc{D.~Janovská} (2013)
  \enquote{Some applications of Filippov’s dynamical systems,} \emph{Journal
  of Computational and Applied Mathematics}, \textbf{254}, pp. 132--143,
  nonlinear Elliptic Differential Equations, Bifurcation, Local Dynamics of
  Parabolic Systems and Numerical Methods.

\bibitem{dox}
\textsc{Tsodyks, M.~V.}, \textsc{W.~E. Skaggs}, \textsc{T.~J. Sejnowski}, and
  \textsc{B.~L. McNaughton} (1997) \enquote{Paradoxical Effects of External
  Modulation of Inhibitory Interneurons,} \textbf{17}(11), pp. 4382--4388.

\bibitem{wang1}
\textsc{Wang, X.~J.} (2002) \enquote{Probabilistic Decision Making by Slow
  Reverberation in Cortical Circuits,} \emph{Neuron}, \textbf{36}(5), pp.
  955--968.

\bibitem{wang2}
\textsc{Wong, K.} and \textsc{X.~J. Wang} (2006) \enquote{A Recurrent Network
  Mechanism of Time Integration in Perceptual Decisions,} \emph{The Journal of
  Neuroscience}, \textbf{26}(4), pp. 1314--1328.

\bibitem{wang3}
\textsc{Wong, K.}, \textsc{A.~Huk}, \textsc{M.~Shadlen}, and \textsc{X.~Wang}
  (2007) \enquote{Neural circuit dynamics underlying accumulation of
  time-varying evidence during perceptual decision making,} \emph{Frontiers in
  Computational Neuroscience}, \textbf{Volume 1 - 2007}.

\bibitem{bcmarchitect}
\textsc{Albantakis, L.} and \textsc{G.~Deco} (2011) \enquote{Changes of Mind in
  an Attractor Network of Decision-Making,} \emph{PLOS Computational Biology},
  \textbf{7}.

\bibitem{cpg}
\textsc{Yuste, R.}, \textsc{J.~MacLean}, \textsc{J.~Smith}, and
  \textsc{A.~Lansner} (2005) \enquote{The cortex as a central pattern
  generator,} \emph{Nature Reviews Neuroscience}, \textbf{6}, pp. 477--483.

\bibitem{diverse}
\textsc{Morrison, K.}, \textsc{A.~Degeratu}, \textsc{V.~Itskov}, and
  \textsc{C.~Curto} (2024) \enquote{Diversity of Emergent Dynamics in
  Competitive Threshold-Linear Networks,} \emph{SIAM Journal on Applied
  Dynamical Systems}, \textbf{23}(1), pp. 855--884.

\bibitem{biomod}
\textsc{Edelstein-Keshet, L.} (2005) \emph{Mathematical Models in Biology},
  chap. Applications of Continuous Models to Population Dynamics, 1 ed.,
  Society for Industrial and Applied Mathematics, pp. 224 -- 230.

\bibitem{lv}
\textsc{Strogatz, S.} \emph{Nonlinear Dynamics and Chaos: With Applications to
  Physics, Biology, Chemistry and Engineering}, chap. Phase Plane, 2 ed.,
  Westview Press, pp. 156--160.

\bibitem{toggle}
\textsc{Lugagne, J.-B.}, \textsc{S.~Sosa~Carrillo}, \textsc{M.~Kirch},
  \textsc{A.~Köhler}, \textsc{G.~Batt}, and \textsc{P.~Hersen} (2017)
  \enquote{Balancing a genetic toggle switch by real-time feedback control and
  periodic forcing,} \emph{Nature Communications}, \textbf{8}(1), p. 1671.

\bibitem{lvsols}
\textsc{Abdelkader, M.~A.} (1974) \enquote{Exact solutions of Lotka-Volterra
  equations,} \emph{Mathematical Biosciences}, \textbf{20}(3), pp. 293--297.

\bibitem{lcexist}
\textsc{Bel, A.}, \textsc{R.~Cobiaga}, \textsc{W.~Reartes}, and \textsc{H.~G.
  Rotstein} (2021) \enquote{Periodic Solutions in Threshold-Linear Networks and
  Their Entrainment,} \emph{SIAM Journal on Applied Dynamical Systems},
  \textbf{20}(3), pp. 1177--1208.

\bibitem{conley}
\textsc{Mischaikow, K.} (1999) \enquote{The Conley index theory: A brief
  introduction,} \emph{Banach Center Publications}, \textbf{47}(1), pp. 9 --
  19.

\bibitem{con2}
\textsc{Gedeon, T.}, \textsc{B.~Cummins}, \textsc{S.~Harker}, and
  \textsc{K.~Mischaikow} (2018) \enquote{Identifying robust hysteresis in
  networks,} \emph{PLOS Computational Biology}, \textbf{14}.

\bibitem{MDL}
\textsc{Ding, J.} and \textsc{A.~Zhou} (2007) \enquote{Eigenvalues of rank-one
  updated matrices with some applications,} \emph{Applied Mathematics Letters},
  \textbf{20}(12), pp. 1223--1226.

\bibitem{vand}
\textsc{Moya-Cessa, H.} and \textsc{F.~Soto-Eguibar} (2012), \enquote{Inverse
  of the Vandermonde and Vandermonde confluent matrices,} \eprint{1211.1566}.
\newline\urlprefix\url{https://arxiv.org/abs/1211.1566}

\bibitem{branch}
\textsc{Langdon, C.}, \textsc{M.~Genkin}, and \textsc{T.~Engel} (2023)
  \enquote{A unifying perspective on neural manifolds and circuits for
  cognition,} \emph{Nature Reviews Neuroscience}, \textbf{24}(6), pp. 363 --
  377.

\bibitem{smform}
\textsc{Bartlett, M.~S.} (1951) \enquote{{An Inverse Matrix Adjustment Arising
  in Discriminant Analysis},} \emph{The Annals of Mathematical Statistics},
  \textbf{22}(1), pp. 107 -- 111.

\bibitem{fark}
\textsc{Bertsimas, D.} and \textsc{J.~Tsitsiklis} (1997) \emph{Introduction to
  Linear Optimization}, chap. Duality Theory, 1 ed., Athena Scientific, p. 165.

\bibitem{binary}
\textsc{Sridhar, V.}, \textsc{L.~Li}, \textsc{D.~Gorbonos}, \textsc{M.~Nagy},
  \textsc{B.~Schell}, \textsc{T.~Sorochkin}, \textsc{N.~Gov}, and
  \textsc{I.~Couzin} (2021) \enquote{The geometry of decision-making in
  individuals and collectives,} \emph{Proc. Natl. Acad. Sci. U.S.A.},
  \textbf{118}(50).

\bibitem{sturm}
\textsc{Mishra, B.} (1993) \emph{Real Algebra}, Springer New York, New York,
  NY, pp. 297--383.

\end{thebibliography}
	\end{singlespace}

%\begin{singlespace}
%\begin{thebibliography}{99}
%\addcontentsline{toc}{chapter}{Bibliography}
%\frenchspacing

%\bibitem{Wisdom87} J. Wisdom, ``Rotational Dynamics of Irregularly Shaped Natural Satellites,'' \emph{The Astronomical Journal}, Vol.~94, No.~5, 1987  pp. 1350--1360.

%\bibitem{G&H83} J. Guckenheimer and P. Holmes, \emph{Nonlinear Oscillations, Dynamical Systems, and Bifurcations of Vector Fields}, Springer-Verlag, New York, 1983.

%\end{thebibliography}
%\end{singlespace}

\backmatter

% Vita
% Place your vita below.
Safaan Sadiq graduated from American University in 2017 with a BS in Mathematics and a BA in International Studies.  He then graduated from Northwestern University in 2018 with an MS in Engineering Sciences and Applied Mathematics.  In 2019, he entered the Ph.D. program at Penn State University where he has worked with Carina Curto.  

\begin{center}
\textbf{Teaching}
\end{center}

Instructor - MATH21 - College Algebra I \hfill FA2020, SP2021

Instructor - MATH22 - College Algebra II \hfill FA2021-SP2023 and SP2024-FA2024

Grader - MATH310 - Combinatorics \hfill SP2020

Grader - MATH311 -  Discrete Mathematics \hfill SP2024

\end{document}